# On Low-Rank Multiplicity-Free Fusion Categories

## Gert Vercleyen

Thesis Presented for the Degree of
Doctor of Philosophy

to

Maynooth University

Department of Theoretical Physics

February 26, 2024

Department Head
Dr. J. K. Slingerland

Research Advisor
Dr. J. K. Slingerland

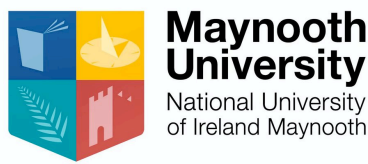

# Declaration

I declare that this thesis has not been submitted in whole, or in part, to this or any other university for any other degree and is, except where otherwise stated, the original work of the author.

Gert Vercleyen, February 10, 2024

# Acknowledgements

First, I want to thank my promoter, Joost Slingerland, who endured my ceaseless firing of questions and crazy ideas and still managed to keep me on the right track. Special thanks go to Colleen Delaney, Eric Samperton, and Catherine Meusburger, whose encouragement has helped me overcome doubt about the value of my work. I also want to thank Eddie Ardonne, Eric Rowell, Andrew Schopieray, and Sebastien Palcoux for their great enthusiasm, without which the AnyonWiki would not have been possible. Several people from the University of Ghent deserve my deepest gratitude. First of all, this project was only possible because of the confidence and support of Frank Verstraete, who kickstarted the project of solving pentagon equations. I would also like to thank Professor Hendrik De Bie and the Clifford research group for their hospitality. The most productive period has been the time spent in their presence.

I would also like to thank the examiners, Eddy Ardonne and Paul Watts, for doing a fantastic job. The many helpful comments and suggestions show that you have spent quite the effort delving into the abstract nonsense that constitutes my thesis.

This work would not have been possible without the encouragement of my family. I want to thank Mom for giving me access to the many puzzle and riddle books when I was young and for being a creative genius who has inspired me in many ways. I want to thank Dad for supporting me all the way and showing me the joy of living a carefree, non-judgemental life. Last but not least, I would like to thank my Brother for accompanying me to the Pokémon TCG world championships in Florida and for the good memories we share. Then there is also my +dad Chris, whose renowned cooking makes my day, and my +mom Esmeralda, who has aided my quest for discovery ever since primary school.

My friends have supported me greatly throughout my career as a doctoral student. Sam and Spoon have been there for me when I needed a safe place and board game evening. Tommeke's support means more than he can imagine. I know he is rooting for me, and I am proud to call him my friend. There are, of course, also Nasha, Denis, and Elena, with whom I can spend hours talking about virtually anything. And then there's Eden, who sneaks out past bedtime and always manages to score a few hugs from the conversing grown-ups before his glorious return to dreamland. You truly are a little angel. Deborah, I'm so happy I met you and hope to keep in touch for many years. You are a great inspiration to keep fighting for what I want. Nadia, you have always believed in me, even at times when I did not. It really means a lot to me. Lastly, Laurent deserves special mention as well. You were an integral part of the best year in high school, and I'm so happy we still keep in touch.

The acknowledgments would not be complete without thanking my Brother's personal dishwasher, our family's favorite footwarmer, the winner of multiple prestigious gardening awards[1], the entity that strikes fear in many cats: Zempo, the crazy dog. Who's a good woof woof? You are!

---

[1] Including award for deepest pit, highest heap, longest distance traveled by dirt, and most bones buried per $m^2$

At last, I want to conclude the acknowledgments by thanking my partner in crime for making the previous year one of the best so far. Your support is boundless, and your encouragement keeps me going until the end. From failing at crossword puzzles to failing at cooking to failing at woodworking in the garage to being successfully trampled by a group of angry mammoths, you are always there to support me.

# Abstract

This thesis explains the methods and algorithms we used to obtain explicit $F$ symbols, $R$ symbols, and pivotal coefficients of all multiplicity-free pivotal fusion categories up to rank 7. The thesis starts by introducing the concept of a unitary modular fusion system via two applications: modeling anyons for topological quantum computation and calculating braid group representations. Next, the notions of a pivotal, spherical, braided, ribbon, and modular fusion system are introduced. These are arithmetic counterparts to fusion categories with the respective structure. Unitarity and its implications on the pivotal structure are discussed as well. The next part of the thesis is devoted to algorithms for finding fusion systems and compatible structures. First, an algorithm to find low-rank fusion rings is explained, and its results are given. Special attention is given to the structure of non-commutative fusion rings and the construction of songs, which are generalizations of the Tambara-Yamagami and Haagerup Izumi fusion rings, is given. Then, the algorithms used to find fusion systems are discussed. Particular attention is given to how the individual steps for solving the consistency equations are done with Anyonica, a software package we developed for working with fusion systems. Gauge and automorphism equivalence are reviewed, and algorithms that put solutions in a unitary gauge and remove redundant solutions are given. Some results on the categorification of all multiplicity-free pivotal fusion rings up to rank 7 are presented. The final part of the thesis is devoted to building models of anyons on graphs and how their behavior differs from those in the plane. The appendices contain a minimal mathematical exposition on fusion categories with their relation to fusion systems, a list of all multiplicity-free fusion rings up to rank 9 with information on categorifiability, a list of all multiplicity-free fusion categories up to rank 7, and data on some graph-braid models.

# Contents

# Part I

# Introduction

# Chapter 1

# Anyons as Unitary Modular Fusion Systems

Anyons are particle-like excitations that only exist in two dimensions [81, 82, 79, 56]. In contrast to particles living in three dimensions, whose exchange is governed by the sign representation of the symmetric group, the exchange of anyons is governed by representations of the braid group. For anyons, it is possible that more exotic phases $e^{i\phi} \neq \pm 1, \phi \in \mathbb{Q}$, are required to describe the swap of two anyons. It might also be that a higher dimensional, non-Abelian representation of the braid group is required. The reason for this strange behavior comes from the fact that the fundamental group of the configuration space of particles in $3 + 1$ dimensions is the symmetric group, while in $2 + 1$ dimensions, it is the braid group [60].

One of the aims of this chapter is to explore two of the consequences of these braid-group statistics of anyons. The first one is of mathematical interest, namely that a system of anyons can be used to obtain representations of the braid group. The other one is of physical interest, namely that anyons play a central role in topological quantum computation.

Another aim of this chapter is to introduce the concept of a unitary modular fusion system by means of the theory of anyons. These unitary modular fusion systems and their generalizations form the main topic of this thesis and are directly related to the more abstract fusion categories (see Section 7.2). This introduction is mainly based on Section E of Kitaev's paper [56], but the normalization and graphical conventions from Bonderson's thesis [10] are used. The exposition is necessarily ad-hoc since building up the mathematical structures starting from field theory would distract from the main story of this thesis. The proofs that are not necessary to understand the diagrammatic language are also left out. The contents of this chapter can be summarized as follows. Section 1.1 introduces the notion of a unitary fusion system and explains how it relates to the fusion and splitting of anyons in the plane using a diagrammatic language. The notion of progressive planar isotopy is discussed, conventions for dealing with bends in diagrams are introduced, and how to assign topological invariants to disjoint loops in diagrams is shown. In Section 1.2 the definition of a unitary modular fusion system is given. Such a system allows us to describe the braiding of anyons and express specific processes with anyons via braid diagrams. The chapter concludes with Section 1.3, which explains how anyons allow us to construct braid group representations and how these can be used for topological quantum computation.

It must be noted that a lot of the physical exposition relies on the fact that we will be working with a *unitary theory*, i.e. the underlying mathematical description is given by a unitary fusion category. Unitarity simplifies matters a lot but also sweeps certain subtleties under the carpet. In Chapter 2, we will drop the unitarity assumption at first and later discuss the consequences of demanding unitarity.

## 1.1 Fusion Theory

For our purposes, we imagine a finite set of anyons on a line. While anyons live in 2 dimensions, one can and should define an order on points in the plane in order to arrange the anyons in a non-ambiguous way, such as in figure 1.1. We start by considering a system of two anyons.

### 1.1.1 Fusion and Splitting of Two Anyons

The state space of two anyons, say $a, b$ with total charge $c$, is a Hilbert space $V_{a,b}^c$. An element of this space corresponds to a process where the two charges $a, b$ fuse to give charge $c$. These should be thought of as operators that map a two-particle state to a one-particle state. The dimension $\dim V_{a,b}^c$ equals the different number of inequivalent ways such a process can happen. Let $\left\{ \langle a, b; c, \alpha | \,\middle|\, \alpha = 1, \ldots, \dim V_{a,b}^c \right\}$ be an orthonormal basis for $V_{a,b}^c$, and $\left\{ | a, b; c, \alpha \rangle \,\middle|\, \alpha = 1, \ldots, \dim V_{a,b}^c \right\}$ a dual basis for the dual space $V_c^{a,b}$. We can use a diagrammatic language that makes it much easier to deal with anyons by using the following conventions.

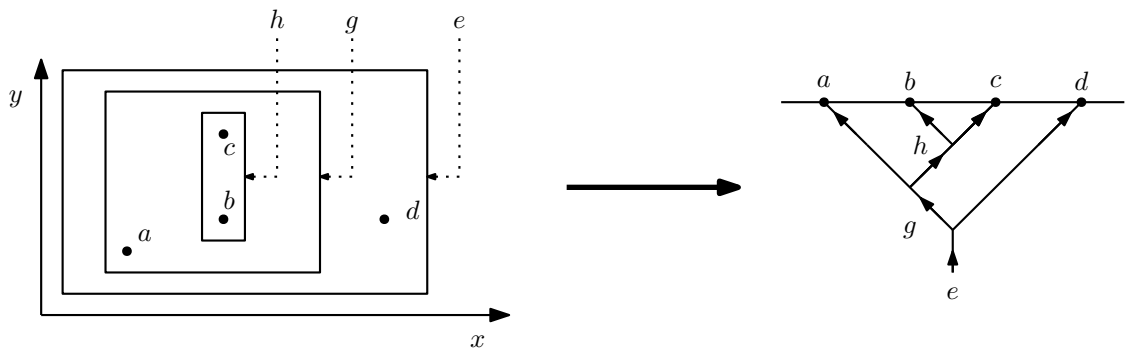

Figure 1.1: By using the lexicographic ordering on the $(x, y)$ coordinates, the positions of the four anyons can be ordered unambiguously. In this case, we have also grouped the anyons in subsystems and assigned labels to these.

$$\begin{gathered}\begin{array}{c}a \quad\quad b\\ \diagdown\;\;\diagup\\ \alpha\\ \big|\\ c\end{array}\end{gathered} = \sqrt[4]{\frac{d_c}{d_a d_b}}\langle a,b;c,\alpha|, \tag{1.1}$$

$$\begin{gathered}\begin{array}{c}c\\ \big|\\ \alpha\\ \diagup\;\;\diagdown\\ a \quad\quad b\end{array}\end{gathered} = \sqrt[4]{\frac{d_c}{d_a d_b}}|a,b;c,\alpha\rangle. \tag{1.2}$$

The labeled lines in these diagrams can be thought of as trajectories in space-time. For now, one may regard the $d_i$ as formal symbols. We will assign these well-chosen non-zero values later on that will simplify computations.

> **Note 1.** Diagrams like (1.1) and (1.2) are modeled via equivalence classes of oriented labeled planar graphs[a] embedded inside a rectangle $[x_1, x_2] \times [y_1, y_2] \subset \mathbb{R}^2$ for which
>
> - the labeled edges always point upwards. In particular, no edge is allowed to have a horizontal tangent at any point (yet). They need not be straight lines but must have a tangent at each point and be non-pathological (see [101] for more detail).
> - Every edge is connected to either a labeled disk or the interior of the bottom or top line of the rectangle.
> - The vertices are either labeled disks that are connected to three edges, with at least one incoming and one outgoing edge, or they are points on the bottom or top line of the rectangle.
> - two diagrams are regarded as equivalent if they can be transformed into each other by moving the vertices (and thus deforming the edges as well) of the labeled graph in such a way that
>   - no vertices overlap,
>   - no edges cross,
>   - no vertex enters or leaves the bottom or top lines of the rectangle,
>   - and no edges have a horizontal tangent
>
> at any point during this process. Graphs like these are called progressive planar graphs, the allowed deformations are called progressive planar deformations, and two graphs that are equivalent under such deformations are called progressive planar isotopic. This is a weaker form than full planar isotopy, where the vertices are allowed to move everywhere, and the edges need not have horizontal tangents, but overlap of vertices and crossing of edges are still not allowed.
>
> ---
> [a]In [53] and [101] the precise definition of these classes are given, but here we will present their definition more intuitively.

Splitting operators from $V^{a,b}_{c'}$ and fusion operators from $V^{c}_{a,b}$ can be composed as follows

$$a \underset{\beta}{\overset{\alpha}{\bigcirc}} b \;\; c \atop c' \;=\; \delta^{c}_{c'} \delta^{\alpha}_{\beta} \sqrt{\frac{d_c}{d_a d_b}} \;\; c, \tag{1.3}$$

where the line labeled by $c$ should be thought of as the identity operator and one is allowed to use equation (1.3) to replace lines in any diagram by a decomposition into a splitting and fusion operator. Note that we do not sum over fusion labels. The appearance of $\delta^{c}_{c'}$ can be considered the equivalent of demanding charge conservation. It might seem strange at first that, algebraically, $\langle a, b; c', \beta | a, b; c, \alpha \rangle$ does not result in a number but rather in a number multiplied with an identity operator. These two are, however, equivalent. One can regard the identity operator as a basis vector in the 1-dimensional space of processes, which keeps the charge $c$ unchanged. Combining two processes from this space, say $\lambda \, \mathrm{id}_c$ and $\tau \, \mathrm{id}_c$, then corresponds to a process $\lambda \tau \, \mathrm{id}_c$.

We can also combine a fusion operator with a splitting operator the other way. The decomposition of the identity operator on two charges is expressed using such a combination as follows

$$\bigg| a \quad \bigg| b \;=\; \sum_{c,\alpha} \sqrt{\frac{d_c}{d_a d_b}} \;\; \overset{a \quad b}{\underset{a \quad b}{\overset{\alpha}{\bigcirc}} c \underset{\alpha}{\bigcirc}} \;. \tag{1.4}$$

### 1.1.2  Fusion and Splitting of Multiple Anyons

For systems with more than two anyons, one must combine multiple splitting and fusion operators in a compatible way. All lines can only have one label, and one can only connect lines of diagrams with matching labels. To describe, e.g., a process where an anyon with charge $d$ splits into two anyons with charges, $e$ and $c$, followed by the process where $e$ splits into charges $a$ and $b$ one would compose the operators $\langle a, b; e, \alpha | \otimes \mathrm{id}_c$ and $\langle e, c; d, \beta |$. Up to a scalar factor, this composition is represented by the following diagram,

$$\overset{a \quad b \quad c}{\underset{d}{\overset{\alpha}{\underset{e \;\; \beta}{\bigcirc}}}} \;. \tag{1.5}$$

Each composition is completely determined by the couple $\langle a, b; e, \alpha | \langle e, c; d, \beta |$ and is thus described by elements of a Hilbert space isomorphic to $V^{a,b}_{e} \otimes V^{e,c}_{d}$. This isomor-

phism is typically used to write $\langle a, b; e, \alpha | \langle e, c; d, \beta |$ instead of the longer $(\langle a, b; e, \alpha | \otimes \mathrm{id}_c)\langle e, c; d, \beta |$.

There might be multiple charges $e$ that allow $d$ to split in this way to $a, b, c$, so the space that describes such processes is given by $\bigoplus_e V_e^{a,b} \otimes V_d^{e,c}$. Now, if we are just given the information that $d$ is the total charge of the anyons with charges $a, b, c$, we might as well describe this space using basis vectors of the following form:

$$
\begin{array}{c}
a \quad b \quad c \\
\diagup\!\!\diagdown\!\!\diagup \\
\phantom{xx}\delta \\
\gamma \quad f \\
\uparrow \\
d
\end{array}
\qquad (1.6)
$$

The two sets of basis vectors are related by a unitary change of basis which is denoted by

$$
\begin{array}{c}
a \quad b \quad c \\
\alpha \\
e \quad \beta \\
\uparrow \\
d
\end{array}
= \sum_{f,\gamma,\delta} [F_d^{abc}]^{(e,\alpha,\beta)}_{(f,\gamma,\delta)}
\begin{array}{c}
a \quad b \quad c \\
\delta \\
\gamma \quad f \\
\uparrow \\
d
\end{array}.
\qquad (1.7)
$$

In what follows, we will assume that the triples $(e, \alpha, \beta), (f, \gamma, \delta)$ are ordered lexicographically, and we regard each triple as a single index. This way, the basis transform is given by unitary matrices $\{[F_d^{abc}]\}$ that are called the *F*-matrices of the theory. Its elements $[F_d^{abc}]^{(e,\alpha,\beta)}_{(f,\gamma,\delta)}$ are called the *F*-symbols, and a basis transform using the *F*-symbols is called an *F*-move. We will denote the inverse $F$ matrices by $\left[\tilde{F}_d^{abc}\right]$.

The fusion theory of systems with any number of anyons is determined by a *unitary fusion system*. A unitary fusion system is a collection of data $(\mathbf{L}, *, \mathbf{N}, \mathbf{V}, \mathbf{F})$ that allows one to consistently deal with processes involving any number of anyons that fuse and split. It is given by

1. A finite list **L** of labels (also called charge types, superselection sectors, particle types, anyon types, or whatever suits your needs) which we will denote by $\{1, ..., r\}$. Each label represents a type of anyon and one of the labels, 1 in our case, represents the vacuum (or trivial charge). The number $r$ will be called the rank of the unitary fusion system.

2. A map $* : \mathbf{L} \to \mathbf{L}$ called the conjugation, that maps anyons to their dual (or conjugate, or antiparticle). The map has to satisfy that $1^* = 1$ and $(a^*)^* = a$.

3. A finite set of natural numbers $\mathbf{N} = \left\{ N_{a,b}^c \,\middle|\, a, b, c = 1, ..., r \right\}$, the dimensions of the fusion spaces $\left\{ V_{a,b}^c \,\middle|\, a, b, c = 1, ..., r \right\}$, that satisfy

    - Fusing with the vacuum is trivial: $N_{a,1}^b = N_{1,a}^b = \delta_a^b$
    - Only fusion with antiparticles can create the vacuum: $N_{a,b}^1 = \delta_{b^*}^a$
    - Fusion is associative: $N_{a,b,c}^d := \sum_e N_{a,b}^e N_{e,c}^d = \sum_f N_{a,f}^d N_{b,c}^f$

- Fusion is commutative: $N_{a,b}^c = N_{b,a}^c$

4. A finite set of Hilbert spaces $\mathbf{V} = \left\{ V_{a,b}^c \,\middle|\, a, b, c = 1, \ldots, r \right\}$, also called fusion spaces, for which $\dim V_{a,b}^c = N_{a,b}^c$ and whose inner product induces an Hermitian adjoint † that satisfies

$$d_a := (|a, a^*; 1, 1\rangle)^\dagger |a, a^*; 1, 1\rangle = \langle a, a^*; 1, 1 | a, a^*; 1, 1 \rangle = \frac{1}{\left|[F_a^{aa^*a}]_{(1,1,1)}^{(1,1,1)}\right|} \quad (1.8)$$

$$d_{a^*} = d_a, \quad (1.9)$$

$\forall a \in \mathbf{L}$. The interpretation of $d_a$ is explained in Section 1.1.3.

5. A finite set $\mathbf{F} = \left\{ [F_d^{abc}] \in \mathrm{Mat}_{N_{a,b,c}^d \times N_{a,b,c}^d}(\mathbb{C}) \,\middle|\, a, b, c, d = 1, \ldots, r \right\}$ of finite-dimensional unitary matrices, whose entries, also called $F$-symbols, are denoted by

$$[F_d^{abc}]_{(f,\gamma,\delta)}^{(e,\alpha,\beta)}, \quad \begin{array}{l} \alpha \in \left\{1, \ldots, N_{a,b}^e\right\} \\ \gamma \in \left\{1, \ldots, N_{a,f}^d\right\} \end{array}, \quad \begin{array}{l} \beta \in \left\{1, \ldots, N_{e,c}^d\right\} \\ \delta \in \left\{1, \ldots, N_{b,c}^f\right\} \end{array}, \quad (1.10)$$

where the labels $(e, \alpha, \beta)$ and $(f, \gamma, \delta)$ are ordered lexicographically and we use the notation $\left[\tilde{F}_d^{abc}\right]$ for the inverse of $[F_d^{abc}]$. These matrices are required to satisfy

- Fusion with the vacuum is irrelevant:

$$[F_d^{1bc}] = \mathbb{1}_{N_{b,c}^d \times N_{b,c}^d}, \quad (1.11)$$

$$[F_d^{a1c}] = \mathbb{1}_{N_{a,c}^d \times N_{a,c}^d}, \quad (1.12)$$

$$[F_d^{ab1}] = \mathbb{1}_{N_{a,b}^d \times N_{a,b}^d}, \quad (1.13)$$

where $[\mathbb{1}_{m \times n}]_j^i = \delta_j^i$ is the $m \times n$ identity matrix.

- The pentagon equations:

$$\sum_\zeta [F_e^{fcd}]_{(l,\zeta,\varepsilon)}^{(g,\beta,\gamma)} [F_e^{abl}]_{(k,\theta,\eta)}^{(f,\alpha,\zeta)} = \sum_{h,\iota,\kappa,\lambda} [F_g^{abc}]_{(h,\kappa,\iota)}^{(f,\alpha,\beta)} [F_e^{ahd}]_{(k,\theta,\lambda)}^{(g,\kappa,\gamma)} [F_k^{bcd}]_{(l,\eta,\varepsilon)}^{(h,\iota,\lambda)}, \quad (1.14)$$

where whenever a zero dimensional matrix is encountered in a term, it is automatically 0. Equivalently, one could demand that the summation index $h$ on the RHS of (1.14) only takes values for which none of the $F$-matrices are zero-dimensional.

- The snake property

$$[F_a^{aa^*a}]_{(1,1,1)}^{(1,1,1)} \neq 0, \quad (1.15)$$

which demands that the $d_a$ from equation (1.8) are well-defined.

The interpretation of the pentagon equations is the following. If we want to describe the splitting of a charge into $n$ charges, we would have a basis for every binary tree with $n$ leaves. Any basis can be transformed into another basis by a sequence of $F$-moves, but if $n > 3$, there are multiple different combinations of $F$-moves that relate one basis to the other. To have a consistent theory, we demand that the $F$-symbols satisfy the pentagon equations. These can be graphically depicted as in figure 1.2. Mac Lane's coherence theorem [67] then states that these demands, on configurations with only 4 anyons, are strong enough to guarantee that any sequence of $F$ moves between two fixed trees of any shape is equal.

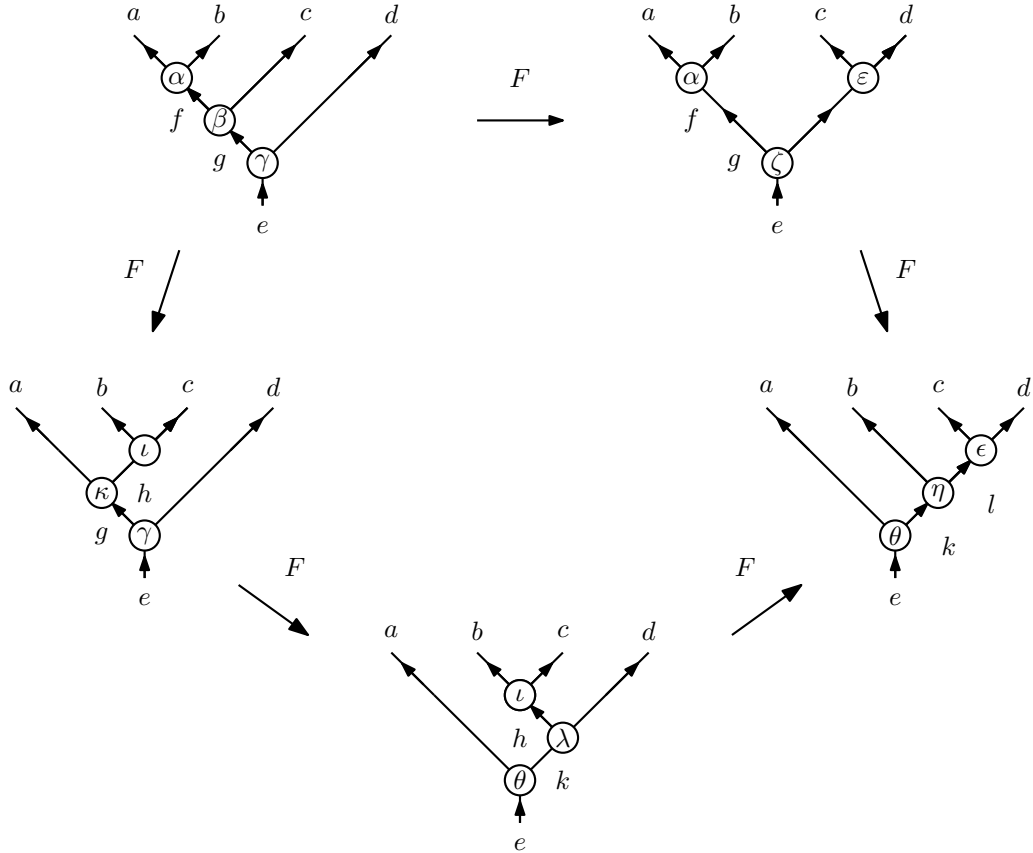

Figure 1.2: The graphical interpretation of the pentagon equations.

For a fusion system, an arbitrary process that starts with $m + 1$ anyons $b_1, ..., b_{m+1}$ and results in $n + 1$ anyons $a_1, ..., a_{n+1}$ (both in that order on the line) can be de decomposed into a superposition of processes that consist of two parts: first all anyons $b_1, ..., b_{m+1}$ are fused and next the fusion product is split into the anyons $a_1, ..., a_{n+1}$. Figure 1.3 shows a possible convention of a standard basis for all such processes.

The following procedure can express any fusion diagram as a linear combination of standard basis vectors. Given a diagram with $m + 1$ bottom strands and $n + 1$ top strands it works as follows

1. Transform this diagram into a top and bottom diagram connected by a single vertical line. This can be done as follows. First, we can draw a horizontal line at any height of the diagram, such that it does not overlap with any vertices. Consider this our split between the two diagrams. If only one vertical line crosses this horizontal line, we are done. If there are multiple, then one can apply the decomposition of the identity (1.4) to reduce the number of vertical lines until only one remains.

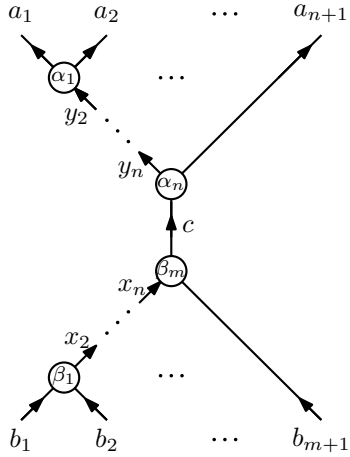

Figure 1.3: A possible standard basis in which any process that starts with *m* anyons and ends with *n* anyons can be expressed.

2. If neither the top nor bottom contains any loops, they must be trees. In this case, a sequence of *F*-moves can be used to recouple the trees to the required form, and we are done. If a diagram contains one or more loops, these can be removed by combining *F*-moves, inverse *F*-moves, and formula (1.3). To do so, pick any loop that contains no inner lines. Then, use the *F*-moves to move all lines attached to this loop to two other lines of choice. The various moves that can be used to do so are described in figures 1.4 and 1.5. Formula (1.3) can then be used to remove this loop. By removing all loops this way we end up with two tree diagrams and a sequence of *F*-moves can be used to recouple the trees to the required form.

### 1.1.3 Vertical Bends and Removal of Vacuum Lines

The interpretation of the snake property is that we would like the following process to have a non-zero expectation value.

$$a \underset{1}{\overset{a}{\diamondsuit}} a^* \overset{1}{\diamondsuit} a = [F_a^{aa^*a}]^{(1,1,1)}_{(1,1,1)} \; a \underset{1}{\overset{a}{\diamondsuit}} a^* \overset{1}{\diamondsuit} a = d_a [F_a^{aa^*a}]^{(1,1,1)}_{(1,1,1)} \uparrow a \quad (1.16)$$

Note that

- we have not included the labels for the basis vectors of the splitting and fusion

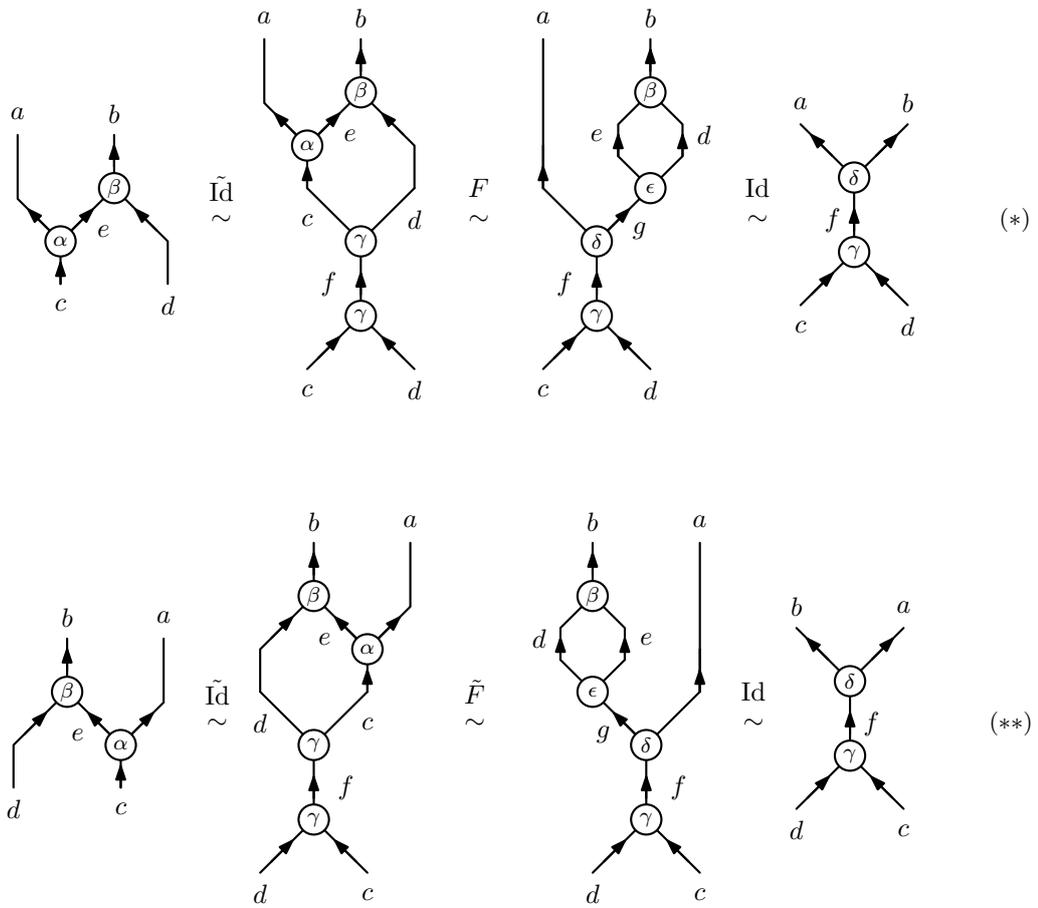

Figure 1.4: By adding a loop, performing an (inverse) *F*-move and removing the loop we can transform '*H*-shaped' diagrams into superpositions of '*I*'-shaped diagrams. Here Id means that we have used the decomposition of the identity (1.4), while $\tilde{\mathrm{Id}}$ means we have used equation (1.3)

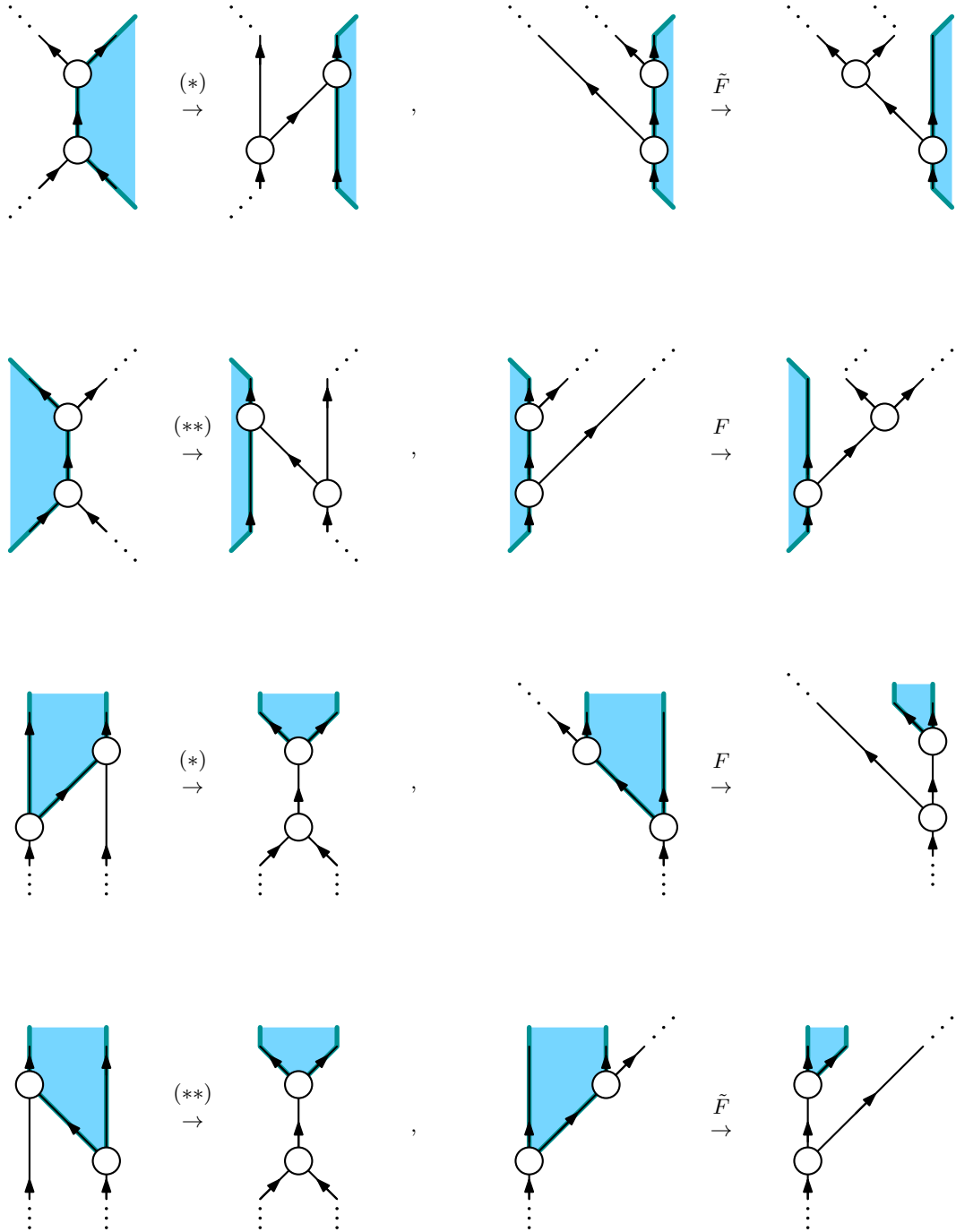

Figure 1.5: Using the equivalences defined in figure 1.4 we can reduce the number of lines attached to a loop (whose edges and interior are drawn in blue)

spaces. This is because, in this case, those spaces are 1-dimensional, so there is no ambiguity in leaving those labels out.

- We used dotted lines to denote vacuum charges. One of the goals of this subsection is to set up conventions such that the vacuum lines can be left out. Diagrams like the one on the LHS of (1.16) will look like a single bent line. Later on, in Section 1.2, when we add over and under crossings of lines, this will enable us to construct braid diagrams. Under the assumption that diagrammatic calculus becomes *isotopy-invariant*, we can compute (framed) knot invariants by reducing these diagrams.

There are two independent snake equations that we need to fix:

$$a \underset{a^*}{\diamondsuit} a = d_a [F_a^{aa^*a}]^{(1,1,1)}_{(1,1,1)} \, a, \quad a \underset{a^*}{\diamondsuit} a = d_a \overline{[F_a^{aa^*a}]^{(1,1,1)}_{(1,1,1)}} \, a. \quad (1.17)$$

Since $d_a = \left|[F_a^{aa^*a}]^{(1,1,1)}_{(1,1,1)}\right|^{-1}$ we can define $\chi_a := d_a [F_a^{aa^*a}]^{(1,1,1)}_{(1,1,1)} \in U(1)$ and we find that

$$a \underset{a^*}{\diamondsuit} a = \chi_a \, a, \quad a \underset{a^*}{\diamondsuit} a = \overline{\chi_a} \, a.$$

$d_a$ is called the quantum dimensions of the particle $a$ and it is invariant under unitary basis transforms of the fusion and splitting spaces. For a unitary fusion system, the quantum dimensions are, by definition, positive and have the following properties. If we denote by $V_{a^n}^b$ the Hilbert space that describes the fusion of $n$ particles $a$ into a particle $b$ then for a unitary fusion system

$$\dim\left(\bigoplus_{b \in L} V_{a^n}^b\right) = \sum_{i_2,\ldots,i_{n-1},b} N_{a,a}^{i_2} N_{i_2,a}^{i_3} \cdots N_{i_{n-1},a}^b \overset{n \to \infty}{\sim} (\text{FPDim}(a))^n = d_a^n, \quad (1.18)$$

where FPDim($a$), the Frobenius-Perron dimension, is the largest positive eigenvalue of the matrix $[N_a]$ (which always exists due to the Frobenius-Perron theorem). For unitary fusion systems $d_a = \text{FPDim}(a)$. The intuition of $d_a$ is thus that, as one increases the number $n$ of anyons $a$, the dimension of the Hilbert space describing all possible fusion processes of the particles $a$ grows as $d_a^n$. Interestingly, $d_a$ can be non-integer for generic anyons. Another interpretation of the number $d_a$ is the following [103]. Given a process (as in equation (1.26)) where an anyon $x$ and its dual $x^*$ split off from the vacuum and immediately fuse back to the vacuum. The chance that that process involves the anyon $a$ is given by $d_a^2/(\sum_{b\in L} d_b^2)$. So, the higher the value of $d_a$, the more likely it will pop up in such a process.

For a self-dual particle $a$, $\chi_a = \pm 1$ and is invariant under unitary basis transforms. For unitary fusion systems, it is called the Frobenius-Schur indicator of $a$. For non-self-dual particles, the value $\chi_a$ depends on the choice of basis for the fusion and splitting spaces and therefore has no invariant meaning.

> **Notes 1.** In chapter 2 we will drop the unitarity assumption. One of the consequences is that the quantum dimensions could be complex numbers. In this case, one can still use the Frobenius-Perron dimensions (which only depend on the list **N**) as scaling exponents. Another consequence of dropping unitarity is that, to even talk about quantum dimensions, we must provide a list of phases $\chi_a$ that satisfy specific properties. In the more general setting, these are called *pivotal coefficients*. Chapter 2 also discusses these in more detail.

By introducing flagged cups and caps as follows

$$\begin{aligned}
\smile_{a\ a^*} &:= \bigvee_{1}^{a\ a^*}, \quad \frown^{a\ a^*} := \left(\smile_{a\ a^*}\right)^\dagger = \bigwedge_{a\ a^*}^{1}, \\
\frown_{a^*\ a} &:= \overline{\chi_a}\bigwedge_{a^*\ a}^{1}, \quad \smile^{a^*\ a} := \left(\frown_{a^*\ a}\right)^\dagger = \chi_a \bigvee^{a^*\ a}_{1},
\end{aligned}$$
(1.19)

we find that

$$a\ \bigcirc\!\!\!\bigcirc\ a = \Big|\,a = a\ \bigcirc\!\!\!\bigcirc\ a.$$
(1.20)

If we swap $a$ with $a^*$ in definitions (1.19) it follows that

$$\frown_{a\ a^*} = \chi_{a^*} \frown_{a\ a^*},$$
(1.21)

$$\smile^{a\ a^*} = \overline{\chi_{a^*}} \smile^{a\ a^*},$$
(1.22)

and thus also

$$a \bigcirc a^* \bigcirc a = a = a \bigcirc a^* \bigcirc a. \tag{1.23}$$

So far, all arrows have been pointing upward. From now on, we will also use downward arrows, which are defined as follows

$$\uparrow a =: \downarrow a^*. \tag{1.24}$$

We will also use the convention that whenever two flags on a line point in the opposite direction, we will not draw those flags. Equation (1.21) in combination with the fact that $\chi_a \in \mathrm{U}(1)$, $\forall a$ implies that this does not result in any inconsistencies. The snake identities become

$$\bigcirc a \bigcirc = a = \bigcirc a \bigcirc, \tag{1.25}$$

so they basically say that we can pull any bends in a line straight.

By combining a cup and a cap, with flags pointing in opposite directions along the line, we find that a loop labeled by particle $a$ evaluates to $d_a$:

$$a \bigcirc = d_a. \tag{1.26}$$

When a loop is a disconnected part of a bigger diagram, we can remove it and multiply the diagram by the loop's value. We can also put it back at any point in the diagram. The reason this can be done is because of the demand that vacuum $F$-matrices are identity matrices.

The results from the above derivations can be summarized as follows:

- In any diagram, we can perform the following substitutions

$$\begin{array}{c} a \quad a^* \\ \vee \\ \vdots \\ 1 \end{array} \mapsto \begin{array}{c} a \quad a^* \\ \smile \end{array}, \qquad \begin{array}{c} 1 \\ \vdots \\ \wedge \\ a \quad a^* \end{array} \mapsto \begin{array}{c} \frown \\ a \quad a^* \end{array}, \tag{1.27}$$

that remove vacuum lines and introduce bent flagged lines whose flags point to the right. Together with the fact that all vacuum $F$-matrices are identity matrices, we can always remove any vacuum line from any diagram and add vacuum lines

as desired.[1]

- When a line contains two opposite flags, these flags may be removed. Likewise, one can always add pairs of opposite flags to any line as desired. Flags may also be moved from one bend to another but generically not through fusion vertices.

- The direction of a flag on a line with label $a$ can be changed at the cost of a phase factor $\chi_a$.

- Double bends in lines can be straightened out. In [56] and [10], it is shown that one can also straighten out single bends at the cost of introducing linear transformations, which are often called $A$ and $B$ matrices. Such transforms allow one to introduce spirals in diagrams, but this lies outside the scope of this thesis.

- If a disconnected loop is encountered at any point in a diagram, it may be replaced by the quantum dimension of its particle label.

## 1.2 Unitary Modular Fusion Theory

One of the interesting properties of anyons is their strange behavior under particle exchange. In contrast to particles in three-dimensional space, the effect of swapping anyons is governed by representations of the braid group. The specific properties of the braiding of anyons are captured by the notion of a *unitary modular fusion system*. A unitary modular fusion system $(\mathbf{L}, *, \mathbf{N}, \mathbf{F}, \mathbf{V}, \mathbf{R})$ is a unitary fusion system together with

- a finite list $\mathbf{R}$ of unitary matrices $\left\{ R_c^{ab} \in \text{Mat}_{N_c^{a,b} \times N_c^{a,b}}(\mathbb{C}) \,\middle|\, a, b, c = 1, ..., r \right\}$, with inverses $\{\tilde{R}_c^{ab}\}$ which satisfy the hexagon equations:

$$\sum_{\lambda,\gamma} [R_e^{ca}]_\lambda^\alpha [F_d^{acb}]_{(g,\mu,\gamma)}^{(e,\lambda,\beta)} [R_g^{cb}]_\nu^\gamma = \sum_{f,\sigma,\delta,\psi} [F_d^{cab}]_{(f,\sigma,\delta)}^{(e,\alpha,\beta)} [R_d^{cf}]_\psi^\sigma [F_d^{abc}]_{(g,\mu,\nu)}^{(f,\delta,\psi)}, \quad (1.28)$$

$$\sum_{\lambda,\gamma} [\tilde{R}_e^{ac}]_\lambda^\alpha [F_d^{acb}]_{(g,\mu,\gamma)}^{(e,\lambda,\beta)} [\tilde{R}_g^{bc}]_\nu^\gamma = \sum_{f,\sigma,\delta,\psi} [F_d^{cab}]_{(f,\sigma,\delta)}^{(e,\alpha,\beta)} [\tilde{R}_d^{fc}]_\psi^\sigma [F_d^{abc}]_{(g,\mu,\nu)}^{(f,\delta,\psi)}, \quad (1.29)$$

and are such that the matrix $\hat{S} \in \text{Mat}_{r \times r}(\mathbb{C})$, whose entries are given by

$$[\hat{S}]_b^a = \sum_{c=1}^r \sum_{i=1}^{N_{a,b^*}^c} \sum_{j=1}^{N_{c,b}^a} \sum_{i'=1}^{N_{b^*,a}^c} \sum_{i''=1}^{N_{a,b^*}^c} [\tilde{F}_a^{ab^*b}]_{(c,i,j)}^{(1,1,1)} [R_c^{b^*a}]_{i'}^i [R_c^{ab^*}]_{i''}^{i'} [F_a^{ab^*b}]_{(1,1,1)}^{(c,i'',j)}, \quad (1.30)$$

is invertible. Note that as a consequence of the hexagon equations, we have that

$$[R_a^{a1}]_1^1 = 1 = [R_a^{1a}]_1^1 \quad, \text{ and } \quad [\tilde{R}_a^{a1}]_1^1 = 1 = [\tilde{R}_a^{1a}]_1^1. \quad (1.31)$$

---

[1]This is because the demand that vacuum $F$-matrices are the identity implies the triangle equations (7.6) for a monoidal category. Mac Lane's coherence theorem then assures that the removal or addition of identity maps is allowed at will.

The action of the *R*-matrices can be graphically represented as

$$\text{(diagram)} = \sum_{\beta} [R_c^{ba}]_{\beta}^{\alpha} \; \text{(diagram)}, \tag{1.32}$$

$$\text{(diagram)} = \sum_{\beta} [\tilde{R}_c^{ab}]_{\beta}^{\alpha} \; \text{(diagram)} \tag{1.33}$$

The diagrammatic interpretation of the hexagon equations is given in figure 1.6.

Similarly to the pentagon equations, there is also a coherence theorem [52] for the hexagon equations. It states that the hexagon equations imply that any two transformations between two fixed fusion trees, using only *F* and *R*-symbols, must be equal. So it does not matter which combination of *F* and *R* one chooses to transform a certain tree *A* into another tree *B*; eventually, they all have the same effect.

Note that any process that starts with *m* anyons and ends with *n* anyons can still be uniquely expressed in a standard basis. To do so one can add a decomposition of the identity (1.4) on the bottom of all braids, and then remove each braid using equations (1.32) and (1.33). The rest of the procedure is then the same as for the reduction of non-braided diagrams.

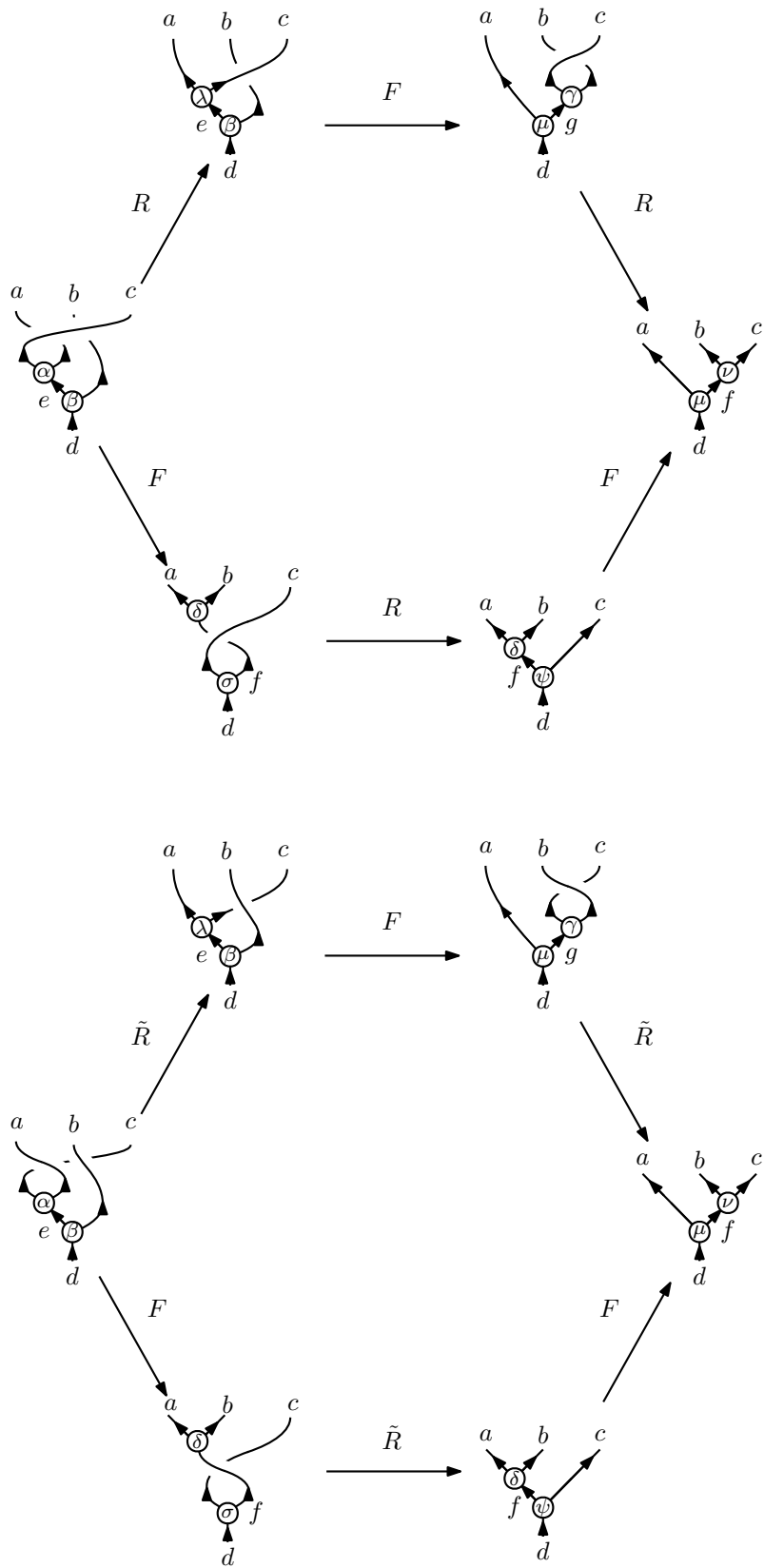

Figure 1.6: The diagrammatic interpretation of the hexagon equations

## 1.3 Braid Group Representations and Topological Quantum Computation

This section aims to give a theoretical, and necessarily naive, introduction to two of the most profound applications of anyons: the construction of braid group representations, and topological quantum computation. The exposition is based on several other, more thorough, introductions such as [91, 20, 103, 82, 104, 87].

### 1.3.1 Constructing Braid Group Representations

Consider $n$ anyons of the same charge $a$ with total charge $b$. Let us assume, for simplicity, that the modular fusion system describing these anyons is multiplicity-free. The state of these $n$ anyons is described by a Hilbert space $V_{a^n}^b$. Take the set of well-formed left-ordered fusion trees, such as in (1.34), as a basis for this space. If fixing the top charges as $a$ completely fixes the tree then dim $V_{a^n}^b = 1$ and we say that $a$ is an Abelian anyon. The reason for this name will become clear soon. If there are multiple fusion trees with top charges $a$ and bottom charge $b$ then the dim $V_{a^n}^b$ depends on the specific anyon model but it grows as dim $V_{a^n}^b \sim d_a^n$. We call anyons of this type non-Abelian anyons, or nonabelions. A nonabelion has the interesting property that its state space carries a non-Abelian representation of the braid group [90], hence the name. The specific representation of the braid group can be computed as follows. Let $a_i$ denote the anyon $a$ at the $i$th leg of the fusion tree and denote the operator that braids $a_i$ around $a_{i+1}$ by $\sigma_i$. The braided tree is still a tree with top charges $a$ and total charge $b$ so $\sigma_i$ is a map from $V_{a^n}^b$ to itself. The matrix representing this map can be calculated by expressing the braided tree in terms of the basis elements of $V_{a^n}^b$, via $F$ and $R$ moves. Since the hexagon equations imply the Yang-Baxter equations (see e.g. prop 8.1.10 in [25]) we have that the representation $\rho$ is automatically a dim $V_{a^n}^b$-dimensional representation of the braid group on $n$ strands.

*Example* 1.3.1. Consider the Fibonacci theory which has two anyons $1, \tau$ where $1$ is the vacuum and the fusion of $\tau$ with itself can result in either $1$ or $\tau$. The space $V_{\tau^4}^1$ is two dimensional and we choose

$$\left\{ \begin{array}{c} \tau \quad \tau \quad \tau \quad \tau \\ \diagup\diagdown\diagup\diagdown \\ x \\ \tau \\ 1 \end{array} \;\middle|\; x \in \{1, \tau\} \right\} \tag{1.34}$$

as its basis. Since $\max N_{a,b}^c = 1$ we will write $[F_d^{abc}]_f^e$ instead of $[F_d^{abc}]_{(f,1,1)}^{(e,1,1)}$ and $R_c^{ab}$

instead of $[R_c^{ab}]_1^1$. For $\sigma_1$ we find that

$$\text{(diagram)} = R_x^{\tau\tau} \text{ (diagram)} \tag{1.35}$$

so $[\rho(\sigma_1)]_y^x = \delta_y^x R_x^{\tau\tau}$. For $\sigma_2$ we find that

$$\text{(diagram)} = \sum_{z \in \{1,\tau\}} [F_\tau^{\tau\tau\tau}]_z^x \text{ (diagram)} \tag{1.36}$$

$$= \sum_{z \in \{1,\tau\}} [F_\tau^{\tau\tau\tau}]_z^x R_z^{\tau\tau} \text{ (diagram)} \tag{1.37}$$

$$= \sum_{z \in \{1,\tau\}, y \in \{1,\tau\}} [F_\tau^{\tau\tau\tau}]_z^x R_z^{\tau\tau} [\tilde{F}_\tau^{\tau\tau\tau}]_y^z \text{ (diagram)} \tag{1.38}$$

so $[\rho(\sigma_2)]_y^x = \sum_{z \in \{1,\tau\}} [F_\tau^{\tau\tau\tau}]_z^x R_z^{\tau\tau} [\tilde{F}_\tau^{\tau\tau\tau}]_y^z$. Lastly for $\sigma_3$ we find that

$$\text{(diagram)} = \sum_{z \in \{1,\tau\}} [F_1^{x\tau\tau}]_z^\tau \text{ (diagram)} \tag{1.39}$$

$$= \sum_{z \in \{1,\tau\}} [F_1^{x\tau\tau}]_z^\tau R_z^{\tau\tau} \text{ (diagram)} \tag{1.40}$$

$$= \sum_{z\in\{1,\tau\},y\in\{1,\tau\}} [F_\tau^{x\tau\tau}]_z^\tau R_z^{\tau\tau} [\tilde{F}_1^{x\tau\tau}]_y^z \quad \begin{array}{c}\tau\ \tau\ \tau\ \tau\\ \diagup\!\!\!\diagup\\ x\quad y\\ 1\end{array} \quad , \quad (1.41)$$

$$= R_y^{\tau\tau} \quad \begin{array}{c}\tau\ \tau\ \tau\ \tau\\ \diagup\!\!\!\diagup\\ x\quad y\\ 1\end{array} \quad , \quad (1.42)$$

so $[\rho(\sigma_3)]_y^x = \delta_y^x R_y^{\tau\tau} = [\rho(\sigma_1)]_y^x$.

### 1.3.2 Topological Quantum Computation

Quantum computation is a term used to describe a computation method based on creating, manipulating, and measuring quantum states [37]. Several algorithms with quite remarkable properties are based on quantum computation. For example,

- factoring integers and computing discrete logarithms can be done "exponentially faster" using Shor's algorithm [93].

- Searching for a specific entry in an unstructured database can be done with Grover's algorithm in $\mathcal{O}(\sqrt{N})$ evaluations as opposed to the classical $\mathcal{O}(N)$ evaluations [45].

- The time required for simulating certain quantum systems, in particular the system that describes the computer itself, can be exponentially faster [35, 66].

- Quantum invariants of three-dimensional manifolds and knots, e.g., the absolute value of the Jones polynomial of a link $L$ at certain roots of unity, can be approximated exponentially fast [37].

One of the main difficulties in quantum computing is the accumulation of errors due to environmental interactions. Indeed, any interaction of a quantum system with the environment can be seen as a measurement and thus causes a collapse of the wave function. Quantum memory can, therefore, be quite volatile. This is where anyons can play a big role. Anyonic systems realize quantum memory that is inherently fault-tolerant [56, 55]. The clue lies in the fact that anyon statistics are governed by the braid group, and thus only depend on the topology of the paths taken during the exchange. These paths may be perturbed, but no errors will occur as long as the perturbations are smaller than the distance between two anyons.

For the system described in example 1.3.1, the Hilbert space associated with the four $\tau$ anyons with total charge 1 is two-dimensional. This means that, in theory, one can use such a system as a qubit. By braiding the $\tau$ particles around each other, one can apply transformations to these qubits. The specific transforms that can be realized on these qubits (via braiding) depend on the type of anyons used. In the case of Fibonacci anyons, it turns out [103] that any unitary transform can be approximated with arbitrary precision. Anyons for which such a property hold are said to be universal for quantum computation. Not all anyonic systems have this property, the Ising anyons being a standard counter-example. Since the outcome of the braiding of qubits depends only on the topology of the space-time trajectories, this scheme of quantum computation is called topological quantum computation (TQC).

The mathematical structure that describes anyons is that of a unitary modular fusion system or, more abstractly, a unitary modular fusion category [56]. The advantage of the categorical description is that it can be used without worrying about the explicit realization of anyons in a lab. This allows us to investigate the properties of anyons without the need for a microscopic description.

There are a myriad of practical challenges that need to be overcome, though, if one wants to use nonabelions for TQC. At the moment of writing, I do not know any experimental setup that has realized nonabelions that can be manipulated in the ways required for TQC. While there are signs of the existence of Ising anyons [86, 22, 76, 14, 36, 2], and proposals for the detection of Fibonacci anyons [74], one still needs to be able to manipulate these anyons in a controlled manner to perform actual quantum computation. That being said, there are some recent breakthroughs in simulating nonabelions on quantum computers [42, 49].

> Notes 2.
> - The treatment given here not only ignores various practical difficulties but also several important theoretical issues. One of these is that it is unclear whether the representations obtained from anyons are irreducible or not. Another one is that it is not obvious which anyons lead to universal quantum computation. The papers cited at the beginning of this section provide a more in-depth discussion of these problems.
>
> - A famous conjecture [78] says that if $\sum_a d_a^2 \notin \mathbb{N}$, there are non-abelian anyons whose braiding allows for universal quantum computing.

# Chapter 2

# From Fusion System to Unitary Modular Fusion System

In the previous chapter, we saw how to describe processes with anyons via diagrams. To do so, we started with a unitary fusion system and very quickly introduced the more specialized unitary modular fusion system. This chapter aims to present the more general fusion systems and to add structures, one by one until we arrive at the notion of a unitary modular fusion system. It will also point out a relation between a fusion system and a fusion category, but leave the exact details for Section 7.2. While not all fusion systems are suitable to describe anyons, they still provide a powerful language that has proven useful in bridging the gap between many research areas, such as knot theory [54, 84, 83, 96], representation theory of weak Hopf algebras and quantum groups [54, 8], low-dimensional topology and topological field theory [97, 16, 8, 18, 98], topological quantum computation [89, 87, 88, 103, 56], subfactor theory and planar algebras [51, 44, 13, 30], vertex operator algebras and conformal field theory [75, 17, 30], etc.

The outline of this chapter is the following. The chapter starts by introducing the notion of a (not necessarily unitary) fusion system in Section 2.1. In Section 2.2, we will start adding extra structures to a fusion system. This way, we will arrive at the notions of pivotal, spherical, braided, ribbon, and modular fusion systems. In Section 2.3, the notion of unitarity of fusion systems will be discussed. The chapter concludes with Section 2.4 that discusses some of the challenges in research in fusion categories and how the rest of this thesis tackles some of these. Most of the material in this chapter comes straight from [19], an excellent resource for people who want to bridge the gap between the abstract fusion categories and the down-to-earth numerical fusion systems.

## 2.1 Fusion Systems

The definition of a fusion system is closely related to that of a unitary fusion system, given in Chapter 1.

**Definition 2.1.1**. A **fusion system** is a collection of data $(\mathbf{L}, *, \mathbf{N}, \mathbf{F})$ where

1. $\mathbf{L} = \{1, ..., r\}$

2. $* : \mathbf{L} \to \mathbf{L}$ is a map for which $1^* = 1$ and $(a^*)^* = a$.

3. $\mathbf{N} = \left\{ N^c_{a,b} \middle| a, b, c = 1, ..., r \right\}$ is a finite set of natural numbers that satisfy

$$N^b_{a,1} = N^b_{1,a} = \delta^a_b, \tag{2.1}$$

$$N^1_{a,b} = \delta^a_{b^*}, \tag{2.2}$$

$$N^d_{a,b,c} := \sum_e N^e_{a,b} N^d_{e,c} = \sum_f N^d_{a,f} N^f_{b,c}. \tag{2.3}$$

4. $\mathbf{F} = \left\{ [F^{abc}_d] \in \text{Mat}_{N^d_{a,b,c} \times N^d_{a,b,c}} \middle| a, b, c, d = 1, ..., r \right\}$ is a finite set of finite-dimensional invertible matrices, with inverses $\left\{ [\tilde{F}^{abc}_d] \right\}$, that satisfy

$$[F^{aa^*a}_a]^{(1,1,1)}_{(1,1,1)} \neq 0, \tag{2.4}$$

$$[F^{1bc}_d] = \mathbb{1}_{N^d_{b,c} \times N^d_{b,c}}, \tag{2.5}$$

$$[F^{a1c}_d] = \mathbb{1}_{N^d_{a,c} \times N^d_{a,c}}, \tag{2.6}$$

$$[F^{ab1}_d] = \mathbb{1}_{N^d_{a,b} \times N^d_{a,b}}, \tag{2.7}$$

$$[F^{fcd}_e]^{(g,\beta,\gamma)}_{(l,\zeta,\varepsilon)} [F^{abl}_e]^{(f,\alpha,\zeta)}_{(k,\theta,\eta)} = \sum_{h,\iota,\kappa,\lambda} [F^{abc}_g]^{(f,\alpha,\beta)}_{(h,\kappa,\iota)} [F^{ahd}_e]^{(g,\kappa,\gamma)}_{(k,\theta,\lambda)} [F^{bcd}_k]^{(h,\iota,\lambda)}_{(l,\eta,\varepsilon)}, \tag{2.8}$$

where whenever a zero dimensional matrix is encountered in a term, it is automatically 0.

> **Note 2.** We did not demand that $N^c_{a,b} = N^c_{b,a}$, i.e. that fusion is commutative. Neither did we demand that the $F$-matrices are unitary. The $N^c_{a,b}$ can still be interpreted as the dimensions of fusion and splitting spaces, but these are now just vector spaces instead of Hilbert spaces. There is no notion of a canonical inner product for a generic fusion system.

Fusion systems are closely related to fusion categories. In [19] it is shown that each fusion system gives rise to a unique fusion category. On the other hand, each fusion category gives rise to an infinite number of fusion systems. This is because a choice of bases, also called a choice of the gauge, for the various fusion spaces $V^{a,b}_c$ is required to determine the $F$ matrices. Let the matrices $[G^{ab}_c]$, with inverses $[\tilde{G}^{ab}_c]$, represent basis transformations (also called gauge transformations) of the various $V^{a,b}_c$. The induced transformations on the $F$-matrices are of the form

$$[F^{abc}_d]^{(e,\alpha,\beta)}_{(f,\gamma,\delta)} \mapsto \sum_{\alpha',\beta',\gamma',\delta'} [G^{ab}_e]^{\alpha'}_\alpha [G^{ec}_d]^{\beta'}_\beta [F^{abc}_d]^{(e,\alpha',\beta')}_{(f,\gamma',\delta')} [\tilde{G}^{af}_d]^{\gamma'}_\gamma [\tilde{G}^{bc}_f]^{\delta'}_\delta. \tag{2.9}$$

For any solution $\left\{ [F^{abc}_d]^{(e,\alpha,\beta)}_{(f,\gamma,\delta)} \right\}$ to the pentagon equations,

$$\left\{ \sum_{\alpha',\beta',\gamma',\delta'} [G^{ab}_e]^{\alpha'}_\alpha [G^{ef}_d]^{\beta'}_\beta [F^{abc}_d]^{(e,\alpha',\beta')}_{(\gamma',\delta',\delta')} [\tilde{G}^{af}_d]^{\gamma'}_\gamma [\tilde{G}^{bc}_f]^{\delta'}_\delta \right\} \tag{2.10}$$

is also a solution to the pentagon equations. So as long as the gauge transforms are chosen such that (2.4), (2.5), (2.6) and (2.7) are satisfied for the new set of $F$-symbols (which is always possible), we obtain a fusion system that corresponds to the same fusion category.

A fusion system is called **multiplicity-free** if $\max\left\{N_{a,b}^c\right\} = 1$. For a multiplicity-free fusion system, we use the notation $[F_d^{abc}]_f^e := [F_d^{abc}]_{(f,1,1)}^{(e,1,1)}$ and the demands on the $F$-matrices simplify to

$$[F_a^{aa^*a}]_1^1 \neq 0, \tag{2.11}$$

$$[F_d^{1bc}] = [F_d^{a1c}] = [F_d^{ab1}] = \mathbb{1}_{1\times 1}, \tag{2.12}$$

$$[F_e^{fcd}]_l^g [F_e^{abl}]_k^f = \sum_h [F_g^{abc}]_h^f [F_e^{ahd}]_k^g [F_k^{bcd}]_l^h. \tag{2.13}$$

For a multiplicity-free fusion system each fusion and splitting space is one-dimensional and, by writing $g_c^{ab} := [G_c^{ab}]_1^1$, the $F$-symbols transform as

$$[F_d^{abc}]_f^e \mapsto \frac{g_e^{ab} g_d^{ec}}{g_d^{af} g_f^{bc}} [F_d^{abc}]_f^e. \tag{2.14}$$

> **Note 3.** In contrast with [19], definition 2.1.1 demands that any vacuum $F$ matrix is the identity. While actually only demand (2.6) is necessary, it is, without loss of generality, always possible to satisfy (2.5) and (2.7) (see proposition 4.1.3 for a proof for the multiplicity-free case). The main difference between demand (2.6) and demands (2.5) and (2.7) is that the latter two are not necessary and only there to simplify calculations, while the first is equivalent to the triangle equations (7.6) and thus necessary. The latter two also restrict the choice of gauge, while the first does not.

The terms fusion system and fusion category will often be used interchangeably in what follows. The latter is the more common term, while the former is, as far as I am aware, only introduced in [19] to point out the relationship between the abstract structure and the more down-to-earth, numerical description of a fusion category. For most of this thesis, understanding the category theory is unnecessary. Therefore, the categorical definition and its relation with the numerical description is postponed to Section 7.2.

## 2.2 From Fusion System to Modular Fusion System

We start by adding a *pivotal structure*.

**Definition 2.2.1**.

- A **pivotal fusion system** $(\mathbf{L}, *, \mathbf{N}, \mathbf{F}, \mathbf{P})$ is a fusion system together with a list of

phases $\mathbf{P} = \{p_a \in \mathrm{U}(1) \,|\, a \in \mathbf{L}\}$, called **pivotal coefficients**, for which

$$p_1 = 1, \tag{2.15}$$

$$p_a = p_{a^*}^{-1}, \tag{2.16}$$

$$\frac{p_c}{p_a p_b} = \sum_{s=1}^{N_{b,c^*}^{a^*}} \sum_{t=1}^{N_{c^*,a}^{b^*}} [F_1^{abc^*}]_{(a^*,1,s)}^{(c,i,1)} [F_1^{bc^*a}]_{(b^*,1,t)}^{(a^*,s,1)} [F_1^{c^*ab}]_{(c,1,i)}^{(b^*,t,1)}, \tag{2.17}$$

for all $i \in \left\{1, \dots, N_{a,b}^c\right\}$.

- The **quantum dimensions** $\{d_a \in \mathbb{C} \,|\, a \in L\}$ of a pivotal fusion system are defined as

$$d_a := \frac{p_a}{[F_a^{aa^*a}]_{(1,1,1)}^{(1,1,1)}}. \tag{2.18}$$

- The pivotal structure $\mathbf{P}$ is called **spherical** if $d_a = d_{a^*}$ for all $a \in \mathbf{L}$. In this case, the pivotal fusion system is called a **spherical fusion system**.

It might seem odd that we need to add an extra structure in order to be able to define the quantum dimensions of a fusion system, whereas, in the previous chapter, we did not make such demands. This is because, in the previous chapter, we worked with a unitary fusion system, where the values of the quantum dimensions are fixed by definition. Section 2.3 expands more on this subtlety.

The pivotal coefficients are also gauge-dependent. A gauge-transform, with matrices $[G_c^{ab}]$, has the following effect

$$p_a \mapsto \frac{[G_1^{aa^*}]_1^1 [G_a^{1a}]_1^1}{[G_a^{a1}]_1^1 [G_1^{a^*a}]_1^1} p_a = \frac{[G_1^{aa^*}]_1^1}{[G_1^{a^*a}]_1^1} p_a, \tag{2.19}$$

where the last equality comes from the fact that the vacuum $F$-symbols are not allowed to change, and thus $[G_a^{1a}]_1^1 / [G_a^{a1}]_1^1 = 1$. The reason this is the correct transform comes from a categorical argument. For a pivotal fusion category, the quantum dimensions $d_a$ are defined in a basis-independent manner (see 7.1.21) and must therefore be gauge-invariant. Since the quantum dimensions are gauge-invariant, $p_a$ must transform the same way as $[F_a^{aa^*a}]_{(1,1,1)}^{(1,1,1)}$, i.e., as in equation 2.19.

We can also add a *braided structure* to a fusion system independent of a pivotal structure.

**Definition 2.2.2**.

- A **braided fusion system** $(\mathbf{L}, *, \mathbf{N}, \mathbf{F}, \mathbf{R})$ is a fusion system together with a finite list $\mathbf{R}$ of invertible matrices $\left\{ [R_c^{ab}] \in \mathrm{Mat}_{N_c^{a,b} \times N_c^{a,b}}(\mathbb{C}) \,\middle|\, a, b, c = 1, \dots, r \right\}$, with in-verses $\{[\tilde{R}_c^{ab}]\}$, that satisfy the hexagon equations:

$$\sum_{\lambda,\gamma}[R_e^{ca}]_\lambda^\alpha[F_d^{acb}]_{(g\mu,\gamma,\gamma)}^{(e,\lambda,\beta)}[R_g^{cb}]_\nu^\gamma = \sum_{f,\sigma,\delta,\psi}[F_d^{cab}]_{(f,\sigma,\delta)}^{(e,\alpha,\beta)}[R_d^{cf}]_\psi^\sigma[F_d^{abc}]_{(g,\mu,\nu)}^{(f,\delta,\psi)}, \quad (2.20)$$

$$\sum_{\lambda,\gamma}[\tilde{R}_e^{ac}]_\lambda^\alpha[F_d^{acb}]_{(g,\mu,\gamma)}^{(e,\lambda,\beta)}[\tilde{R}_g^{bc}]_\nu^\gamma = \sum_{f,\sigma,\delta,\psi}[F_d^{cab}]_{(f,\sigma,\delta)}^{(e,\alpha,\beta)}[\tilde{R}_d^{fc}]_\psi^\sigma[F_d^{abc}]_{(g,\mu,\nu)}^{(f,\delta,\psi)}. \quad (2.21)$$

- A **ribbon fusion system** is a spherical braided fusion system.

- A **modular fusion system** (**L**, ∗, **N**, **F**, **P**, **R**) is a ribbon fusion system for which the matrix $\hat{S} \in \mathrm{Mat}_{r \times r}(\mathbb{C})$, whose entries are given by

$$[\hat{S}]_b^a = \sum_{c=1}^r \sum_{i=1}^{N_{a,b^*}^c} \sum_{j=1}^{N_{c,b}^a} \sum_{i'=1}^{N_{b^*,a}^c} \sum_{i''=1}^{N_{a,b^*}^c} [\tilde{F}_a^{ab^*b}]_{(c,i,j)}^{(1,1,1)}[R_c^{b^*a}]_{i'}^i[R_c^{ab^*}]_{i''}^{i'}[F_a^{ab^*b}]_{(1,1,1)}^{(c,i'',j)}, \quad (2.22)$$

is invertible.

Each of the systems above defines a fusion category, with the corresponding adjectives, that is unique up to equivalence. See Section 7.2 for the exact correspondences.

Just like the $F$-symbols and pivotal coefficients, the $R$-symbols are gauge-dependent. A gauge-transform, with matrices $[G_c^{ab}]$, has the following effect

$$[R_c^{ab}]_\beta^\alpha \mapsto \sum_{\gamma,\delta}[\tilde{G}_c^{ba}]_\gamma^\alpha[R_c^{ab}]_\beta^\alpha[G_c^{ab}]_\delta^\beta. \quad (2.23)$$

If the underlying fusion system is multiplicity-free, then several formulas are simplified. The pivotal equations lose the summation on the RHS and are simplified to

$$\frac{p_c}{p_a p_b} = [F_1^{abc^*}]_{a^*}^c[F_1^{bc^*a}]_{b^*}^{a^*}[F_1^{c^*ba}]_c^{b^*}. \quad (2.24)$$

Let $R_c^{ab} := [R_c^{ab}]_1^1$ then the hexagon equations simplify to

$$R_e^{ca}[F_d^{acb}]_g^e R_g^{cb} = \sum_f [F_d^{cab}]_f^e R_d^{cf}[F_d^{abc}]_g^f, \quad (2.25)$$

$$\tilde{R}_e^{ac}[F_d^{acb}]_g^e \tilde{R}_g^{bc} = \sum_f [F_d^{cab}]_f^e \tilde{R}_d^{fc}[F_d^{abc}]_g^f. \quad (2.26)$$

Equations (2.34) and (2.35) respectively simplify to

$$\frac{\theta_c}{\theta_a \theta_b} = R_c^{ab} R_c^{ba}, \quad \forall a,b,c \in \mathbf{L}, \text{ with } N_{a,b}^c \neq 0 \quad (2.27)$$

$$\theta_a = \frac{1}{d_a}\sum_{c=1}^r d_c R_c^{aa}, \quad (2.28)$$

and the matrix $\hat{S} \in \mathrm{Mat}_{r \times r}(\mathbb{C})$, for a multiplicity-free ribbon category has entries

$$[\hat{S}]_b^a = \sum_{c=1}^r [\tilde{F}_a^{ab^*b}]_c^1 R_c^{b^*a} R_c^{ab^*}[F_a^{ab^*b}]_1^c. \quad (2.29)$$

For the multiplicity-free case, the gauge transforms (2.19) and (2.23) also simplify to

$$p_a \mapsto \frac{g_1^{aa^*}}{g_1^{a^*a}} p_a \tag{2.30}$$

$$R_c^{ab} \mapsto \frac{g_c^{ab}}{g_c^{ba}} R_c^{ab}. \tag{2.31}$$

**Notes 3.**

- The origin of the equations (2.15), (2.16), and (2.17) is described in [19]. They stem from the fact that $x^{**} \cong x$ in the categorical language rather than $x^{**} = x$. For several formulas, including the one to calculate quantum dimensions, a set of isomorphisms $\psi_x : x^{**} \to x$ is required to match the input and output of various morphisms properly. Requiring that a given set of isomorphisms $\{\psi_x\}$ behaves naturally with respect to the tensor product comes down to demanding the pivotal equations.

- For each ribbon fusion system, there exists a canonical list of gauge-independent phases, also called twists, $\mathbf{T} = \{\theta_i \in U(1) | i \in \mathbf{L}\}$ that satisfy equations

$$\theta_1 = 1, \tag{2.32}$$

$$\theta_a = \theta_{a^*}, \tag{2.33}$$

$$\frac{\theta_c}{\theta_a \theta_b} = \sum_{\beta,\gamma} [R_c^{ab}]_\beta^\gamma [R_c^{ba}]_\gamma^\beta, \tag{2.34}$$

$$\theta_a = \frac{1}{d_a} \sum_{c=1}^{r} \sum_\alpha d_c [R_c^{aa}]_\alpha^\alpha. \tag{2.35}$$

Often, the definition of a ribbon structure on a monoidal category (see, e.g., [25]) only demands that the first three equalities (2.32), (2.33), and (2.34) hold. These are independent of the pivotal structure, and there could be multiple ribbon structures that one can put on a braided category. One could, therefore, also investigate the notion of non-spherical ribbon categories. Such ribbon categories do not admit the twist's interpretation as a combination of a braid with caps and cups (which is, in essence, what equation (2.35) demands). The authors of some books, including [25], therefore identify ribbon fusion categories with braided spherical categories. Note that the twists are gauge-independent.

- The $S$-matrix of a modular fusion system is defined as

$$[S]_b^a = d_a d_b [\hat{S}]_b^a. \tag{2.36}$$

It can be shown that it is, up to scaling, always a unitary matrix [27]. The rank of $S$ equals that of $\hat{S}$, but typically, $\hat{S}$ has a nicer form. Moreover, $\hat{S}$ can be defined independent of the specific pivotal structure, so to calculate the $S$-matrix for multiple pivotal fusion systems with the same $F$ and $R$-symbols, one only has to calculate $\hat{S}$ once.

- Let $T$ be an $r \times r$ matrix with entries

$$[T]^a_b := \delta^a_b \theta_a, \tag{2.37}$$

then, for a modular fusion system, $S$ and $T$ satisfy

$$(ST)^3 = \lambda S^2, \tag{2.38}$$
$$S^4 = 1 \tag{2.39}$$

for some $\lambda \in \mathbb{C}^\times$ [25]. They define a projective representation of the modular group $SL_2(\mathbb{Z})$, hence the origin of the adjective 'modular'.

- Via a construction known as the center of a category, one can construct modular fusion categories from spherical fusion categories [77]. This is yet another reason why it is interesting to study more general fusion categories than just modular ones.

## 2.3 Unitary Fusion Systems

Not demanding unitarity from the start makes it easier to see several subtleties that were otherwise hidden. One of these subtleties is the existence (and choice) of different pivotal and possibly non-spherical structures. In the previous chapter, we did not choose a pivotal structure for the fusion system because the fusion system corresponded to a unitary fusion category. This is a fusion category that comes with a *Hermitian structure*, †, that satisfies several properties (see, e.g., [40, 48] for a definition). The Hermitian structure defines the notion of unitary maps and can, in particular, be used to define cups and caps that can be stacked on each other to calculate the quantum dimensions $d_a$ [50, 106, 105]. The quantum dimensions, arising this way, satisfy $d_a = \text{FPDim}(a)$. The dagger also fixes a spherical pivotal structure $\mathbf{P}$, which, for the corresponding fusion system, is given by $\left\{ p_a = d_a [F_a^{aa^*a}]^{(1,1,1)}_{(1,1,1)} \,\middle|\, a \in \mathbf{L} \right\}$. This spherical structure is also called the canonical spherical structure of the unitary fusion category. While a priori, it looks like there might be multiple Hermitian structures that one can put on a fusion category, it was recently shown in [85] that any unitarizable fusion category admits a unique unitary structure (up to unitary monoidal equivalence). Therefore, the canonical spherical structure can be implicitly included in the definition of a unitary fusion category.

The link between a unitary fusion category and a fusion system is the following. If a fusion system has a gauge for which its $F$-matrices are unitary, then its category can always be made into a unitary category via the choice of an appropriate Hermitian structure (see [106], part 4). Now, define a unitary fusion system as follows.

**Definition 2.3.1**. A **unitary fusion system** is a spherical fusion system (**L**, ∗, **N**, **F**, **P**) for which the *F*-matrices are unitary and the quantum dimensions satisfy $d_a = \text{FPDim}(a)$.

For a unitary category, there always exists a gauge choice for which the *F*-symbols of the corresponding spherical fusion system are unitary. So, the notion of a unitary fusion system is equivalent to that of a unitary fusion category.

> **Note 4.** In the previous chapter, we already introduced a unitary fusion system as a fusion system with unitary *F*-symbols and a specific Hermitian structure that gives rise to positive quantum dimensions. That definition is equivalent to definition 2.3.1 but not as practical so we will use definition 2.3.1 from now on.

It is important to remember that if the *F*-matrices of a fusion system are unitary, one could still choose a (spherical) pivotal structure that is not canonical. For example, any fusion system with a fusion ring corresponding to the group algebra $\mathbb{Z}_3$ has unitary *F*-symbols, but there are multiple non-spherical pivotal structures for which the quantum dimensions are not positive. We will not call these fusion systems unitary but rather say that they admit a unitary gauge.

The definition of unitarity of a fusion category can be extended to braided, ribbon, and modular fusion categories by demanding that the braided and ribbon structures are *compatible with the unitary structure*. In the paper [40], these demands are written out, and the following statements are proven:

- If a unitary fusion category admits a braiding, it is automatically a unitary braided fusion category.

- every unitary braided fusion category admits a unique unitary ribbon structure. This ribbon structure is the canonical ribbon structure that comes from the canonical spherical structure in combination with the braiding.

In particular, a unitary braided fusion category is immediately a ribbon fusion category. We will therefore use the following definitions

**Definitions 2.3.2**.

- A **unitary braided fusion system**, or equivalently a **unitary ribbon fusion system**, is a ribbon fusion system (**L**, ∗, **N**, **F**, **P**, **R**) for which (**L**, ∗, **N**, **F**, **P**) form a unitary fusion system and the *R*-matrices are unitary matrices.

- A **unitary modular fusion system** is a unitary ribbon fusion system for which the matrix $\hat{S}$ is invertible.

We end this chapter by noting that not all fusion categories are unitary. Typically, the non-unitary categories lead to a much richer variety of examples. This is because the cascade of canonical structures that follow from the Hermitian structure of a unitary category is not there for a generic category. Therefore, one can put many more combinations of different structures on non-unitary categories.

> **Note 5.** The demand that the *F*-and *R*-symbols of a unitary fusion system are unitary is out of convenience rather than necessity. Since such a basis always exists and it is not hard to find (see Section 4.1.9 for an algorithm) this should cause no issues in practice.

## 2.4 The Landscape of Fusion Categories

As we have seen, every fusion category is fully determined by a fusion system (**L**, ∗, **N**, **F**). The mathematical structure that captures the properties of **L**, ∗, and **N** is that of a fusion ring [25], which is defined as follows.

**Definitions 2.4.1**. A $\mathbb{Z}_+$-ring is an associative unital ring $\mathcal{R} \equiv (\mathcal{R}, B, 1, \times, +)$, finitely generated and free as a $\mathbb{Z}$-module, which is equipped with a distinguished basis $B$ such that $1 \in B$ and for which the structure constants $\{N_{ab}^c\}$ are non-negative.

A **fusion ring** $\mathcal{R} \equiv (\mathcal{R}, B, 1, \times, +, *)$ is a $\mathbb{Z}_+$ ring with basis $B$ and a linear involution $\cdot^* : a \mapsto a^*$ such that $N_{a,b^*}^1 = \delta_{ab}$ and $N_{a,b}^c = N_{a^*,c}^b$ for all $a, b, c \in B$. The size of $B$ is called the **rank of the fusion ring**, and the number $\max\left\{N_{a,b}^c\right\}$ is called the **multiplicity of the fusion ring**. If the multiplicity of a fusion ring is 1 we say the ring is **multiplicity-free**. We also say that $B$ generates $\mathcal{R}$ and write $\mathcal{R} = \langle B \rangle \equiv \langle \psi_1, \dots, \psi_r \rangle$.

> **Notes 4.**
>
> - From now on, we will reserve the notation $\mathcal{R}$ for a fusion ring, $B$ for its basis, $r$ for its rank, $m$ for its multiplicity, and $N_{a,b}^c$ for its structure constants. We will always use the convention that the basis of a fusion ring is ordered so that its unit is the first element in this ordering. By a fusion ring automorphism, we mean a map $\sigma : B \to B$ that satsifies $N_{\sigma(a),\sigma(b)}^{\sigma(c)} = N_{a,b}^c, \forall a, b, c \in B$ is meant.
>
> - Since a fusion ring is entirely determined by a finite set of integer structure constants $\left\{ 0 \leq N_{a,b}^c \leq m \,\middle|\, a, b, c = 1, \dots, r \right\}$, there are only a finite number of fusion rings for a given rank $r$ and multiplicity $m$.

A fusion ring is the only required information to set up the necessary constraints (2.4), (2.5), (2.6), (2.7), and (2.8) on the *F*-symbols in order to obtain a fusion system. These constraints form a system of polynomial equations and inequalities. Such a system is, in theory, always algorithmically solvable, but it could be that the solution set is empty. Solving these equations is also known as categorifying the fusion ring.

Most known fusion categories are of some '*standard* form', a combination of these (via, e.g., tensor products), modifications of these (via, e.g., zesting [21]) or extensions of these (via, e.g., constructions such as the Tambara-Yamagami or Haagerup-Izumi categories (see Section 3.4.1) ). By a *standard* form I[1] mean that the fusion ring is one

---
[1]Different authors might have different opinions on what should be regarded as *standard*.

of the following:

- A group ring of a finite group.

- A ring of finite dimensional irreps of a finite group with the product being the tensor product.

- A fusion ring coming from the representation theory of quantum groups at roots of unity. (see [92] for a gentle introduction)

To what extent these standard categories populate the landscape of fusion categories is unknown. In order to get an overview of a small part of the landscape of fusion categories one could theoretically, apply the following approach:

1. Fix a multiplicity $m$, and a rank $r$.

2. Find all fusion rings with this rank and multiplicity.

3. Solve the pentagon equations per fusion ring found.

4. Solve the pivotal and hexagon equations per solution to the pentagon equations.

5. For each triple of $F$-symbols, $R$-symbols (possibly empty), and pivotal coefficients (possibly empty but not for any of the categories we considered), calculate all other properties such as the quantum dimensions, twists, and the $S$-matrix.

If one fixes the multiplicity $m$ and lists fusion rings by their rank $r$, one gets a tree of possible fusion categories per fixed value of $m$. While the standard constructions typically provide several categories for each rank $r$ and thus descend deep into these trees, the approach above produces all categories per rank, i.e., it goes wide into the tree (see Figure 2.1).

There are several caveats when applying this strategy, however.

1. Firstly, due to gauge symmetry, there are an infinite number of solutions to the consistency equations that, in essence, describe the same fusion system. Without breaking this gauge symmetry beforehand, solving the consistency equations is only possible for the simplest examples. For fusion rings with multiplicity, i.e., with $N_{a,b}^c > 1$ for some labels, I failed to find a general way to break this symmetry. This made it very hard for me to solve the pentagon equations for the case with multiplicity, even for the smallest ranks. If the fusion ring is multiplicity-free, though, the symmetry transformations have a simple form, and we can get rid of this symmetry before solving the pentagon equations. I, therefore, decided to restrict my attention to the categorification of multiplicity-free fusion rings.

2. Secondly, while theoretically, any set of polynomial equations can be solved algorithmically, there are some serious limits on the systems for which this can be done in practice. A system of polynomial equations is algorithmically solved by finding a suitable Gröbner basis that brings the system of equations into an "upper triangular" form. By this, we mean a new, typically smaller, system with the same roots as the original one, but from which one can find all roots by only solving polynomial equations in one variable. Most computer algebra systems have standard built-in algorithms to calculate Gröbner bases, such as Buchberger's original algorithm [11], or the more recent and more optimized F4 [34] and F5 [33] algorithms. We found that none of these algorithms is capable of solving the pentagon equations, without some serious pre-processing of the system. The number of equations and *F*-symbols just grows too fast with the rank of the fusion ring (see figure 2.2 for the exact numbers).

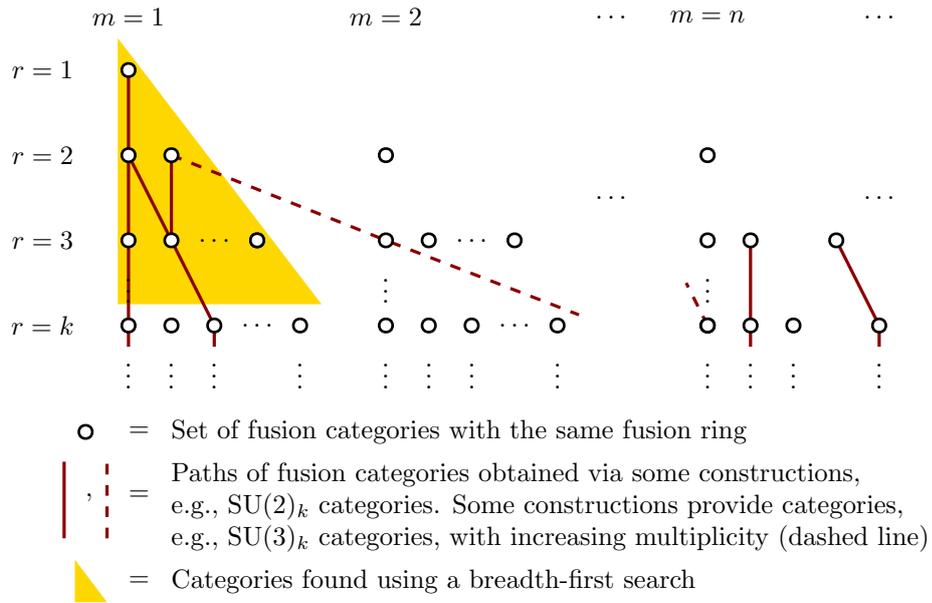

Figure 2.1: By fixing a multiplicity *m* and listing fusion rings by rank *r*, one can visualize the landscape of fusion categories as a series of trees. While the standard constructions cut out deep paths in these trees, our approach cuts out a wide area in the $m = 1$ tree.

By using methods specifically designed for solving pentagon and hexagon equations, I believe[2] we found all multiplicity-free pivotal[3] fusion categories up to rank seven, however. Much of the progress in solving these equations comes from methods and heuristics aimed at reducing sparse systems of polynomial equations. While these methods were written to solve pentagon and hexagon equations, it soon became clear that they are more widely applicable. Therefore we decided to create the Anyonica package [100].

Anyonica has multiple use cases. For one, it is meant to ease research in the theory of anyon models and, more generally, fusion categories. To this end, it contains lists of fusion rings and categories together with all the relevant data we could find. Moreover, many functions have been implemented to aid in finding properties of fusion rings and categories and transforming and or combining them. The second way in which the package is useful is via the many specialized functions it provides for reducing large systems of polynomial equations. Different reduction techniques and heuristics can and have been applied for solving both the pentagon and hexagon equations as well as for other systems of polynomial equations, such as the equations governing anyons on wire networks [68]. The package, written in the Wolfram Language, a.k.a. Mathematica, is freely available on https://github.com/gert-vercleyen/Anyonica.

---

[2] See the notes at the end of this Section for why this is a statement of informed faith, rather than one of absolute truth

[3] If all fusion categories with $PSU(2)_{12}$ fusion rules are pivotal then we found all multiplicity-free fusion categories up to rank seven.

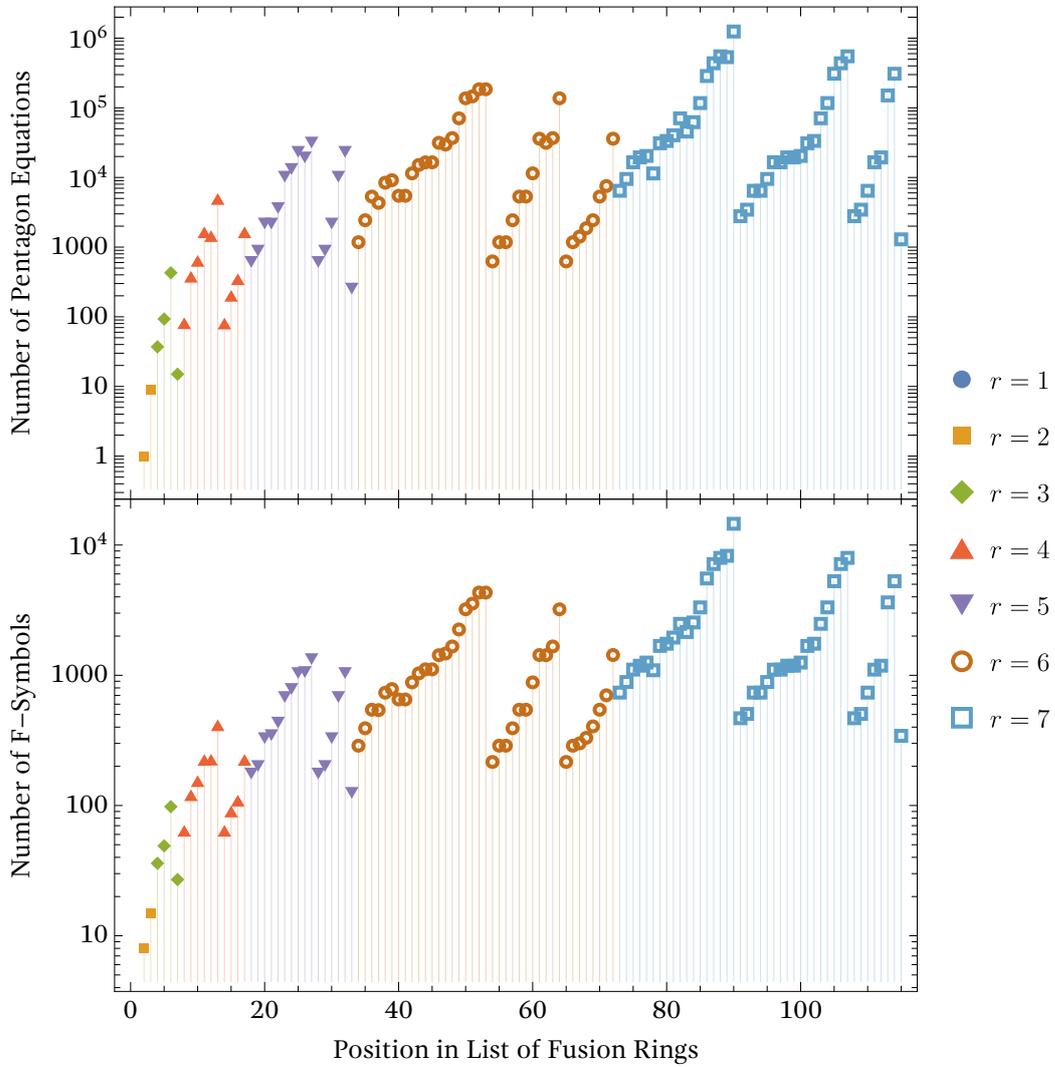

Figure 2.2: Number of pentagon equations and $F$-symbols per fusion ring. The numbers on the horizontal axis correspond to the position of the fusion ring in the list of multiplicity-free fusion rings in Section 8. Here, we assumed that the vacuum $F$-symbols are already known and the trivial pentagon equations have already been removed. This is also why the trivial ring (at position 1) has no data.

The next part of the thesis explains the various algorithms developed and how they helped find fusion categories. Also, small examples of code will be provided to illustrate how the Anyonica package can be used. Most of the examples will focus on the fusion ring $FR_2^{3,1,0} = \text{Rep}(D_3)$ which has the following fusion rules

$$\begin{array}{|ccc|} \hline 1 & 2 & 3 \\ 2 & 1 & 3 \\ 3 & 3 & 1+2+3 \\ \hline \end{array}.$$

These examples should be regarded as if they belong to the same Mathematica session, so if the variable `repD3` is defined in one example, it must be assumed that it is accessible in the other examples as well. Although the programming style and keywords are those of the Wolfram Language, the code should be as readable as any pseudo-code.

> **Notes 5.**
> - Ideally, one would first look at the classification of fusion rings before classifying fusion categories. The defining properties of a fusion ring are only a slight generalization of those that define a group. Indeed, any group ring is a fusion ring, and for any fusion ring $\mathscr{R}$, demanding that $a \times a^* = 1$ for all $a \in \mathscr{R}$ comes down to demanding that $\mathscr{R}$ is a group ring. This means that classifying fusion rings is at least as hard as classifying finite groups, and there is little hope that all fusion rings will be classified soon.
>
> - There is no reason to believe that there won't be a generic algorithm that can tackle the pentagon equations sometime in the future. Often, a direct Gröbner basis calculation is thought to be undoable since theoretically (see [24]) the largest degree of an element of a Gröbner basis is bounded above by
> $$2\left(\frac{d^2}{2} + d\right)^{2^{n-2}}, \qquad (2.40)$$
> with $d$ the highest degree of the input polynomials (always 3 in our case), and $n$ the number of variables (easily larger than 100 for fusion rings of rank 5 or higher as can be seen in figure 2.2). While there exist polynomial systems for which such a degree is obtained [72], we found that the pentagon equations typically can be reduced to a Gröbner basis of a degree of the order of $r$. The double exponential asymptotic scaling in time and memory of Gröbner basis algorithms might be correct for arbitrary sys-

- tems of polynomial equations but the pentagon and hexagon equations are in no way arbitrary. There is a lot of structure behind these equations that is not taken into account in an expression for a bound like (2.40).

- Some other efforts at achieving similar results have also been made. These will be reviewed in Section 4.6.

- Besides the gauge symmetry, one would ideally also like to break permutation symmetry, coming from the fusion ring automorphisms. However, we did not attempt to break this symmetry while solving the consistency equations and only removed equivalent solutions afterward. One of the reasons this workflow was chosen is that it is very hard to get rid of this symmetry beforehand. The complexity of the workflow would also increase a lot by adding intermediate steps to hunt down these redundancies. It might be interesting or even necessary to add this functionality in order to categorify larger rings with bigger groups of automorphisms.

- The completeness of the classification of all multiplicity-free fusion categories up to rank seven depends on symbolic computer algorithms. These algorithms have been tested over several years, and up to rank six, the results look in line with theory. Each time the consistency equations were solved, the solver created log files. These log files contain reports on when and why intermediate solutions to parts of the systems were refused. They also allow us to check the calculations without the need to redo them again. Even though the algorithms rely on exact methods, they are still programmed by a non-perfect human (me) in a high-level language (the Wolfram Language) that can and does contain bugs. If there is anything I have learned throughout my Ph.D., then it is that one should never unquestioningly trust a computer.

# Part II

# Finding Fusion Categories

# Chapter 3

# Finding Fusion Rings

We start our quest to find fusion categories by searching for fusion rings. In this chapter we take the viewpoint that fusion rings are interesting structures in their own right. We will, therefore, also search for fusion rings of higher multiplicity despite the fact that the categories of interest are multiplicity-free. As a side quest, we will also delve a bit deeper in the structure of non-commutative fusion rings and introduce a recipe for creating so called SONGs (Single Orbit Normal Group fusion rings).

The explanation of the algorithm, together with a quick overview of the results it brought forth, are described in section 3.1. Section 3.2 discusses a canonical naming scheme for fusion rings, and Section 3.3 deals with methods for finding fusion ring characters and modular data for commutative fusion rings. In section 3.4 we introduce a method to create generalizations of the Tambara-Yamagami and Haagerup Izumi fusion rings, which we call songs. Furthermore, the structure of non-commutative fusion rings with a subgroup is reviewed, and the one-and two-particle extensions of groups are classified.

## 3.1 Algorithm

Any fusion ring $\mathscr{R}$ of rank $r$ and multiplicity $m$ is entirely determined by a set of structure constants $\left\{ 0 \leq N_{a,b}^c \leq m \,\middle|\, a, b, c = 1, \ldots, r \right\}$. Therefore, the search for fusion rings can be reduced to filling three-dimensional tables with natural numbers such that the defining properties, like associativity, unitality, etc, are apparent.

Several algorithms for doing this exist. One could, e.g., try using brute force to generate all integer rings of a certain rank and multiplicity and filter those that do not satisfy the requirements of a fusion ring. Even after breaking symmetry and reducing the number of variables (see sections 3.1.1 and 3.1.2), this method quickly becomes unfeasible, as can be seen in table 3.1.

Another strategy (see [41]) consists of simultaneously diagonalising the fusion matrices of the particles, which is always possible if all fusion matrices commute. For non-commutative fusion rings one can only guarantee a block diagonal form and this method becomes quite cumbersome to implement. Instead, we built an algorithm based on a backtracking approach or tree search. This is a classical method that is also used to, e.g., solve sudokus, a problem that is of exactly the same nature as ours. Before delving into the details of the algorithm, we first present some general results and techniques that were applied to make the task more tractable.

|   | 3 | 4 | 5 | 6 | 7 | 8 | 9 |
|---|---|---|---|---|---|---|---|
| 1 | $2.0 \times 10^1$ | $1.2 \times 10^3$ | $1.1 \times 10^6$ | $3.5 \times 10^{10}$ | $7.3 \times 10^{16}$ | $1.9 \times 10^{25}$ | $1.3 \times 10^{36}$ |
| 2 | $9.0 \times 10^1$ | $6.1 \times 10^4$ | $3.5 \times 10^9$ | $5.0 \times 10^{16}$ | $5.2 \times 10^{26}$ | $1.2 \times 10^{40}$ | $1.8 \times 10^{57}$ |
| 3 | $2.7 \times 10^2$ | $1.1 \times 10^6$ | $1.1 \times 10^{12}$ | $1.2 \times 10^{21}$ | $5.2 \times 10^{33}$ | $3.7 \times 10^{50}$ | $1.8 \times 10^{72}$ |
| 4 | $6.5 \times 10^2$ | $9.8 \times 10^6$ | $9.6 \times 10^{13}$ | $2.9 \times 10^{24}$ | $1.4 \times 10^{39}$ | $5.2 \times 10^{58}$ | $7.5 \times 10^{83}$ |
| 5 | $1.3 \times 10^3$ | $6.1 \times 10^7$ | $3.7 \times 10^{15}$ | $1.7 \times 10^{27}$ | $3.8 \times 10^{43}$ | $2.3 \times 10^{65}$ | $2.4 \times 10^{93}$ |
| 6 | $2.5 \times 10^3$ | $2.8 \times 10^8$ | $8.0 \times 10^{16}$ | $3.8 \times 10^{29}$ | $2.1 \times 10^{47}$ | $9.7 \times 10^{70}$ | $2.6 \times 10^{101}$ |
| 7 | $4.2 \times 10^3$ | $1.1 \times 10^9$ | $1.2 \times 10^{18}$ | $4.1 \times 10^{31}$ | $3.7 \times 10^{50}$ | $7.2 \times 10^{75}$ | $2.3 \times 10^{108}$ |
| 8 | $6.6 \times 10^3$ | $3.5 \times 10^9$ | $1.2 \times 10^{19}$ | $2.5 \times 10^{33}$ | $2.7 \times 10^{53}$ | $1.4 \times 10^{80}$ | $3.2 \times 10^{114}$ |

Table 3.1: Size of the search space of fusion rings of rank $r$ (columns) and multiplicity $m$ (rows) with two significant digits after reduction of the number of variables and symmetry breaking.

### 3.1.1 Reducing the Number of Variables

There are relations between the structure constants that can be used to reduce the number of variables greatly. From the definition of a fusion ring, it follows that

$$N_{a,b}^c = \sum_{e=1}^r N_{a,b}^e N_{e,c^*}^1 = \sum_{f=1}^r N_{a,f}^1 N_{b,c^*}^f = N_{b,c^*}^{a^*}. \tag{3.1}$$

Combined with the relations $N_{a,b}^c = N_{a^*,c}^b$ we obtain

$$N_{a,b}^c = N_{a^*,c}^b = N_{c,b^*}^a = N_{b,c^*}^{a^*} = N_{c^*,a}^{b^*} = N_{b^*,a^*}^{c^*}, \tag{3.2}$$

which we will call *pivotal relations*. A reduced set of fusion coefficients can be obtained using the pivotal relations. Since these relations depend on the number $s$ of self-dual particles, we will assume from here on that we are searching for fusion rings with a fixed value for $s$. One only needs to apply the algorithm to each of the $\left\lfloor \frac{r+1}{2} \right\rfloor$ values of $s$ to find all fusion rings of rank $r$ and multiplicity $m$.

### 3.1.2 Breaking Permutation Symmetry

When expressing structure constants using tables, a labelling of the elements of the basis $B$ is implicitly made. Any relabeling of the elements of $B$ results in a table of structure constants that describe the same ring. In particular, for every fusion ring of rank $r$, there are up to $r!$ different, yet equivalent, tables of structure constants. This redundancy dramatically increases the work of searching for fusion rings by constructing multiplication tables.

One way to break this symmetry slightly is by numbering the basis elements and demanding that the first element is the unit element. We can break the symmetry further by requiring that all self-dual elements appear before the non-self-dual elements, and all non-self-dual elements are grouped in pairs. To break the symmetry even further, we added a set of constraints on the structure constants to the set of associativity relations. To explain the idea behind these constraints, we will first assume that all particles are self-dual and later generalise to generic fusion rings. The constraints are built up by looking at a particle that is not the unit, demanding it will be the 2nd particle and sorting the other particles *based on their fusion* with this particle. Doing so gives a candidate for the 3rd particle. Then we demand that all particles, apart from the 1st, 2nd, and 3rd particle, whose positions are not uniquely fixed by their fusion with the 2nd particle, are sorted by their fusion with the 3rd particle. A candidate for the 4th particle is then given, and we can continue this way until all particle labels are fixed. To apply this scheme, we need to define an ordering of a set of particles $\{\psi_i\}$ that is solely based on their fusion with a given particle, say $\psi_a$. We used the function $\iota_a : i \mapsto N^i_{a,i}$ for this. Practically this means that we demand that for particle $\psi_2$ the following inequalities

$$\iota_2(j) \leq \iota_2(j+1), \qquad (3.3)$$

for $j = 3, 4, \ldots, r-1$, hold. For some particles $\psi_i, \psi_k$ we still might have that $\iota_2(i) = \iota_2(k)$. Those particles are then sorted further using the values of $\iota_3$, which yields the following inequalities

$$\neg(\iota_2(j) = \iota_2(j+1)) \vee (\iota_3(j) \leq \iota_3(j+1)), \qquad (3.4)$$

for $j = 4, 5, \ldots, r-1$. Applied to all particles $\{\psi_2, \ldots, \psi_r\}$ we get

$$\neg\left(\bigwedge_{n=2}^{i-1} \iota_n(j) = \iota_n(j+1)\right) \vee (\iota_i(j) \leq \iota_i(j+1)), \qquad (3.5)$$

for $i = 2, \ldots, r, j = i+1, \ldots, r-1$. Although equations (3.5) break some symmetry, it is clear that there is still redundancy in the choice of the 2nd particle. Furthermore, there might also be multiple choices for the 3rd particle since the order of the particles, determined by fusion with the 2nd particle, is not strict. The same is true for the 4th particle and so on. To reduce this redundancy, we demand that the 2nd particle should be such that

$$\sum_{i=1}^{r} \iota_2(i) \geq \sum_{i=1}^{r} \iota_k(i), \qquad (3.6)$$

for $k = 3, \ldots, r$. For the 3rd, 4th, and other particles, a similar set of inequalities can be constructed, but we can only compare with particles that could not be distinguished by all previous particles. These extra constraints become so tedious that considering them only increases computation time. Therefore we only added constraints (3.6) to the set of constraints.

Now assume not all particles are self-dual. In this case, we can still apply symmetry

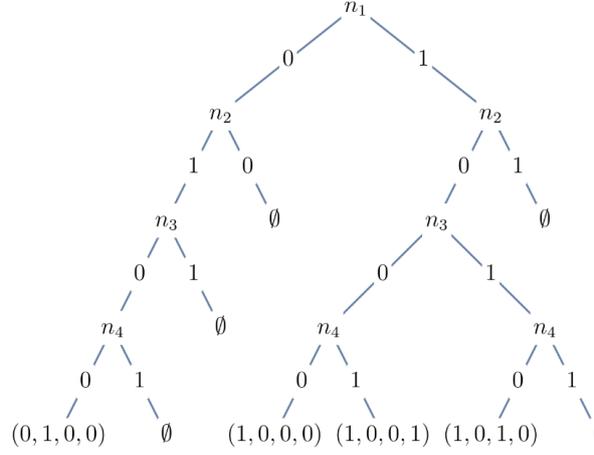

**Equations**
$$\begin{cases} n_1^2 + n_2^2 = 1 \\ n_1 n_2 + n_1 n_3 = n_3^2 \\ n_1 n_2 + n_2 n_3 + n_3 n_4 + n_4 n_1 = n_4^2 \end{cases}$$

**Variables** $n_i \in \{0,1\},\ i = 1,\ldots,4$

**Pseudocode**

```
for n1 in 0:1
    for n2 in 0:1
        if(n1^2 + n2^2 == 1)
            for n3 in 0:1
                if(n1*n2+n1*n3 == n3^2)
                    for n4 in 0:1
                        if(n1*n2+n2*n3+n3*n4+n4*n1 == n4^2)
                            save(n1,n2,n3,n4)
```

Figure 3.1: Procedure for finding solutions to a system of constraints applied to a system of 3 equations in 4 unknowns with values in $\{0,1\}$. Note that the order in which the constraints are checked is vital for the performance.

breaking but should do so separately for self-dual and non-self-dual particles. This must be done so that no comparison occurs between the structure constants of self-dual particles and non-self-dual particles. Note that for the non-self-dual particles, the constraints above will not make an ordering between dual particles because of the pivotal relations. Therefore, only pairs of particles can be sorted this way.

### 3.1.3 Backtracking

Under the assumption that the techniques described above have been applied, one should now have (a) a reduced set of unknowns, say $\{n_i\}$, and (b) an enlarged set of constraints consisting of a set of associativity relations together with a set of inequalities (forcing a non-strict order on the particles). The remaining procedure then consists of finding solutions to the system of constraints. Here, a backtracking approach naturally arises since, for any constraint, only partial information about the values of the $n_i$ is needed to check its validity. We can therefore do a tree search for admissible solution sets. The branches of the tree at level $j$ correspond to a choice of $n_j$. The leaves at all but the last level correspond to invalidated constraints, and those at the last level correspond to valid solutions. Figure 3.1 shows a graphical depiction of this process.

Whenever a constraint is violated, $m^K$ configurations (where $K$ denotes the number of remaining unknowns) are ruled out. The order in which the unknowns are given values is thus of great importance. To cut off branches of the search tree as soon as possible, we sorted the unknowns in the following way. First, we look for the constraints with the fewest number of different unknowns. If there are multiple, pick any as a first constraint. Then regard all the unknowns in this constraint as known and choose a second constraint with the least number of unknowns (thus, after removing the unknowns from the first constraint from all other constraints). Keep repeating this procedure until no constraints remain. Now group the constraints in sets $C[i]$, $i = 1, \ldots, k$, such that every constraint in $C[i]$ requires the same set of new unknowns to be validated. For each set $C[i]$ construct a set $V[i] = \{V[i,1], V[i,2], \ldots, V[i,l_i]\}$ that contains the $l_i$ unknowns in $C[i]$, given that all the unknowns in $V[i-1], \ldots, V[1], \varnothing$ are known. The following code then finds all fusion rings of rank $r$, multiplicity $m$, and number of self-dual particles $s$:

```
for V[1,1] in 0:m, ..., V[1,l_1] in 0:m
  if( all constr in C[1] are verified )
    for V[2,1] in 0:m, ..., V[2,l_2] in 0:m
      if( all constr in C[2] are verified )
        ...
          for V[k,1] in 0:m, ..., V[k,l_k] in 0:m
            if( constraints in C[k] are verified )
              saveSol({ V[1,1], V[1,2], ..., V[k,l_k] })
```

A few remarks are in place.

- Because of the nature of the above algorithm, different code must be created for each rank, multiplicity, and number of self-dual particles separately. To accommodate this need, we used the Wolfram Language to generate the polynomial constraints, reduce the number of variables, break the symmetry, sort the equations and unknowns, and create, compile, and execute the corresponding C source code. The link to the Wolfram Language code we used to find the fusion rings can be found as an attachment to the paper [102].

- It is important to note that adding other specific constraints on the fusion rings is very easy. Since the constraints are sorted purely based on the number of variables, any constraint on the structure constants can easily be added and sorted with the rest. One could, e.g., add constraints on the maximum number of non-zero structure constants per particle or put a bound on several structure constants.

- Just like [58] pointed out, it is very hard to benchmark the code since the performance crucially depends on how efficiently cache memory is used. Sometimes the CPU decides to swap between cache and RAM, and computation times get multiplied by a factor that can go up to 100 or even higher.

### 3.1.4 Results

Using the algorithm described in Subsection 3.1.3, 28451 fusion rings were found, of which 353 are multiplicity-free, and 118 are non-commutative. These results were obtained without the use of a high-end machine. A summary of the number of rings per rank and multiplicity is given in table 3.2. A more detailed overview of fusion rings with $r \geq 6$ is shown in table 3.3.

|   | Rank |   |   |   |   |   |   |   |   |
|---|---|---|---|---|---|---|---|---|---|
| Multiplicity | | 1 | 2 | 3 | 4 | 5 | 6 | 7 | 8 | 9 |
| | 1 | 1 | 2 | 4 | 10 | 16 | 39 | 43 | 96 | 142 |
| | 2 | 0 | 1 | 3 | 17 | 37 | 154 | 319 | 874+ | |
| | 3 | 0 | 1 | 4 | 24 | 82 | 384 | 562+ | | |
| | 4 | 0 | 1 | 6 | 45 | 134 | 872 | 1236+ | | |
| | 5 | 0 | 1 | 5 | 55 | 209 | 533+ | | | |
| | 6 | 0 | 1 | 9 | 81 | 336 | 872+ | | | |
| | 7 | 0 | 1 | 6 | 92 | 477 | 976+ | | | |
| | 8 | 0 | 1 | 10 | 137 | 733 | 1672+ | | | |
| | 9 | 0 | 1 | 12 | 151 | 1463 | | | | |
| | 10 | 0 | 1 | 9 | 186 | 1794 | | | | |
| | 11 | 0 | 1 | 10 | 238 | 2283 | | | | |
| | 12 | 0 | 1 | 20 | 291 | 3049 | | | | |
| | 13 | 0 | 1 | 9 | 246 | 1300+ | | | | |
| | 14 | 0 | 1 | 13 | 340 | 1323+ | | | | |
| | 15 | 0 | 1 | 16 | 349 | 1550+ | | | | |
| | 16 | 0 | 1 | 25 | 525 | 1925+ | | | | |

Table 3.2: Table of total number of fusion rings per rank and multiplicity. The grey numbers with a + indicate partial results from an incomplete search.

A complete list of all the fusion rings with multiplication tables and other properties can be found as a part of the Anyonica package. A list fusion matrices, "FusionRing-MultiplicationTables", can be also be found as an attachment to the paper [102]. The number of fusion rings per rank and multiplicity and least-squares fits are shown in figure 3.2.

*Remark* 3.1.1. It seems that the number of fusion rings for a given multiplicity $m$ can be approximated well by a function of the form $a_m \times (m+1)^{br}$, where $a_m$ is a constant and $b \approx 1$. Whether the number of fusion rings with multiplicity $m$ grows asymptotically as $(m+1)^r$ for all $m$ (or as $m$ becomes larger as well) is unknown to the authors.

|   | $6^0$ | $6^2$ | $6^4$ | $7^0$ | $7^2$ | $7^4$ | $7^6$ | $8^0$ | $8^2$ | $8^4$ | $8^6$ | $9^0$ | $9^2$ | $9^4$ | $9^6$ | $9^8$ |
|---|---|---|---|---|---|---|---|---|---|---|---|---|---|---|---|---|
| 1 | 20 | 9, 2 | 8 | 18 | 14, 3 | 7 | 1 | 38 | 17, 13 | 3, 15 | 7, 3 | 46 | 34, 11 | 12, 21 | 13, 3 | 2 |
| 2 | 13 | 37, 2 | 102 | 2 | 32 | 86, 5 | 194 | | | | | | | | | |
| 3 | 16 | 81, 1 | 286 | | | | | | | | | | | | | |
| 4 | 17 | 151, 1 | 703 | | | | | | | | | | | | | |

Table 3.3: Table of fusion rings per multiplicity (rows) and rank (columns), where the rank is subdivided into sets $r^i$ with $i$ the number of non-self-dual particles. Numbers separated by a comma, $a, b$, indicate that there are $a$ commutative rings and $b$ non-commutative rings.

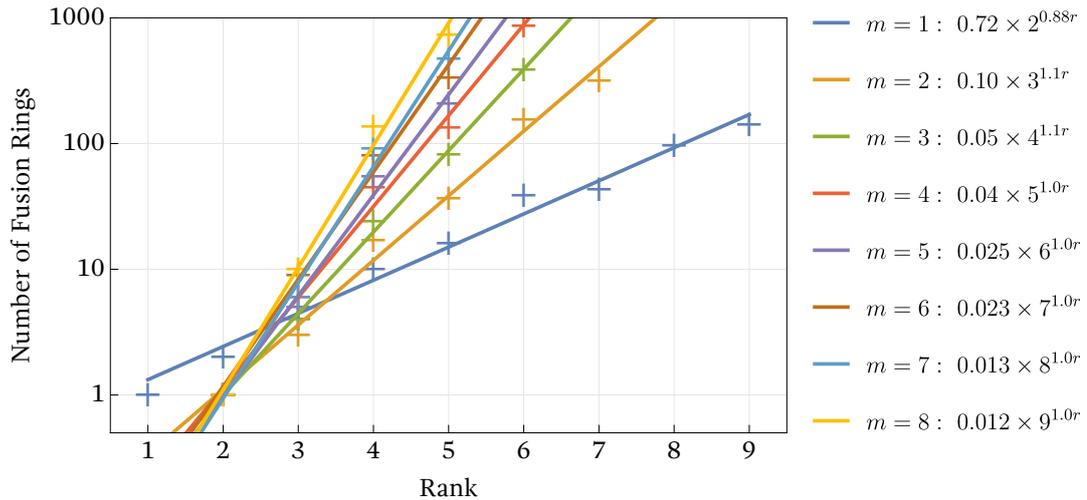

Figure 3.2: The number of fusion rings per rank. Each line represents a least squares fit of the function $a(m+1)^{br}$.

## 3.2 Naming of Fusion Rings

Many fusion rings are still unnamed, which is inconvenient for referencing specific rings. To resolve this issue, we constructed a naming scheme that uniquely characterises a fusion ring using 3 to 4 natural numbers. Every fusion ring is denoted by a formal name $\text{FR}_i^{r,m,n}$ where

- $r$ denotes the rank of the fusion ring,
- $m$ equals the multiplicity of the fusion ring, i.e. the largest structure constant, and
- $n$ denotes the number of particles that are not self-dual.

Every triple $(r, m, n)$ determines a finite list of fusion rings that can be sorted canonically. The number $i$ then denotes the position of the fusion ring in such a list. The canonical order we implemented is based on a unique number assigned to each fusion ring that can be computed as follows. First, we sort the fusion rings by multiplicity, rank, number of non-self-dual particles, and number of non-zero structure constants. If there is any remaining ambiguity, we sort further by calculating a unique number for each fusion ring as follows.

1. First, the elements of the rings themselves are sorted. All self-dual particles are grouped, and the non-self-dual particles are grouped in dual pairs. The self-dual particles appear before the pairs and are sorted by increasing quantum dimension. The pairs are sorted by increasing maximum quantum dimension.

2. Let $N_{a,b}^c$ denote the structure constants of a multiplication table of a fusion ring. For each permutation of the elements of the ring (where the identity is kept fixed, and the previous ordering is kept fixed), a list of digits can be created by ordering the structure constants $N_{a,b}^c$ lexicographically on $a, b, c$:

$$N_{1,1}^1, N_{1,1}^2, ..., N_{1,2}^1, N_{1,2}^2, ..., N_{r,r}^r.$$

3. By regarding each list as a number, in base $m + 1$, with digits given by the list elements, a unique number is assigned to each permutation of the same ring. Moreover, these numbers are unique for each different fusion ring.

4. Taking the maximum value of the numbers generated per ring, a unique number is assigned to each fusion ring.

If $m = 1$, we omit this value and just write $\mathrm{FR}_i^{r,n}$. The nice thing about formal names is that it allows one to easily look up a fusion ring. For example the list of small mulitplicity-free fusion rings on the anyonwiki uses these names and the Anyonica package has a function `FusionRingByCode` which takes a list of four numbers and returns the respective fusion ring.

## 3.3 Characters and Modular Data of Commutative Fusion Rings

### 3.3.1 Characters of Commutative Fusion Rings

Finding a character table of commutative fusion rings comes down to finding a matrix $V$ that simultaneously diagonalises the fusion matrices $[N_k]_i^j$ [39]. A simple way to do this is by taking a linear combination $M = \sum_{k=1}^r c_k [N_k]$, where the $c_k$ are random real numbers in an interval $[a, b]$, and finding the eigenspace of $M$. There are values of $c_k$ for which this method provides an incorrect answer, but it is always easy to test whether this is the case by performing the diagonalisation. When a wrong result is returned, a new set of $c_k$ can be chosen at random until a set that works is found. The following lemma shows that the set of vectors $\vec{c}$ for which the procedure above fails is a subset of a strict sub-vector space of $\mathbb{C}^r$.

**Proposition 3.3.1.** *Let $\{M_k\}$ be a set of simultaneously diagonalisable $n \times n$-matrices. Let $c \in \mathbb{C}^n$ be a vector for which there exists a matrix $V$ that diagonalises $M := \sum_{k=1}^n c_k M_k$ but does not diagonalise each $M_k$ individually. The set of vectors $\{c\}$ for which this property holds is a subset of a strict subspace of $\mathbb{C}^n$.*

*Proof.* First we note that for any $c \in \mathbb{C}^n$ there always exists a matrix $V$ that diagonalises all $M_k$ and therefore $M$ as well. Now assume there exists a $V$ that diagonalises $M$ but not every $M_k$. We then have that

$$[V^{-1}MV]_j^i = \sum_{k=1}^n c_k [V^{-1}M_k V]_j^i = 0, \text{ for } i \neq j. \tag{3.7}$$

Since not all $[V^{-1}M_k V]^i_j$ are 0 (3.7) is a system of linear equations in the variables $c_k$ with at least 1 non-trivial equation. ∎

Whether the character table contains

1. symbolic expressions, or

2. approximate floating point numbers

depends on the technique for finding eigenspaces. For the first case, we found exact character tables for 28227 out of the 28333 commutative rings. In the second case, we found character tables for all commutative fusion rings expressing each character with 99 significant digits. In both cases, we checked diagonalisation by performing matrix multiplications such that the matrix elements in the final results were correct up to 99 significant digits.

### 3.3.2  Modular Data

Not every fusion ring can be categorified into a modular fusion category (MFC). If they can, then some of the structure of their categories can be extracted without the need for categorification. Since categorification of a fusion ring can be quite hard to do, any information that can be extracted beforehand is more than welcome. In particular, any MFC has an associated $S$ and $T$-matrix that define a representation of the modular group. The possible (normalized) $S$ and $T$-matrices belonging to an MFC can be constructed using only information from the fusion ring. In particular, if no such $S$ or $T$-matrices exist, we know there can be no MFC associated with the fusion ring. Note that the $S$ and $T$-matrices found this way do not necessarily appear as the modular data of some category. There exists at least one modular fusion ring ($FR_1^{7,1,2}1$) that has no categorifications. Whether each categorifiable modular fusion ring has at least one modular fusion category is not known to us. The code we used to find the possible $S$-matrices and $T$-matrices can be found in the file "CharactersAndModularData.wl", attached to the paper [102].

### 3.3.3  Finding $S$-matrices

An $S$-matrix associated with a fusion ring $\mathcal{R}$ is a square, symmetric, invertible matrix that diagonalises the set of matrices $\{[N_i]\}_{i=1}^r$, and satisfies

$$\left[S^2\right]_{ij} = N^1_{i,j} \tag{3.8}$$

$$\left[S^2\right]_{11} = 1 \tag{3.9}$$

Once the character table of a commutative fusion ring is found, it is easy to construct $S$-matrices belonging to the ring. Indeed: we know that the rows of the character table are the simultaneous eigenvectors of the fusion matrices $[N_i]$. Likewise, the rows and columns of the S-matrix consist of simultaneous eigenvectors of the $[N_i]$. By permuting the rows of the character table and rescaling them such that $S_{1i} = S_{i1}$, one can find a set of possible S-matrices starting from the character table.

### 3.3.4 Finding T-Matrices

Recall that a T-matrix belonging to a fusion ring $\mathscr{R}$ is a square diagonal matrix for which there exists a $\lambda \in \mathbb{C}\backslash\{0\}$ such that

$$(ST)^3 = \lambda S^2, \qquad (3.10)$$

i.e. the S- and T matrices form a projective representation of the modular group. Instead of solving equations (3.10) directly we can use a theorem from Vafa [99]:

**Theorem 3.3.2.** *Let T be a T-matrix, with diagonal entries $\theta_i$, belonging to a fusion ring $\mathscr{R}$ then the following (Vafa) equations hold*

$$\theta_1 = 1 \qquad (3.11)$$

$$\left(\theta_i \theta_j \theta_k \theta_l\right)^{\sum_{n=1}^{r} N_{i,j}^{n^*} N_{k,l}^{n}} = \prod_{n=1}^{r} \theta_n^{N_{i,j}^{n} N_{l,n}^{k^*} + N_{j,k}^{n} N_{l,n}^{i^*} + N_{i,k}^{n} N_{l,n}^{j^*}}, \qquad (3.12)$$

where $i, j, k, l = 1, \ldots, r$. To find admissible T-matrices, one can first solve the Vafa equations and then check whether the modularity constraint (3.10) holds. To solve the Vafa equations, one can take a logarithm of both sides of the equations to obtain

$$\sum_{n=1}^{r} N_{i,j}^{n^*} N_{k,l}^{n} t_i + N_{i,j}^{n^*} N_{k,l}^{n} t_j + N_{i,j}^{n^*} N_{k,l}^{n} t_k + N_{i,j}^{n^*} N_{k,l}^{n} t_l \qquad (3.13)$$

$$- \left(N_{i,j}^{n} N_{l,n}^{k^*} + N_{j,k}^{n} N_{l,n}^{i^*} + N_{i,k}^{n} N_{l,n}^{j^*}\right) t_n \in \mathbb{Z} \qquad (3.14)$$

where $t_i := \ln(\theta_i)/(2\pi i)$. This is a system of linear equations with integer coefficients, which can, e.g. be solved using a Smith decomposition.

From the collection of commutative fusion rings we obtained, 54 have matching S- and T-matrices.

## 3.4 Some Comments on Non-commutative Fusion Rings

Out of the 28451 fusion rings we have found, 118 are non-commutative. Apart from 4 exceptions (see 3.4.3), all non-commutative fusion rings contain a non-trivial subgroup.

### 3.4.1 Song Extensions of Groups

Fusion rings that contain a subgroup are called generalised near-group fusion rings [95]. The most notable of such fusion rings are the Tambara-Yamagami fusion rings [94] and the Haagerup-Izumi fusion rings of groups. The structure of both of these rings can be generalised as follows.

**Definition 3.4.1.** Let $G$ be a finite group, $T$ a finite set, and $\sigma_l : G \times T \to T : (g, t) \mapsto \sigma_l(g, t) =: g \cdot t$ a left action of $G$ on $T$ such that $T = G \cdot t_1$ for some $t_1 \in T$ and the left stabilizer of $t_1$, $G_{t_1}^l$, obeys $G_{t_1}^l = H \trianglelefteq G$. Let $\tilde{g} \in G$, $n \in \mathbb{N}$ and

- $A : G/H \to G/H$ be an automorphism such that

$$A^2([g]) = [\tilde{g}^{-1} g \tilde{g}], \ \forall g \in G, \text{ and} \quad (3.15)$$
$$A([\tilde{g}]) = [\tilde{g}], \quad (3.16)$$

  where $[\cdot]$ denotes the canonical projection from $G$ to $G/H$,

- $\Phi : T \to G/H$ be such that $\Phi(g \cdot t_1) = [g]$, i.e. $\Phi$ maps a $t \in T$ to the class in $G/H$ that maps $t_1$ to $t$, and

- $\lambda : G/H \to G$ be a lift of the elements of $G/H$ into $G$, i.e. $[\lambda(gH)] = g, \forall g \in G$.

The set $G \sqcup T$ with the following product

$$g \times g' = gg', \quad (3.17)$$
$$g \times t = g \cdot t \quad (3.18)$$
$$t \times g = \lambda(\Phi(t) A([g])) \cdot t_1, \quad (3.19)$$
$$t \times t' = \lambda(\Phi(t) A(\Phi(t'))) \tilde{g}^{-1} \sum_{h \in H} h + n \sum_{t \in T} t, \quad (3.20)$$

$\forall g, g' \in G, \forall t, t' \in T$, is called the $n$'th **single orbit normal group** (or **song**) extension of $G$ featuring $H$, $A$, and $\tilde{g}$ and we denote it by $[H \trianglelefteq G]_{\tilde{g}|n}^A$.

The proof that songs are well-defined and the rings they produce are fusion rings can be found in the paper [102]. We have the following

*Examples* 3.4.2.  1. Let $T = \{t\}$, $\tilde{g} = 1$, and $n \in \mathbb{N}$. Then $H = G$, $A$ is trivial, and the fusion rules become

$$g_i \times g_j = g_i g_j, \quad (3.21)$$
$$g_i \times t = t \times g_i = t, \quad (3.22)$$
$$t \times t = \sum_{g \in G} g + nt. \quad (3.23)$$

This ring is called a near-group fusion ring, and in particular, when $n = 0$, this ring is called the Tambara-Yamagami fusion ring of the group $G$: TY($G$). Such a ring is non-commutative iff the the group $G$ is non-commutative. Corollary 3.4.4 tells us that these songs capture all extensions of any group by one particle.

2. Let $G = \{g_1, \ldots, g_n\}$ be a commutative group, $T = \{t_1, g_2 \cdot t_1, \ldots, g_n \cdot t_1\}$ (and

thus $H = \{1\}$), $\tilde{g} = 1$, and $A : g \mapsto g^{-1}$ then the fusion rules become

$$g_i \times g_j = g_i g_j, \tag{3.24}$$
$$g_i \times t_j = (g_i g_j) t_1, \tag{3.25}$$
$$t_i \times g_j = (g_i g_j^{-1}) t_1, \tag{3.26}$$
$$t_i \times t_j = g_i g_j^{-1} g_0^{-1} + n \sum_{t \in T} t. \tag{3.27}$$

If $n = 1$, this ring is called the Haagerup-Izumi fusion ring of the commutative group $G$: HI($G$). Such a ring is non-commutative iff the group $G$ (seen as a group fusion ring) contains non-self-dual particles.

3. Let $G = D_3$ be the dihedral group with 6 elements, $T = \{t_1, t_2\}$, $H = \mathbb{Z}_3$, $\tilde{g} = 1$, and $A$ trivial. The song $[\mathbb{Z}_3 \trianglelefteq D_3]_{1|0}^{\text{Id}}$ (= $\text{FR}_2^{8,1,2}$) is a non-commutative categorifiable fusion ring that is not of the type TY($G$) or HI($G$) for any group $G$. This ring has categorifications since it is the Grothendieck ring of the crossed product category $\mathscr{C}_{\text{TY}(\mathbb{Z}_3)} \rtimes \mathbb{Z}_2$. [29]

4. Let $\alpha : \mathbb{Z}_3 \to \mathbb{Z}_3 : g \mapsto g^{-1}$. The song $[\mathbb{Z}_2 \trianglelefteq \mathbb{Z}_6]_{1|0}^{\alpha}$ (= $\text{FR}_3^{9,1,4}$) also has categorifications but is neither of the types TY($G$) or HI($G$) for any group $G$, nor is it the Grothendieck ring of a crossed product category. The links to the categorifications for this and the previous example can be found as attachments to the paper [102].

### 3.4.2 Generic Non-commutative Fusion Rings

Not all non-commutative fusion rings are of this type, however. Table 3.4 gives an overview of the multiplicity-free fusion rings per subgroup and rank.

To get some insight into the properties of generic non-commutative rings and understand the structure of table 3.4, it is interesting to look at the consequences of having a subgroup structure.

Let $G = \langle g_1 = 1, \ldots, g_n \rangle$ be a group fusion ring and consider extending it by adding elements from a finite set, say $t_1, \ldots, t_m \in T$. Now consider a product, say $\times$, on $G \sqcup T$ that coincides with the product on $G$, i.e. $g \times h = gh, \forall g, h \in G$. Since $G$ is a group, and a fusion ring is associative and has a unique unit, the left and right action $\sigma_g^l : t \mapsto g \times t$ and $\sigma_g^r : t \mapsto t \times g$ of $G$ on $T$ must permute the elements of $T$. Requiring that the ring $G_T := \langle g_1 = 1, \ldots, g_n, t_1, \ldots t_m \rangle$ with product $\times$ is a fusion ring puts further restrictions on $\times$:

**Proposition 3.4.3.** *Let $G$ and $T$ be as above and write $G_t^l$ ($G_t^r$) for the left (right) stabilizer of $t$ under $G$, $[t]^l$ ($[t]^r$) for the left (right) orbit of $t$ under $G$ and $H_i^l$ ($H_i^r$) for the elements of $G/G_t^l$ ($G/G_t^r$) (with $H_1^l = G_t^l$ and $H_1^r = G_t^r$) then for all $t \in T, g \in G$*

1.
$$G_t^l = G_{t^*}^r, \tag{3.28}$$

|  | 6 | 7 | 8 | 9 |
|---|---|---|---|---|
| $\mathbb{Z}_2$ |  |  | $FR_{11}^{8,1,2}$, $FR_{29}^{8,1,2}$, $FR_{30}^{8,1,2}$ | $FR_{32}^{9,1,2}$, $FR_{38}^{9,1,2}$, $FR_{41}^{9,1,2}$, $FR_{44}^{9,1,2}$, $FR_{20}^{9,1,4}$, $FR_{31}^{9,1,4}$, $FR_{33}^{9,1,4}$ |
| $\mathbb{Z}_3$ | $HI(\mathbb{Z}_3)$ | $FR_{15}^{7,1,2}$ |  | $FR_{27}^{9,1,4}$ |
| $\mathbb{Z}_4$ |  |  | $FR_8^{8,1,2}$, $HI(\mathbb{Z}_4)$, $FR_8^{8,1,4}$, $FR_7^{8,1,6}$, $[I \trianglelefteq \mathbb{Z}_4]_{2;1}^{\alpha}$ | $FR_{36}^{9,1,2}$, $FR_9^{9,1,4}$, $FR_{17}^{9,1,4}$, $FR_{13}^{9,1,6}$ |
| $\mathbb{Z}_2 \times \mathbb{Z}_2$ |  |  | $FR_6^{8,1,2}$, $FR_9^{8,1,2}$, $FR_{10}^{8,1,2}$, $[I \trianglelefteq \mathbb{Z}_2 \times \mathbb{Z}_2]_{1;1}^{\alpha}$ | $FR_7^{9,1,2}$, $FR_{22}^{9,1,2}$, $FR_{37}^{9,1,2}$, $FR_4^{9,1,4}$, $FR_8^{9,1,4}$, $FR_{22}^{9,1,4}$ |
| $\mathbb{Z}_6$ |  |  |  | $[\mathbb{Z}_2 \trianglelefteq \mathbb{Z}_6]_{1;0}^{\alpha}$, $[\mathbb{Z}_2 \trianglelefteq \mathbb{Z}_6]_{1;1}^{\alpha}$ |
| $D_3$ | $D_3$ | $TY(D_3)$, $[D_3 \trianglelefteq D_3]_{1;1}^{Id}$ | $[\mathbb{Z}_3 \trianglelefteq D_3]_{1;0}^{Id}$, $[\mathbb{Z}_3 \trianglelefteq D_3]_{2;1}^{Id}$, $[\mathbb{Z}_3 \trianglelefteq D_3]_{1;1}^{Id}$, $[\mathbb{Z}_3 \trianglelefteq D_3]_{2;0}^{Id}$, $FR_3^{8,1,2}$ | $FR_4^{9,1,2}$, $FR_6^{9,1,4}$ |
| $D_4$ |  |  | $D_4$ | $TY(D_4)$, $[D_4 \trianglelefteq D_4]_{1;1}^{Id}$ |
| $Q$ |  |  | $Q$ | $TY(Q)$, $[Q \trianglelefteq Q]_{1;1}^{Id}$ |

Table 3.4: Non-commutative multiplicity-free fusion rings per non-trivial largest subgroup (rows) and rank (columns). Here $I$ stands for the trivial group, $HI(G)$ for the Haagerup-Izumi fusion ring of $G$ and $TY(G)$ for the Tambara-Yamagami fusion ring of $G$. $\alpha$ denotes the, up to isomorphism, only non-trivial group automorphism, **1** the unit element, and **2** the, up to isomorphism, only non-trivial group element. For the meaning of the names $FR_i^{r,m,n}$ see Section 3.2

2. for all $a, b \in G \sqcup T : N^g_{a,b} \in \{0, 1\}$ and

$$N^g_{t,t^*} = 1 \Leftrightarrow g \in G^l_t, \text{ and} \qquad (3.29)$$
$$N^g_{t^*,t} = 1 \Leftrightarrow g \in G^r_t \qquad (3.30)$$

3. every $\tau \in [t]^l$ can be labelled by a unique $H^l_i$ (say $\tau_{H^l_i} \equiv \tau_i$) in such a way that for all $g$ in $G$

$$N^g_{\tau_i,t^*} = 1 \Leftrightarrow g \in H^l_i, \qquad (3.31)$$

4. if $|T| < |G|$ then there exists a $t \in T$ with non-trivial $G^l_t$ such that $\left|[t]^l\right|$ divides $|G|$,

5. if $|T| = |G|$ and there exists a $t \in T$ with trivial $G^l_t$, then $[t]^l = T$ and $G^l_t$ is trivial for all $t \in T$, and

6. if $|T| > |G|$ then at least $\mathrm{mod}(|G|, |T|)$ elements of $T$ have a non-trivial left stabilizer.

*Proof.* Let $e$ denote the unit of $G$.

1. For all $h$ in $G^l_t$ : $e \in t^* \times t = t^* \times (h \times t) = (t^* \times h) \times t$ but every element of a fusion ring has a unique dual so $t^* \times h = t^*$ for all $h$ in $G^l_t$.

2. Since $e \in t \times t^*$, $g \in g \times t \times t^*$ so $g \in G^l_t \Rightarrow g \in t \times t^*$. Now assume that $g \notin G^l_t$ and $g \in t \times t^*$. Then $e \in g^{-1} \times t \times t^*$ so $g^{-1} \times t = (t^*)^* = t$ which is impossible by assumption. Similar reasoning can be applied to prove the second formula.

3. For all $h_i$ in $H^l_i$: $\tau_i \times t^* = h_i \times t \times t^* \ni h_i$

The last three statements follow directly from the orbit stabiliser theorem. ∎

The following corollary then classifies all 1-particle extensions of groups to fusion rings.

**Corollary 3.4.4.** *Let $G$ and $T$ be as above. If $|T| = 1$ then for each $k \in \mathbb{N}$ there exists only one extension $\mathscr{R}$ of $G$ by $T$. Moreover $\mathscr{R}$ is commutative if and only if $G$ is commutative and for $t \in T$ and all $g \in G$ (seen as elements of $\mathscr{R}$)*

$$g \times t = t \times g = t, \qquad (3.32)$$
$$t \times t = \sum_{g \in G} g + k\, t. \qquad (3.33)$$

*Proof.* Since $G$ stabilizes $t$ and $t$ is self-dual it follows from (3.30) that $N^g_{t,t} = N^g_{t,t^*} = 1$ for all $g$ in $G$. The only remaining degree of freedom is the value of $N^t_{t,t} \in \mathbb{N}$, which can be chosen freely as verified by checking the associativity constraints. ∎

For $|T| = 2$ commutativity of an extension also depends solely on the commutativity of the group, and all extensions can be classified as well.

**Proposition 3.4.5.** *If $\mathcal{R}$ is an extension of a non-trivial group $G = \langle g_1, \ldots, g_n \rangle$ by $T = \{t_1, t_2\}$ then we have for all $a, b, c, d \in \mathbb{N}$ that*

1. *$\mathcal{R}$ is a commutative fusion ring iff $G$ is a commutative group,*

2. *if $G = G_t^l$ $\forall t \in T$ and $t_i = t_i^*$*

$$t_1 \times t_1 = \sum_{g \in G} g + at_1 + bt_2, \tag{3.34}$$

$$t_2 \times t_2 = \sum_{g \in G} g + ct_1 + dt_2, \tag{3.35}$$

$$t_1 \times t_2 = t_2 \times t_1 = bt_1 + ct_2, \tag{3.36}$$

   *where: $b(b - d) + c(c - a) = |G|$,*

3. *if $G = G_t^l$ $\forall t \in T$ and $t_i \neq t_i^*$*

$$t_1 \times t_1 = at_1 + bt_2, \tag{3.37}$$

$$t_2 \times t_2 = bt_1 + at_2, \tag{3.38}$$

$$t_1 \times t_2 = t_2 \times t_1 = \sum_{g \in G} g + at_1 + at_2, \tag{3.39}$$

   *where: $b^2 - a^2 = |G|$,*

4. *if $G \neq G_t^l$ $\forall t \in T$ and $t_i = t_i^*$*

$$t_1 \times t_1 = t_2 \times t_2 = \sum_{g \in G_t^l} g + at_1 + bt_2, \tag{3.40}$$

$$t_1 \times t_2 = t_2 \times t_1 = \sum_{g \notin G_t^l} g + bt_1 + at_2. \tag{3.41}$$

5. *if $G \neq G_t^l$ $\forall t \in T$ and $t_i \neq t_i^*$*

$$t_1 \times t_1 = t_2 \times t_2 = \sum_{g \notin G_t^l} g + at_1 + at_2, \tag{3.42}$$

$$t_1 \times t_2 = t_2 \times t_1 = \sum_{g \in G_t^l} g + at_1 + at_2. \tag{3.43}$$

6. *if $|G| > 2$ there are no multiplicity-free extensions $\mathcal{R}$ of $G$ by $T$ where $G$ stabilizes every element of $T$,*

7. *if $\mathcal{R}$ is multiplicity-free then $|G|$ is even.*

*Proof.* (1) By setting $a = c^*$ in the pivotal relation $N_{a,b}^c = N_{b,c^*}^{a^*}$ we obtain $N_{a,b}^{a^*} = N_{b,a}^{a^*}$. For both the case where $t_1 = t_1^*$ and the case where $t_1 = t_2^*$ this results in the relations $N_{t_1,t_2}^{t_1} = N_{t_2,t_1}^{t_1}$ and $N_{t_1,t_2}^{t_2} = N_{t_2,t_1}^{t_2}$. Therefore $t_i \times t_j = t_j \times t_i$, $i = 1, 2$. We also have that $G_t^l = G_t^r$, $\forall t \in T$ (and therefore $t_i \times r = r \times t_i \forall r \in \mathcal{R}$). Indeed: for the case where $t_i = t_i^*$ this follows from equation (3.28). If $t_i \neq t_i^*$ and there would exist a $t \in T, g \in G$ with $g \in G_t^l$ but $g \notin G_t^r$ then $t_1 \times t_1 = t_1 \times g \times t_1 = t_2 \times t_1 \Rightarrow t_1 = t_1^*$, a contradiction.

(2)…(5) Follows directly from applying equations (3.30) and (3.31) in combination with the pivotal relations (3.2), and the associativity constraints.

(6) If $G$ stabilizes every element of $T$ then either $|G| = b(b-d) + c(c-a) \leq 2$ or $|G| = b^2 - a^2 \leq 1$.

(7) If $G$ stabilizes every element of $T$ then, since $G$ is non-trivial, statement nr. 6 implies that $|G| = 2$. If $G$ does not stabilise every element of $T$, then for any $t \in T$, $|[t]| = |T|$ and the result follows directly from statement nr. 4 of proposition 3.4.3.

∎

A lot can be said about fusion rings with subgroups by looking at the cardinality of the stabilisers. Indeed, in proposition 3.4.5, we found several relations between the size of stabiliser subgroups (in that case, the whole group) and the fusion coefficients. These are useful tools for finding obstructions to extend groups into fusion rings. The following lemma and its corollaries will prove useful to exclude extensions of groups to fusion rings for more general $T$.

**Lemma 3.4.6.** *Let $\tau \in T$, then for all $a \in \mathcal{R}$*

$$\left|G_\tau^l \cap G_a^l\right| = \sum_{g \in G} (N_{\tau,a}^g)^2 + \sum_{t \in T} \left(N_{\tau,a}^t\right)^2 - \sum_{t \in T} N_{\tau,t}^\tau N_{t,a}^a. \tag{3.44}$$

*Proof.* For any particle $\tau \in T$ and $a \in \mathcal{R}$ we have that

$$\tau^* \times (\tau \times a) = \sum_{e \in \mathcal{R}} \sum_{d \in \mathcal{R}} N_{\tau,a}^d N_{\tau^*,d}^e e, \tag{3.45}$$

while

$$(\tau^* \times \tau) \times a = \sum_{g \in G_\tau^l} g \times a + \sum_{t \in T} N_{\tau^*,\tau}^t t \times a \tag{3.46}$$

$$= \sum_{g \in G_\tau^l \cap G_a^l} a + \sum_{g \in G_\tau^l \setminus G_a^l} g \times a + \sum_{t \in T} N_{\tau^*,\tau}^t t \times a \tag{3.47}$$

$$= \left|G_\tau^l \cap G_a^l\right| a + \sum_{g \in G_\tau^l \setminus G_a^l} g \times a + \sum_{e \in \mathcal{R}} \sum_{t \in T} N_{\tau^*,\tau}^t N_{t,a}^e e \tag{3.48}$$

By looking at the $a$'th component and using the pivotal relation $N_{a,b}^c = N_{a^*,c}^b$ we find that

$$\left|G_\tau^l \cap G_a^l\right| = \sum_{g \in G} (N_{\tau,a}^g)^2 + \sum_{t \in T} \left(N_{\tau,a}^t\right)^2 - \sum_{t \in T} N_{\tau,t}^\tau N_{t,a}^a. \tag{3.49}$$

∎

This bound becomes stronger when one or multiple particles are fixed by the group $G$.

**Corollary 3.4.7.** *If $|T| \geq 2$ and $\exists \tau \in T$ for which $G_\tau^l = G$ then for any $a \in T$ with $a \neq \tau^*$*

$$\left|G_a^l\right| = \sum_{t \in T} \left(N_{\tau,a}^t\right)^2 - \sum_{t \in T} N_{\tau,t}^\tau N_{t,a}^a \leq |T| m^2, \tag{3.50}$$

*where m is the multiplicity of the fusion ring.*

We can now generalize proposition 3.4.5(6):

**Proposition 3.4.8.** *If $2 \leq |T| < |G|$ and $G_t^l = G$ for all $t \in T$ then there exists no multiplicity-free extension $\mathcal{R}$ of G by T. In particular, if $|G|$ is prime and $2 \leq |T| < |G|$, there exists no multiplicity-free-extensions of G by T.*

We are now in a position where several patterns in table 3.4 can easily be explained. From corollary 3.4.4 we conclude that the only groups with rank less than 9 and non-commutative 1-particle extensions are $D_3$, $D_4$ and $Q_8$. For each group, there are 2 such extensions. Corollary 3.4.4 also explains the absence of the table's 3 commutative groups of order 8. Proposition 3.4.8 implies that both $\mathbb{Z}_5$ and $\mathbb{Z}_7$ have no multiplicity-free non-commutative extensions with rank less than 10 hence their absence in the table.

### 3.4.3 Non-Commutative Fusion Rings Without Non-Trivial Subgroup

Only 4 non-commutative rings were found that contain no non-trivial subgroup. Three of these ($FR_{312}^{6,6,2}$, $FR_{115}^{6,7,2}$, and $FR_{48}^{6,8,2}$) are simple, i.e. contain no subring at all. The sole non-commutative non-simple fusion ring without a proper subgroup has the following multiplication table

| 1 | 2     | 3     | 4                 | 5         | 6         |
|---|-------|-------|-------------------|-----------|-----------|
| 2 | 1 + 2 | 6     | 4 + 5             | 4         | 3 + 6     |
| 3 | 5     | 1 + 3 | 4 + 6             | 2 + 5     | 4         |
| 4 | 4 + 6 | 4 + 5 | 1 + 2 + 3 + 2 4 + 5 + 6 | 3 + 4 + 5 + 6 | 2 + 4 + 5 + 6 |
| 5 | 3 + 5 | 4     | 2 + 4 + 5 + 6     | 4 + 6     | 1 + 3 + 4 |
| 6 | 4     | 2 + 6 | 3 + 4 + 5 + 6     | 1 + 2 + 4 | 4 + 5     |

The elements **2** and **3** generate Fibonacci subrings. Elements $5 = 3 \times 2$ and $6 = 2 \times 3$ can be regarded as couples of Fibonacci particles and $4 = 2 \times 3 \times 2 = 3 \times 2 \times 3$ as a triple of Fibonacci particles. Indeed, the above fusion ring is completely determined by the generators **1**, **2** and **3** together with the relations $2^2 = 1 + 2, 3^2 = 1 + 3$, and $2 \times 3 \times 2 = 3 \times 2 \times 3$. The structure of the ring thus corresponds to that of a Hecke algebra with generators **2** and **3**.

# Chapter 4

# Categorifying Fusion Rings: Solving Consistency Equations

Now that we have an extensive list of multiplicity-free fusion rings, it is time to turn to the quest of categorification and the search for braided and pivotal data. All data we are after are fully determined by finite sets of variables that satisfy polynomial constraints. The workflow per fusion ring is the following

1. First, we set up the pentagon equations and combine these with the constraints that each $F$-matrix must be invertible, i.e., $\left\{ \det([F_d^{abc}]) \neq 0 \,\middle|\, a,b,c,d = 1, \ldots, r \right\}$. The other constraints, (2.12), on the $F$-matrices can be implemented beforehand to reduce the number of variables. The solutions to the consistency relations are sets of $F$-matrices. Next, the sets of $F$-matrices for which a basis exists in which they are unitary are expressed in such a basis. Some of the sets of $F$-symbols might be equivalent due to a gauge transform, possibly in combination with a fusion ring automorphism. By removing redundant solutions, we obtain a final set of inequivalent fusion systems, say $\{\mathscr{F}_i := (\mathbf{L}, *, \mathbf{N}, \mathbf{F}_i)\}_{i=1}^{n_F}$.

2. For each fusion system, the hexagon equations, (2.25) and (2.26), are set up and combined with the constraints that none of the $R$-symbols can be 0. If there are solutions to the hexagon equations, then these are combined with the fusion system into a braided fusion system. Here, we also might have equivalent braided fusion systems, and the redundant ones have to be removed as well. The result is a set of possibly braided fusion systems $\left\{(\mathscr{F}_i, \mathscr{R}_{i,j})\right\}_{j=1}^{n_R(i)}$ where $\mathscr{R}_{i,j}$ might be empty in the case of a non-braided fusion system.

3. For each braided fusion system, the pivotal constraints, (2.24), are solved. The solutions to these equations are combined with the (braided) fusion systems. Here, we also might have equivalent (braided) pivotal fusion systems, and redundant systems have to be removed as well. The result is a set of (braided) pivotal fusion systems $\left\{(\mathscr{F}_i, \mathscr{R}_{i,j}, \mathscr{P}_{i,k})\right\}_{k=1}^{n_P(i)}$ where $\mathscr{R}_j$ might be empty. Although it is a famous open question whether the pivotal equations always have a solution, this is the case for all the rings we investigated.

We will, despite the fact that we also take other constraints into account, from now on refer to step 1 as "solving the pentagon equations", step 2 as "solving the hexagon equations" and step 3 as "solving the pivotal equations".

In order to solve these equations, several techniques must be applied to reduce the systems of consistency equations. This chapter aims to explain the various methods used to reduce and eventually solve these sets of consistency equations. In particular, Section 4.1 explains the various techniques developed and their application to solving pentagon equations. Most, but not all, of the techniques used also find their application in solving the hexagon equations. The minor differences between the strategy used to solve the hexagon equations and that to solve the pentagon equations are explained in Section 4.2. Section 4.3 explains how the pivotal equations were solved. The solutions are discussed in Section 4.4. Section 4.5 discusses the naming scheme we use for fusion categories, and Section 4.6 concisely discusses some predecessors to Anyonica and alternative software.

All categorifiable rings up to rank 6 have already been classified in [63]. By using Anyonica, I believe we classified all pivotal categorifiable multiplicity-free fusion rings of rank 7. If the $\text{PSU}(2)_{12}$ fusion ring only has pivotal solutions, then we classified all of the multiplicity free categorifiable fusion rings up to rank 7. Moreover, all $F$-symbols, $R$-symbols, and pivotal coefficients of the fusion categories have been found and are now available as well. For each ring that could not be categorified, there is a Mathematica notebook that explains why this is the case. The only place where an undetectable error could occur is during the procedure to find zero values of $F$-symbols. Since logging intermediate results for this step causes enormous slowdowns, we did not let the computer do this.

> Note 6. Steps 2 and 3 in the process of finding fusion categories are independent and can be interchanged.

## 4.1 Solving Pentagon Equations

This section explains the procedures for solving the pentagon equations in the order they are applied when one evaluates the function `SolvePentagonEquations` from the Anyonica package. The only procedures not a standard part of the `SolvePentagonEquations` function are checking for obstructions to categorification, trying to fix a unitary gauge, and removing duplicate solutions. This was done manually since the SageMath software package needed to be used for some categories.

### 4.1.1 Obstructions To Categorification

The quickest way to categorify a fusion ring is not categorifying it at all. Before starting the process of categorification, it is interesting to check whether there might be obstructions from categorification. For a given fusion ring, one can sometimes rule out categorification using specific criteria. The criteria used all come from the papers

| Criterion | Abbreviation | Only Commutative | Rules out |
|---|---|---|---|
| Zero-Spectrum | ZS | No | FC |
| Commutative Schur Product | CSP | Yes | UFC |
| $d$-number | DN | Yes | FC |
| Pivotal Drinfeld Center | PDC | Yes | PFC |
| Lagrange | L | No | FC |
| Extended Cyclotomic | EC | Yes | FC |

Table 4.1: The various categorifiability criteria that were used. Here "Only Commutative" means that the criterium only applies to commutative fusion rings. FC, UFC, and PFC respectively stand for fusion category, unitary fusion category, and pivotal fusion category.

[65], [63], and [64]. Table 4.1 gives an overview of the various criteria.

The ZS criterion is implemented in Anyonica as `ZSCriterion` and was able to rule out general categorification for 267 of the 28451 fusion rings.

The CSP criterion is implemented via the function `CSPCValue`, whose output is a floating point number $p$. If $p < 0$, then the fusion ring has no unitary fusion category. We could not implement an exact test since the CSP criterion uses fusion ring characters, and we do not have symbolic versions of all character tables. This is also why a number is returned rather than a boolean: it is up to the user to determine whether $|p|$ is big enough to be convinced that the test result is correct. For an accuracy of 64-digits, the CSP criterion ruled out unitary categorification for 19471 of the 28333 commutative fusion rings.

Some of the more decisive criteria, such as the DN, PDC, L, and EC criteria, require algorithms from number field theory (see [63] for the source code). These algorithms could be highly demanding, to the point of needing more than 128GB of RAM. These were, therefore, only applied to the multiplicity-free fusion rings, some of which were still too hard to crack. The DN, PD, and L criteria are implemented in Anyonica as `DNCriterion`, `PDCCriterion`, and `LCriterion`. Mathematica does not have the required functionality that I[1] needed to implement the EC criterion, so I used the SageMath code from the original paper to calculate its outcome.

*Example* 4.1.1. The following code

```
rings = FusionRingList[[ { 5, 16, 22, 29 } ]];

Print @
TableForm @
Prepend[ { "ZS", "L", "CSP", "PDC", "DN" } ] @
Table[
  {
    ZSCriterion[r], LCriterion[r], CSPCValue[10][r],
    PDCCriterion[r], DNCriterion[r]
```

---
[1]Several functions for working with field extensions are required, and I do not have the skills to implement these myself.

```
  },
  { r, rings }
]
```

returns

```
ZS      L       CSP             PDC     DN
False   False   ∞               False   False
False   True    ∞               True    True
True    True    -0.96226071     True    True
False   True    ∞               True    True
```

So for the first ring, Rep($D_3$), none of the criteria exclude categorification, while for each of the other rings, there is at least one criterion that rules out categorification. The numerical value, different from $\infty$, indicates that the third ring has no unitary categorifications due to the CSP criterion. The other criteria already made the stronger statement that this ring has no categorifications at all, however. As a matter of fact, all multiplicity-free fusion rings up to rank 7 for which the CSP criterion rules out unitary categorification are already ruled out from general categorification due to some other criteria.

In Section 8, the categorifiability data of the multiplicity-free fusion rings up to rank 9 is listed. All rings of rank 8 or greater, except for $FR_4^{8,2}$, $FR_2^{8,4}$, $FR_5^{8,4}$, $FR_2^{9,2}$, and $FR_3^{9,6}$, that are listed as not categorifiable were ruled out because of the DN, PDC, L, or EC criteria. The rings $FR_4^{8,2}$, $FR_2^{8,4}$, $FR_5^{8,4}$, $FR_2^{9,2}$, and $FR_3^{9,6}$ are ruled out from categorification by using Anyonica to reduce the pentagon equations to the point it was clear that no non-singular solution exists. The rings listed as categorifiable, but for which we don't have explicit data, are rings with a special structure. For example fusion rings arising from finite groups, the representation theory of quantum groups at roots of unity, or products of these are automatically categorifiable.

Note that there is one fusion ring, $FR_{38}^{9,0}$, that has no pivotal categorifications due to the PDC criterium but is not ruled out from general categorification by any of the other criteria. This ring is, therefore, the only multiplicity-free fusion ring of rank 9 or less that might be a counterexample to the famous conjecture that all fusion categories have a pivotal structure.

### 4.1.2 Determining Which *F*-Symbols Could Be Zero

Multiple steps in solving the pentagon equations assume that the variables to be solved for are non-zero. Therefore, we must determine which variables could be zero beforehand. The main steps that we performed, in the order they are applied, are the following:

1. Translate the pentagon equations and invertibility constraints to a logical proposition.

2. Reduce the proposition via tetrahedral symmetries, pattern matching, and conversion between normal forms.

3. Find all lists of booleans that satisfy the proposition.

#### 4.1.2.1 Constructing the logical proposition

The pentagon equations and the demands that all *F*-matrices have full rank can be translated to a boolean proposition in the following way. First, every *F*-symbol $F_I$ (where *I* is some admissible label) is replaced by a boolean variable $b_I$ for which

$$b_I = \begin{cases} \text{FALSE} & \text{iff } F_I = 0 \\ \text{TRUE} & \text{iff } F_I \neq 0 \end{cases}. \tag{4.1}$$

Next, for every pentagon equation, a proposition is constructed that demands that it is not allowed for exactly one term to be non-zero. For example, for the following equation

$$F_I F_J = F_K F_L F_M \tag{4.2}$$

this translates to the proposition

$$b_I \wedge b_J \Leftrightarrow b_K \wedge b_L \wedge b_M, \tag{4.3}$$

while for

$$F_I F_J = F_K F_L F_M + F_N F_O F_P \tag{4.4}$$

this translates to

$$\neg \left( (p_1 \wedge \neg p_2 \wedge \neg p_3) \vee (\neg p_1 \wedge p_2 \wedge \neg p_3) \vee (\neg p_1 \wedge \neg p_2 \wedge p_3) \right), \tag{4.5}$$

with

$$p_1 = b_I \wedge b_J, \tag{4.6}$$

$$p_2 = b_K \wedge b_L \wedge b_M, \tag{4.7}$$

$$p_3 = b_N \wedge b_O \wedge b_P. \tag{4.8}$$

The demand that the *F*-matrices are invertible would typically translate to the determinant of each matrix being non-zero. There is no way to translate such a demand to a proposition in the variables $b_I$ since it depends on the exact values of the *F*-symbols. Therefore, we weaken this demand to the demand that at least one of the terms of the determinant should be non-zero. If this demand is violated, the determinant is equal to zero, but the opposite is clearly not true.

Let $p(b_{I_1}, \ldots, b_{I_n})$ be the logical conjunction of all propositions constructed for a certain fusion ring. Then, the task of finding all admissible sets of zero values of *F*-symbols comes down to finding lists of booleans for which $p$ evaluates to TRUE. This problem is a specific case of an ALLSAT problem in theoretical computer science.

However, the proposition $p$ is often too big to handle directly, so some heuristics were implemented to reduce its complexity.

#### 4.1.2.2 Reducing the Proposition

Several methods were applied to reduce the proposition $p$. Firstly, we made use of so-called tetrahedral symmetries, which allow us to identify several of the boolean variables as follows

$$[b_d^{abc}]_f^e = [b_{d^*}^{cba^*}]_{f^*}^e = [b_{b^*}^{a^*dc}]_f^{e^*} = [b_{a^*}^{e^*cf^*}]_{d^*}^{b^*}. \tag{4.9}$$

These symmetries come from a group of natural transformations in the fusion category (see definition 5.1. in [64]) that map $F$-symbols to other $F$-symbols up to a non-zero scalar factor. In the physics literature, sometimes a stronger form of these symmetries, namely

$$[F_d^{abc}]_f^e = [F_{d^*}^{cba^*}]_{f^*}^e = [F_{b^*}^{a^*dc}]_f^{e^*} = [F_{a^*}^{e^*cf^*}]_{d^*}^{b^*}\sqrt{\frac{d_e d_f}{d_b d_d}}, \tag{4.10}$$

is demanded (see [61] for examples and applications). This helps reduce the pentagon equations, but there are fusion systems for which the stricter version (4.10) do not hold. There is, for example, no gauge for the fusion systems belonging to the $HI(\mathbb{Z}_3)$ fusion ring, introduced in [43], for which (4.10) holds.

> **Note 7.** The factors that appear in front of the $F$-symbols after applying the natural transformations come from the fact that in the categorical language $x^{**} \cong x$ rather than $x^{**} = x$. The translation to a language where $x^{**} = x$ must involve the isomorphisms $(x^{**} \to x) \in \hom(x^{**}, x)$ which cause nonzero numerical factors to appear. The group of natural transformations is a group of linear transformations, so they must map non-zero $F$-symbols to non-zero $F$-symbols and vice versa. In particular, they hold exactly on the variables $b_I$.

The reduction in the number of variables is quite significant. For example, for the fusion ring $PSU(2)_6$, the number of variables in the proposition gets reduced from 163 to 44, while for $HI(\mathbb{Z}_3)$ the number of variables gets reduced from 1259 to 396. Once the tetrahedral symmetries are exhausted, one can still reduce the system by using pattern matching and recursion. The invertibility constraints on the 1$D$ $F$-symbols cause $p$ to contain sub-propositions of the form

(a) $b_{I_1} \wedge \ldots \wedge b_{I_k}$, and

(b) $b_I \Leftrightarrow b_J$

that can be factored out. Factors of form (a) can be used to conclude that $b_{I_1} =$ TRUE$, \ldots, b_{I_k} =$ TRUE and this new information can be used to update $p$. Factors of form (b) can be used to set up equivalence relations between certain variables and, therefore, reduce the number of variables appearing in *p*. After updating *p* in this way, new factors of the form (a) and (b) can appear, and the process is repeated until no factors are left.

One last procedure can be applied to reduce the proposition. Propositions can be expressed in certain normal forms, such as

- CNF (Conjunctive Normal Form): the proposition is written as an AND of OR's,

- DNF (Disjunctive Normal Form): the proposition is written as an OR of AND's.

For example, proposition (4.3) looks like

$$(\neg F_I \vee \neg F_J \vee F_K) \wedge (\neg F_I \vee \neg F_J \vee F_L) \wedge (\neg F_I \vee \neg F_J \vee F_M) \wedge \\ (F_I \vee \neg F_K \vee \neg F_L \vee \neg F_M) \wedge (F_J \vee \neg F_K \vee \neg F_L \vee \neg F_M)$$
(4.11)

in CNF, but like

$$(F_I \wedge F_J \wedge F_K \wedge F_L \wedge F_M) \vee (\neg F_I \wedge \neg F_K) \vee (\neg F_I \wedge \neg F_L) \vee \\ (\neg F_I \wedge \neg F_M) \vee (\neg F_J \wedge \neg F_K) \vee (\neg F_J \wedge \neg F_L) \vee (\neg F_J \wedge \neg F_M)$$
(4.12)

in DNF. By expressing the proposition in CNF form, sometimes extra factors of the form (a) can be identified. When the proposition is expressed in DNF form, sometimes variables appear in all AND clauses, which implies they have to be TRUE. The method used to find zero values converts back and forth between CNF and DNF forms to extract these extra pieces of information until no extra info can be extracted.

### 4.1.2.3 Finding All Instances

Mathematica has a function `SatisfiabilityInstances`, which can be used to find all boolean vectors that satisfy a proposition. It is also possible to use the built-in function `Reduce` or our implementation of an ALLSAT solver that writes, compiles, and executes C code without leaving Mathematica. This solver uses the same technique (see 3.1.3) as used to find fusion matrices.

All of the above functionality has been implemented in the `FindZeroValues` function of Anyonica. This function takes two arguments `equations` and `variables` and several optional arguments. The option `"FindZerosBy"` allows you to change the algorithm to solve the ALLSAT problem, while `"InvertibleMatrices"` allows you to add matrices of variables that are required to be invertible. By default, not all equations containing more than two terms are included in the boolean proposition. This is because these equations are often much weaker than those with only two terms. Also, the execution time and memory pressure of the algorithm increases drastically by including all such equations. So by default, `FindZeroValues` only includes 30% of the *least complex* equations with more than two terms, a value that can be changed via the option `"SumSubSetParameter"`. At the end of the algorithm, however, solutions incompatible with the other 70% of these equations are removed, so this value has no effect on the output itself. The default measure of complexity that

`ZeroValues` uses is the built-in `LeafCount` function, which returns the number of leaves that the formal tree form of an expression has. A custom function can be used by setting the `"SumSubsetFunction"` option.

*Example* 4.1.2. The following code

```
repD3 = FusionRingList[[5]];
pEqns = PentagonEquations @ repD3;
fSymb = FSymbols   @ repD3;
fMats = FMatrices  @ repD3;
FindZeroValues[ pEqns, fSymb, "InvertibleMatrices" -> fMats ]
```

returns

```
{ { }, { [F_3^{333}]_3^3 -> 0 } }
```

which means that there are two possible sets of zero values. Either no $F$-symbol is zero, or only $[F_3^{333}]_3^3 = 0$. For both of these possibilities, the pentagon equations need to be solved. It turns out that the first configuration, with no zero $F$-symbols, does not lead to any solutions, and $\text{Rep}(D_3)$ is the smallest multiplicity-free fusion ring with a non-trivial zero $F$-symbol.

> **Note 8.** Since code listings can take up much space, LaTeX typesetting will be used to display the variables in these environments. In particular, $[F_d^{abc}]_f^e$ will be used instead of the standard output `F[ a, b, c, d, e, f ]` from the package. Likewise, triple dots `...` will be used to condense all but a few elements of long lists whenever their content is similar to the elements that will be shown.

For all but three multiplicity-free fusion rings with rank up to seven there is a unique configuration of zero values for the $F$-symbols. The three exceptions are the following.

- $\text{FR}_1^{7,1,0} = \text{Adj}(\text{SO}(16)_2)$ admits 56 (braided) pivotal fusion categories of which 32 have 192 $F$-symbols equal to 0, 16 have 96 zero-$F$-symbols equal to 0, and 8 have 24 $F$-symbols equal to 0.

- $\text{FR}_6^{7,1,0} = \text{Adj}(\text{SO}(11)_2)$ has 5 categorifications of which 3 have 135 $F$-symbols that are 0 and 2 have 45 $F$-symbols that are zero.

- $\text{FR}_4^{7,1,2}$ has 22 categorifications of which 16 have 192 $F$-symbols that are zero and 6 have 24 $F$-symbols that are 0.

For all three cases, the smaller sets of $F$-symbols that are zero are included in all larger sets of $F$-symbols that are zero.

### 4.1.3 Fixing the Gauge

The pentagon and hexagon equations never have a finite number of solutions because they have gauge freedom associated with each distinct vertex that amounts to the choice of basis vectors. Let $\left\{ g_c^{ab} \in \mathbb{C}\backslash\{0\} \,\middle|\, N_{a,b}^c \neq 0 \right\}$ be a set of numbers that denote the change of basis of the 1D vector spaces $V_c^{ab}$. By applying such a basis transform, the values of the $F$-symbols change as

$$[F_d^{abc}]_f^e \mapsto \frac{g_e^{ab} g_d^{bc}}{g_d^{af} g_f^{bc}} [F_d^{abc}]_f^e. \tag{4.13}$$

Since the pentagon equations are derived in a basis-independent way, they are symmetric under such transforms. In particular, for every solution to the pentagon equations given by a set $\left\{[F_d^{abc}]_f^e\right\}$, the set $\left\{\frac{g_e^{ab} g_d^{bc}}{g_d^{af} g_f^{bc}} [F_d^{abc}]_f^e \,\middle|\, g_c^{ab} \in \mathbb{C}\backslash 0\right\}$ is also a valid solution. This allows us, even without knowing any of the values *of the* $F$-symbols a priori, to fix the values of certain $F$-symbols before solving the pentagon equations. In effect, we are choosing specific representatives for the orbits of the solutions under such gauge transforms. This is why we call this process fixing the gauge. Interestingly, a result of Ocneanu tells us that after fixing the gauge, the number of solutions to the pentagon (and hexagon) equations is finite [27]. Completely fixing the gauge is often not possible, though. What we can do, however, is fix the gauge up to a discrete group of gauge transforms. After we solve the pentagon (and hexagon) equations, we will need to check for equivalent solutions and remove these.

The definition of a fusion system already has several gauge demands in disguise. Indeed, the demand that any $F$-symbol $[F_d^{abc}]_f^e$ with $a = 1$, or $c = 1$ equals 1 has no effect on the categorical level. Here, we show that these demands are artificial in that it is always possible to fix a gauge for which they are true. Moreover, if such a gauge is chosen, one also has that $[F_d^{a1c}]_f^e = 1$, so any vacuum $F$-symbol, i.e., for which $a = 1, b = 1$, or $c = 1$, must equal 1.

**Proposition 4.1.3.** *For any set of F-symbols of a multiplicity-free fusion category it is possible to choose a gauge such that any vacuum F-symbol equals* 1.

*Proof.* First we note that any vacuum $F$-matrix must be a $1 \times 1$ invertible matrix. In particular any vacuum $F$-symbol is non-zero. Now assume a set $\left\{[F_d^{abc}]_f^e\right\}$ of values of $F$-symbols is given. We will show that there exists a set of symbols $\left\{g_c^{ab} \in \mathbb{C}\backslash\{0\} \,\middle|\, N_{a,b}^c \neq 0\right\}$ parametrizing a gauge transform $\varphi$ such that $\varphi([F_d^{abc}]_f^e) = 1$ whenever $a = 1$, $b = 1$, or $c = 1$. This will be shown by listing all cases of vacuum symbols in order of decreasing number of vacuum labels and refining the gauge transform as we go along.

Denote by $E(a, b, c, d, e, f, g, h, i)$ the pentagon equation in the gauge transformed variables:

$$\varphi([F_d^{abc}]_f^e)\varphi([F_d^{ghf}]_i^a) = \sum_j \varphi([F_e^{ghb}]_j^a)\varphi([F_d^{gjc}]_i^e)\varphi([F_i^{hbc}]_f^j). \tag{4.14}$$

This is nothing but equation (4.14), where the indices are relabeled to make the arguments easier to follow.

**Case 1:** $a = b = c = 1$

In this case also $d = e = f = 1$ and there is only one such symbol: $[F_1^{111}]_1^1$. This symbol is gauge-invariant, i.e. $\varphi([F_1^{111}]_1^1) = [F_1^{111}]_1^1$ but from the pentagon equation

$$E(1,1,1,1,1,1,1,1,1) : \varphi([F_1^{111}]_1^1)^2 = \varphi([F_1^{111}]_1^1)^3 \tag{4.15}$$

it follows immediately that $\varphi([F_1^{111}]_1^1) = 1$.

**Case 2:** $a = b = 1, c \neq 1$

In this case $c = d = f, e = 1$ and we find that

$$\varphi([F_c^{11c}]_c^1) = \frac{g_1^{11}}{g_c^{1c}}[F_c^{11c}]_c^1, \tag{4.16}$$

so if we set $g_c^{1c} = g_1^{11}/[F_c^{11c}]_c^1$ for all $c \neq 1$ then $\varphi([F_c^{11c}]_c^1) = 1$.

**Case 3:** $a = c = 1, b \neq 1$

In this case $b = d = e = f$ and we find that the symbol is gauge invariant. Because of the choice of $\varphi$ from case 2 we find that $E(1, b, 1, b, b, b, 1, 1, b) : \varphi([F_b^{1b1}]_b^b)\varphi([F_b^{11b}]_b^1) = \varphi([F_b^{11b}]_b^1)\varphi([F_b^{1b1}]_b^b)\varphi([F_b^{1b1}]_b^b)$ simplifies to $\varphi([F_b^{1b1}]_b^b) = \varphi([F_b^{1b1}]_b^b)^2$ and thus $[F_b^{1b1}]_b^b = 1$ for all $b \neq 1$.

**Case 4:** $b = c = 1, a \neq 1$

In this case $d = e = a, f = 1$ and we find that

$$\varphi([F_a^{a11}]_1^a) = \frac{g_a^{a1}}{g_1^{11}}[F_a^{a11}]_1^a, \tag{4.17}$$

so if we set $g_a^{a1} = [F_a^{a11}]_1^a/g_1^{11}$ for all $a \neq 1$ then $\varphi([F_a^{a11}]_1^a) = 1$.

**Case 5:** $a = 1, b \neq 1, c \neq 1$

In this case $e = b, f = d$ and we find that

$$\varphi([F_d^{1bc}]_d^b) = \frac{g_b^{1b}}{g_d^{1d}}[F_d^{1bc}]_d^b = \frac{[F_d^{11d}]_d^1}{[F_b^{11b}]_b^1}[F_d^{1bc}]_d^b \tag{4.18}$$

and thus its value is fixed because we already put constraints on the relevant gauge factors. By choosing suitable labels, however, we find that

$$E(1, b, c, d, b, d, 1, 1, d) : \varphi([F_d^{1bc}]_d^b)\varphi([F_d^{11d}]_d^1) = \varphi([F_b^{11b}]_b^1)\varphi([F_d^{1bc}]_d^b)\varphi([F_d^{1bc}]_d^b) \tag{4.19}$$

or, because of our choice of $\varphi$,

$$\varphi([F_d^{1bc}]_d^b) = \varphi([F_d^{1bc}]_d^b)^2 \tag{4.20}$$

and thus $\varphi([F_d^{1bc}]_d^b) = 1$.

**Case 6:** $a \neq 1, b = 1, c \neq 1$

In this case $e = a, f = c$ and from

$$E(a, 1, c, d, a, c, a, 1, c) : \varphi([F_d^{a1c}]_c^a)^2 = \varphi([F_d^{a1c}]_c^a) \tag{4.21}$$

we find that $\varphi([F_d^{a1c}]_c^a) = 1$.

**Case 7:** $a \neq 1, b \neq 1, c = 1$

In this case we choose our labels such that the second factor in the pentagon equation equals $[F_d^{ab1}]_b^d$,

$$E(d, 1, 1, d, d, 1, a, b, b) : \varphi([F_d^{ab1}]_b^d) = \varphi([F_d^{ab1}]_b^d)^2 \tag{4.22}$$

and we find that $\varphi([F_d^{ab1}]_b^d) = 1$. ∎

In what follows, we will assume that vacuum $F$-symbols have already been set to 1, and no gauge transform is allowed to change these values.

One can apply the following procedure to fix the gauge for the other $F$-symbols. Let $\left\{ \varphi([F_d^{abc}]_{ef}) \right\}$ be a set of $F$-symbols obtained after applying a gauge transform with variables $\left\{ g_c^{ab} \in \mathbb{C}\backslash\{0\} \,\middle|\, N_{a,b}^c \neq 0 \right\}$ on $\left\{ [F_d^{abc}]_f^e \right\}$, i.e.

$$\varphi(F_d^{abc}]_{ef}) = \frac{g_e^{ab} g_d^{bc}}{g_d^{af} g_f^{bc}} [F_d^{abc}]_f^e. \tag{4.23}$$

Pick an $F$-symbol whose transform has a gauge variable, say $g_z^{xy}$, that appears linearly (to the power 1 or $-1$) in Eq. (4.23). Set the transformed $F$-symbol equal to 1, and solve this equation for $g_z^{xy}$. Then, substitute this value of $g_z^{xy}$ into the other gauge equations and repeat this procedure with another $F$-symbol. Sometimes, this procedure can be carried through until no more free gauge factors appear in the equations. When this happens, the gauge freedom of the $F$-symbols is completely exhausted. However, we may generally run out of equations containing linear gauge factors before the gauge is fully fixed. In such cases, one can continue the process using quadratic or higher-order equations in the gauge factors. Such equations do not have unique solutions for the gauge factors, which results in a partly broken gauge. The continuous symmetry is broken, but a discrete symmetry from a residual finite group of gauge transforms remains. This is not a problem when solving the pentagon equations since the number of solutions will still be finite, but afterward, duplicate equivalent solutions should be deleted.

The Anyonica package contains several functions that help deal with multiplicative symmetries, i.e., symmetry transforms that act via multiplication with specific variables. The function `BreakMultiplicativeSymmetry` takes a set of symmetries (that can be created using the `GaugeSymmetries` function) and returns a couple of (1) remaining symmetries and (2) a list of values of fixed symbols.

For example

*Example* 4.1.4. For the Rep($D_3$) fusion ring initialized in example 4.1.2, evaluating

```
symmetries = GaugeSymmetries[ FSymbols @ repD3, g ];
brokenSymmetries = BreakMultiplicativeSymmetry[ symmetries ];

Print[symmetries];
Print[brokenSymmetries];
```

returns

```
<|
  "Transforms" -> {
    [F_1^{111}]_1^1 -> [F_1^{111}]_1^1, ...,
    [F_3^{333}]_2^3 -> ((g_3^{33})^2 [F_3^{333}]_2^3)/(g_3^{32} g_2^{33}), [F_3^{333}]_3^3 -> [F_3^{333}]_3^3
  },
  "Symbols" -> { g }
|>

{
  <|
    "Transforms" -> {
      [F_1^{111}]_1^1 -> [F_1^{111}]_1^1, ..., [F_3^{333}]_2^3 -> [F_3^{333}]_2^3, [F_3^{333}]_3^3 -> [F_3^{333}]_3^3
    },
    "Symbols" -> { g }
  |>
  ,
  {
    [F_2^{112}]_2^1 -> 1, [F_3^{113}]_3^1 -> 1, [F_2^{211}]_1^2 -> 1, [F_3^{223}]_3^1 -> 1,
    [F_1^{233}]_2^3 -> 1, [F_3^{311}]_1^3 -> 1, [F_3^{322}]_1^3 -> 1, [F_3^{333}]_3^1 -> 1
  }
}
```

It is possible to set desired $F$-symbols to specific values by setting the option `"GaugeDemands"` in which case these symbols will be changed to the desired value as soon as possible in the process, to the extent that they are free to choose. Likewise, one can exclude variables from being fixed by using the `"ExcludedVariables"` option. While the symmetries in the example were calculated via the `GaugeSymmetries` function that recognizes $F$-symbols and $R$-symbols, the function `BreakMultiplicativeSymmetry` accepts arbitrary symbols and transforms as long as the symmetries are multiplicative.

### 4.1.4 Solving Binomial Equations

Now that the gauge is fixed and none of the remaining variables are 0, the pentagon equations can be greatly reduced using the subset of equations with two terms. These equations have the specific form $m_{i_1} = m_{i_2}$ where the $m_{i_j}$ are monomials in non-zero variables. Since none of the variables can be 0, one can apply a logarithm to transform these equations to a linear system of the form $A.\vec{x} = \vec{y} \mod 2\pi i \mathbb{Z}^n$, where $n$ is the number of $F$-symbols. Such a system can be solved by taking a Smith decomposition of the matrix $A$.

The general form of the solution to a system of binomial equations in the variables $f_j$ is the following:

$$f_j = v_j \prod_{k=1}^{\text{rk}(A)} e^{2\pi i [Q]_j^k m_k} \prod_{k=1}^{n-\text{rk}(A)} z_k^{[Z]_j^k}, \tag{4.24}$$

where $v$ is a constant vector, $Q$ a matrix of fractions, $Z$ a matrix of integers, the $m_k$ are integer parameters, the $z_k$ are complex parameters, and rk($A$) is the rank of $A$. From the RHS in (4.24), we see that only a finite number of lists of integers $\vec{m} = (m_1, \ldots, m_r)$ give rise to different solutions. For each $\vec{m}$, we get a parametrization of the $F$-symbols in fewer variables than before. These parametrizations can be substituted in the remaining non-binomial equations, which become equations in the parameters $z_k$. Parametrizations for which an $F$-matrix becomes singular or for which the updated set of equations is inconsistent are removed at this point.

Typically, a new subset of binomial equations arises after the parametrization, and thus, we can repeat the process recursively until no new binomial equations are found. This procedure results in a finite number of couples consisting of (a) a system with fewer equations and fewer variables and (b) a parametrization of the $F$-symbols in these variables.

The reduction in the number of variables and equations is often quite drastic. For example, for the Rep($D_3$) and PSU(2)$_6$ fusion rings, all equations are solved by this procedure. For the HI($\mathbb{Z}_3$) fusion ring, we get a reduction of 36018 equations to 3319 and 1212 variables to 5.

In Anyonica, the function `SolveBinomialSystem` solves a binomial system of equations using the Smith decomposition. It takes three arguments, `equations`, `variables`, and `parameter`, and returns a list of solutions parametrized by `parameter`. Several options can be set, such as

- `"NonSingular"` which is set to `True` if no variable is allowed to be zero,

- `"ZeroValues"` which is set to a list of sets of variables that are known to be 0,

- `"Symmetries"` which is set to a list of symmetries that should be broken,

- `"InvertibleMatrices"` which is set to a list of matrices of variables that should be invertible,

- `"PolynomialConstraints"` which can be set to a list of more general polynomial constraints (such as inequalities) that should be satisfied by the solutions,

- `"UseDatabaseOfSmithDecompositions"` which can be set to `True` in order to reuse previously calculated Smith decompositions, and

- `"StoreDecompositions"` which can be set to `False` in order not to waste time storing decompositions.

*Example* 4.1.5. After fixing the gauge for Rep($D_3$) and substituting $\{[F_3^{333}]_3^3 \to 0\}$, the pentagon equations become a system of 82 polynomial equations in 19 variables. 72 of these equations are binomial. If we solve these as follows

```
binEqns =
  {
    1 == [F_2^{222}]_1^1^2, 1 == [F_1^{332}]_3^2 [F_2^{332}]_3^1, ...,
    [F_3^{333}]_2^3 == [F_3^{323}]_3^3 [F_3^{333}]_1^3, [F_3^{333}]_1^3 == -[F_3^{323}]_3^3 [F_3^{333}]_3^2 [F_3^{333}]_2^3
```

```
    }

  vars =
    { [F_2^{222}]_1^1, [F_3^{232}]_3^3, ..., [F_3^{333}]_2^3 }

  invMats =
    {
      {{[F_2^{222}]_1^1}}, {{[F_3^{232}]_3^3}}, {{[F_2^{233}]_1^3}}, ..., {{[F_2^{333}]_3^3}},
      {{[F_3^{333}]_1^1,[F_3^{333}]_2^1,1},{[F_3^{333}]_1^2,[F_3^{333}]_2^2,[F_3^{333}]_3^2},{[F_3^{333}]_1^3,[F_3^{333}]_2^3,0}}
    }

  SolveBinomialSystem[
    binEqns, vars, z, "InvertibleMatrices" -> invMats
  ]
```

then we obtain 6 parametrizations of the $F$-symbols in terms of one parameter $z_1$

```
{
  { [F_2^{222}]_1^1 -> 1, ..., [F_3^{333}]_1^3 -> z_1,           [F_3^{333}]_2^3 -> -z_1            },
  { [F_2^{222}]_1^1 -> 1, ..., [F_3^{333}]_1^3 -> z_1 e^{-iπ/3}, [F_3^{333}]_2^3 -> -e^{2iπ/3} z_1  },
  { [F_2^{222}]_1^1 -> 1, ..., [F_3^{333}]_1^3 -> z_1 e^{-2iπ/3},[F_3^{333}]_2^3 -> -e^{-2iπ/3} z_1 },
  { [F_2^{222}]_1^1 -> 1, ..., [F_3^{333}]_1^3 -> -z_1,          [F_3^{333}]_2^3 -> -z_1            },
  { [F_2^{222}]_1^1 -> 1, ..., [F_3^{333}]_1^3 -> z_1 e^{2iπ/3}, [F_3^{333}]_2^3 -> -e^{2iπ/3} z_1  },
  { [F_2^{222}]_1^1 -> 1, ..., [F_3^{333}]_1^3 -> z_1 e^{iπ/3},  [F_3^{333}]_2^3 -> -e^{-2iπ/3} z_1 }
}
```

By substituting these parametrizations into the 10 remaining equations, we obtain 6 new, greatly simplified systems of equations.

To repeatedly reduce a system of equations with more than two terms, the function `ReduceByBinomials` can be used. It takes four arguments, `sumEquations`, `twoTermEquations`, `variables`, and `parameter`. Besides all the options it inherits from `SolveBinomialSystem`, it also takes the option `"SimplifyIntermediateResultsBy"` (with default `Identity`), which can be set to a function that simplifies intermediate solutions before substituting them. This is primarily useful if the polynomial equations contain non-integer coefficients. While this situation can occur when solving the pentagon equations, the non-integer coefficients are always cyclotomic integers at this point in the calculation. Mathematica typically simplifies combinations of these coefficients by default. However, the coefficients that appear in the hexagon equations are solutions to the pentagon equations and can be pretty wild. Since the solutions to the hexagon equations are always phases, it is often desirable to express intermediate coefficients as phases.

### 4.1.5 Reduction Via Linear Polynomials

When no more binomial equations are present we can reduce the system of equations even further by using equations that are linear in a certain variable, i.e. of the form

$$zp + q = 0, \tag{4.25}$$

where $p$ and $q$ are polynomials that are independent of the variable $z$. From this equation it is clear that either

$$z = -q/p, \quad \text{or} \quad p = 0. \tag{4.26}$$

The process of reduction is the following. Let $D = (l_P, l_A, l_Z)$ be a triple of lists, to which we will refer to as 'the data', with the following purpose:

- $l_P$ is the list of polynomials that we want to reduce. During the reduction process the polynomials in this list will be updated, and polynomials equal to 0 will be deleted. Initially $l_P$ contains the list of all polynomials we want to reduce.

- $l_A$ is the list of assumptions that are made during the reduction process. Here polynomials will be stored that are assumed to be non zero. $l_A$ is initially empty.

- $l_Z$ is the list of deduced values of variables. During the reduction process, more values will be added and previous values will be updated. $l_Z$ is initially empty.

If at any point during the reduction $l_P$ contains a non-zero number, $l_A$ contains 0, or any of the values in $l_Z$ becomes zero then the data at that point is regarded as invalid. The following procedure is then applied to reduce the list of polynomials.

Let $L_{\text{reduced}}$ be an empty stack.

1. Find the *simplest* linear polynomial in $l_P$.

2. If there is no linear polynomial, throw $D$ on $L_{\text{reduced}}$.
   If there is, then

   (a) if $p$ is monomial, solve for $z$ and update $D$ to $D_{\text{new}}$ by substituting the value of $z$ in $D$, and adding its value to $l_Z$. If $D_{\text{new}}$ is valid then remove all 0 polynomials from it, convert all equations with rational functions to polynomial equations, and repeat step 1 with the $D_{\text{new}}$ as input.

   (b) If $p$ is not monomial then create two copies of $D$, say $D_1$ and $D_2$, and

      i. solve for $z$ under the assumption that $p \neq 0$. Update $D_1$ to $D_{\text{new},1}$ by substituting the value of $z$ in $D_1$, adding $z$ to $l_{P_1}$, and adding $p$ to $l_{A_1}$. If $D_{\text{new},1}$ is valid, remove all 0 polynomials from it, convert all equations with rational functions to polynomial equations, and repeat step 1 with $D_{\text{new},1}$.

      ii. Update $D_2$ to $D_{\text{new},2}$ by reducing all polynomials in $D_2$ modulo $p$. If $D_{\text{new},2}$ is valid, remove all 0 polynomials from it and repeat step 1 with the updated data.

In each step where a linear polynomial is encountered the list of polynomials gets either reduced by at least 1 element or, in the case of an invalid system, the process quits. The algorithm, therefore, always stops in a finite time.

After performing these steps one should have a stack of triples

$$\{\{l_{P;1}, l_{A;1}, l_{Z;1}\}, \ldots, \{l_{P;n}, l_{A;n}, l_{Z;n}\}\},$$

all corresponding to reduced systems of equations. Note that while the systems of equations have fewer equations and variables, the size (in bits) can become quite big.

However, the systems created this way are easier to tackle using Gröbner bases algorithms combined with some tricks.

The function `ReduceByLinearity` performs the reduction of a system of polynomial equations by linear equations. It takes two standard arguments `"polynomial-List"` and x, where the variables of the polynomials should be of the form `x[1],...,x[n]`. The *simplest* linear equation is based on (1) the lexicographic weight of the function `{ MonomialQ[p], MonomialQ[q] }` which prefers fractions of monomials but prioritizes fraction where only $p$ is monomial over those where only $q$ is monomial, and (2) `LeafCount[p/q]` which provides a simple measure of the complexity of an expression. With the option `"LinearReductionWeight"`, a custom weight function can be given to decide the simplest polynomial. This can be necessary since the execution time, maximum memory used, the size, and the form of the result depend greatly on the order of the variables that are solved for. Extra options can also be set to parallelize evaluation and control the memory pressure.

Notes 6.
- This procedure can also be used to reduce the system binomial equations to a great extent (which is what J.K. Slingerland's code does). I still decided to use F. Verstraete's idea to use the Smith decomposition since these decompositions can be stored. This allows one to check and redo calculations much faster. The Smith decomposition does have some serious drawbacks, however. Firstly, the required memory for the algorithm scales quite badly with the size of the system. Secondly, the Smith decomposition returns all solutions to the binomial system in one go. For example, for the $SO(7)_2$ fusion ring, the outcome of `ReduceByBinomials` is a list of 64 reduced systems. Many of these intermediate solutions do not result in an actual solution and are thus a waste of space and time. Lastly, if a crash occurs during the computation of the Smith decomposition, it is very hard to recover the computations performed so far. While the linear reduction can also suffer from memory blowup, this can often be prevented by choosing a different order in which the variables are substituted. Because it is easy to keep track of the steps of the linear reduction, it is easier to avoid intermediate memory blowups. The linear reduction also avoids the blowup in the number of solutions since quadratic and higher order terms are not substituted for. For bigger systems of pentagon equations, it might be more beneficial to use the linear reduction directly after fixing the gauge instead of reducing the system via the Smith decomposition.

> - While $l_A$ is initialized as an empty list when solving all systems, it might be beneficial to initialize it to a list of determinants of the *F*-matrices. This does not affect the eventual solutions found since, in the last step, all singular solutions are removed. It might affect computation speed, however.

### 4.1.6 Incremental and Parallel Gröbner Bases

The penultimate step in solving the pentagon and hexagon equations is the computation of Gröbner bases. A Gröbner basis of a set of polynomials is a set of polynomials that has the same roots as the original system, but from which it is easier to extract certain properties of the system. Both the form of the basis and the properties that can be extracted depend on a given monomial ordering. A Gröbner basis in "Lexicographic order" is especially interesting for our purpose. It has a polynomial upper triangular form: the first polynomial contains only one variable, the second contains only two variables, one shared with the first polynomial, and so on. While the number of variables at this stage is quite low, the number and size of equations could still be high. Typically, though, a lot of the equations are redundant. Therefore, we implemented the function `IncrementalGroebnerBasis`, which creates such a basis incrementally. It takes two arguments, `polynomials` and `variables`, and works as follows. Let $l_P$ be the list of polynomials given by `polynomials`.

1. First find several of the *simplest* polynomials and compute a Gröbner basis, say $\mathcal{G} = \{p_1, \ldots, p_n\}$, for these.

2. The system $l_P$ is then reduced via $\mathcal{G}$ as follows. Each polynomial $q$ in $l_P$ can be written as a linear combination $q = a_{q;1} p_1 + \cdots a_{q;n} p_n + r_q$ where $a_{q;1}, \ldots, a_{q;n}$, and $b(q)$ are polynomials such that $b(q)$ is indivisible by any of the $p_i$, i.e. is the remainder of the division of $q$ by $\{p_1, \ldots, p_n\}$. Since the $p_i$ are assumed to be zero, one can replace all polynomials $q$ by $r_q$.

3. Next all 0 polynomials and all duplicate polynomials are removed from $l_P$. If no polynomials remain then $\mathcal{G}$ is a Gröbner basis for the system. If some polynomials remain then we add $\mathcal{G}$ to the reduced $l_P$ and repeat steps 1, 2, 3 with the updated $l_P$.

The Gröbner bases are calculated using the standard `GroebnerBasis` function that comes with Mathematica. While the implementation of this function uses algorithms that are somewhat dated, I found no alternative algorithm that could properly deal with polynomial systems with non-integer symbolic coefficients.

Just like `ReduceByLinearity`, the memory consumption of `IncrementalGroebnerBasis` can easily blow up. To reduce the risk of memory overflows, we made it possible to set certain options by hand. The option `"Cutoff"` can be set to a number between 0 and 1, determining which fraction of all polynomials to consider in step 1. The option `"GroebnerWeightFunction"` determines how *simple* a polynomial is, and by default, it is the function `PolynomialDegree`. In some cases, it might be beneficial to choose another measure of complexity, such as e.g. the number of terms or a measure based on the complexity of the coefficients. If the polynomial system contains symbolic roots of polynomials as coefficients, it might be interesting to set the option `"ReduceRoots"` to `True`. This implies that the algorithm will reduce algebraic expressions containing roots to roots of single polynomials after each reduction step. Without intermediate simplification of this kind, large algebraic expressions without variables often linger in the system while they should either be 0 or, if not, allow us to conclude that the system has no roots. A custom function to perform simplifications can also be given by setting the `"SimplifyIntermediateResultsBy"` option.

The order of the variables given to the `GroebnerBasis` function also plays an important role. Different permutations of the list of variables given to the function can result in considerable differences in execution time, memory consumption, and output complexity. The function `ParallelGroebnerBasis` was constructed to deal with this issue. For every available parallel kernel, the function chooses a permutation of the variables and starts the task of computing a Gröbner basis. The moment one of the tasks finishes, it returns the Gröbner basis, and all other tasks are killed. This way, one can obtain speedups of factors 10000 or higher using only a few extra cores.

For the pentagon and hexagon equations we tackled, such a basis typically contains less than 20 polynomials and less than 5 variables.

> **Note 9.** For some systems, it is not beneficial to compute Gröbner bases in the first place. If the results from the linear reduction are systems in one variable, it is often much faster to find the lowest degree polynomial in that variable, solve it, and remove solutions that do not satisfy the assumptions from $l_A$ or are incompatible with the other equations in $l_P$. It might even be beneficial to do this for systems with two or more variables (which might or might not have equations containing only one variable), but at the moment, only the check for systems with one variable is implemented.

### 4.1.7 Solving the Pentagon Equations

To find roots of Gröbner bases, the built-in function `Reduce` was used. This function returns a proposition that provides the solutions for a polynomial system with a finite number of solutions. Sometimes, a closed form cannot be found or is too complicated, so Mathematica solves the system in terms of `Root` expressions. These are symbolic placeholders, in our case of the form `Root[p(x), n]` where $p$ is a polynomial and $n$ denotes the index of the root according to a specific ordering. Such an expression contains all the information to specify the root it represents exactly. One can, in particular, obtain the numeric value of a `Root` expression to arbitrary precision and combine and simplify root objects symbolically.

All the functionality described so far is bundled in the function `SolvePentagonEquations`. By default, `SolvePentagonEquations` takes only a single argument: a fu-sion ring. There are, however, a large number of options one can set. Each option for the functions discussed so far is also valid for `SolvePentagonEquations`. Other options come from built-in functions such as `GroebnerBasis`, `Reduce`, and there are also some other options which can looked up in the documentation of the package. For multiplicity-free fusion rings up to rank 6, only several of the options are necessary. To categorify a generic fusion ring of higher rank, however, one likely needs to experiment with various options to avoid bottlenecks and crashes. It is even quite likely that more fine-tuned control is required and that, for example, it is necessary to swap the order of specific steps, substitute solutions to subsystems, or manually find and solve certain polynomial equations. Various extra functions were implemented to accommodate this need. These include functions that manipulate polynomial expressions, gauge symmetries, fusion rings, fusion categories, and more. Anyonica contains more than 170 different documented functions at the moment of writing. By evaluating `?Anyonica`*` in Mathematica, one gets a list of all functions with clickable hyperlinks that reveal usage information.

*Example* 4.1.6. The result from evaluating the following code

```
solutions = SolvePentagonEquations[repD3];

Print[ Length @ solutions ];
Print[ First @ solutions ];
```

is

```
6

{
  [F_1^{111}]_1^1 -> 1, [F_2^{112}]_2^1 -> 1, [F_3^{113}]_3^1 -> 1, [F_2^{121}]_2^2 -> 1, [F_1^{122}]_1^2 -> 1, ...
  [F_3^{333}]_3^2 -> -1, [F_3^{333}]_1^3 -> 1/2, [F_3^{333}]_2^3 -> -(1/2), [F_3^{333}]_3^3 -> 0
}
```

### 4.1.8 Removing Equivalent Solutions

We say that two solutions to the pentagon equations are equivalent if they can be transformed into each other via

- a gauge transform: $\left\{[F_d^{abc}]_f^e\right\} \mapsto \left\{\frac{g_e^{ab} g_d^{ec}}{g_d^{af} g_f^{bc}}[F_d^{abc}]_f^e\right\}$ for some set of numbers

$$\left\{g_c^{ab} \in \mathbb{C}\backslash\{0\} \,\middle|\, N_{a,b}^c \neq 0\right\},$$

- a fusion ring automorphism: $\left\{[F_d^{abc}]_f^e\right\} \mapsto \left\{[F_{\sigma(d)}^{\sigma(a)\sigma(b)\sigma(c)}]_{\sigma(f)}^{\sigma(e)}\right\}$ where $\sigma \in S_r$ is a permutation that leaves the multiplication table of the fusion ring invariant, or

- a combination of the previous two transforms.

After the pentagon equations are solved, there are typically several equivalent solutions, and it is desirable to remove redundant solutions. To do so, we used a 'gauge-split set' of rational functions in formal $F$-symbols to check whether two solutions are gauge equivalent. By this, we mean a list of rational monomials (RMs) in formal $F$-symbols (i.e., without assigned values) for which the first $m$ RMs form a generating set of the ring for all gauge-invariant rational functions in the $F$-symbols, and the last $n$ RMs are completely gauge-dependent. The generating set for the gauge-invariant RMs is constructed as follows. Let $I_1, ..., I_k$ be all admissible labels for which $F_{I_j} \neq 0$, $j = 1, ..., k$, and define

$$G_{I_i} := \frac{g_{e_i}^{a_i b_i} g_{d_i}^{e_i c_i}}{g_{d_i}^{a_i f_i} g_{f_i}^{b_i c_i}} \tag{4.27}$$

as the overall multiplicative gauge factor that appears in the gauge transform $F_{I_i} \mapsto G_{I_i} F_{I_i}$. An arbitrary RM in the $F$-symbols of a fusion category can be written as

$$(F_{I_1})^{p_1} (F_{I_2})^{p_2} \cdots (F_{I_k})^{p_k}, \quad p_i \in \mathbb{Z}. \tag{4.28}$$

The demand that it is gauge invariant then comes down to

$$(G_{I_1} F_{I_1})^{p_1} (G_{I_2} F_{I_2})^{p_2} \cdots (G_{I_k} F_{I_k})^{p_k} = (F_{I_1})^{p_1} (F_{I_2})^{p_2} \cdots (F_{I_k})^{p_k}, \tag{4.29}$$

$$\iff \left(G_{I_1}\right)^{p_1} \left(G_{I_2}\right)^{p_2} \cdots \left(G_{I_k}\right)^{p_k} = 1, \tag{4.30}$$

$$\iff \left(\frac{g_{e_1}^{a_1 b_1} g_{d_1}^{e_1 c_1}}{g_{d_1}^{a_1 f_1} g_{f_1}^{b_1 c_1}}\right)^{p_1} \cdots \left(\frac{g_{e_k}^{a_k b_k} g_{d_k}^{e_k c_k}}{g_{d_k}^{a_k f_k} g_{f_k}^{b_k c_k}}\right)^{p_k} = 1, \tag{4.31}$$

$\forall g_c^{ab} \in \mathbb{C}\setminus\{0\}$. Now assume that the $g_c^{ab}$ are ordered lexicographically on $(a, b, c)$ in a list of length $l$ and write $g_j$ for the $j$th variable that appears in this list. By collecting factors $g_j$ in equation 4.31 we get that

$$g_1^{\sum_{i_1} [Z]_{i_1}^1 p_{i_1}} g_2^{\sum_{i_2} [Z]_{i_2}^2 p_{i_2}} \cdots g_l^{\sum_{i_l} [Z]_{i_1}^l p_{i_l}} = 1, \quad \forall g_1, ..., g_l \in \mathbb{C}\setminus\{0\}, \tag{4.32}$$

where $[Z]$ is a matrix with integer coefficients. Equation (4.32) can only be satisfied if each exponent of each $g_i$ is identically zero, i.e.,

$$Z.\vec{p} = \vec{0} \tag{4.33}$$

with $\vec{p} := (p_1, p_2, ..., p_k)$. The space of vectors that produce gauge-invariant RMs is isomorphic to the kernel of $Z$, while span of the vectors that give rise to gauge-dependent RMs is isomorphic to the co-kernel of $Z$. In particular $m = \dim(\ker Z)$, $n = \dim(\text{CoKer } Z)$. The kernel and co-kernel of $Z$ can be constructed via a Smith decomposition. Indeed, if we write $Z = U.D.V$, with $U, V$ orthogonal matrices and $D$ diagonal, then the dimension of the kernel equals the number of non-zero rows of $D$ and is spanned by the first $m$ columns of $V$. The remaining columns of $V$ span the co-kernel. The matrix $V$ thus allows us to transform a lexicographically ordered list of $F$-symbols into a gauge-split basis. Since $V$ is orthogonal, any non-zero (and trivially any zero)

*F*-symbol can be written as a unique multiplication of powers of gauge-invariant and gauge-dependent RMs. This implies, in particular, that two solutions to the pentagon equations are gauge-equivalent if and only if their gauge-invariant RMs evaluate to the same values.

Several functions are implemented in Anyonica to play around with gauge invariants.

- `GaugeSplitTransform` takes a fusion ring and returns a couple of (1) the matrix *V* and (2) the number *m*.

- `GaugeSplitBasis` takes a fusion ring and returns a couple of lists. The first list corresponds to the gauge-invariant RMs and the second to the remaining gauge-dependent RMs.

- `GaugeInvariants` takes a fusion ring and returns only the first list from `GaugeSplitBasis`.

For all these functions, one can set option `"Zeros"` to a list of symbols that are 0. By default, these functions also include *R*-symbols in the gauge invariants (which has no meaning if the categories are non-braided), but by setting the option `"IncludeOnly"` to `"FSymbols"`, it is possible to return invariants in *F*-symbols. These functions then form the core of the function `GaugeEquivalentQ`, which takes a fusion ring and two sets of *F*-symbols and returns true if the solutions are gauge equivalent.

*Example* 4.1.7. The following code

```
gInv =
  GaugeInvariants[
    repD3, "IncludeOnly" -> "FSymbols", "Zeros" -> { [F_3^{333}]_3^3 }
  ];
Print[gInv];
```

returns

```
{
  [F_3^{333}]_3^3, [F_1^{111}]_1^1, [F_2^{121}]_2^2, ...,
  ([F_3^{113}]_3^1 [F_3^{311}]_1^3 [F_3^{322}]_1^3 [F_3^{333}]_3^1^2 [F_3^{333}]_2^3^2)/([F_3^{223}]_3^1 [F_1^{233}]_2^3^2)
}
```

Using these invariants one can easily see which solutions are gauge equivalent. The following code

```
gaugeValues[i_] := ReplaceAll[ solutions[[i]] ] @ gInv;
DeleteDuplicates[ Range @ Length @ solutions, gaugeValues ]
```

returns the indices of the unique solutions, in this case

```
{ 1, 2, 3, 4, 5, 6 }
```

Since we started with six solutions, we find that none of the solutions are gauge equivalent. We will see that only three unique solutions are left when we delete solutions based on full equivalence, however.

To check whether two solutions are equivalent up to a gauge transform in combination with a fusion ring automorphism, one can check gauge equivalence between the first solution and a permutation of the second solution for each pair of solutions and each fusion ring automorphism. The function `SymmetryEquivalentQ` can be used to check for such general equivalence. However, removing multiple redundant solutions by pairwise comparison with `SymmetryEquivalentQ` is not efficient. This is mainly because this algorithm computes and checks permutations one by one. If no matches are found, it will compute the same permutations again, wasting time. Instead, one can use the more efficient `DeleteEquivalentSolutions`. This function creates all permutations of each solution and calculates the values of the gauge invariants using these permutations. This way, each solution has a corresponding orbit of lists of gauge invariant values. Two solutions are then equivalent if their orbits intersect[2]. Since the function checks for equality to remove solutions, it is by default conservative: it might return solutions that are still equivalent but whose equivalence is not immediately clear.

*Example* 4.1.8. Any computer algebra system has to deal with how much it should automatically simplify symbolic expressions. Mathematica is quite conservative, and for all but the most trivial cases, the user must choose when to simplify. The following, almost trivial, equality is, for example, not simplified by default

```
Exp[ -2 I Pi / 3 ] == ( -1 - I Sqrt[3] ) / 2
```

However, applying a function such as `RootReduce` transforms this equality to `True`.

To avoid this situation, it is possible to set the option `"PreEqualCheck"` to a function of choice that will be evaluated on any two arguments before they are checked for equality. Note that this way it becomes possible that too many solutions are removed. To avoid this from happening, built-in functions such as `FullSimplify` or `N` (which gives a numerical approximation) should be avoided, if possible, and replaced by safe variants such as `RootReduce`. It is also possible to delete redundant solutions using numerical methods, which are almost always faster. By setting the option `"Numeric"` to `True` each gauge invariant is calculated with an accuracy of (by default) 64 digits and infinite precision (which means that Mathematica uses as many extra digits required in intermediate calculations to ensure that those 64 digits are 100% correct). A different accuracy can be set if desired using the `"Accuracy"` option.

*Example* 4.1.9. The following code

```
pentSoln = DeleteEquivalentSolutions[ solution, repD3 ];
Length @ pentSoln
```

returns

```
3
```

which means that checking for gauge invariance alone is not enough to remove redundant solutions.

---

[2]Actually, if they are equal, but since the orbits are not sorted, it is faster to check whether they intersect.

We removed redundant solutions using both the symbolic and numeric methods for each set of solutions to the pentagon equations and checked whether these agreed. This was always the case for an accuracy of 1000 digits.[3]

> **Notes 7.**
> - Computing gauge invariant monomials can be done almost completely with information from only the fusion ring. The only information the $F$-symbols provide is which $F$-symbols are 0. If these zeros are not considered up-front, they might appear in a RM denominator, resulting in problems when checking gauge equivalence.
> 
> - The algorithm above was already described in [47], but I only recently came across this paper. The paper [47] views the problem from the more general viewpoint of algebraic geometry, however.

### 4.1.9 Fixing a Unitary Gauge

The specific values of the $F$-symbols are gauge-dependent, and thus, we can manipulate the gauge to put solutions to the pentagon equations, as provided by the `SolvePentagonEquations`, in a more appealing form. A particularly interesting gauge is one for which all $F$-matrices are unitary. Not all solutions permit such a unitary gauge, though, but if they do, we would like to represent them in such a gauge.

If a set of $F$-symbols is in a unitary gauge then we know that

$$[F_d^{abc}]_f^e [\tilde{F}_d^{abc}]_e^f = \left|[F_d^{abc}]_f^e\right|^2 \tag{4.34}$$

for all admissible $F$-symbols, and vice versa. Since the left hand side of this equation is gauge-invariant we already know its value, even for a set of non-unitary $F$-symbols.

Now assume that we have a set of $F$-symbols that is not in a unitary gauge. Finding a unitary gauge then comes down to finding a set of gauge variables $\left\{g_c^{ab} \in \mathbb{C}\backslash\{0\} \,\middle|\, N_{a,b}^c \neq 0\right\}$ such that

$$[F_d^{abc}]_f^e [\tilde{F}_d^{abc}]_e^f = \left|\frac{g_e^{ab} g_d^{ec}}{g_d^{af} g_f^{bc}} [F_d^{abc}]_f^e\right|^2. \tag{4.35}$$

Since we know all the values of the $F$-symbols involved, this is a set of equations in the variables $h_c^{a,b} := \left|g_c^{ab}\right|^2$. These equations can be tackled one by one in a similar manner that we fixed the gauge before solving the pentagon equations: we pick the simplest equation, solve it, substitute the solution into the rest of the equations, and continue. At some point, equations might appear that only contain higher powers of the variables $h_c^{a,b}$ but, as was the case with fixing the gauge, we are free to choose any solution for the $h_c^{a,b}$ variable in consideration. The difference between the various solutions for the variable $h_c^{a,b}$ is that certain $F$-symbols will differ by a phase factor when comparing solutions. However, if we look at (4.35), we see that the right-hand side does not care about phases, and the left-hand side does not care about gauge transforms. Therefore, this redundancy is harmless, and we can use the first solution for the $h_c^{a,b}$ that the `Solve` function returns. It is always possible to solve for all $\left\{h_c^{a,b}\right\}$, but this only implies that the specific equations we used to find such a solution are satisfied. The final step in the algorithm consists of checking whether all equations of the form (4.35) are satisfied. If so then we can apply a gauge transform with gauge variables $g_c^{ab} = \sqrt{\left|h_c^{a,b}\right|}$ to the original set of $F$-symbols $\left\{[F_d^{abc}]_f^e\right\}$ to put it into a unitary gauge. Note that also here there is a redundancy of the phases of the various $g_c^{ab}$, but this does not matter for the same reason that the redundancy in the phases for the various $h_c^{a,b}$ do not matter.

---

[3]While the symbolic method should, in principle, always give a correct result, it is still important to check the results by other means, whenever possible

## 4.2 Solving Hexagon Equations

The hexagon equations (2.25) and (2.26) have a structure that is very similar to the pentagon equations. In particular, per solution to the pentagon equations, there are, up to gauge equivalence, only a finite number of solutions to the hexagon equations (see remark 2.33 of [27]). The same tools we used to solve the pentagon equations were also used to solve the hexagon equations. There are a few important differences, however. Since we solve the hexagon equations per solution to the pentagon equations, the only variables that appear are 1-dimensional invertible $R$-matrices. The $R$ symbols are, in particular, never 0. To solve the pentagon equations, we already fixed the gauge up to a certain degree for the $F$-symbols, and therefore, we can now only fix the remaining degrees of freedom that leave the $F$-symbols invariant. To do so, the function `RestrictMultiplicativeSymmetries` was implemented. It takes a list of symmetry transforms `symmetries`, a set of variables `variables`, and a symbol `g` that denote the gauge variables `{g[a,b,c]}` as arguments and returns a set of gauge transforms that leave the set `variables` invariant. The remaining gauge can then be fixed the same way as for the $F$-symbols. For most fusion rings, once the gauge for the $F$-symbols was fixed, no gauge freedom was left for the values of the $R$-symbols. Thus, one should not expect a significant reduction in variables. Once the gauge is fixed we can apply the usual functions such as `ReduceByBinomials`, `ReduceByLinearity`, and `IncrementalGroebnerBasis`. All this functionality is implemented in the function `SolveHexagonEquations`, which takes one argument, the fusion ring, and several options. The most important option is `"Knowns"`, which can be set to a solution to the pentagon equations. Without this option, `SolveHexagonEquations` would assume that the $F$-symbols are variables and try to find their value as far as possible. The other important options overlap with those of `SolvePentagonEquations`.

*Example* 4.2.1. The following code

```
hexSol =
  Table[
    SolveHexagonEquations[ repD3, "Knowns" -> sol ],
    { sol, pentSol }
  ];
```

```
  Map[Length] @ hexSol
```

returns

```
  { 3, 0, 0 }
```

which means that the first solution to the pentagon equations has at most three compatible braided structures, and the other ones have none. It turns out there are actually exactly three braided structures since the following code

```
  uniqueHexSol = DeleteEquivalentSolutions[ First @ hexSol, repD3 ];
  Length @ uniqueHexSol
```

returns

```
  3
```

There is one catch, however. While the function `SolveBinomialSystem` (that forms the core of `ReduceByBinomials`) works quite fine for pentagon equations, it does not perform nearly as well for hexagon equations. The cause of this problem lies with the Smith decomposition. On the one hand, the solutions to the pentagon equations often contain root expressions, i.e., expressions containing $n$th roots or even expressions containing formal roots of the form `Root[`$p(x), n$`]`. On the other hand, the matrices that result from taking a Smith decomposition often contain big integers. Eventually, these integers are the exponents with which the root expressions appear in the solution to a binomial system. If the root expressions were just integers, then Mathematica has no issues canceling fractions of gigantic integers. Fractions of root objects, raised to huge powers, form a big problem, however. Moreover, even if the binomial system can be solved, it often contains many solutions. In contrast to the pentagon equations, the main restriction to the number of solutions of the binomial equations comes from the equations containing more than two terms. Eventually, most of the solutions to the binomial equations fail to satisfy the other equations, but they do cause a blowup of required memory and time.

For some fusion rings, solving the hexagon equations by first reducing the system via binomial equations was impossible. In these cases, it was, however, quite easy to solve the equations via the `ReduceByLinearity` function in combination with the option `"SimplifyIntermediateResultsBy"` to simplify intermediate results as soon and frequently as possible.

After solutions for the hexagon equations have been obtained, we need to remove the equivalent solutions again since some discrete gauge freedom might remain for the $R$-symbols. This is done the same way as for the $F$-symbols, but now we use gauge invariants that include $R$-symbols to distinguish solutions.

## 4.3  Solving Pivotal Equations

To find pivotal structures for a given (braided) fusion system $\mathscr{C}$, we must solve the pivotal equations (2.24). While Anyonica can solve these equations, they are so simple that the built-in `Solve` function is much more efficient. As always, once a set of solutions has been found, we must eliminate redundant solutions. Since the values of the quantum dimensions $d_a = \frac{p_a}{[F_a^{aa^*a}]_1^1}$ are gauge invariants, and different for every solution, only fusion ring automorphisms can cause redundancy in the solutions. These automorphisms must be such that, up to gauge transforms, they have no effect on the $F$ and $R$-symbols and map the quantum dimensions of one solution to the quantum dimensions of another. Once this is done, the pivotal structures can be added to the data of the (braided) fusion system $\mathscr{C}$. In order to ease calculations with a graphical language, I decided to apply a gauge transform such that every pivotal coefficient of a non-self-dual particle equals 1. This implies that in the list of stored fusion categories in Anyonica, there are pivotal fusion systems that correspond to the same solution to the pentagon equations but have slightly different $F$-symbols.

## 4.4 Results

The pentagon, hexagon, and pivotal equations for all multiplicity-free fusion rings up to rank 7 were solved, and all redundant solutions were removed. For all but two rings, namely $\mathrm{PSU}(2)_{12}$ and $\mathrm{PSU}(2)_{13}$, Anyonica's functionality was sufficient. An alternative approach for the two exceptions was necessary because the systems of pentagon equations were too big to handle. The categorifications of $\mathrm{PSU}(2)_{2n+1}, n \in \mathbb{N}$ are interestingly already classified [38], and the $F$ and $R$-symbols can be easily calculated by the alatc package from E. Ardonne (see 4.6 for more info). For $\mathrm{PSU}(2n), n \in \mathbb{N}$ only the pivotal categorifications are classified (see appendix of [28]). While the same paper argues that it is most likely the case that all fusion categories are pivotal and therefore most likely the all fusion categories with $\mathrm{PSU}(2)_k, k \in \mathbb{N}$ are classified, we will not make such an assumption. We therefore have the slightly weaker result that all pivotal multiplicity-free fusion categories up to rank 7 are now classified and this classification extends to all multiplicity-free fusion categories up to rank 7 if the $\mathrm{PSU}(2)_{12}$ fusion ring only has pivotal categorifications. The $F$ and $R$-symbols for the pivotal categories with $\mathrm{PSU}(2)_{12}$ fusion rules were also obtained using alatc.

Out of the 115 multiplicity-free fusion rings of rank 7 or less, 59 are categorifiable, leading to a total of 225 inequivalent fusion systems. All of these fusion systems have solutions to the pivotal equations, and 108 of the fusion systems admit a braided structure. In total, there are 977 pivotal and possibly braided fusion systems. The number of inequivalent solutions to the pentagon equations per rank can be seen in figure 4.1. Figure 4.3 gives a complete overview of the number of pivotal and braided fusion systems per fusion ring, and figure 4.2 shows a Venn diagram of the number of systems that have a combination of any of the five properties: braided, unitary, spherical, ribbon, and modular. In Section 9, one can find a list of all multiplicity-free fusion categories with indicators for braided, unitary, spherical, ribbon, and modular structures. All categorifiable fusion rings up to rank 7 have at least one fusion category that is either braided or unitary.

From figure 4.3, it is clear that the number of (pivotal) (braided) fusion systems of the product of two fusion rings is not necessarily equal to the product of the number of (pivotal) (braided) fusion systems for each ring separately. For, e.g., the $\mathbb{Z}_2 \times$ Ising fusion ring one finds 6 solutions to the pentagon equations while the $\mathbb{Z}_2$ and Ising fusion rings have 2 solutions each. Moreover, the number of braided fusion systems and the number of pivotal fusion systems for $\mathbb{Z}_2 \times$ Ising varies per solution to the pentagon equations, which is not the case for the individual fusion rings. There are even fusion systems for $\mathbb{Z}_2 \times$ Ising that cannot be braided at all, even though all fusion systems are braided for $\mathbb{Z}_2$, and Ising separately!

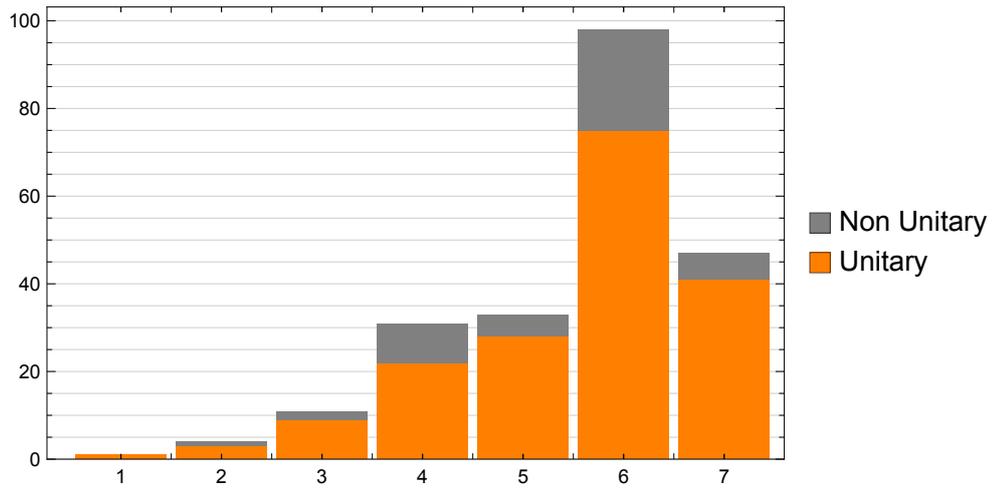

Figure 4.1: Number of solutions to the pentagon equations for multiplicity-free fusion rings per rank (horizontal axis). Unitary means that a gauge exists in which the *F*-symbols are unitary.

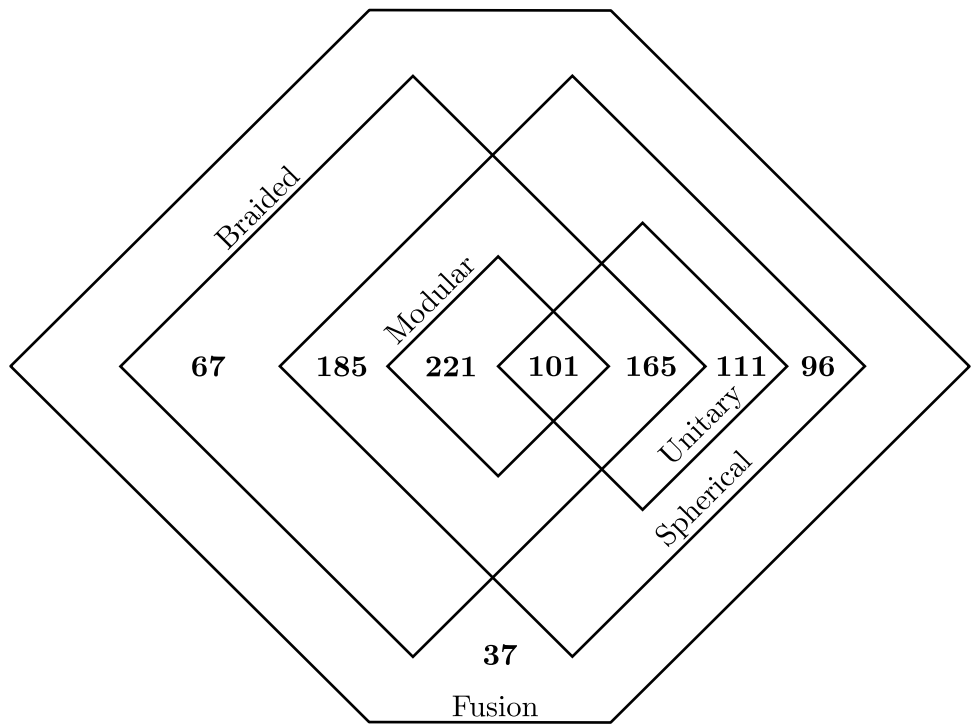

Figure 4.2: Venn diagram of the number of multiplicity-free fusion systems up to rank 7 with certain properties. By definition, the ribbon fusion systems belong to the intersection of the braided and spherical squares. Each number indicates the number of fusion systems that lie in the connected region of the Venn diagram. There are thus, e.g., 37 fusion systems with no extra properties, 185 ribbon fusion systems that are neither modular nor unitary, 101 fusion systems that are modular and unitary, 221 fusion systems that are modular but non-unitary and thus 322 modular fusion systems in total.

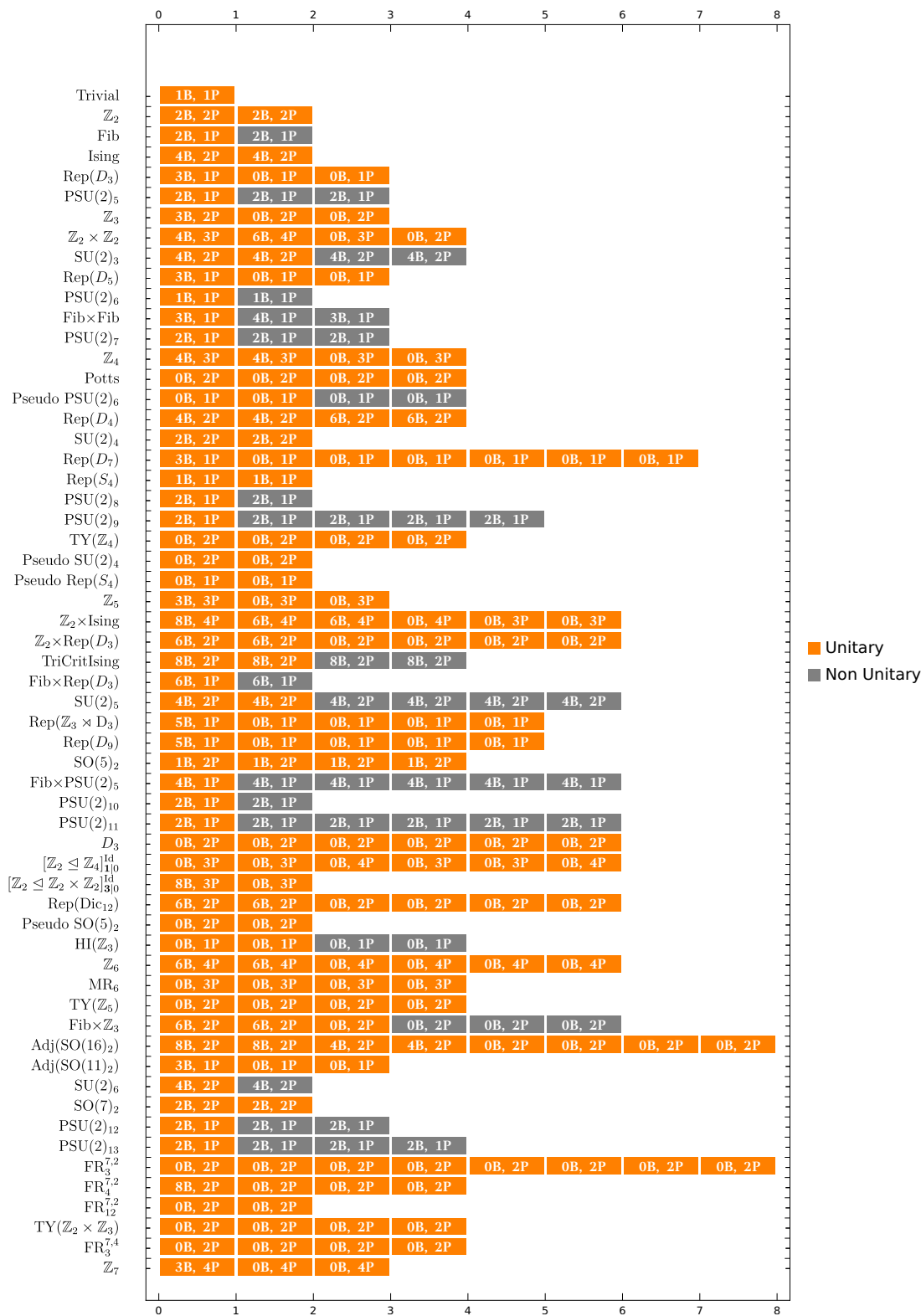

Figure 4.3: Number of solutions to the pentagon equations per fusion ring. Each rectangle corresponds to a unique fusion system. The first number in a rectangle denotes the number of braided fusion systems corresponding to that fusion system while the second denotes the number of pivotal fusion systems. Unitary (orange color) means a gauge exists in which the *F*-symbols are unitary, and **not** that all braided and pivotal fusion systems are unitary.

Almost all categorifiable fusion rings up to rank 7 all belong to one of the following groups (note that these might overlap)

- Finite group rings:
  - $\mathbb{Z}_k$ for $k = 1, ..., 7$, $\mathbb{Z}_2 \times \mathbb{Z}_2$, $\mathbb{Z}_2 \times \mathbb{Z}_3$
  - $D_3$.

- Representation rings of finite groups:
  - $\text{Rep}(D_k)$ for $k = 3, 4, 5, 7, 9$,
  - $\text{Rep}(S_4)$,
  - $\text{Rep}(\mathbb{Z}_3 \rtimes D_3)$,
  - $\text{Rep}(\text{Dic}_{12})$.

- Fusion rings coming from the representation theory of quantum groups at roots of unity:
  - $\text{PSU}(2)_k$ for $k = 3, ..., 13$ (note that Fib = $\text{PSU}(2)_3$),
  - $\text{SU}(2)_k$ for $k = 2, ..., 6$ (note that Ising = $\text{SU}(2)_2$),
  - $\text{SO}(2k+1)_2$ for $k = 1, 2, 3$

- Products of rings in one of the groups above:
  - $\mathbb{Z}_2 \times \mathscr{R}$ with $\mathscr{R}$ any of the rings from the groups above with rank up to 3,
  - Fib $\times \mathscr{R}$ with $\mathscr{R}$ any of the rings from the groups above with rank up to 3 (note that TriCritIsing = Fib $\times$ Ising).

- Extensions of rings in one of the groups above:
  - TY($G$) with $G$ any commutative group ring up to rank 6 (note that Potts = TY($\mathbb{Z}_3$)),
  - HI($\mathbb{Z}_3$).

- Zestings (see [21] for a definition) of rings in one of the groups above:
  - According to [63], Pseudo $\text{SU}(2)_4 = \text{Zest}(\text{SU}(2)_4)$, Pseudo $\text{SO}(5)_2 = \text{Zest}(\text{SO}(5)_2)$, and the three rings $[\mathbb{Z}_2 \trianglelefteq \mathbb{Z}_4]^{\text{Id}}_{1|0}$, $[\mathbb{Z}_2 \trianglelefteq \mathbb{Z}_2 \times \mathbb{Z}_2]^{\text{Id}}_{3|0}$, and $\text{MR}_6$ are zestings of $\mathbb{Z}_2 \times$ Ising.

- Adjoint fusion rings (see [25] Chapter 3) of fusion rings coming from quantum groups at roots of unity: Adj(SO(16)$_2$), Adj(SO(11)$_2$)

The remaining fusion rings are the following: Pseudo PSU(2)$_6$, Pseudo Rep($S_4$), FR$_3^{7,2}$, FR$_4^{7,2}$, FR$_{12}^{7,2}$, and FR$_3^{7,4}$.

According to [63], Pseudo PSU(2)$_6$ comes from the even part of the 1-supertransitive subfactor of index $3 + 2\sqrt{2}$ with non-self-adjoint objects [62].

The structure of the multiplication tables and the Frobenius-Perron dimensions (and private conversations with E. Rowell) indicate that FR$_{12}^{7,2}$ might be a zesting of SO(7)$_2$, and FR$_4^{7,2}$ might be a zesting of Adj(SO(11)$_2$). From a conversation with S. Palcoux on Stackoverflow, we know that according to the paper [26] Pseudo Rep($S_4$), FR$_3^{7,2}$, and FR$_3^{7,4}$ should be *weakly group-theoretical*.

### 4.4.1 Correctness of the Results

For each fusion ring, a notebook[4] is kept with the code we used to solve the equations, together with a log file that can be used to access intermediate results (see figures 4.4, 4.5, and 4.6). The log file is a notebook designed to be readable and contains clickable hyperlinks to the notebooks with intermediate results. For almost any function, there is the possibility to create log files. To do so, one wraps the code for which a log file is desired inside the `PrintLog` function. The function `PrintLog` takes the options `"Directory"` and `"FileName"`, which are by default set to respectively a temporary directory and a filename containing the date and time of execution.

*Example* 4.4.1. To create a log of the steps taken during the evaluation of `SolvePentagonEquations[repD3]` one can use the following code

```
PrintLog[ SolvePentagonEquations[repD3] ]
```

---

[4]These are available upon request rather than online since they take up quite a lot of memory.

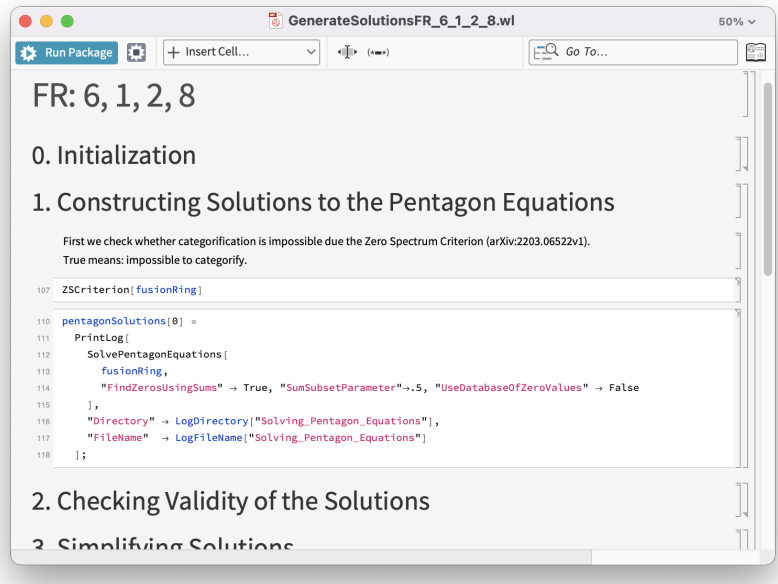

Figure 4.4: For each fusion ring, a notebook that contains all steps from the categorification process is kept. This is an example of such a notebook for the $FR_8^{6,1,2} = HI(\mathbb{Z}_3)$ fusion ring. We only knew about the zero-spectrum criterion when this ring was categorified.

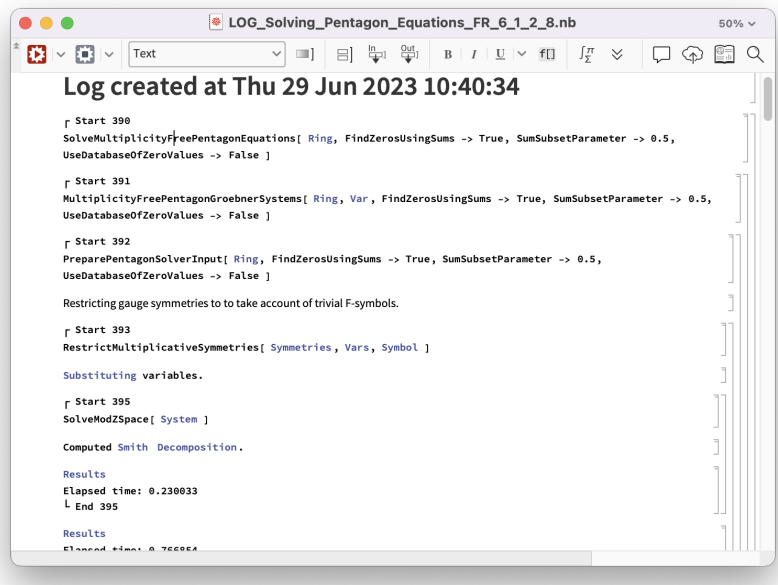

Figure 4.5: For each fusion ring, several log files were kept. These files log the steps taken by various functions and give warnings whenever decisions are made based on the equality of symbolic expressions. This is an example of such a log file for the function `SolvePentagonEquations`, applied to the $FR_8^{6,1,2} = HI(\mathbb{Z}_3)$ fusion ring

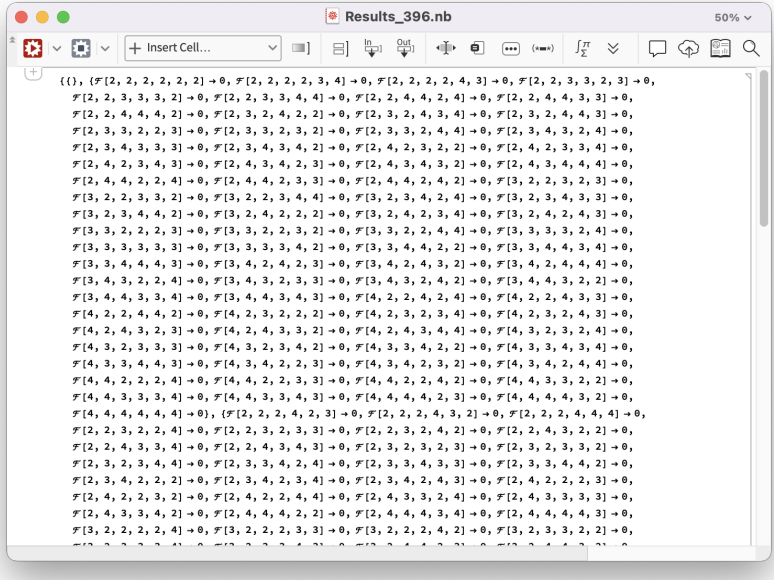

Figure 4.6: The log files contain clickable hyperlinks that open notebooks with the relevant data. This way, it is easy to check the arguments and results of intermediate computations. The figure shows the results of the `FindZeroValues` function for the pentagon equations corresponding to the $\mathrm{FR}_8^{6,1,2} = \mathrm{HI}(\mathbb{Z}_3)$ fusion ring

## 4.5 Naming of Fusion Categories

To be able to refer to fusion systems or categories, the following naming scheme is used. A fusion system is written down as follows

$$[\text{RingName}]_{i,j}^k$$

where RingName is the name of its fusion ring, and $i, j, k$ are natural numbers that distinguish between the solutions to the pentagon, hexagon, and pivotal equations, respectively. If there are no solutions to the hexagon equations, then $j = 1$. For example, $[\text{Rep}(D_3)]_{1,2}^1$ is the fusion system with fusion ring $\text{Rep}(D_3)$, corresponding to the *first* solution to the pentagon equations, the *second* solution to the hexagon equations, and the *first* solution to the pivotal equations. This scheme only makes sense if there is a canonical ordering on the solutions to the consistency equations. The ordering applied on the pentagon solutions is based on the list of data $(b_B, b_U, b_S, b_R, b_M, w_F)$, where

- $b_B, b_U, b_S, b_R, b_M$ are 1 if the fusion system can be expanded to respectively a braided, unitary, spherical, ribbon, modular fusion system, and 0 otherwise.

- $w_F$ is a canonical weight function that calculates the list of values of the gauge-invariant rational monomials (in a specific order) and rewrites every gauge-invariant number $re^{i\phi}, \phi \in [0, 2\pi[$ as a couple $(-r, -\phi)$.

The values of these lists then sort the solutions in decreasing lexicographic order. In particular, the solutions that admit braiding have a lower $i$ index than those that do not. Within these classes, the unitary solutions have a lower index than the non-unitary ones, and so on.

The following is why the weight $w_F$ can be considered unique. For every fusion ring, there exists a canonical basis, determined as in Section 3.2. This ordering implies a lexicographic ordering on the labels of the $F$-symbols. In our case we let $[F^{a_1 b_1 c_1}_{d_1}]^{e_1}_{f_1} < [F^{a_2 b_2 c_2}_{d_2}]^{e_2}_{f_2} \Leftrightarrow (a_1, b_1, c_1, d_1, e_1, f_1) < (a_2, b_2, c_2, d_2, e_2, f_2)$ in the list of formal $F$-symbols (i.e. $F$-symbols without values). Each gauge-invariant rational monomial can be expressed via a list $l$ of $n$ exponents, where $n$ is the number of $F$-symbols and $l_m \in \mathbb{Z}$ denotes the exponent of the $m$th symbol for the rational monomial. For example, for Fib we have that $[F^{112}_2]^1_2 [F^{122}_1]^2_1$ is a gauge-invariant rational monomial and in the list of $F$-symbols $[F^{112}_2]^1_2, [F^{122}_1]^2_1$ have positions $2, 4$ out of $15$. The list $l$ for this rational monomial would thus be $(0, 1, 0, 1, 0, \ldots, 0)$.

The gauge-invariant monomials are then sorted lexicographically by the values of $(\sum_m |l_m|, \text{reverse}(l))$, where $(\text{reverse}(l))_m = l_{n-m}$, in increasing order[5] and the resulting list is a canonical list of gauge-invariant rational monomials.

For all fusion systems with equivalent $F$-symbols, the solutions to the hexagon equations are sorted via the list of data $(b_S, b_R, b_M, w_R)$ where $w_R$ is a weight function that works analogous to $w_F$, except for the fact that it uses gauge-invariant rational monomials that include $R$-symbols.

For all fusion systems with equivalent $F$ and $R$-symbols, the solutions to the pivotal equations were sorted lexicographically by $(b_S, b_R, b_M, w_P)$ in decreasing order (where the booleans now say that the fusion system is respectively spherical, ribbon, modular). Here $w_P$ is a weight function that calculates the quantum dimensions $d_a$ and rewrites every $d_a = re^{i\phi}, \phi \in [0, 2\pi[$ as a couple $(-r, -\phi)$.

## 4.6 Other Software to Work With Fusion Rings and Fusion Categories

### 4.6.1 Predecessors

Anyonica is partly based on software that has not been published. P. Bonderson's thesis [10] describes in some detail basic techniques for solving multiplicity-free pentagon and hexagon equations on a computer, and we are using several of those. J. K. Slingerland wrote Mathematica code based on those techniques, and by using their code, they produced the tables of pentagon/hexagon solutions in the back of Bonderson's thesis.

Subsequently, J. K. Slingerland found all multiplicity-free fusion rings up to rank 6 with a search algorithm similar to that used by Gepner [41] (but not specialized to rings with modular properties). He also found all solutions to the pentagon and hexagon equations for all categorifiable rings up to rank 6, although it was not clear that these were all solutions because unitarity was assumed in finding the zero $F$-symbols.

---

[5]The reverse function is used to ensure that the rational monomials that also contain $R$-symbols are sorted behind the ones without $R$-symbols.

Non-unitary systems with those same zeros patterns were also found, but there could have been systems with another configuration of zero $F$-symbols. Pivotal and spherical structures were not systematically considered. Neither the code nor the full results were published.

F. Verstraete also programmed software in Matlab to solve the pentagon and hexagon equations. Instead of fixing the gauge separately, a Smith decomposition was used to solve the logarithm of the binomial equations and to find a subspace of the solution set that is orthogonal to the space of gauge transforms. Under the assumption that a zero value can only appear for one $F$-symbol, the pentagon equations could also be solved numerically. To find a unitary gauge, the solver searched for a solution to the algebraic equations describing the unitarity of $F$ matrices. As far as I know, neither the code nor the results were published.

While Anyonica is written from scratch, it is still based on several techniques of P. Bonderson, J. K. Slingerland, and F. Verstraete. The method used for gauge fixing is the same as that of J. K. Slingerland and P. Bonderson. The reduction of systems of linear equations is based on that of P. Bonderson and J. K. Slingerland but has been generalized in the sense that Anyonica also divides by non-monomials. This requires more bookkeeping since assumptions need to be saved and checked recursively, but there is a payoff in the level of reduction that can be achieved. Anyonica's method for solving binomial equations is the same as that of F. Verstraete, except that Anyonica fixes the gauge before solving these equations. While doing both in one go is more elegant, it is more efficient not to do so. Solving the eventual reduced system of equations is done the same way by every author, namely by using a built-in solver based on Gröbner basis calculations.

Some of the things of which we think Anyonica supersedes its predecessors are

- the fact that no assumptions (in particular no unitarity) on the solutions are made,

- all configurations of 0 $F$-symbols can be found,

- finding a unitary gauge, which used to be very hard, is now almost trivial,

- any operation can be logged so it is easy to see when and why solutions are rejected,

- the code consists of modular pieces that can be used for any system of polynomial equations. The programming style is almost completely functional, so running the same code twice should result in the same outcome,

- all fusion rings and categories (including pivotal, spherical, and modular data) are available, easy to access, easy to work with due to various extra functions, and combined into a single package.

There is still lots of functionality left to implement such an algorithm to compute the center of a category, check whether two categories are Morita-equivalent, find possi-ble equivariantizations and de-equivartiantizations of a category, compute zestings of fusion categories, etc.

### 4.6.2 Other Packages

Anyonica is not the only software that provides tools to work with fusion rings and fusion categories. Some of the other software packages are the following

- Recently, W. Aboumrad released a software package that solves the pentagon and hexagon equations for specific fusion rings [1]. The functionality is more oriented towards applications in topological quantum computation and, at the moment, seems only to categorify multiplicity-free fusion rings coming from the representation theory of quantum groups at roots of unity. The nice thing about the software is that it is entirely open source, uses parallelization, and can be paused and resumed in a user-friendly way at any time. It also cleverly handles the creation of the pentagon equations: these are constructed at the same time they are being solved. At the moment, Anyonica creates all the equations in one go, and then it starts simplifying. This leads to unnecessary memory consumption, so I hope to implement a more clever method soon. The way that pentagon and hexagon equations are being dealt with is clever, but some assumptions, such as the existence of braiding and orthogonality of the $F$-symbols, do not hold for general fusion categories.

- There is another package for finding fusion categories related to the representation theory of quantum groups at roots of unity: alatc (affine Lie algebra and tensor categories), written by E. Ardonne [6, 5]. This package finds fusion categories and braided structures based on calculating Clebsch-Gordan coefficients. It can handle fusion rings with multiplicity but comes with the downside that, just like Anyonica, it is implemented in the Wolfram Language. It can find $F$-symbols for fusion rings that are much larger than Anyonica is capable of handling. There are plans to incorporate alatc into Anyonica.

- F. Meuser and U. Thiel are developing tensorcategories.jl [69, 70, 71], a package that provides a framework as well as examples for computations in the realm of categories.

- T. Hagge and M. Titsworth constructed a Mathematica package (referenced in the introduction of [47]) that contains data on various categories, including some of rank 8 and higher. It also contains data for rigidity, pivotal, and braiding structures, calculates $S$ and $T$ matrices, and can check for monoidal equivalence of two categories.

- J. C. Bridgeman created the GitHub repository "smallRankUnitaryFusionData" [12], which contains data on all multiplicity-free unitary fusion categories (and their braidings) up to rank six. This data has also been incorporated in the Ten-

sorKit.jl [46] package from J. Haegeman via the extension CategoryData.jl [23] from L. Devos.

- Kac is a software package that computes fusion rules for rational conformal field theories based on affine Lie algebras. It can be accessed at `https://www.nikhef.nl/~t58/Site/Kac.html`.

- The SageMath code to check various categorifiablity criteria listed in 4.1.1 can be found in the paper [63].

- For the classification of modular data of integral fusion rings up to rank 12 [3] the Normaliz software was used. The integrality constraint puts bounds on the sums of certain fusion coefficients, and Normaliz is very fast at finding all integer lattice points within a polytope defined by such bounds.

# Part III

# Anyons on Graphs

# Chapter 5

# Anyons on Graphs

In Chapter 1 we saw how braiding anyons in a plane could be useful for topological quantum computation. A significant advantage of topological quantum computing is its protection against decoherence. There is another way in which topological protection can be acquired, though. Recently several experiments have demonstrated evidence for the existence of so-called Majorana modes at opposite ends of a nanowire [76, 86, 22, 14, 36, 2]. To measure the properties of one of the modes, one needs to measure those of the other one at the same time. The longer the wire, the lower the chance this accidentally happens via interaction with the environment. These wires could, therefore, provide a way to store quantum information. Sadly, there is no way to braid these states around each other on a single wire. But what if one connects multiple wires to create a network? The modes then still move in one dimension, but depending on the shape of the network, they might actually be braided around each other. While planar braiding has been studied for quite some time [7], graph-braiding has received less attention. Only recently have people started investigating the mathematical theory of graph braiding [31, 32, 59, 4]. The idea of describing anyons on graphs is even more recent. In 2023 A. Conlon and J. K. Slingerland published a paper discussing the braiding of anyons on a tri and tetrajunction [15]. The scope of this paper got expanded to braiding on more general networks by T. Maciazek, A. Conlon, J. K. Slingerland, and me [68]. In this chapter, a shorter account of that paper will be presented.

In particular, we will set up the theory for several of the simplest graphs, and we will see that consistency equations differ from one graph to another. Braiding on the two simplest graphs, the circle and the trijunction, will be considered respectively in Sections 5.1 and 5.2, while braiding on a lollipop graph is discussed in Section 5.3. The chapter concludes with Section 5.4 that shows how to solve the graph-braid equations for the circle, trijunction, and lollipop graphs, and discusses theirs solutions for certain fusion systems.

In the following, we will assume that we are always working with a fusion category whose fusion ring is commutative. We will still call the 'particles' in question anyons (and their respective fusion systems anyon models), but this terminology should be taken with a grain of salt because no planar braiding may exist.

**Notes 8.**
- The exposition will be very naive and there are a myriad of questions one should ask themselves, such as
  - Why are the fusion systems for graph-braiding are the same as for planar braiding? Isn't it possible that one also needs new pentagon equations?
  - How do we know that these consistency equations are all consistency equations?
  - Why are the paths of particles denoted by lines instead of ribbons, and why do some of the paths contain seemingly unnecessary bends?
  - What about bigger graphs?
  - What about non-planar graphs?
  - Etc.

  Discussions on these topics are delibirately avoided since they are quite technical and can be read in the paper [68] of T. Maciazek, A. Conlon, J. K. Slingerland, and me.

- On a graph it might actually make sense to look at non-commutative fusion because the particles are bound to move in one dimension. We have not managed to develop this idea into a proper theory of non-commutative fusion, though.

## 5.1 Anyons on a Circle

Let us start by considering braiding when anyons are restricted to move on a circle $S^1$. On the circle, the only possible permutations of anyons that can be performed are cyclic permutations. If one would be braiding in the plane then for two particles, $a, b$ with total charge $c$, such a permutation would be described by the $R$-symbol $R_c^{ab}$. For movement of particles that is restricted to a circle, we will represent this permutation by the $D$-symbol $D_c^{ab}$ (see figure 5.1). We will assume that these $D$-symbols are unitary, i.e. $D_c^{ab} \in \mathrm{U}(1)$.

Similar to the planar case we demand that braiding around a circle behaves naturally with respect to the fusion of particles. By this we mean that if we move particles around one by one then this should have the same effect as moving the fusion product of those particles around. This comes down to demanding that the diagrams in figures 5.2 and 5.3 commute.

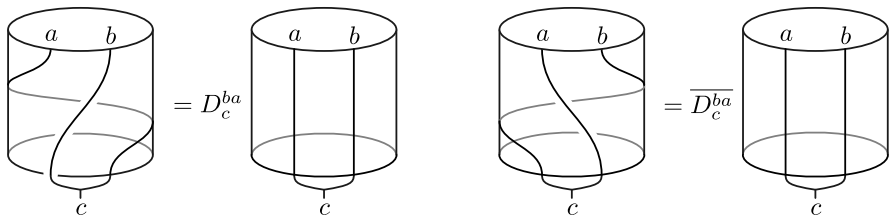

Figure 5.1: The $D$-symbol represents the braiding of a single anyon around the circle.

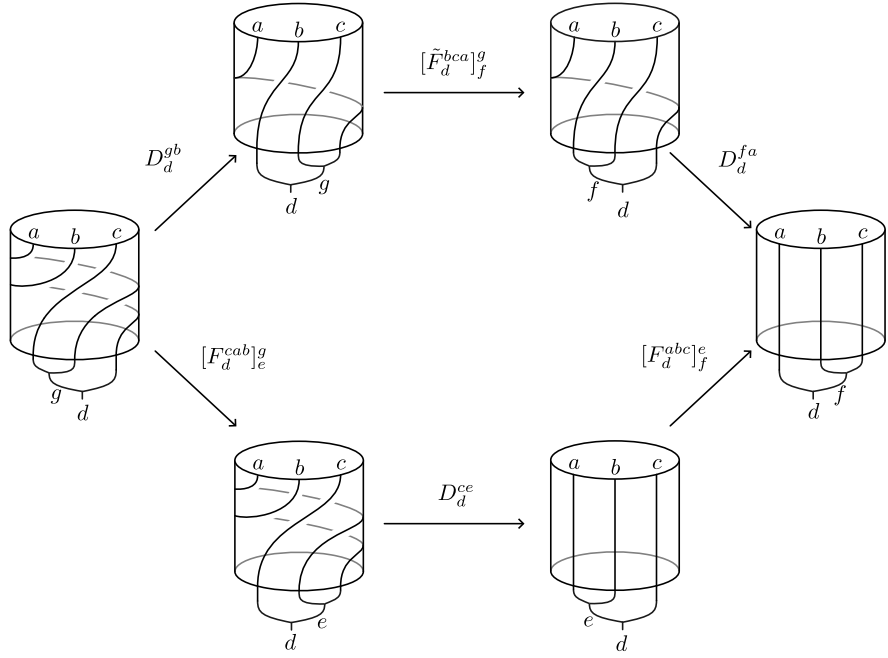

Figure 5.2: The naturality condition for braiding clockwise around the circle

This leads to two sets of consistency relations which will be called the *D*-hexagon equations, or circle equations:

$$D_d^{gb}[\tilde{F}_d^{bca}]_f^g D_d^{fa} = \sum_e [F_d^{cab}]_e^g D_d^{ce} [F_d^{abc}]_f^e, \tag{5.1}$$

$$\overline{D_d^{gb}}[F_d^{cab}]_e^g \overline{D_d^{ec}} = \sum_f [\tilde{F}_d^{bca}]_f^g \overline{D_d^{af}} [\tilde{F}_d^{abc}]_e^f. \tag{5.2}$$

Equations (5.2) are equivalent to equations (5.1), though. Indeed, let $c = 1$ in (5.1). Then we find that

$$D_e^{ab} D_e^{ba} = D_e^{1e}. \tag{5.3}$$

By setting $b \to f, e \to d$ and inverting two of the *D*-symbols in (5.3) we obtain that $\overline{D_d^{af}} = D_d^{fa} \overline{D_d^{1d}}$. By substituting $\overline{D_d^{af}}$ in the RHS of (5.2) and inserting $1 = D_d^{gb} \overline{D_d^{gb}}$ we see that

$$\sum_f [\tilde{F}_d^{bca}]_f^g \overline{D_d^{af}} [\tilde{F}_d^{abc}]_e^f = \overline{D_d^{gb}} \sum_f \left( D_d^{gb} [\tilde{F}_d^{bca}]_f^g D_d^{fa} \right) \overline{D_d^{1d}} [\tilde{F}_d^{abc}]_e^f, \tag{5.4}$$

where the factor in parenthesis is nothing but the LHS of (5.1). By using (5.1) we see

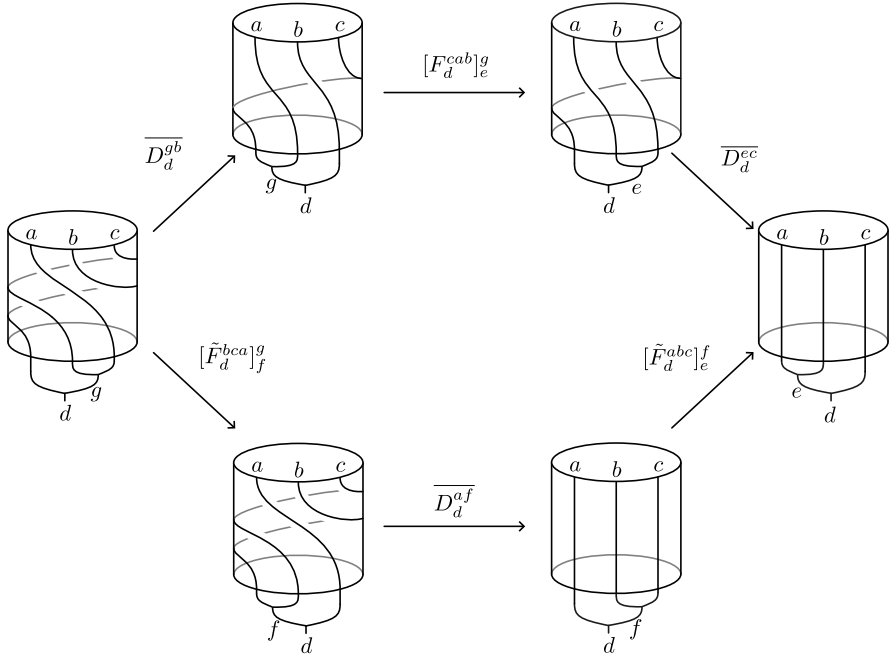

Figure 5.3: The naturality condition for braiding anti-clockwise around the circle

that the RHS of (5.1) is nothing but

$$\overline{D_d^{gb}} \overline{D_d^{1d}} \sum_{f,e'} [F_d^{cab}]_{e'}^g D_d^{ce} [F_d^{abc}]_f^{e'} [\tilde{F}_d^{abc}]_e^f = \overline{D_d^{gb}} \overline{D_d^{1d}} \sum_{e'} \delta_{e'}^e [F_d^{cab}]_{e'}^g D_d^{ce'} \quad (5.5)$$

$$= \overline{D_d^{gb}} [F_d^{cab}]_e^g D_d^{ce} \overline{D_d^{1d}} \quad (5.6)$$

$$= \overline{D_d^{gb}} [F_d^{cab}]_e^g \overline{D_d^{ec}} \quad (5.7)$$

which equals the LHS of (5.2).

Interestingly, in contrast to the $R$-symbols, we only have that $D_a^{a1} = 1, \forall a$. $D_a^{1a}$ can take other values because the particle's frame also makes a full twist upon moving around the circle. The $D_a^{1a}$-symbols are the topological twists for the graph braid models. These are more general than the classical topological twists $\theta_a$ since any solution to the planar hexagon equations is automatically a solution to the $D$-hexagon equations. For e.g., anyons with $\mathbb{Z}_3$ fusion rules the traditional twists $\theta_a$ are always third roots of unity, while the solutions to the $D$-hexagon equations for these anyons also allow twists that are a ninth root of unity.

In [68] it is shown that the $D$-hexagon equations are sufficient to provide coherence, i.e. the $D$-hexagons imply that any combination of $F$ and $D$ moves between two given graph-braided fusion trees on a cirle must always be the same.

In Section 5.4.2 we present solutions of the $D$-hexagon equations for some low-rank anyon models. Interestingly, we always found a finite number of solutions. Whether there exists an Ocneanu-type result for these equations is not known to us.

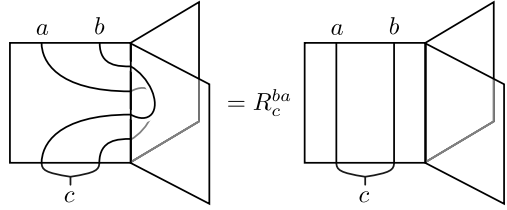

Figure 5.4: The exchange of two particles on a trijunction is governed by a set of $R$-symbols.

## 5.2 Anyons on a Trijunction

When considering anyons that are restricted to move on a trijunction the situation is more complex. The main reason is that extra symbols appear whenever we increase the number of particles. This is not necessary for planar and circle braiding since the coherence theorem identifies all combinations of simple braids, that have the same outcome, with each other. We have not managed to prove coherence for the trijunction so we need to investigate the theory of trijunction braiding per number of particles. The situation is not as bad as it sounds, though. Just like it is possible to express any planar braid operation on $n$ particles via a sequence of braids on 2 particles, there are also relations that relate the braiding operations on, e.g., 4 particles to the braiding operators on 3 particles. T. Maciazek, A. Conlon, J. K. Slingerland and me hypothesize that as the particle number increases, these relations will be strong enough to provide coherence for any particle number, given that we have coherence for 4 particles. For a more in-depth discussion on this topic we refer to [68]. We will now investigate the theory of anyons on a trijunction for up to four particles. Here, some extra assumptions will be made in order to keep the exposition more readable. In particular, an edge of the trijunction will be fixed as the base edge where all particles reside, and we will assume that after a braid, all particles return to that base edge. This way, we avoid the subtle question of how to represent states of particles living on a different edge. We also assume that the Hilbert space of these particles is the same as if the particles would reside in the plane (which are interpreted as lying on a line as in figure 1.1). Section 5.2.1, 5.2.2 and 5.2.3 consider a system of respectively two, three, and four particles. Each time an extra particle is added, new symbols and consistency equations arise.

### 5.2.1 Two Particles on a Trijunction

For this section, we will assume that the trijunction is configured as in figure 5.4: the base edge points to the left, the back edge is called edge (1), the front edge is called edge (2), latin characters label particles, and the vertical dimension represents time, which flows from bottom to top. For two anyons $a, b$ with total charge $c$ on the base edge the situation is quite simple. There are two (inverse) ways in which these particles can exchange. We can move $a$ to edge (1), $b$ to edge (2), $a$ back to the base, and then $b$ back to the base, or we can do it the other way around. These are represented by unitary $R$-symbols as shown in figure 5.4.

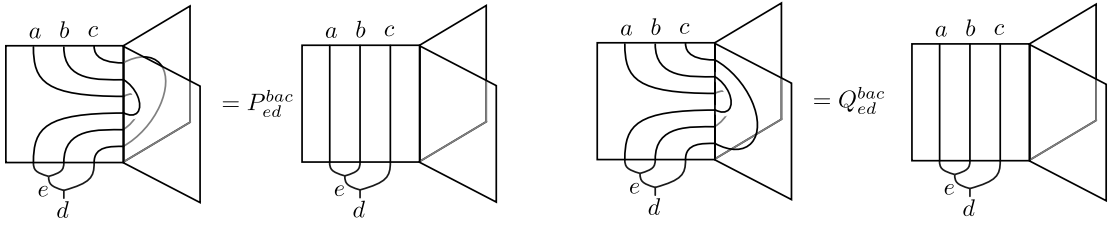

Figure 5.5: The exchange of particles *a* and *b* can be done in two topologically different ways. These moves are represented by *P* and *Q*-symbols.

> **Note 10.** We chose to have the initial configuration of the particles on the bottom in such a way that after the braid is applied, we end up with the labels in lexicographic order at the top. This will be the convention for the rest of the chapter.

These *R*-symbols should not be confused with the *R*-symbols for planar braiding. As a matter of fact, for two particles we don't even have consistency equations that the *R*-symbols need to satisfy. This changes when we consider systems with three anyons.

### 5.2.2 Three Particles on a Trijunction

Let $a, b, c$ be three particles on the base vertex of a trijunction, such that $c$ lies closest to, and $a$ lies furthest from the vertex. There are now three distinct braid operations. Firstly, we can still exchange $b$ and $c$ the same way as in the two particle case. To exchange $a$ and $b$, however, we will need to move $c$ out of the way. As shown in figure 5.5, this can be done in two inequivalent ways. We will represent such exchanges by the symbols $P, Q$. Here, we choose to allow their values to depend on all particles in the fusion tree. In [68], we also mention the simplified case where each symbol that represents a braid on a system with $n$ particles depends on only four labels: two labels for the exchanged particles, one for the total charge of these particles, and one for the total charge of the particles moving out of the way. While this simplifies matters, we will not do this here since the theory without such simplifications is more general and still not too involved.

For three particles on a trijunction, we can now demand that fusion interacts naturally with braiding, i.e., braiding a product of particles should be equal to braiding the individual particles one by one. By demanding that the diagram in figure 5.6 commutes we arrive at the *Q*-hexagon equations,

$$R_g^{ca}[\tilde{F}_d^{bac}]_e^g Q_{ed}^{bac} = \sum_f [\tilde{F}_d^{bca}]_f^g R_d^{fa}[\tilde{F}_d^{abc}]_e^f. \tag{5.8}$$

Just like for the circle, there is another diagram for the *Q* braid but it turns out that the equations it produces are equivalent to (5.8). Via a similar deduction the *P*-hexagon equations,

$$P_{gd}^{cab}[F_d^{acb}]_f^g R_f^{cb} = \sum_e [F_d^{cab}]_e^g R_d^{ce}[F_d^{abc}]_f^e, \tag{5.9}$$

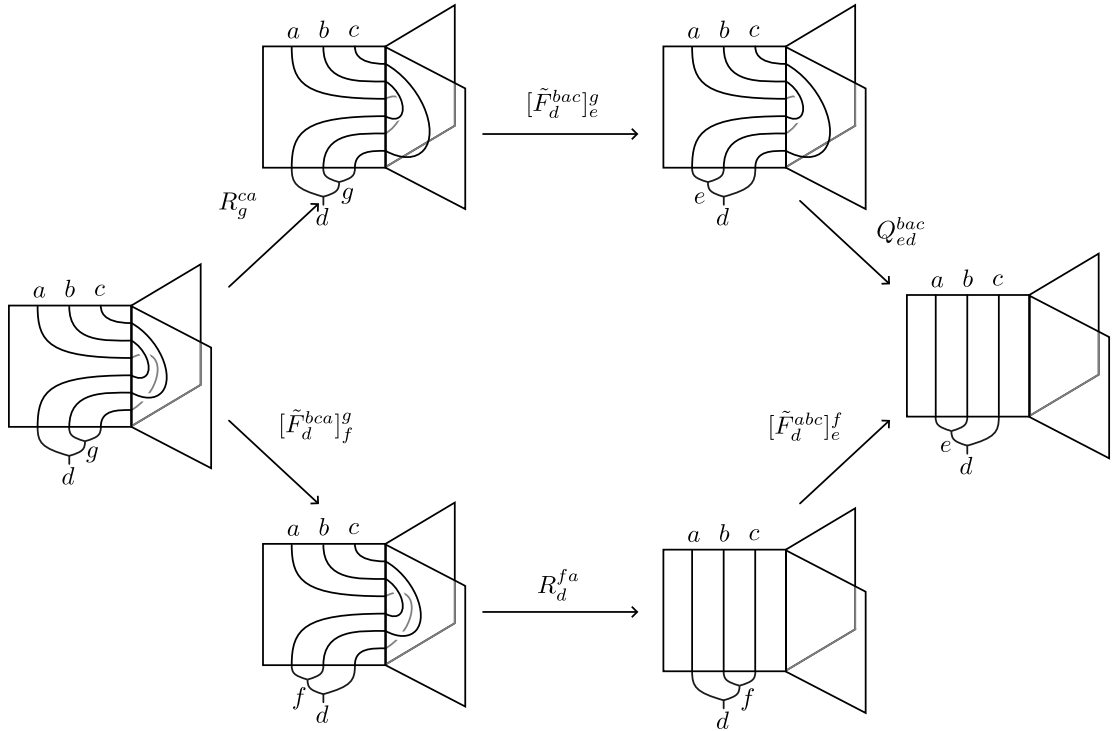

Figure 5.6: One of the diagrams that demands naturality for the Q braid.

can be derived.

The *P* and *Q* hexagon equations often have an infinite number of solutions because there are no constraints on the *R*-symbols besides being elements of U(1).

> **Note 11.** Although we are working with a trijunction, similar arguments apply to a star graph with more edges. With an increasing number of edges we get an increasing number of new variables but as was shown in [15] these do not interact with each other. The solutions for the different sets of equations are all the same as those for a single set of *P* and *Q*-hexagon equations.

### 5.2.3 Four Particles on a Trijunction

For four particles, the situation becomes more complex. When we exchange two particles we define the symbols using a basis in which a common fusion vertex joins them. In the case of three particles there is only one fusion tree that allows this situation. For four particles, however, there might be multiple fusion trees. If e.g., *a* and *b* exchange, a common fusion vertex must join them. However, then *c* and *d* can either be fused one by one with this vertex, or first fused together and only then fused with this vertex. The former allows us to move *c* and *d* to different vertices, while the latter only allows us to move *c* and *d* to the same vertex. We will, therefore, define the new set of braid symbols via braids on a left-ordered fusion basis, where the other particles have the freedom to move to different edges.

If we assume that the the *R*, *P,* and *Q*-symbols govern the braiding of the three particles closest to the vertex, then we only need to define symbols that exchange the two particles furthest from the vertex. There are four ways in which the two particles closest to the vertex can be moved out of the way, and thus four new types of braid operations. The action of these operations, which we denote by the symbols $X$, $Y$, $A$, and $B$, on the particles $b$, $a$, $c$, $d$ are the following:

$X$: moves $d$ and $c$ to edge (1), applies an $R$ move on $b$ and $a$, and moves $c$ and $d$ back,

$Y$: moves $d$ and $c$ to edge (2), applies an $R$ move on $b$ and $a$, and moves $c$ and $d$ back,

$A$: moves $d$ to edge (1), $c$ to edge (2), applies an $R$ move on $b$ and $a$, and moves $c$ and $d$ back (without braiding $c$ and $d$),

$B$: moves $d$ to edge (2), $c$ to edge (1), applies an $R$ move on $b$ and $a$, and moves $c$ and $d$ back (without braiding $c$ and $d$).

In [68] it is shown that demanding naturality comes down to demanding that at least the following equations hold

$$X^{bacd}_{fge}\delta^g_{g'} = \sum_l [F^{fcd}_e]^g_l P^{bal}_{fd} [\tilde{F}^{fcd}_e]^l_{g'}, \tag{5.10}$$

$$Y^{bacd}_{fge}\delta^g_{g'} = \sum_l [F^{fcd}_e]^g_l Q^{bal}_{fe} [\tilde{F}^{fcd}_e]^l_{g'}, \tag{5.11}$$

$$A^{badc}_{fje}\delta^j_{j'} = \sum_{g,l,l'} [\tilde{F}^{fdc}_e]^{l'}_{j'} (R^{dc}_{l'})^{-1} [F^{fdc}_e]^j_l R^{dc}_l [\tilde{F}^{fcd}_e]^l_g B^{bacd}_{fge} [F^{fcd}_e]^g_{l'}, \tag{5.12}$$

$$B^{cabd}_{nge}\delta^n_{n'}\delta_{gg'} = \sum_{f,h,k} [F^{cab}_g]^n_f Q^{cfd}_{ge} [F^{abc}_g]^f_h [F^{ahd}_e]^g_k (Q^{cbd}_{hk})^{-1} [\tilde{F}^{ahd}_e]^k_{g'} [\tilde{F}^{acb}_{g'}]^h_{n'}. \tag{5.13}$$

These look asymmetric, but that is mainly because these four equations come from a much larger set of consistency equations. Many of the equations in this more extensive set turned out to be equivalent, though, and eventually only equations (5.10), (5.11), (5.12), and (5.13) are independent. We do not include the commutative diagrams for these equations since they are too big to depict properly.

It is interesting to note the particular way in which the consistency equations split into two parts. If, for an equation, the LHS is zero, then the equation becomes a consistency equation for the three particle symbols (for equation (5.12) we assume that the value of $B$ from (5.13) has been substituted). If the LHS is non-zero, the equation can be seen as the definition of the symbols $X$, $Y$, $A$, or $B$ appearing on the LHS. The same trick can be applied for the $P$ and $Q$-hexagon equations: by moving all the $F$-symbols to the RHS, these equations split in, on the one hand, consistency relations for the $R$-symbols, and on the other hand, in definitions for the $P$ and $Q$-symbols. In this sense, there are no new symbols at all. However, we get new consistency equations each time we add a particle. If it can be shown that the validity of such equations for any particle number follows from that of those for a certain finite number of particles, then we would have a coherence theorem.

The trijunction equations for three and four particles were solved for several fusion systems. An overview of the solutions from Section 5.4.2, and an in-depth discussion for the fusion systems based on the fusion rings Ising, $\mathbb{Z}_2 \times \mathbb{Z}_2$, and $\text{TY}(\mathbb{Z}_3)$ can be found, respectively, in Sections 10.1, 10.2, and 10.3.

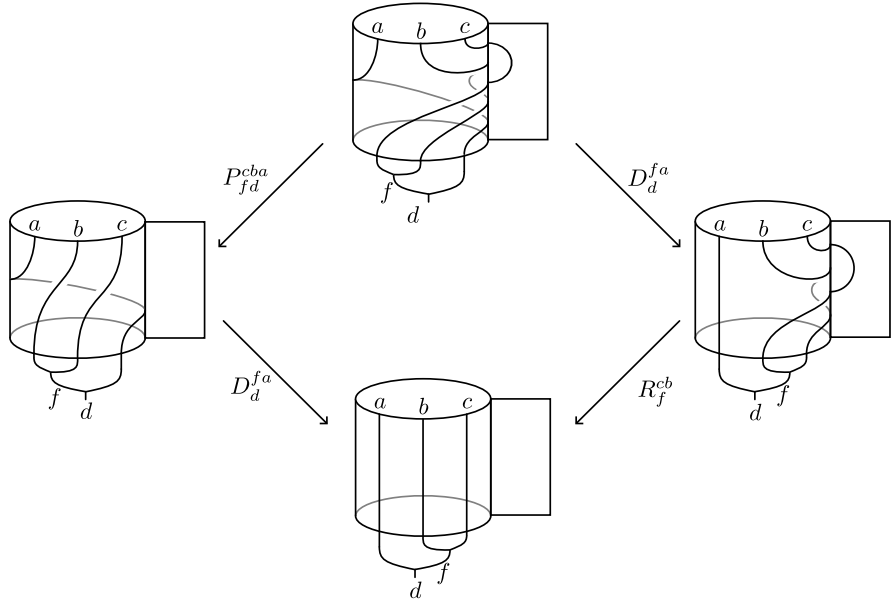

Figure 5.7: The two different paths that lead from the top diagram to the bottom diagram must be identical

## 5.3 Three Particles on a Lollipop Graph

The last graph we will consider is the lollipop graph. This graph is a trijunction graph where the base edge is joined with edge (1). Since the lollipop is a combination of the trijunction with the circle, the consistency equations include both the *P*-hexagon relations, the *Q*-hexagon relations, and the circle equations. There is also some interaction between moves on the trijunction and circle subgraphs. Demanding that such interaction is natural comes down to demanding that the diagram in figure 5.7 commutes. Since $D_d^{fa} \in U(1)$ this is equivalent to demanding that

$$D_d^{fa} P_{fd}^{cba} = R_f^{cb} D_d^{fa} \tag{5.14}$$
$$\iff \quad P_{fd}^{cba} = R_f^{cb} \tag{5.15}$$

Note that if we substitute this result in the *P*-hexagon equation, we recover one of the planar hexagon equations. So, by 'adding' a loop to the trijunction, we obtain restrictions that start to look more like the planar hexagon equations. The more connected a planar graph becomes, the more its braiding starts to resemble the usual planar braiding. As a matter of fact, in [68], it is shown that braiding on a triconnected graph, i.e., a connected graph such that deleting any two vertices (and incident edges) results in a graph that is still connected, is already equivalent to braiding on the plane.

We will see that for some models $Q_{fd}^{cba} = R_f^{cb}$ as well. In this case, the solutions to the graph braid equations are the same as those to the planar hexagon equations.

There is one final move on the lollipop that we have not considered yet. It consists of moving the particle closest to the vertex onto the stick and swapping the other two particles by using the circle. This move does not introduce new consistency relations, though, and we will not include it in this discussion. The interested reader is referred to part 6.4.1 of the original paper [68].

## 5.4 Solving Graph-Braid Equations

The consistency equations for braiding on a graph are similar to the pentagon and hexagon equations. These were, therefore, tackled by the functions provided by Anyonica package. While there is no function `SolveGraphBraidEquations`, we do have the function `SolvePolynomialSystem`, which is specifically designed to solve large sparse systems of polynomial equations with multiplicative gauge symmetries. The function `DeleteDuplicateSolutions` is also capable of using the custom gauge symmetries for the graph-braid symbols.

Instead of just generating the consistency equations and feeding those to `SolvePolynomialSystem` it is interesting to look at some heuristics that we used to save time. For both the lollipop and trijunction graphs, equations (5.8) and (5.9) need to be satisfied. It is, therefore, beneficial to start by searching for admissible sets of $P$, $Q$, and $R$-symbols for each given set of $F$-symbols. By inverting some of the arrows and going around the whole hexagon, these equations can be re-expressed in terms of $P$ and $Q$ as follows.

$$P_{ed}^{cab}\delta_{e'}^{e} = \sum_{f,g}^{r} [\tilde{F}_d^{acb}]_{e'}^{g}[F_d^{cab}]_f^{e} R_d^{cf}[F_d^{abc}]_g^{f}(R_g^{cb})^{-1}, \tag{5.16}$$

$$Q_{ed}^{bac}\delta_{e'}^{e} = \sum_{f,g}^{r} [F_d^{bac}]_g^{e'}(R_g^{ca})^{-1}[\tilde{F}_d^{bca}]_f^{g} R_d^{fa}[\tilde{F}_d^{abc}]_e^{f}. \tag{5.17}$$

Here the $\delta_{e'}^{e}$ appears as a consequence of the fact that we demand $P$ and $Q$ to preserve the charge $e$. As mentioned in Section 5.2.3 these equations split into two sets.

- If $e \neq e'$ we might get a consistency equation on the $R$-symbols which can be solved in terms of $R$.

- If $e = e'$ we get a definition for the $P$ and $Q$-symbols in terms of the $R$-symbols we solved for. In particular we don't need to solve for these symbols.

> **Note 12.** The consistency equations on $R$ might be trivially satisfied. This is always the case for Abelian anyons and for these theories all the $R$-symbols remain completely free.

For three particles on the trijunction, these are all the equations we need to solve.

For four particles, we need to take into account equations (5.10), (5.11), (5.12), and (5.13). These split in the same manner as the $P$ and $Q$ hexagons do. It is best to tackle equations (5.10) and (5.11) first. Typically, the solutions to the $P$ and $Q$ hexagon equations have continuous degrees of freedom. By substituting their values into the consistency equations coming from the $X$ and $Y$ moves, this solution set might get restricted somewhat. By substituting these restricted solutions into the equations for the *A* and *B*-symbols we may obtain further restrictions on the values of the *P* and *Q*-symbols.

For the lollipop graph, extra constraints on the *P* and *R*-symbols (5.14) together with equations 5.1 for the *D*-symbols, need to be added. The equations for the *D*-symbols can be solved separately.

If there is no gauge freedom left after fixing the *F*-symbols then the solutions to the lollipop equations consist of all possible combinations of solutions to the circle equations 5.1 with solutions to equations (5.8), (5.9), and (5.14). Of the anyon models we investigated, only one model has gauge freedom left after fixing the *F*-symbols: $\mathbb{Z}_2 \times \mathbb{Z}_2$. The method with which we constructed solutions to the lollipop equations for $\mathbb{Z}_2 \times \mathbb{Z}_2$ is described in 10.2.3. Once the lollipop equations were solved we checked the planarity of the solutions by checking whether $R_e^{ab} \equiv P_{ed}^{abc} \equiv Q_{ed}^{abc}$ and $D_e^{ab} \equiv R_e^{ab} D_b^{1b}$.

### 5.4.1 Removing Equivalent Solutions

Similar to the case of planar braiding, given a solution to the graph braid equations, one can create an infinite set of other solutions by applying gauge transforms. Apart from the *F*-symbols, all graph-braid symbols $S$ $(= R, Q, P, A, B, X, Y)$ correspond to exchanging two anyons with each other and therefore transform in the same way:

$$S \mapsto \frac{g_c^{ab}}{g_c^{ba}} S. \quad (5.18)$$

For all sets of *F*-symbols we considered, except for $\mathbb{Z}_2 \times \mathbb{Z}_2$, the gauge was already completely fixed. For $\mathbb{Z}_2 \times \mathbb{Z}_2$, the remaining gauge transforms form a $\mathbb{Z}_2$ group, and we removed this symmetry after solving the equations.

For the rings of type $SU(2)_k$ and $PSU(2)_k$, we did not have access to all *F*-symbols of these rings at the time of the project and made do with a single unitary solution per ring. We obtained these using the methods in [6]. Moreover, the specific form of the solutions for *k* an odd number were too complicated to derive the solutions symbolically. Eventually, we did find symbolic solutions by solving the systems numerically and reverting the numeric solutions to roots of polynomial equations. All solutions obtained this way are correct with an accuracy of 1000 decimal digits and infinite precision (meaning the computer used as many internal extra digits as needed to ensure all 1000 digits are correct).

### 5.4.2 Results

We solved the graph braiding equations for the circle, the trijunction (with three and four particles), and the lollipop graph for the following anyon models: $\mathbb{Z}_2$, Fib, Ising, Rep($D_3$), PSU(2)$_5$, $\mathbb{Z}_3$, $\mathbb{Z}_2 \times \mathbb{Z}_2$, SU(2)$_3$, $\mathbb{Z}_4$, TY($\mathbb{Z}_3$), Rep($D_4$), and SU(2)$_4$.

Some of these anyon models have different properties when braiding is confined to a graph rather than the plane. There exist, in particular, several fusion categories that have solutions for the graph-braid equations despite having no solutions to the planar hexagon equations. For the anyon models we studied, we observed the following:

- The equations (5.1) for anyons on a circle, like the planar hexagon equations, lead to discrete sets of solutions. There are always at least as many solutions as the planar hexagons allow. The equations for a circle sometimes admit solutions for models for which the planar hexagon equations don't. The $\text{TY}(\mathbb{Z}_3)$ fusion model (see 10.3.2 for the solutions) is such an example.

- As was pointed out in [15], solutions to the trijunction equations for three particles sometimes contain free parameters. If we add the equations for four particles, then, depending on the model, this freedom either remains unaltered (e.g., for Abelian anyons), gets partially restricted (e.g., for Ising anyons), or disappears completely (e.g., for $\text{Rep}(D_3)$ anyons). For the models we investigated, we found that if a model has solutions for the three-particle equations, it also has solutions for the four-particle equations. Specific results on the number of free variables and solutions to the trijunction equations can be found in table 5.1.

- The equations for the lollipop graph consist of (a) the trijunction equations (5.9) and (5.8), (b) equations demanding equality between the $P$ and $R$-symbols (5.14), and (c) equations for anyons on a circle (5.1). We will call the combined set of (a) and (b) the lollipop trijunction equations. The lollipop trijunction equations are sufficient to fix all degrees of freedom in the standard trijunction solutions. Since the equations on a circle give rise to a discrete set of solutions, all investigated models have a discrete set of solutions to the full lollipop equations. Let $n_c, n_t, n_l$ denote the number of gauge-inequivalent solutions to the circle equations, lollipop trijunction equations, and full lollipop equations, respectively. Although the equations for a circle graph are independent of the lollipop trijunction equations, $n_l$ need not be equal to $n_c n_t$. This happens when there is still some gauge freedom left after fixing the values of the $F$-symbols. In this case, the number of solutions to each set of equations gets reduced by the same factor. This implies that the number of gauge-inequivalent solutions to the combined set of equations will be greater than the product of the number of solutions of the individual equations. For the cases studied, only the $\mathbb{Z}_2 \times \mathbb{Z}_2$ model has remaining gauge symmetry. More information on the number of solutions to the planar hexagon equations, the circle equations, lollipop trijunction equations, and full lollipop equations can be found in tables 5.2 and 5.3.

If all the anyons are Abelian (i.e., the fusion algebra is a group algebra), then:

- The trijunction equations are trivially fulfilled for 3 and 4 particles. All non-trivial $R$- symbols are thus free variables for the trijunction. In particular, each set of trijunction equations admits an infinite set of solutions for any set of $F$-symbols. This is not the case for the planar hexagon equations. For, e.g., $\mathbb{Z}_3$ anyons only the trivial $F$-symbols admit a braided structure and for $\mathbb{Z}_2$ and $\mathbb{Z}_2 \times \mathbb{Z}_2$ only half of the sets of $F$-symbols admit a braided structure.

- For the circle, Lollipop trijunction, and full lollipop equations, we find that, for

| Fusion Ring | Solutions to the trijunction equations per set of unitary $F$-symbols | | | | |
|---|---|---|---|---|---|
| | $N = 3$ | | $N = 4$ | | |
| | # Solutions | # Free Variables | # Solutions | # Free Variables | Planar |
| Fib | 2 | None | 2 | None | Always |
| Ising | 2 | 2 | 2 | 1 | UCC |
| $PSU(2)_5$ | 2* | None | 2 | None | Always |
| $Rep(D_3)$ | 1** | 2** | 3** | 0** | Always |
| $SU(2)_3$ | 2* | 2 | 2 | 1 | UCC |
| $SU(2)_4$ | 2* | 2 | 2 | 1 | UCC |
| $TY(\mathbb{Z}_3)$ | 0 | | | | |
| $Rep(D_4)$ | 4 | 10 | 4 | 1 | UCC |

Table 5.1: Generic properties of solutions to the trijunction equations for three and four particles for various non-Abelian anyon models. Here UCC means that the solutions are planar under certain conditions on the free $R$-symbols. All solutions listed are gauge-inequivalent. Note that the number of solutions corresponds to the number of gauge-inequivalent families of solutions, possibly parametrized by some free variables. *For these models we only obtained solutions for 1 set of unitary $F$-symbols per model. **For $Rep(D_3)$ it looks like there are more solutions to the equations for $N = 4$, but this is only due to the fact that for $N = 4$ all free parameters are fixed and thus instead of 2 continuous families of solutions we find 3 solutions without freedom.

a fixed fusion ring, each set of $F$-symbols gives rise to the same number of solutions. If the $F$-symbols allow solutions to the planar hexagon equations, then some solutions to the lollipop equations are also planar. Because of the $D$-symbols (which are also gauge-dependent), the number of planar solutions to the lollipop equations is always greater than the number of solutions to the planar hexagon equations. For more information on the number of solutions to the lollipop equations for Abelian anyons, see table 5.3.

If some of the anyons are not Abelian, then:

- The solutions to the trijunction equations without free variables are always planar, and the solutions with free variables are planar for a discrete set of values of the free variables.

- All solutions to the lollipop equations are planar. The number of planar solutions to the lollipop equations is always greater than the number of solutions to the planar hexagon equations.

The exact solutions for the the Ising model, the quantum double of $\mathbb{Z}_2$, and $TY(\mathbb{Z}_3)$ can be found respectively in sections 10.1, 10.2, and 10.3.

| Fusion Ring | Number of solutions per type of equations (3 particles) per set of unitary $F$-symbols | | | |
|---|---|---|---|---|
| | Planar Hexagon | Circle | Lollipop Trijunction | Full Lollipop |
| Fib | 2 | 2 | 2 | $2^2$ |
| Ising | $2^2$ | $2^4$ | $2^2$ | $2^6$ |
| $PSU(2)_5$ | $2^*$ | $2^2$ | 2 | $2^3$ |
| $Rep(D_3)$ | 3, 0, 0 | 3, 3, 3 | 3, 0, 0 | $3^2$, 0, 0 |
| $SU(2)_3$ | $2^*$ | $2^6$ | 2 | $2^7$ |
| $TY(\mathbb{Z}_3)$ | 0 | 3 | 0 | 0 |
| $SU(2)_4$ | $2^*$ | $2^8$ | 2 | $2^9$ |
| $Rep(D_4)$ | $2^3$ | $2^7$ | $2^3$ | $2^{10}$ |

Table 5.2: Number of gauge inequivalent solutions to the consistency equations for various non-Abelian anyon models. Except for the planar hexagon equations all equations were constructed for systems with only three anyons. All of the solutions to the lollipop trijunction equations in this table are planar i.e., $P = Q = R$. For $Rep(D_3)$ a different number of solutions was found for the different solutions to the pentagon equations and so we used a notation where the $i^{th}$ number in each column corresponds to data regarding the $i^{th}$ solution to the pentagon equations. *For these models we only obtained solutions for 1 set of unitary $F$-symbols per case.

| Fusion Ring | Number of solutions per type of equations (3 particles) per set of equivalent $F$-symbols | | | | |
|---|---|---|---|---|---|
| | Planar Hexagon | Circle | Lollipop Trijunction | Full Lollipop | Lollipop but non-planar |
| $\mathbb{Z}_2$ | 2 | $2^2$ | 2 | $2^3$ | 0 |
| $\mathbb{Z}_3$ | 3 | $3^3$ | $3^2$ | $3^5$ | $\left(\frac{2}{3}\right)3^5$ |
| | 0 | $3^3$ | $3^2$ | $3^5$ | $3^5$ |
| $\mathbb{Z}_2 \times \mathbb{Z}_2$ | $2^3$ | $2^7$ | $2^5$ | $2^{13}$ | $\left(\frac{3}{4}\right)2^{13}$ |
| | 0 | $2^7$ | $2^5$ | $2^{13}$ | $2^{13}$ |
| $\mathbb{Z}_4$ | $2^2$ | $2^8$ | $2^6$ | $2^{14}$ | $\left(\frac{15}{4}\right)2^{14}$ |
| | 0 | $2^8$ | $2^6$ | $2^{14}$ | $2^{14}$ |

Table 5.3: Number of gauge inequivalent solutions to the consistency equations for various Abelian anyon models. Here we say two sets of $F$-symbols are equivalent iff they both have solvable planar hexagon equations. We chose to do this because, within each equivalence class, all members give rise to identical rows.

# Part IV

# Summary and Outlook

# Chapter 6

# Summary and Outlook

## 6.1 Summary

This thesis shows the methods we used to get an overview of the landscape of fusion categories. A fast algorithm is presented to find low-rank fusion rings via a tree search. The results of the search is a list of 28451 fusion rings that, in particular, contains all multiplicity-free fusion rings up to rank 9. We also found that out of the 118 non-commutative fusion rings in this list, all but 4 contain a subgroup. We investigated rings with a subgroup structure and classified all extensions of group rings by two elements. We also derived several other structure theorems that can be used to explain specific patterns in the table of non-commutative fusion rings per rank and subgroup. The notion of a song (single-orbit normal group) extension, a generalization of the Tambara-Yamagami (TY) and Haagerup-Izumi (HI), was introduced, and we showed that there exist categorifiable songs that are not of TY or HI type. The thesis continues by explaining the methods used to obtain the *F*-symbols, *R*-symbols, and pivotal coefficients of all multiplicity-free fusion categories up to rank 7. A total of 977 pivotal (possibly braided) fusion categories have been found. The results we found have been summarized in various ways, and a complete table of fusion categories can be found in Section 9. Most of the fusion categories can be identified as being of a standard structure, but such identification is not immediately apparent for some categories. It turns out that not all fusion rings with modular data are categorifiable. For the multiplicity-free fusion rings up to rank 7, we have the result that if they can be categorified, a modular categorification also exists. Interestingly, not all fusion categories belonging to the same fusion ring have the same configuration of zero *F*-symbols. The first three examples for which this is the case appear at rank 7. All the data on fusion rings and categories are part of the Anyonica package. This package also contains all methods for finding fusion categories and useful functions for working with these rings and categories. The package has also been used to export all data to the AnyonWiki website. In the last part of the thesis, the theory of graph braiding is presented in a somewhat naive fashion. The equations governing the braiding of anyons on the circle, trijunction, and lollipop graphs are derived, and a summary of their solutions is presented for several fusion systems. For some fusion systems, graph braiding is possible despite the lack of solutions to the planar hexagon equations. We also find that some fusion systems have more planar solutions on the lollipop graph than in the plane due to the appearance of extra gauge-dependent variables. For Abelian anyons, we also find that the graph-braid equations are always trivially satisfied on a trijunction, so there are always infinite solutions. While not all graphs lead to a finite number of solutions, it must be noted that, for the models we investigated, this is always the case for the circle, and one may wonder whether there is an Ocneanu-type theorem for this graph as well.

## 6.2 Outlook

Now that the Anyonica package is completed and there is an extensive database of multiplicity-free fusion categories, the way to several other interesting projects has opened.

1. The first project that comes to mind is the classification of multiplicity-free fusion categories up to rank 8 or even higher. To do so, Anyonica could use some updates. Some bottlenecks can still be dealt with, especially regarding the Smith decomposition and memory usage. The package would, in particular, benefit from making the Smith decomposition optional, having a smarter method of setting up equations (such as in [1]) and parallelizing some of the most time-consuming steps. It would also be nice if assumptions on the categories could be added in order to optimize the search. For example, if one assumes the fusion category is modular, then it must be braided, and one can immediately identify some values of the $R$-symbols. These can be used to simplify the hexagon equations, whose solutions, in turn, can be used to simplify the pentagon equations. Likewise, if one assumes a fusion category is unitary, then several extra constraints (such as orthogonality [1]) can be demanded that simplify the pentagon equations. There are many other heuristics that can be applied and are already implemented in J. K. Slingerland's Mathematica code to solve pentagon equations; therefore, it should not take too much time to incorporate these in Anyonica. Such functionality would especially be useful for the next project.

2. Recently, a paper that classifies all modular data up to rank 11 has been published [80]. We would like to find solutions to the pentagon, hexagon, and pivotal equations for these modular fusion rings.

3. The equations to find (bi)module categories over fusion categories are similar to the pentagon equations. I want to try to use Anyonica to solve these equations as well. The examples found can then be investigated to, e.g., get insight into the existence of new finite index subfactors or how the braiding on a bimodule category relates to that of the fusion categories that work on it. Fusion categories and bimodule categories are also interesting to compute state sum invariants of 3-manifolds, such as the Turaev-Viro(-Barret-Westubury) and Reshetikhin-Turaev state sums[9, 97]. If Anyonica could be connected to an on-line database of 3-manifolds (e.g., "The Atlas of 3-manifolds", hosted at http://www.matlas.math.csu.ru/), then hopefully it should not be too hard to compute these invariants. One could even try to go further and calculate state sums for manifolds with defects, which have applications in topological quantum field theory [73, 18, 57].

4. As shown in Chapter 1, braided fusion categories directly provide the required data to compute braid group representations. It should not be hard to implement this functionality. If the category in question is a ribbon fusion category, then one can use the quantum trace to find knot invariants as well [54, 84, 83, 96]. The theory to perform these calculations has been worked out numerous times in the literature, and it would be nice if Anyonica provided functions to perform these calculations. I would, in particular, like to program a function (or user interface) that simplifies fusion trees and thus calculates amplitudes of processes with anyons.

5. Solving generic pentagon equations for $m > 1$ is very hard. Most known examples come from the representation theory at roots of unity. These can be calculated by using the alatc package of E. Ardonne. We also know that the zesting construction [21] transforms fusion rings into other ones that do not belong to the quantum group picture. By implementing the zesting procedure in Anyonica on the level of the $F$- and $R$-symbols and incorporating alatc's functionality, we may be able to find new examples of fusion categories with multiplicity.

6. One complaint about Anyonica is that it requires Mathematica. Both the data and the source code are freely available, though, and I would love it if some of the techniques and tables were imported into other languages. For example, the TensorCategories package of F. Maurer and U. Thiel shows great potential, and maybe one day I might develop in Julia rather than Mathematica.

Besides these projects, I plan to maintain the AnyonWiki as long as possible. To do so, I could use all the help that the community can offer.

I hope that the data and some of the techniques described in this thesis might also be useful to other people. If you are one of those people, then I wish you the best of luck, and who knows, we might even meet and talk fusion cats over a cup of tea.

Gert Vercleyen

# Part V

# Appendix

# Chapter 7

# The Mathematics of Anyons

In this chapter we introduce the concept of a fusion category and add extra structures until we arrive at the concept of a unitary modular category. This is the notion that is most useful for physicists since such categories provide an algebraic description of systems with anyons. We also describe two different viewpoints of fusion categories. On the one hand there is the approach that treats fusion categories as strict categories, which simplifies abstract treatments and proofs considerably. On the other hand there is a more concrete aproach that describes fusion catgories via their fusion data; sets of numbers satisfying certain algebraic equations. While our approach for most part of the thesis is the second one, the first approach is essential to understand some of the subtle points of fusion systems, such as the need for a pivotal structure. Therefore, the exact way that one can translate one approach into the other will also be explained.

## 7.1 From Category to Modular Category

In order to settle some notational conventions we will revise some of the main definitions that are necessary to arrive to the concept of a fusion category. This section is not intended as an introduction to fusion categories. Thinking categorically is a skill that can only be obtained by practice and endurance.

### 7.1.1 From Category to Fusion Category

We start by defining the usual notions of a category, functor, natural transformation, and equivalence of categories in order to settle the notation.

**Definitions 7.1.1**.

- A **category** $\mathscr{C}$ consists of

    - a collection $\mathrm{Obj}\,\mathscr{C}$ of **objects**;
    - for any two objects $a, b \in \mathrm{Obj}\,\mathscr{C}$, a collection $\hom_{\mathscr{C}}(a, b)$ (or just $\hom(a, b)$ if it is clear with which category we're working) of **morphisms** with **domain** $a$ and **codomain** $b$;

- for each object $a \in \operatorname{Obj}\mathscr{C}$, an **identity morphism** $1_a \in \hom(a,a)$, (or just 1 if it is clear on which object it acts);

- for any objects $a, b, c \in \operatorname{Obj}\mathscr{C}$, an associative **composition operator** $\circ$ for morphisms $\hom(b,c) \times \hom(a,b) \xrightarrow{\circ} \hom(a,c)$, sending $(g,f)$ to $g \circ f = gf$ for which $1_b \circ f = f$ and $g \circ 1_a = g$.

- A **functor** $F : \mathscr{C} \to \mathscr{D}$ is a map from a category $\mathscr{C}$ to a category $\mathscr{D}$ that maps objects to objects, i.e., $F : \operatorname{Obj}\mathscr{C} \longrightarrow \operatorname{Obj}\mathscr{D}; a \longmapsto Fa$, morphisms to morphisms, i.e., $F : \hom(a,b) \longrightarrow \hom(Fa, Fb), \quad f \longmapsto Ff$, and satisfies

  - Preservation of Identity: $\forall a \in \operatorname{Obj}\mathscr{C}$

  $$F(1_a) = 1_{Fa} \tag{7.1}$$

  and

  - Preservation of Composition: $\forall (g,f) \in \hom(b,c) \times \mathscr{C}(a,b)$

  $$F(g \circ f) = Fg \circ Ff. \tag{7.2}$$

  The functor $\operatorname{id}_\mathscr{C}$ that is the identity on objects and morphisms is called the **identity functor**.

- A **natural transformation** $\theta : F \Rightarrow G$ between two functors $F : \mathscr{C} \to \mathscr{D}$ and $G : \mathscr{C} \to \mathscr{D}$ consists of a set of morphisms $\left\{ \theta_a \in \hom_\mathscr{D}(Fa, Ga) \,\middle|\, a \in \operatorname{Obj}\mathscr{C} \right\}$ such that $\forall a, b \in \operatorname{Obj}\mathscr{C}, \forall \psi \in \hom_\mathscr{C}(a,b)$ the following diagram commutes

  $$\begin{array}{ccc} Fa & \xrightarrow{F\psi} & Fb \\ \downarrow{\theta_a} & & \downarrow{\theta_b} \\ Ga & \xrightarrow{G\psi} & Gb \end{array} \tag{7.3}$$

- If $\theta : F \Rightarrow G$ and $\eta : G \Rightarrow H$ are natural transformations, their **composition** is the natural transformation $\eta\theta : F \Rightarrow H$ with structure morphisms $(\eta\theta)_a = \eta_a \circ \theta_a$, $\forall a \in \operatorname{Obj}\mathscr{C}$.

- A **natural isomorphism** is a natural transformation in which every structure morphism is an isomorphism.

- We say that a functor $F : \mathscr{C} \to \mathscr{D}$ is an **equivalence** if there is a functor $F^{-1} : \mathscr{D} \to \mathscr{C}$, called a quasi-inverse of $F$, such that $F^{-1} \circ F \cong \operatorname{id}_\mathscr{C}$ and $F \circ F^{-1} \cong \operatorname{id}_\mathscr{D}$. We say that the categories $\mathscr{C}$ and $\mathscr{D}$ are **equivalent** if there exists an equivalence between them.

The other definitions and propositions in this section will follow the EGNO [25] but are adapted to the goal of understanding the following definition as quickly as possible.

**Definition 7.1.2.** A fusion category over $\mathbb{C}$ is a category which is **[FC1]** $\mathbb{C}$–linear, **[FC2]** Abelian, **[FC3]** indecomposable, **[FC4]** semi-simple, **[FC5]** finite, **[FC6]** monoidal with unit **1** and product $\otimes$ that is bilinear on morphisms, **[FC7]** rigid, and for which **[FC8]** $\hom(\mathbf{1}, \mathbf{1}) \cong k$

In particular we will only revise the notions that are necessary to understand fusion categories over $\mathbb{C}$ and will use more specific definitions whenever possible.

**Definition 7.1.3. [FC1]** A category is $\mathbb{C}$-**linear** if

1. Every set $\hom(a, b)$ is equipped with a structure of a complex vector space such that composition of morphisms is bilinear.

2. There exists a zero object $0 \in \mathscr{C}$ such that $\hom(0, 0) = \mathbf{0}$ (the zero-dimensional vector space).

3. (Existence of direct sums) For any objects $a_1, a_2 \in \mathscr{C}$ there exists an object $b \in \mathscr{C}$ and morphisms $p_1 : b \to a_1, p_2 : b \to a_2, i_1 : a_1 \to b, i_2 : a_2 \to b$ such that $p_1 \circ i_1 = 1_{a_1}, p_2 \circ i_2 = 1_{a_2}$, and $i_1 \circ p_1 + i_2 \circ p_2 = 1_b$.

A functor between two $\mathbb{C}$-linear categories is called $\mathbb{C}$-**linear** if its action on morphisms is $\mathbb{C}$-linear.

Note that $b$ is often written as $a_1 \oplus a_2$ and moreover $\oplus$ can be extended to a bifunctor (whose action on morphisms is defined via the maps $p_j, i_j$).

**Definitions 7.1.4.** Let $\mathscr{C}$ be $\mathbb{C}$-linear and $\varphi : a \to b$ a morphism in $\mathscr{C}$.

- The **kernel** $\ker(\varphi)$ of $\varphi$ (if it exists) is an object $k$ together with a morphism $\kappa : k \to a$ such that $\varphi \circ \kappa = 0$, and if $\kappa' : k' \to a$ is such that $\varphi \circ \kappa' = 0$ then there exists a unique morphism $\lambda : k' \to k$ such that $\kappa \circ \lambda = \kappa'$.

- If $\ker(\varphi) = (0, 0)$ then $\varphi$ is called a **monomorphism**.

- An object $a$ together with a monomorphism $\iota : a \to b$ is called a **subobject** of $b$. We also write $a \subseteq b$ if there exists such a $\iota$.

Dually one defines

- The **cokernel** $\mathrm{CoKer}(\varphi)$ of a morphism $\varphi : a \to b$ in $\mathscr{C}$ (if it exists) is an object $c$ together with a morphism $\gamma : b \to c$ such that $\gamma \circ \varphi = 0$, and if $\gamma' : b \to c'$ is such that $\gamma' \circ \varphi = 0$ then there exists a unique morphism $\lambda : c \to c'$ such that $\lambda \circ \gamma = \gamma'$.

- If $\mathrm{CoKer}(\varphi) = (0, 0)$ then $\varphi$ is called an **epimorphism**.

- An object $c$ together with an epimorphism $\varepsilon : b \to c$ is called a **quotient object** of $b$

**Definition 7.1.5. [FC2]** A $k$-linear category $\mathscr{C}$ is called **Abelian** if for every morphism $\varphi : a \to b$ there exists a sequence

$$k \xrightarrow{\kappa} a \xrightarrow{i} \text{Im}(\varphi) \xrightarrow{j} b \xrightarrow{\gamma} c \qquad (7.4)$$

such that

1. $j \circ i = \varphi$,

2. $(k, \kappa) = \ker(\varphi)$, $(c, \gamma) = \text{CoKer}(\varphi)$,

3. $(\text{Im}(\varphi), i) = \text{CoKer}(\kappa)$, $(\text{Im}(\varphi), j) = \ker(\gamma)$.

The abelian category with one object 0 is called the **zero category**.

We can use the $\oplus$ functor to define the direct sum of abelian categories

**Definition 7.1.6.** Let $\mathscr{C}_i, i \in I$, be a family of $\mathbb{C}$-linear categories. The **direct sum** $\mathscr{C} = \oplus_{i \in I} \mathscr{C}_i$ is the category whose objects are sums $a = \oplus_{i \in I} a_i$, $a_i \in \mathscr{C}_i$, such that almost all $a_i$ are zero, with $\text{hom}(a, b) = \oplus_{i \in I} \text{hom}_{\mathscr{C}_i}(a_i, b_i)$ for $a = \oplus_{i \in I} a_i$ and $b = \oplus_{i \in I} b_i$.

**Definition 7.1.7. [FC3]** An abelian category $\mathscr{C}$ is **indecomposable** if it is not equivalent to a direct sum of two nonzero categories.

The definition of a subobject can be used to define the fundamental notion of simplicity.

**Definition 7.1.8. [FC4]** A nonzero object $a$ in $\mathscr{C}$ is called **simple** if its only subobjects are $\mathbf{0}$ and $a$. An object $a$ in $\mathscr{C}$ is called semisimple if it is a direct sum of simple objects, and a category $\mathscr{C}$ is called **semisimple** if every object of $\mathscr{C}$ is semisimple.

**Definition 7.1.9.** Let $a$ be an object in an Abelian category $\mathscr{C}$. We say that $a$ has **finite length** if there exists a finite list of objects $\{a_0 = 0, a_1, ..., a_n = a\}$, called a Jordan-Holder series, such that $a_{i-1} \subseteq a_i, i = 1, ..., n$ and $a_i/a_{i-1}$ is simple for all $i = 1, ..., n$.

Due to a theorem by Jordan-Holder any Jordan-Holder series for an object has the same length and any simple object, $a_i/a_{i-1}$, in a Jordan-Holder series of $a$ appears the same number of times in any other Jordan-Holder series of $a$.

For a fusion category any object has finite length and admits a unique decomposition into a direct sum of simple objects.

**Definitions 7.1.10.** A sequence of morphisms

$$\ldots \longrightarrow a_{i-1} \xrightarrow{\varphi_{i-1}} a_i \xrightarrow{\varphi_i} a_{i+1} \longrightarrow \ldots$$

in an Abelian category is called **exact in degree** $i$ if $\text{Im}\,\varphi_{i-1} = \ker \varphi_i$. The sequence is called **exact** if it is exact in every degree. An exact sequence

$$0 \longrightarrow a \longrightarrow b \longrightarrow c \longrightarrow 0$$

is called a **short exact sequence**.

**Definitions 7.1.11.** An object $p$ in an Abelian category $\mathscr{C}$ is called **projective** if the functor $\hom(p, \cdot)$ maps short exact sequences to short exact sequences. A **projective cover** of an object $a$ is a projective object $p_a$ together with an epimorphism $\pi : p_a \to a$ such that if $\pi' : q \to a$ is an empimorphism form a projective object $q$ to $a$, there exists an epimorphism $\lambda : q \to p_a$ with $\pi \circ \lambda = \pi'$.

**Definition 7.1.12. [FC5]** A $\mathbb{C}$-linear Abelian category $\mathscr{C}$ is called **finite** if

- $\dim \hom(a, b) < \infty$ for all $a, b \in \mathrm{Obj}\,\mathscr{C}$,

- there are finitely many isomorphism classes of simple objects, and

- every object $a \in \mathrm{Obj}\,\mathscr{C}$ has finite length and has a projective cover.

A monoidal category is a category with an associative unital tensor product. More specifically:

**Definition 7.1.13. [FC6]** A **monoidal category** is a six-tuple $(\mathscr{C}, \otimes, \mathbf{1}, \alpha, \lambda, \rho)$ where

- $\mathscr{C}$ is a category;

- $\mathbf{1}$ is an object of $\mathscr{C}$, called the **unit object** of $\mathscr{C}$;

- $\otimes : \mathscr{C} \times \mathscr{C} \to \mathscr{C}$ is a bifunctor, called the **tensor (or monoidal) product**;

- $\alpha : (\cdot_1 \otimes \cdot_2) \otimes \cdot_3 \to \cdot_1 \otimes (\cdot_2 \otimes \cdot_3)$ is a natural isomorphism;

- $\lambda : \mathbf{1} \otimes \cdot \to \cdot$ and $\rho : \cdot \otimes \mathbf{1} \to \cdot$ are natural isomorphisms;

such that for all $a, b, c, d \in \mathscr{C}$ the following diagrams commute

$$\begin{array}{c}
((a \otimes b) \otimes c) \otimes d \\
{\alpha_{a,b,c} \otimes \mathrm{id}_d \swarrow \qquad \searrow \alpha_{a \otimes b, c, d}} \\
(a \otimes (b \otimes c)) \otimes d \qquad\qquad (a \otimes b) \otimes (c \otimes d) \\
{\alpha_{a, b \otimes c, d} \searrow \qquad \swarrow \alpha_{a, b, c \otimes d}} \\
a \otimes ((b \otimes c) \otimes d) \xrightarrow{\mathrm{id}_a \otimes \alpha_{b,c,d}} a \otimes (b \otimes (c \otimes d))
\end{array} \qquad (7.5)$$

$$\begin{array}{c}
(a \otimes \mathbf{1}) \otimes b \xrightarrow{\alpha_{a,1,b}} a \otimes (\mathbf{1} \otimes b) \\
{\rho_a \otimes 1_b \searrow \qquad \swarrow 1_a \otimes \lambda_b} \\
a \otimes b
\end{array} \qquad (7.6)$$

Equation (7.5) is called the pentagon equation(s) and equation (7.6) is called the triangle equation(s). The map $\alpha$ is called the **associator** and the maps $\lambda$ and $\rho$ are called the **left and right unit constraints**. If $\alpha, \lambda, \rho$ are the identity maps then $\mathscr{C}$ is called a **strict monoidal category**.

**Definitions 7.1.14.** Let $(\mathscr{C}, \otimes, \mathbf{1}, \alpha, \lambda, \rho)$ be a monoidal category. An object $a^*$ in $\mathscr{C}$ is said to be a **left dual** of $a$ if there exist morphisms $e_a : a^* \otimes a \to \mathbf{1}$ and $c_a : \mathbf{1} \to a \otimes a^*$ such that the compositions

$$a \xrightarrow{\lambda_a^{-1}} \mathbf{1} \otimes a \xrightarrow{c_a \otimes 1_a} (a \otimes a^*) \otimes a \xrightarrow{\alpha_{a,a^*,a}} a \otimes (a^* \otimes a) \xrightarrow{1_a \otimes e_a} a \otimes \mathbf{1} \xrightarrow{\rho_a} a \ , \quad (7.7)$$

$$a^* \xrightarrow{\rho_{a^*}^{-1}} a^* \otimes \mathbf{1} \xrightarrow{1_{a^*} \otimes c_a} a^* \otimes (a \otimes a^*) \xrightarrow{\alpha_{a^*,a,a^*}^{-1}} (a^* \otimes a) \otimes a^* \xrightarrow{e_a \otimes 1_{a^*}} \mathbf{1} \otimes a^* \xrightarrow{\lambda_{a^*}} a^* \quad (7.8)$$

are the identity maps. Given a morphism $f \in \mathrm{hom}(a, b)$, its **left dual** $f^* \in \mathrm{hom}(b^*, a^*)$ is defined as

$$f^* := \lambda_a \circ (e_b \otimes 1_{a^*}) \circ ((1_{b^*} \otimes f) \otimes 1_{a^*}) \circ \alpha_{b^*,a,a^*}^{-1} \circ (1_{b^*} \otimes c_a) \circ \rho_{b^*}^{-1} \quad (7.9)$$

An object $^*a$ in $\mathscr{C}$ is said to be a **right dual** of $a$ if there exist morphisms $e'_a : a \otimes {}^*a \to \mathbf{1}$ and $c'_a : \mathbf{1} \to {}^*a \otimes a$ such that the compositions

$$a \xrightarrow{\rho_a^{-1}} a \otimes \mathbf{1} \xrightarrow{1_a \otimes c'_a} a \otimes ({}^*a \otimes a) \xrightarrow{\alpha_{a,{}^*a,a}^{-1}} (a \otimes {}^*a) \otimes a \xrightarrow{e'_a \otimes 1_a} \mathbf{1} \otimes a \xrightarrow{\lambda_a} a^* \ , \quad (7.10)$$

$$^*a \xrightarrow{\lambda_{*a}^{-1}} \mathbf{1} \otimes {}^*a \xrightarrow{c'_a \otimes 1_a} ({}^*a \otimes a) \otimes {}^*a \xrightarrow{\alpha_{*a,a,*a}} {}^*a \otimes (a \otimes {}^*a) \xrightarrow{1_{*a} \otimes e'_a} {}^*a \otimes \mathbf{1} \xrightarrow{\rho_{*a}} {}^*a \quad (7.11)$$

are the identity maps. Given a morphism $f \in \mathrm{hom}(a, b)$, its **right dual** $^*f \in \mathrm{hom}(^*b, {}^*a)$ is defined as

$$^*f := \rho_b \circ (1_{*a} \otimes e'_b) \circ (1_{*a} \otimes (f \otimes 1_{*b})) \circ \alpha_{*a,a,*b} \circ (c'_a \otimes 1_{*b}) \circ \rho_{b^*}^{-1} \quad (7.12)$$

**Definition 7.1.15. [FC7]** An object in a monoidal category is called **rigid** if it has left and right duals. If every object in a monoidal category $\mathscr{C}$ is rigid then we say that $\mathscr{C}$ is **rigid**.

Now that we have defined all necessary notions to make sense of a fusion category it is interesting to see how these interact with each other. The following proposition is a combination of various propositions found throughout the EGNO [25] with the added assumption that we are working with a fusion category. The number in front of each statement equals the number of that statement in the EGNO.

**Proposition 7.1.16.** *Let $(\mathscr{C}, \otimes, \mathbf{1}, \alpha, \lambda, \rho)$ be a $\mathbb{C}$-linear fusion category. Then the following*

*properties hold.*

**Property of simple objects**

> *1.5.2. (Schur's Lemma) Any morphism $\varphi : a \to b$ between two simple objects is either 0 or an isomorphism. In particular, for all simple $a, b$, $\hom(a,a) \cong \mathbb{C}$ and $\hom(a,b) = \mathbf{0}$ if $a$ is not isomorphic to $b$.*

**Properties of $\oplus$ and $\otimes$**

> *1.2.4. For any $\mathbb{C}$-linear functor $F : \mathscr{C} \to \mathscr{D}$ there exists a natural isomorphism $F(\cdot_1) \oplus F(\cdot_2) \Rightarrow F(\cdot_1 \oplus \cdot_2)$.*

> *4.2.1. The functors $a \otimes \cdot$ and $\# \otimes a$ map short exact sequences to short exact sequences.*

> *4.2.8. For any two morphisms $\varphi, \psi$ one has $\mathrm{Im}(\varphi \otimes \psi) \cong \mathrm{Im}\, \varphi \otimes \mathrm{Im}\, \psi$*

**Properties of the unit 1**

> *4.3.8. $\mathbf{1}$ is simple.*

> *2.2.6. $\mathbf{1}$ is unique up to unique isomorphism.*

**Properties of duals**

> *2.10.3. $^*(a^*) \cong a \cong (^*a)^*$, $\mathbf{1}^* \cong \mathbf{1} \cong {}^*\mathbf{1}$, $e'_{a^*} = e_a$, $c'_{a^*} = c_a$ and vice versa $e_{*a} = e'_a$, $c_{*a} = c'_a$.*

> *2.10.5. Left (respectively, right) duals of $a$ are unique up to unique isomorphism*

> *2.10.7. Let $\psi : b \to c, \varphi : a \to b$ then $(\psi \circ \varphi)^* = \varphi^* \circ \psi^*$, $^*(\psi \circ \varphi) = {}^*\varphi \circ {}^*\psi$, $(a \otimes b)^* \cong b^* \otimes a^*$, and $^*(a \otimes b) \cong {}^*b \otimes {}^*a$*

> *2.10.8. $\hom(a \otimes b, c) \cong \hom(a, c \otimes b^*)$, and $\hom(a \otimes b, c) \cong \hom(b, {}^*a \otimes c)$*

> *4.3.9. $e$ and $e'$ are monomorphisms and $c$ and $c'$ are epimorphisms.*

> **Note 13.** The duality between $e', e$ and $c', c$ given in 2.10.3. of proposition 7.1.16 is the reason why some authors only define and use the maps $e, c$.

### 7.1.2 From Fusion Category To Modular Category

There are two independent extra structures that one can put on a fusion category. On the one hand there is a pivotal structure that relates objects to their double duals. On the other hand there is a braiding that relates products $a \otimes b$ to $b \otimes a$. Depending on the availability and properties of these structures, fusion categories are given different adjectvies and find different applications.

From now on we will always assume that $\mathscr{C}$ is a fusion category over $\mathbb{C}$.

### 7.1.2.1 Pivotal and Spherical Fusion Categories

We start by defining *quantum traces*.

**Definitions 7.1.17.** For any object $a \in \mathscr{C}$ and any morphisms $\varphi \in \hom(a, a^{**})$, $\psi \in \hom(^{**}a, a)$ we define

- the **left quantum trace** of $\varphi$ as

$$\mathrm{Tr}^L(\varphi) : \mathbf{1} \xrightarrow{c_a} a \otimes a^* \xrightarrow{\varphi \otimes 1_{a^*}} a^{**} \otimes a^* \xrightarrow{e_{a^*}} \mathbf{1}, \qquad (7.13)$$

and

- the **right quantum trace** of $\psi$ as

$$\mathrm{Tr}^R(\psi) : \mathbf{1} \xrightarrow{c_{*a}} {^*a} \otimes a \xrightarrow{1_{*a} \otimes \psi} {^*a} \otimes {^{**}a} \xrightarrow{e_{**a}} \mathbf{1}. \qquad (7.14)$$

Since $\hom(\mathbf{1}, \mathbf{1}) \cong \mathbb{C}$ we can identify left and right traces of morphisms with complex numbers. In order to introduce a pivotal structure we need a few more definitions.

**Definitions 7.1.18.** Let $(\mathscr{C}, \otimes, \mathbf{1}, \alpha, \lambda, \rho)$ and $(\mathscr{C}', \otimes', \mathbf{1}', \alpha', \lambda', \rho')$ be two monoidal categories. A **monoidal functor** $F : \mathscr{C} \to \mathscr{C}'$ is a triple $(F_0, F_1, F_2)$ where $F : \mathscr{C} \to \mathscr{C}'$ is a functor, $F_0 : \mathbf{1}' \to F_1(\mathbf{1})$ is an isomorphism, and $(F_2)_{a,b} : F_1(a) \otimes' F_1(b) \to F_1(a \otimes b)$ is a natural isomorphism such that the following diagrams commutes for all $a, b, c \in \mathscr{C}$

$$\begin{array}{ccc}
(F_1(a) \otimes' F_1(b)) \otimes' F_1(c) & \xrightarrow{\alpha'_{F_1(a), F_1(b), F_1(c)}} & F_1(a) \otimes' (F_1(b) \otimes' F_1(c)) \\
{\scriptstyle (F_2)_{a,b} \otimes' \mathrm{id}_{F_1(c)}} \downarrow & & \downarrow {\scriptstyle \mathrm{id}_{F_1(a)} \otimes' (F_2)_{b,c}} \\
F_1(a \otimes b) \otimes' F_1(c) & & F_1(a) \otimes' F_1(b \otimes c) \\
{\scriptstyle (F_2)_{a \otimes b, c}} \downarrow & & \downarrow {\scriptstyle (F_2)_{a, b \otimes c}} \\
F_1((a \otimes b) \otimes c)) & \xrightarrow{F_1(\alpha_{a,b,c})} & F_1(a \otimes (b \otimes c))
\end{array} \qquad (7.15)$$

$$\begin{array}{ccccccc}
\mathbf{1}' \otimes' F_1(a) & \xrightarrow{\lambda'_{F_1(a)}} & F_1(a) & & F_1(a) \otimes' \mathbf{1}' & \xrightarrow{\rho'_{F_1(a)}} & F_1(a) \\
{\scriptstyle F_0 \otimes' 1_{F_1(a)}} \downarrow & & \downarrow {\scriptstyle F_1(\lambda_x)^{-1}}, & & {\scriptstyle 1_{F_1(a)} \otimes' F_0} \downarrow & & \downarrow {\scriptstyle F_1(\rho_x)^{-1}} \\
F_1(\mathbf{1}) \otimes' F_1(a) & \xrightarrow{(F_2)_{1,a}} & F_1(\mathbf{1} \otimes a) & & F_1(a) \otimes' F_1(\mathbf{1}) & \xrightarrow{(F_2)_{a,1}} & F_1(a \otimes \mathbf{1})
\end{array} \qquad (7.16)$$

Likewise, a natural transformation of monoidal functors needs to take account of the product structure.

**Definition 7.1.19.** Let $(\mathscr{C}, \otimes, \mathbf{1}, \alpha, \lambda, \rho)$ and $(\mathscr{C}', \otimes', \mathbf{1}', \alpha', \lambda', \rho')$ be two monoidal categories and $F, G$ be monoidal functors from $\mathscr{C}$ to $\mathscr{C}'$. A **natural transformation of monoidal functors** $\eta : (F_0, F_1, F_2) \to (G_0, G_1, G_2)$ is a natural transformation $\eta : F_1 \to$

$G_1$ such that $\eta_1$ is an isomorphism and the following diagram commutes for all $a, b \in \mathscr{C}$:

$$\begin{array}{ccc} F_1(a) \otimes' F_1(b) & \xrightarrow{(F_2)_{a,b}} & F_1(a \otimes b) \\ \downarrow{\eta_a \otimes' \eta_b} & & \downarrow{\eta_{a \otimes b}} \\ G_1(a) \otimes' G_1(b) & \xrightarrow{(G_2)_{a,b}} & G_1(a \otimes b) \end{array} \qquad (7.17)$$

If there exist two monoidal functors $F : \mathscr{C} \to \mathscr{D}$ and $G : \mathscr{D} \to \mathscr{C}$ between fusion categories, such that $G \circ F$ is monoidally isomorphic to the identity functor $1_{\mathscr{C}}$ and $F \circ G$ is monoidally isomorphic to the identity functor $1_{\mathscr{D}}$, then $\mathscr{C}$ and $\mathscr{D}$ are called **fusion equivalent**.

**Definitions 7.1.20.** A **pivotal structure** on $\mathscr{C}$ is an isomorphism of monoidal functors $p : 1_{\mathscr{C}} \to \cdot^{**}$. A fusion category with a pivotal structure is called a **pivotal fusion category**.

This definition implies that the pivotal structure needs to satisfy stronger constraints than a natural transformation between non-monoidal functors. In particular we need to have that $p_a \otimes p_b = p_{a \otimes b}$.

Using the pivotal structure, we can define the quantum dimensions of objects, and the notion of a spherical category

**Definitions 7.1.21.** Let $\psi$ be a pivotal structure on $\mathscr{C}$, the **quantum dimension** of $a \in \mathrm{Obj}\,\mathscr{C}$ is $\dim(a) := \mathrm{Tr}^L_{\psi_a}$. We say that the fusion category $\mathscr{C}$ is **spherical** if $\dim(a) = \dim(a^*)$ for all $a \in \mathrm{Obj}\,\mathscr{C}$.

#### 7.1.2.2 Braided Fusion Categories

**Definition 7.1.22.** A **braided fusion category** is fusion category together with a natural isomorphism $\beta : \cdot_1 \otimes \cdot_2 \Rightarrow \cdot_2 \otimes \cdot_1$ that satisfies the hexagon equations:

$$\begin{array}{ccc} & (b \otimes a) \otimes c \xrightarrow{\alpha_{b,a,c}} b \otimes (a \otimes c) & \\ {}^{\beta_{a,b} \otimes 1}\nearrow & & \searrow{}^{1 \otimes \beta_{a,c}} \\ (a \otimes b) \otimes c & & b \otimes (c \otimes a) \\ {}_{\alpha_{a,b,c}}\searrow & & \nearrow{}_{\alpha_{b,c,a}} \\ & a \otimes (b \otimes c) \xrightarrow{\beta_{a,b \otimes c}} (b \otimes c) \otimes a & \end{array} \qquad (7.18)$$

$$\begin{array}{ccc}
& (b \otimes a) \otimes c \xrightarrow{\alpha_{b,a,c}} b \otimes (a \otimes c) & \\
\beta_{b,a}^{-1} \otimes 1 \nearrow & & \searrow 1 \otimes \beta_{c,a}^{-1} \\
(a \otimes b) \otimes c & & b \otimes (c \otimes a) \\
\alpha_{a,b,c} \searrow & & \nearrow \alpha_{b,c,a} \\
& a \otimes (b \otimes c) \xrightarrow{\beta_{b \otimes c,a}^{-1}} (b \otimes c) \otimes a &
\end{array} \quad (7.19)$$

The notion of a functor between braided categories that is *natural* is the following

**Definitions 7.1.23.** Let $(\mathscr{C}, \otimes, \mathbf{1}, \alpha, \lambda, \rho, \beta)$ and $(\mathscr{C}', \otimes', \mathbf{1}', \alpha', \lambda', \rho', \beta')$ be braided monoidal categories whose braidings are denoted $\beta$ and $\beta'$, respectively. A monoidal functor $(F_0, F_1, F_2)$ from $\mathscr{C}$ to $\mathscr{C}'$ is called **braided** if the following diagram commutes:

$$\begin{array}{ccc}
F_1(a) \otimes' F_1(b) & \xrightarrow{\beta'_{F_1(a), F_1(b)}} & F_1(b) \otimes' F_1(a) \\
\downarrow{(F_2)_{a,b}} & & \downarrow{(F_2)_{b,a}} \\
F_1(a \otimes b) & \xrightarrow{F_1(\beta_{a,b})} & F_1(b \otimes a)
\end{array} \quad (7.20)$$

for all $a, b \in \mathrm{Obj}\,\mathscr{C}$. A **braided equivalence** of braided categories is a braided monoidal functor which is also an equivalence of categories.

> **Note 14.** Being braided is a property of a monoidal functor, rather than an extra structure.

The following proposition (see, e.g., [27]) is one of the reasons braided categories are so interesting.

**Proposition 7.1.24.** *Let $\mathscr{C}$ be a strict monoidal category with braiding $\beta$. For all $a, b, c \in \mathscr{C}$ the Yang-Baxter equation*

$$\left(\beta_{b,c} \otimes 1_a\right) \circ \left(1_b \otimes \beta_{a,c}\right) \circ \left(\beta_{a,b} \otimes 1_c\right) = \left(1_c \otimes \beta_{a,b}\right) \circ \left(\beta_{a,c} \otimes 1_b\right) \circ \left(1_a \otimes \beta_{b,c}\right) \quad (7.21)$$

*holds.*

### 7.1.2.3 Ribbon and Modular Fusion Categories

We finally arive at the notion of a modular category

**Definitions 7.1.25.**

- A spherical braided fusion category $\mathscr{C}$ is called a **ribbon fusion category**. Let $\psi, \beta$ be respectively the spherical and braided structure of a ribbon fusion category, and define the natural isomorphism $u : 1_{\mathscr{C}} \Rightarrow \cdot^{**}$ as follows

$$u_a = \lambda_a \circ (e_a \otimes 1_{a^{**}}) \circ (\beta_{a,a^*} \otimes 1_{a^{**}}) \circ (1_a \otimes c_{a^*}) \circ \rho_{a^*}. \tag{7.22}$$

The maps

$$\theta_a := \psi_a^{-1} \circ u_a, \tag{7.23}$$

are called the **twists** of $\mathscr{C}$ and the natural isomorphism $\theta : 1_{\mathscr{C}} \Rightarrow 1_{\mathscr{C}}$, defined by these maps, is called the **ribbon structure** of $\mathscr{C}$. Two ribbon fusion categories $\mathscr{C}$ and $\mathscr{D}$ are called **ribbon equivalent** if there exists a braided equivalence $F = (F_0, F_1, F_2)$ such that $F_1(\theta_a) = \theta_{F_1(a)}, \forall a \in \mathscr{C}$

- The *S*-**matrix** of a ribbon category is defined as

$$[S]_b^a = \text{Tr}(\beta_{a,b} \circ \beta_{b,a}) \tag{7.24}$$

where the trace is taken with respect to the spherical structure $\psi$.

- A ribbon category whose *S*-matrix is invertible is called a **modular category**. Two modular categories are called **modular equivalent** if they are ribbon equivalent.

## 7.2 From Fusion System to Fusion Category and Back

This section gives the explicit construction used to obtain a fusion system from a fusion category and vice versa. These constructions come from the paper [19] that also proves the following two propositions

**Proposition 7.2.1**. *Given a fusion category $\mathscr{C}$ over $\mathbb{C}$, one can construct a fusion system $FS(\mathscr{C})$ from it. Given a fusion system $\mathscr{F} = (\mathbf{L}, *, \mathbf{N}, \mathbf{F}, \mathbf{P}, \mathbf{R})$, one can construct a fusion category $Cat(\mathscr{F})$ from it, such that $Cat(FS(\mathscr{C}))$ and $\mathscr{C}$ are fusion equivalent.*

**Proposition 7.2.2**. *Given a modular fusion category $\mathscr{C}$ over $\mathbb{C}$, one can construct a modular fusion system $MFS(\mathscr{C})$ from it. Given a modular fusion system $\mathscr{F} = (\mathbf{L}, *, \mathbf{N}, \mathbf{F})$, one can construct a modular fusion category $MCat(\mathscr{F})$ from it, such that $MCat(MFS(\mathscr{C}))$ and $\mathscr{C}$ are modular equivalent.*

> **Note 15.** In [19] these propositions are stated for the more general case of fusion categories over a field $k$. In that case proposition 7.2.2 needs more refinement since the modular category is defined over a different field than the modular fusion system.

The paper [19] proves these propositions in a constructive manner, i.e. by constructing the (modular) category from the (modular) fusion system and vice versa. The constructions are basically the following.

## From Fusion System to Fusion Category

Let $(\mathbf{L}, *, \mathbf{N}, \mathbf{F})$ be a fusion system of rank $r$ and denote the elements of $\mathbf{L}$ by $1, \ldots, r$. Define a category $\mathscr{C}$ as follows. The objects are $r$ tuples of natural numbers

$$\operatorname{Obj} \mathscr{C} = \{(n_1, \ldots, n_r) \mid n_1, \ldots, n_r \in \mathbb{N}\} \tag{7.25}$$

and for every two objects $a, b$ the set of morphisms is the following set of blockdiagonal matrices

$$\hom(a, b) = \bigoplus_{i \in L} \operatorname{Mat}_{a_i \times b_i}(\mathbb{C}). \tag{7.26}$$

Composition of morphisms is given by matrix multiplication and the identity morphism is $1_a = \oplus_{i \in L} \mathbb{1}_{a_i \times a_i}$. The simple objects of $\mathscr{C}$ are the lists $\delta_a$ that have a 1 on the $a$th spot and are zero everywhere else.

The monoidal product $\otimes$ is defined on objects as

$$(a \otimes b)_k = \sum_{i=1, j=1}^{r} a_i b_j N_{i,j}^k, \quad a, b \in \operatorname{Obj} \mathscr{C}, \tag{7.27}$$

and on morphisms $A : a \to a', B : b \to b'$ as

$$A \otimes B = \left( \bigoplus_{i \in L} A_i \right) \otimes \left( \bigoplus_{j \in L} B_j \right) = \bigoplus_{k \in L} \left( \bigoplus_{i,j \in L} A_i \otimes B_j \otimes \mathbb{1}_{N_{i,j}^k \times N_{i,j}^k} \right) \tag{7.28}$$

where $A_i \in \operatorname{Mat}_{a_i \times a_i'}$, $B_i \in \operatorname{Mat}_{b_i \times b_i'}$. The neutral object is $\mathbf{1} = \delta_1$ and the unitor maps $\lambda_a : \mathbf{1} \otimes a \to a$ and $\rho_a : a \otimes \mathbf{1} \to a$ are taken to be the identity. Let $a = \delta_i, b = \delta_j, c = \delta_k$ be simple objects. The associator $\alpha_{a,b,c} : a \otimes (b \otimes c) \to (a \otimes b) \otimes c$ is then defined as

$$\alpha_{a,b,c} := \bigoplus_{l=1}^{r} \left[ \tilde{F}_l^{ijk} \right] \in \bigoplus_{l=1}^{r} \operatorname{Mat}_{N_{i,j,k}^l \times N_{i,j,k}^l}(\mathbb{C}). \tag{7.29}$$

The $*$ function on $\mathbf{L}$ is a permutation on $r$ elements. Let $M$ be the corresponding $r \times r$ matrix, then for any object $a$ its right dual is defined as $a.M$. For any simple object $a$ the evaluation and co-evaluations morphisms $e_a$ and $c_a$ are scalars, given by

$$e_a = 1, \quad c_a = \frac{1}{[F_a^{aa*a}]_{(1,1,1)}^{(1,1,1)}}. \tag{7.30}$$

> **Note 16.** In [19] other conventions are used for several structures:
>
> - The associator is defined the opposite way, namely $\alpha_{a,b,c} : (a \otimes b) \otimes c \to a \otimes (b \otimes c)$. Therefore they use the regular $F$ matrix, rather than the inverse, in definition 7.29.
>
> - The conventions for the evaluation and coevaluation maps are different to ours. This has no effect on the values $e_a$ and $c_a$, since the derivation of these values uses both opposite conventions, whose effects cancel in the

> end.
>
> - The pivotal structure in [19] is also defined as a map $a^{**} \to a$ which is opposite to our definition. Therefore our formula for computing pivotal coefficients 2.24 has an inverse LHS compared to the one in [19].

**Additional Structure**

If the fusion system is pivotal with pivotal structure $\mathbf{P} = (p_1, ..., p_r)$ then the pivotal structure $\psi$ is defined as the set of scalars $\{\psi_a = p_a | a = 1, ..., r\}$. If the fusion system is braided with braiding $\mathbf{R} = \{[R_c^{ab}] | a, b, c = 1, ..., r\}$ then the braiding $\sigma$ on $\mathscr{C}$ is determined by

$$\sigma_{a,b} = \bigoplus_{k=1}^{r} [R_k^{ij}] \in \bigoplus_{k=1}^{r} \mathrm{Mat}_{N_{i,j}^k \times N_{i,j}^k}(\mathbb{C}), \qquad (7.31)$$

where $a = \delta_i$ and $b = \delta_j$.

**From Fusion Category to Fusion System**

Let $\mathscr{C}$ be a fusion category and let $\mathscr{L}$ denote a set of representatives of equivalence classes of simple objects, and $l : \mathscr{L} \to \{1, ..., |\mathscr{L}| =: r\}$ a labeling function which maps the unit of $\mathscr{C}$, $\mathbf{1}$, to 1. Let $a^\star$ denote the dual of any object $a \in \mathrm{Obj}\,\mathscr{C}$. For any simple object $a$ we have that $a^{\star\star} \cong a$ and moreover $\mathbf{1}^\star \cong \mathbf{1}$. Let $[\cdot]_\mathscr{L}$ denote the projection of any simple object onto its representative in $\mathscr{L}$, then the set $\mathbf{L} = \{1, ..., r\}$ together with the map $\cdot^* : \mathbf{L} \to \mathbf{L} = l([(l^{-1}(\cdot))^\star]_\mathscr{L})$ are the first two elements of the desired fusion system $(\mathbf{L}, *, \mathbf{N}, \mathbf{F})$.

Since all hom sets are finite dimensional vector spaces, we can use these to describe the category by means of linear algebra. For all $a, b, c \in \mathscr{L}$ choose a basis $\{|a, b; c, i\rangle\}$ for the hom set $\hom(a \otimes b, c)$ such that:

- the sets of maps $\{\lambda_a\}$, and $\{\rho_a\}$ are respectively the bases for the spaces $\hom(\mathbf{1} \otimes a, a)$ and $\hom(a \otimes \mathbf{1}, a)$, and

- the set of maps $\{e_a\}$ is the basis for the space $\hom(a^* \otimes a, \mathbf{1})$.

We have that $\mathbf{N} := \left\{ N_{a,b}^c := \dim \hom(a \otimes b, c) \big| a, b, c \in \mathscr{L} \right\}$ are the structure constants of a fusion ring and thus satisfy all the required relations for a fusion system.

Let $a, b, c \in \mathscr{L}$ then $\alpha_{a,b,c} : (a \otimes b) \otimes c \to a \otimes (b \otimes c)$ defines a pullback

$$\alpha'_{a,b,c} : \hom(a \otimes (b \otimes c), d) \to \hom((a \otimes b) \otimes c, d) \qquad (7.32)$$

for every $d \in \mathscr{L}$. The choice of bases for the spaces $\hom(i \otimes j, k)$, with $i, j, k \in \mathscr{L}$ implies a choice of basis for higher tensor products as well. In particular the following are bases for the hom spaces appearing in (7.32)

$$\{|a, f; d, i\rangle \circ (1_a \otimes |b, c; f, j\rangle) | i = 1, ..., N_{a,f}^d, j = 1, ..., N_{b,c}^f\} \subset \hom(a \otimes (b \otimes c), d), \quad (7.33)$$

$$\{|e, c; d, j\rangle \circ (|a, b; e, i\rangle \otimes 1_c) | i = 1, ..., N_{a,b}^e, j = 1, ..., N_{e,c}^d\} \subset \hom((a \otimes b) \otimes c, d). \quad (7.34)$$

The matrices $[F_d^{abc}]$ are then by definition the matrices representing $\alpha'_{a,b,c}$ in these bases, i.e.

$$[F_d^{abc}]^{(e,i,j)}_{(f,i',j')} := [\alpha'_{a,b,c}(|e,c;d,j\rangle \circ (|a,b;e,i\rangle \otimes 1_c))]_{(f,i',j')} \tag{7.35}$$

**Additional Structure**

If $\mathscr{C}$ is pivotal with pivotal structure $\psi$ then $\mathbf{P} = (p_1, \ldots, p_r)$, with

$$p_a = \operatorname{Tr}^L(\psi_a)[F_a^{aa^*a}]^{(1,1,1)}_{(1,1,1)}, \tag{7.36}$$

is a set of pivotal coefficients. If $\mathscr{C}$ is braided then the $R$-matrices are found in a similar way as the $F$-symbols: via the pullback of the braiding map $\sigma$. In particular we have that, for all $a, b, c \in \mathscr{L}$,

$$|b,a;c,i\rangle \circ \sigma_{a,b} = \sum_j [R_c^{ab}]^i_j |a,b;c,j\rangle \tag{7.37}$$

## Chapter 8

# List of multiplicity-free fusion rings up to rank 9

A list of multiplicity-free fusion rings of rank up to 9 is given in the table below. $\mathscr{D}^2_{FP}$ denotes the sum of the squares of the quantum dimensions of the basis elements of the ring, and for the last five columns, a value of ✓ indicates that at least one way to categorify the fusion ring to a category with the respective structure exists. The abbreviations FC, PFC, UFC, BFC, and MFC stand for Fusion Category, Pivotal Fusion Category, Unitary Fusion Category, Braided Fusion Category, and Modular Fusion Category. Multiple categories can stem from the same fusion ring with multiple (possibly disjunct) properties. A ✗ indicates that it is known that the ring does not categorify to the respective category, and an empty cell indicates that the authors don't know whether the ring has a category with the respective structure.

Note that:

- The table list the data of which the authors have knowledge. There might be some fusion rings, of which more is known than listed in this table.

- The table on the AnyonWiki is derived from this paper but might contain more information since it gets updated now and then by volunteers.

- Not all fusion rings that have a common name are categorifiable. Some names, such as the Tambara-Yamagami (TY) fusion rings, the Haagerup-Izumi (HI) fusion rings, and the songs, are derived from a construction of a fusion ring from another ring. For e.g., the TY fusion rings, Tambara and Yamagami, proved[94] that if they are based on a non-commutative group, there can be no categorifications. Likewise, a lot of the songs have no categorifications.

Table 8.1: List of multiplicity-free fusion rings up to rank 9

| Name | Common Name | Rank | $\mathcal{D}_{FP}^2$ | Comm. | FC | PFC | UFC | BFC | MFC |
|---|---|---|---|---|---|---|---|---|---|
| $FR_1^{1,0}$ | Trivial | 1 | 1. | ✓ | ✓ | ✓ | ✓ | ✓ | ✓ |
| $FR_1^{2,0}$ | $\mathbb{Z}_2$ | 2 | 2. | ✓ | ✓ | ✓ | ✓ | ✓ | ✓ |
| $FR_2^{2,0}$ | Fib | 2 | 3.618 | ✓ | ✓ | ✓ | ✓ | ✓ | ✓ |
| $FR_1^{3,0}$ | Ising | 3 | 4. | ✓ | ✓ | ✓ | ✓ | ✓ | ✓ |
| $FR_2^{3,0}$ | $\mathrm{Rep}(D_3)$ | 3 | 6. | ✓ | ✓ | ✓ | ✓ | ✓ | ✗ |
| $FR_3^{3,0}$ | $PSU(2)_5$ | 3 | 9.295 | ✓ | ✓ | ✓ | ✓ | ✓ | ✓ |
| $FR_1^{3,2}$ | $\mathbb{Z}_3$ | 3 | 3. | ✓ | ✓ | ✓ | ✓ | ✓ | ✓ |
| $FR_1^{4,0}$ | $\mathbb{Z}_2 \times \mathbb{Z}_2$ | 4 | 4. | ✓ | ✓ | ✓ | ✓ | ✓ | ✓ |
| $FR_2^{4,0}$ | $SU(2)_3$ | 4 | 7.236 | ✓ | ✓ | ✓ | ✓ | ✓ | ✓ |
| $FR_3^{4,0}$ | $\mathrm{Rep}(D_5)$ | 4 | 10. | ✓ | ✓ | ✓ | ✓ | ✓ | ✗ |
| $FR_4^{4,0}$ | $PSU(2)_6$ | 4 | 13.656 | ✓ | ✓ | ✓ | ✓ | ✓ | ✗ |
| $FR_5^{4,0}$ | Fib × Fib | 4 | 13.090 | ✓ | ✓ | ✓ | ✓ | ✓ | ✓ |
| $FR_6^{4,0}$ | $PSU(2)_7$ | 4 | 19.234 | ✓ | ✓ | ✓ | ✓ | ✓ | ✓ |
| $FR_1^{4,2}$ | $\mathbb{Z}_4$ | 4 | 4. | ✓ | ✓ | ✓ | ✓ | ✓ | ✓ |
| $FR_2^{4,2}$ | $TY(\mathbb{Z}_3)$ | 4 | 6. | ✓ | ✓ | ✓ | ✓ | ✗ | ✗ |
| $FR_3^{4,2}$ | $\mathrm{Fib}(\mathbb{Z}_3)$ | 4 | 8.3027 | ✓ | ✗ | ✗ | ✗ | ✗ | ✗ |
| $FR_4^{4,2}$ | Pseudo $PSU(2)_6$ | 4 | 13.656 | ✓ | ✓ | ✓ | ✓ | ✗ | ✗ |
| $FR_1^{5,0}$ | $\mathrm{Rep}(D_4)$ | 5 | 8. | ✓ | ✓ | ✓ | ✓ | ✓ | ✗ |
| $FR_2^{5,0}$ | Fib $(\mathbb{Z}_2 \times \mathbb{Z}_2)$ | 5 | 10.561 | ✓ | ✗ | ✗ | ✗ | ✗ | ✗ |
| $FR_3^{5,0}$ | $SU(2)_4$ | 5 | 12. | ✓ | ✓ | ✓ | ✓ | ✓ | ✓ |
| $FR_4^{5,0}$ | $\mathrm{Rep}(D_7)$ | 5 | 14. | ✓ | ✓ | ✓ | ✓ | ✓ | ✗ |
| $FR_5^{5,0}$ | | 5 | 16.605 | ✓ | ✗ | ✗ | ✗ | ✗ | ✗ |
| $FR_6^{5,0}$ | $\mathrm{Rep}(S_4)$ | 5 | 24. | ✓ | ✓ | ✓ | ✓ | ✓ | ✗ |
| $FR_7^{5,0}$ | $PSU(2)_8$ | 5 | 26.180 | ✓ | ✓ | ✓ | ✓ | ✓ | ✗ |
| $FR_8^{5,0}$ | | 5 | 31.092 | ✓ | ✗ | ✗ | ✗ | ✗ | ✗ |
| $FR_9^{5,0}$ | | 5 | 30.142 | ✓ | ✗ | ✗ | ✗ | ✗ | ✗ |

Table 8.1: List of multiplicity-free fusion rings up to rank 9 (Continued)

| Name | Common Name | Rank | $\mathcal{D}_{FP}^2$ | Comm. | FC | PFC | UFC | BFC | MFC |
|---|---|---|---|---|---|---|---|---|---|
| $FR_{10}^{5,0}$ | $PSU(2)_9$ | 5 | 34.646 | ✓ | ✓ | ✓ | ✓ | ✓ | ✓ |
| $FR_1^{5,2}$ | $TY(\mathbb{Z}_4)$ | 5 | 8. | ✓ | ✓ | ✓ | ✓ | ✗ | ✗ |
| $FR_2^{5,2}$ | $Fib(\mathbb{Z}_4)$ | 5 | 10.561 | ✓ | ✗ | ✗ | ✗ | ✗ | ✗ |
| $FR_3^{5,2}$ | Pseudo $SU(2)_4$ | 5 | 12. | ✓ | ✓ | ✓ | ✓ | ✗ | ✗ |
| $FR_4^{5,2}$ | Pseudo $Rep(S_4)$ | 5 | 24. | ✓ | ✓ | ✓ | ✓ | ✗ | ✗ |
| $FR_5^{5,2}$ | | 5 | 31.092 | ✓ | ✗ | ✗ | ✗ | ✗ | ✗ |
| $FR_1^{5,4}$ | $\mathbb{Z}_5$ | 5 | 5. | ✓ | ✓ | ✓ | ✓ | ✓ | ✓ |
| $FR_1^{6,0}$ | $\mathbb{Z}_2 \times$ Ising | 6 | 8. | ✓ | ✓ | ✓ | ✓ | ✓ | ✓ |
| $FR_2^{6,0}$ | $\mathbb{Z}_2 \times Rep(D_3)$ | 6 | 12. | ✓ | ✓ | ✓ | ✓ | ✓ | ✗ |
| $FR_3^{6,0}$ | | 6 | 18.928 | ✓ | ✗ | ✗ | ✗ | ✗ | ✗ |
| $FR_4^{6,0}$ | TriCritIsing | 6 | 14.472 | ✓ | ✓ | ✓ | ✓ | ✓ | ✓ |
| $FR_5^{6,0}$ | Fib $\times Rep(D_3)$ | 6 | 21.708 | ✓ | ✓ | ✓ | ✓ | ✓ | ✗ |
| $FR_6^{6,0}$ | $SU(2)_5$ | 6 | 18.591 | ✓ | ✓ | ✓ | ✓ | ✓ | ✓ |
| $FR_7^{6,0}$ | $Rep(\mathbb{Z}_3 \rtimes D_3)$ | 6 | 18. | ✓ | ✓ | ✓ | ✓ | ✓ | ✗ |
| $FR_8^{6,0}$ | $Rep(D_9)$ | 6 | 18. | ✓ | ✓ | ✓ | ✓ | ✓ | ✗ |
| $FR_9^{6,0}$ | $SO(5)_2$ | 6 | 20. | ✓ | ✓ | ✓ | ✓ | ✓ | ✓ |
| $FR_{10}^{6,0}$ | | 6 | 25.582 | ✓ | ✗ | ✗ | ✗ | ✗ | ✗ |
| $FR_{11}^{6,0}$ | | 6 | 28.392 | ✓ | ✗ | ✗ | ✗ | ✗ | ✗ |
| $FR_{12}^{6,0}$ | | 6 | 28.392 | ✓ | ✗ | ✗ | ✗ | ✗ | ✗ |
| $FR_{13}^{6,0}$ | | 6 | 33.798 | ✓ | ✗ | ✗ | ✗ | ✗ | ✗ |
| $FR_{14}^{6,0}$ | Fib $\times ExtRep(D_3)$ | 6 | 33.632 | ✓ | ✓ | ✓ | ✓ | ✓ | ✓ |
| $FR_{15}^{6,0}$ | | 6 | 36.779 | ✓ | ✗ | ✗ | ✗ | ✗ | ✗ |
| $FR_{16}^{6,0}$ | $PSU(2)_{10}$ | 6 | 44.784 | ✓ | ✓ | ✓ | ✓ | ✓ | ✗ |
| $FR_{17}^{6,0}$ | | 6 | 55.144 | ✓ | ✗ | ✗ | ✗ | ✗ | ✗ |
| $FR_{18}^{6,0}$ | $PSU(2)_{11}$ | 6 | 56.746 | ✓ | ✓ | ✓ | ✓ | ✓ | ✓ |
| $FR_{19}^{6,0}$ | | 6 | 63.147 | ✓ | ✗ | ✗ | ✗ | ✗ | ✗ |

Table 8.1: List of multiplicity-free fusion rings up to rank 9 (Continued)

| Name | Common Name | Rank | $\mathscr{D}_{FP}^2$ | Comm. | FC | PFC | UFC | BFC | MFC |
|---|---|---|---|---|---|---|---|---|---|
| $FR_{20}^{6,0}$ | | 6 | 63.147 | ✓ | ✗ | ✗ | ✗ | ✗ | ✗ |
| $FR_1^{6,2}$ | $D_3$ | 6 | 6. | ✗ | ✓ | ✓ | ✓ | ✗ | ✗ |
| $FR_2^{6,2}$ | $[\mathbb{Z}_2 \trianglelefteq \mathbb{Z}_4]_{1\vert 0}^{\text{Id}}$ | 6 | 8. | ✓ | ✓ | ✓ | ✓ | ✗ | ✗ |
| $FR_3^{6,2}$ | $[\mathbb{Z}_2 \trianglelefteq \mathbb{Z}_2 \times \mathbb{Z}_2]_{3\vert 0}^{\text{Id}}$ | 6 | 8. | ✓ | ✓ | ✓ | ✓ | ✓ | ✗ |
| $FR_4^{6,2}$ | $\text{Rep}(\text{Dic}_{12})$ | 6 | 12. | ✓ | ✓ | ✓ | ✓ | ✓ | ✗ |
| $FR_5^{6,2}$ | $[\mathbb{Z}_2 \trianglelefteq \mathbb{Z}_4]_{1\vert 1}^{\text{Id}}$ | 6 | 18.928 | ✓ | ✗ | ✗ | ✗ | ✗ | ✗ |
| $FR_6^{6,2}$ | $[\mathbb{Z}_2 \trianglelefteq \mathbb{Z}_2 \times \mathbb{Z}_2]_{3\vert 1}^{\text{Id}}$ | 6 | 18.928 | ✓ | ✗ | ✗ | ✗ | ✗ | ✗ |
| $FR_7^{6,2}$ | Pseudo $SO(5)_2$ | 6 | 20. | ✓ | ✓ | ✓ | ✓ | ✗ | ✗ |
| $FR_8^{6,2}$ | $HI(\mathbb{Z}_3)$ | 6 | 35.725 | ✗ | ✓ | ✓ | ✓ | ✗ | ✗ |
| $FR_9^{6,2}$ | | 6 | 33.798 | ✓ | ✗ | ✗ | ✗ | ✗ | ✗ |
| $FR_{10}^{6,2}$ | | 6 | 36.779 | ✓ | ✗ | ✗ | ✗ | ✗ | ✗ |
| $FR_{11}^{6,2}$ | | 6 | 55.144 | ✓ | ✗ | ✗ | ✗ | ✗ | ✗ |
| $FR_1^{6,4}$ | $\mathbb{Z}_6$ | 6 | 6. | ✓ | ✓ | ✓ | ✓ | ✓ | ✓ |
| $FR_2^{6,4}$ | $MR_6$ | 6 | 8. | ✓ | ✓ | ✓ | ✓ | ✗ | ✗ |
| $FR_3^{6,4}$ | $TY(\mathbb{Z}_5)$ | 6 | 10. | ✓ | ✓ | ✓ | ✓ | ✗ | ✗ |
| $FR_4^{6,4}$ | $[\mathbb{Z}_5 \trianglelefteq \mathbb{Z}_5]_{1\vert 1}^{\text{Id}}$ | 6 | 12.791 | ✓ | ✗ | ✗ | ✗ | ✗ | ✗ |
| $FR_5^{6,4}$ | $\text{Fib} \times \mathbb{Z}_3$ | 6 | 10.854 | ✓ | ✓ | ✓ | ✓ | ✓ | ✓ |
| $FR_6^{6,4}$ | $[\mathbb{Z}_2 \trianglelefteq \mathbb{Z}_4]_{3\vert 1}^{\text{Id}}$ | 6 | 18.928 | ✓ | ✗ | ✗ | ✗ | ✗ | ✗ |
| $FR_7^{6,4}$ | | 6 | 20.485 | ✓ | ✗ | ✗ | ✗ | ✗ | ✗ |
| $FR_8^{6,4}$ | $[I \trianglelefteq \mathbb{Z}_3]_{1\vert 1}^{\text{Id}}$ | 6 | 35.725 | ✓ | ✗ | ✗ | ✗ | ✗ | ✗ |
| $FR_1^{7,0}$ | | 7 | 16. | ✓ | ✓ | ✓ | ✓ | ✓ | ✗ |
| $FR_2^{7,0}$ | | 7 | 21.123 | ✓ | ✗ | ✗ | ✗ | ✗ | ✗ |
| $FR_3^{7,0}$ | | 7 | 27.153 | ✓ | ✗ | ✗ | ✗ | ✗ | ✗ |
| $FR_4^{7,0}$ | | 7 | 28.944 | ✓ | ✗ | ✗ | ✗ | ✗ | ✗ |
| $FR_5^{7,0}$ | | 7 | 29.46 | ✓ | ✗ | ✗ | ✗ | ✗ | ✗ |

Table 8.1: List of multiplicity-free fusion rings up to rank 9 (Continued)

| Name | Common Name | Rank | $\mathscr{D}_{FP}^2$ | Comm. | FC | PFC | UFC | BFC | MFC |
|---|---|---|---|---|---|---|---|---|---|
| $FR_6^{7,0}$ | | 7 | 22. | ✓ | ✓ | ✓ | ✓ | ✓ | ✗ |
| $FR_7^{7,0}$ | $SU(2)_6$ | 7 | 27.3137 | ✓ | ✓ | ✓ | ✓ | ✓ | ✓ |
| $FR_8^{7,0}$ | $SO(7)_2$ | 7 | 28. | ✓ | ✓ | ✓ | ✓ | ✓ | ✓ |
| $FR_9^{7,0}$ | | 7 | 34.3852 | ✓ | ✗ | ✗ | ✗ | ✗ | ✗ |
| $FR_{10}^{7,0}$ | | 7 | 43.3137 | ✓ | ✗ | ✗ | ✗ | ✗ | ✗ |
| $FR_{11}^{7,0}$ | | 7 | 36.9706 | ✓ | ✗ | ✗ | ✗ | ✗ | ✗ |
| $FR_{12}^{7,0}$ | | 7 | 42. | ✓ | ✗ | ✗ | ✗ | ✗ | ✗ |
| $FR_{13}^{7,0}$ | | 7 | 52.93 | ✓ | ✗ | ✗ | ✗ | ✗ | ✗ |
| $FR_{14}^{7,0}$ | $PSU(2)_{12}$ | 7 | 70.6848 | ✓ | ✓ | ✓ | ✓ | ✓ | ✗ |
| $FR_{15}^{7,0}$ | | 7 | 81.6695 | ✓ | ✗ | ✗ | ✗ | ✗ | ✗ |
| $FR_{16}^{7,0}$ | | 7 | 87.0937 | ✓ | ✗ | ✗ | ✗ | ✗ | ✗ |
| $FR_{17}^{7,0}$ | $PSU(2)_{13}$ | 7 | 86.7508 | ✓ | ✓ | ✓ | ✓ | ✓ | ✓ |
| $FR_{18}^{7,0}$ | | 7 | 118.138 | ✓ | ✗ | ✗ | ✗ | ✗ | ✗ |
| $FR_1^{7,2}$ | $TY(D_3)$ | 7 | 12. | ✗ | ✗ | ✗ | ✗ | ✗ | ✗ |
| $FR_2^{7,2}$ | $[D_3 \trianglelefteq D_3]_{\mathbf{1}\vert 1}^{\text{Id}}$ | 7 | 15. | ✗ | ✗ | ✗ | ✗ | ✗ | ✗ |
| $FR_3^{7,2}$ | | 7 | 16. | ✓ | ✓ | ✓ | ✓ | ✗ | ✗ |
| $FR_4^{7,2}$ | | 7 | 16. | ✓ | ✓ | ✓ | ✓ | ✓ | ✗ |
| $FR_5^{7,2}$ | | 7 | 21.1231 | ✓ | ✗ | ✗ | ✗ | ✗ | ✗ |
| $FR_6^{7,2}$ | | 7 | 27.1537 | ✓ | ✗ | ✗ | ✗ | ✗ | ✗ |
| $FR_7^{7,2}$ | | 7 | 27.1537 | ✓ | ✗ | ✗ | ✗ | ✗ | ✗ |
| $FR_8^{7,2}$ | | 7 | 28.9443 | ✓ | ✗ | ✗ | ✗ | ✗ | ✗ |
| $FR_9^{7,2}$ | | 7 | 28.9443 | ✓ | ✗ | ✗ | ✗ | ✗ | ✗ |
| $FR_{10}^{7,2}$ | | 7 | 29.46 | ✓ | ✗ | ✗ | ✗ | ✗ | ✗ |
| $FR_{11}^{7,2}$ | | 7 | 27.3137 | ✓ | ✗ | ✗ | ✗ | ✗ | ✗ |
| $FR_{12}^{7,2}$ | | 7 | 28. | ✓ | ✓ | ✓ | ✗ | ✗ | ✗ |
| $FR_{13}^{7,2}$ | | 7 | 43.3137 | ✓ | ✗ | ✗ | ✗ | ✗ | ✗ |

Table 8.1: List of multiplicity-free fusion rings up to rank 9 (Continued)

| Name | Common Name | Rank | $\mathscr{D}_{FP}^2$ | Comm. | FC | PFC | UFC | BFC | MFC |
|---|---|---|---|---|---|---|---|---|---|
| $FR_{14}^{7,2}$ | | 7 | 52.93 | ✓ | ✗ | ✗ | ✗ | ✗ | ✗ |
| $FR_{15}^{7,2}$ | | 7 | 71.0118 | ✗ | ✗ | ✗ | ✗ | ✗ | ✗ |
| $FR_{16}^{7,2}$ | | 7 | 81.6695 | ✓ | ✗ | ✗ | ✗ | ✗ | ✗ |
| $FR_{17}^{7,2}$ | | 7 | 87.0937 | ✓ | ✗ | ✗ | ✗ | ✗ | ✗ |
| $FR_1^{7,4}$ | $TY(\mathbb{Z}_2 \times \mathbb{Z}_3)$ | 7 | 12. | ✓ | ✓ | ✓ | ✓ | ✗ | ✗ |
| $FR_2^{7,4}$ | $[\mathbb{Z}_6 \trianglelefteq \mathbb{Z}_6]_{1|1}^{Id}$ | 7 | 15. | ✓ | ✗ | ✗ | ✗ | ✗ | ✗ |
| $FR_3^{7,4}$ | | 7 | 16. | ✓ | ✓ | ✓ | ✗ | ✗ | ✗ |
| $FR_4^{7,4}$ | | 7 | 27.1537 | ✓ | ✗ | ✗ | ✗ | ✗ | ✗ |
| $FR_5^{7,4}$ | | 7 | 28.9443 | ✓ | ✗ | ✗ | ✗ | ✗ | ✗ |
| $FR_6^{7,4}$ | | 7 | 57.2354 | ✓ | ✗ | ✗ | ✗ | ✗ | ✗ |
| $FR_7^{7,4}$ | | 7 | 71.0118 | ✓ | ✗ | ✗ | ✗ | ✗ | ✗ |
| $FR_1^{7,6}$ | $\mathbb{Z}_7$ | 7 | 7. | ✓ | ✓ | ✓ | ✓ | ✓ | ✓ |
| $FR_1^{8,0}$ | $\mathbb{Z}_2 \times \mathbb{Z}_2 \times \mathbb{Z}_2$ | 8 | 8. | ✓ | ✓ | ✓ | ✓ | ✓ | ✓ |
| $FR_2^{8,0}$ | $Fib \times \mathbb{Z}_2 \times \mathbb{Z}_2$ | 8 | 14.4721 | ✓ | ✓ | ✓ | ✓ | ✓ | ✓ |
| $FR_3^{8,0}$ | $Rep(D_5) \times \mathbb{Z}_2$ | 8 | 20. | ✓ | ✓ | ✓ | ✓ | ✓ | ✗ |
| $FR_4^{8,0}$ | | 8 | 24. | ✓ | | | | | ✗ |
| $FR_5^{8,0}$ | $PSU(2)_6 \times \mathbb{Z}_2$ | 8 | 27.3137 | ✓ | ✓ | ✓ | ✓ | ✓ | ✗ |
| $FR_6^{8,0}$ | | 8 | 30. | ✓ | ✗ | ✗ | ✗ | ✗ | ✗ |
| $FR_7^{8,0}$ | $Fib \times Fib \times \mathbb{Z}_2$ | 8 | 26.1803 | ✓ | ✓ | ✓ | ✓ | ✓ | ✓ |
| $FR_8^{8,0}$ | | 8 | 38.583 | ✓ | ✗ | ✗ | ✗ | ✗ | ✗ |
| $FR_9^{8,0}$ | | 8 | 26. | ✓ | | | | | ✗ |
| $FR_{10}^{8,0}$ | | 8 | 42.4585 | ✓ | ✗ | ✗ | ✗ | ✗ | ✗ |
| $FR_{11}^{8,0}$ | $Fib \times Rep(D_5)$ | 8 | 36.1803 | ✓ | ✓ | ✓ | ✓ | ✓ | ✗ |
| $FR_{12}^{8,0}$ | | 8 | 47.6333 | ✓ | ✗ | ✗ | ✗ | ✗ | ✗ |
| $FR_{13}^{8,0}$ | $SO(9)_2$ | 8 | 36. | ✓ | ✓ | ✓ | ✓ | ✓ | ✓ |
| $FR_{14}^{8,0}$ | $Rep(D(D_3))$ | 8 | 36. | ✓ | | | | | |

| Name | Common Name | Rank | $\mathscr{D}_{FP}^2$ | Comm. | FC | PFC | UFC | BFC | MFC |
|---|---|---|---|---|---|---|---|---|---|
| $FR_{15}^{8,0}$ | $SU(2)_7$ | 8 | 38.4688 | ✓ | ✓ | ✓ | ✓ | ✓ | ✓ |
| $FR_{16}^{8,0}$ | Fib × $PSU(2)_6$ | 8 | 49.411 | ✓ | ✓ | ✓ | ✓ | ✓ | ✗ |
| $FR_{17}^{8,0}$ | | 8 | 43.0828 | ✓ | ✗ | ✗ | ✗ | ✗ | ✗ |
| $FR_{18}^{8,0}$ | | 8 | 43.0828 | ✓ | ✗ | ✗ | ✗ | ✗ | ✗ |
| $FR_{19}^{8,0}$ | | 8 | 68.6639 | ✓ | ✗ | ✗ | ✗ | ✗ | ✗ |
| $FR_{20}^{8,0}$ | | 8 | 52.6491 | ✓ | ✗ | ✗ | ✗ | ✗ | ✗ |
| $FR_{21}^{8,0}$ | | 8 | 52.6491 | ✓ | ✗ | ✗ | ✗ | ✗ | ✗ |
| $FR_{22}^{8,0}$ | Fib × Fib × Fib | 8 | 47.3607 | ✓ | ✓ | ✓ | ✓ | ✓ | ✓ |
| $FR_{23}^{8,0}$ | $HI(\mathbb{Z}_2 \times \mathbb{Z}_2)$ | 8 | 75.7771 | ✓ | | | | | ✗ |
| $FR_{24}^{8,0}$ | | 8 | 48.0685 | ✓ | ✗ | ✗ | ✗ | ✗ | ✗ |
| $FR_{25}^{8,0}$ | | 8 | 58.1168 | ✓ | ✗ | ✗ | ✗ | ✗ | ✗ |
| $FR_{26}^{8,0}$ | | 8 | 58.1168 | ✓ | ✗ | ✗ | ✗ | ✗ | ✗ |
| $FR_{27}^{8,0}$ | Fib × $PSU(2)_7$ | 8 | 69.5908 | ✓ | ✓ | ✓ | ✓ | ✓ | ✓ |
| $FR_{28}^{8,0}$ | | 8 | 72. | ✓ | | | | | ✗ |
| $FR_{29}^{8,0}$ | | 8 | 72. | ✓ | | | | | ✗ |
| $FR_{30}^{8,0}$ | | 8 | 78.1637 | ✓ | ✗ | ✗ | ✗ | ✗ | ✗ |
| $FR_{31}^{8,0}$ | $PSU(2)_{14}$ | 8 | 105.097 | ✓ | ✓ | ✓ | ✓ | ✓ | ✗ |
| $FR_{32}^{8,0}$ | | 8 | 126.522 | ✓ | ✗ | ✗ | ✗ | ✗ | ✗ |
| $FR_{33}^{8,0}$ | | 8 | 126.522 | ✓ | ✗ | ✗ | ✗ | ✗ | ✗ |
| $FR_{34}^{8,0}$ | | 8 | 128.169 | ✓ | ✗ | ✗ | ✗ | ✗ | ✗ |
| $FR_{35}^{8,0}$ | | 8 | 122.573 | ✓ | ✗ | ✗ | ✗ | ✗ | ✗ |
| $FR_{36}^{8,0}$ | $PSU(2)_{15}$ | 8 | 125.874 | ✓ | ✓ | ✓ | ✓ | ✓ | ✓ |
| $FR_{37}^{8,0}$ | | 8 | 140.586 | ✓ | ✗ | ✗ | ✗ | ✗ | ✗ |
| $FR_{38}^{8,0}$ | | 8 | 201.126 | ✓ | ✗ | ✗ | ✗ | ✗ | ✗ |
| $FR_1^{8,2}$ | $D_4$ | 8 | 8. | ✗ | ✓ | ✓ | ✓ | ✗ | ✗ |
| $FR_2^{8,2}$ | $[\mathbb{Z}_3 \trianglelefteq D_3]_{1\mid 0}^{\mathrm{Id}}$ | 8 | 12. | ✗ | ✓ | | | | ✗ |

Table 8.1: List of multiplicity-free fusion rings up to rank 9 (Continued)

| Name | Common Name | Rank | $\mathscr{D}_{FP}^2$ | Comm. | FC | PFC | UFC | BFC | MFC |
|---|---|---|---|---|---|---|---|---|---|
| $FR_3^{8,2}$ | | 8 | 16.6056 | ✗ | | | | | ✗ |
| $FR_4^{8,2}$ | $[\mathbb{Z}_3 \trianglelefteq D_3]_{1\|1}^{\text{Id}}$ | 8 | 24. | ✗ | ✗ | ✗ | ✗ | ✗ | ✗ |
| $FR_5^{8,2}$ | | 8 | 20. | ✓ | | | | | ✗ |
| $FR_6^{8,2}$ | | 8 | 20. | ✗ | | | | | ✗ |
| $FR_7^{8,2}$ | | 8 | 24. | ✓ | | | | | ✗ |
| $FR_8^{8,2}$ | | 8 | 27.3137 | ✗ | | | | | ✗ |
| $FR_9^{8,2}$ | | 8 | 27.3137 | ✗ | | | | | ✗ |
| $FR_{10}^{8,2}$ | | 8 | 27.3137 | ✗ | | | | | ✗ |
| $FR_{11}^{8,2}$ | | 8 | 26.1803 | ✗ | | | | | ✗ |
| $FR_{12}^{8,2}$ | | 8 | 38.583 | ✓ | ✗ | ✗ | ✗ | ✗ | ✗ |
| $FR_{13}^{8,2}$ | | 8 | 42.4585 | ✓ | ✗ | ✗ | ✗ | ✗ | ✗ |
| $FR_{14}^{8,2}$ | | 8 | 36. | ✓ | | | | | ✗ |
| $FR_{15}^{8,2}$ | | 8 | 36. | ✓ | | | | | ✗ |
| $FR_{16}^{8,2}$ | | 8 | 68.6639 | ✓ | ✗ | ✗ | ✗ | ✗ | ✗ |
| $FR_{17}^{8,2}$ | | 8 | 52.6491 | ✓ | ✗ | ✗ | ✗ | ✗ | ✗ |
| $FR_{18}^{8,2}$ | | 8 | 52.6491 | ✓ | ✗ | ✗ | ✗ | ✗ | ✗ |
| $FR_{19}^{8,2}$ | $HI(\mathbb{Z}_4)$ | 8 | 75.7771 | ✗ | | | | | ✗ |
| $FR_{20}^{8,2}$ | $[I \trianglelefteq \mathbb{Z}_2 \times \mathbb{Z}_2]_{1\|1}^{(3\ 4)}$ | 8 | 75.7771 | ✗ | | | | | ✗ |
| $FR_{21}^{8,2}$ | | 8 | 48.0685 | ✓ | ✗ | ✗ | ✗ | ✗ | ✗ |
| $FR_{22}^{8,2}$ | | 8 | 58.1168 | ✓ | ✗ | ✗ | ✗ | ✗ | ✗ |
| $FR_{23}^{8,2}$ | | 8 | 58.1168 | ✓ | ✗ | ✗ | ✗ | ✗ | ✗ |
| $FR_{24}^{8,2}$ | | 8 | 72. | ✓ | | | | | ✗ |
| $FR_{25}^{8,2}$ | | 8 | 72. | ✓ | | | | | ✗ |
| $FR_{26}^{8,2}$ | | 8 | 126.522 | ✓ | ✗ | ✗ | ✗ | ✗ | ✗ |
| $FR_{27}^{8,2}$ | | 8 | 126.522 | ✓ | ✗ | ✗ | ✗ | ✗ | ✗ |
| $FR_{28}^{8,2}$ | | 8 | 128.169 | ✓ | ✗ | ✗ | ✗ | ✗ | ✗ |

| Name | Common Name | Rank | $\mathscr{D}_{FP}^2$ | Comm. | FC | PFC | UFC | BFC | MFC |
|---|---|---|---|---|---|---|---|---|---|
| $FR^{8,2}_{29}$ | | 8 | 122.573 | ✗ | | | | | ✗ |
| $FR^{8,2}_{30}$ | | 8 | 140.586 | ✗ | | | | | ✗ |
| $FR^{8,4}_{1}$ | $\mathbb{Z}_2 \times \mathbb{Z}_4$ | 8 | 8. | ✓ | ✓ | ✓ | ✓ | ✓ | ✓ |
| $FR^{8,4}_{2}$ | $[\mathbb{Z}_3 \trianglelefteq D_3]^{\mathrm{Id}}_{2\mid 0}$ | 8 | 12. | ✗ | ✗ | ✗ | ✗ | ✗ | ✗ |
| $FR^{8,4}_{3}$ | $\mathbb{Z}_2 \times \mathrm{TY}(\mathbb{Z}_3)$ | 8 | 12. | ✓ | ✓ | ✓ | ✓ | ✗ | ✗ |
| $FR^{8,4}_{4}$ | $\mathbb{Z}_2 \times \mathrm{Fib}(\mathbb{Z}_3)$ | 8 | 16.6056 | ✓ | ✗ | ✗ | ✗ | ✗ | ✗ |
| $FR^{8,4}_{5}$ | $[\mathbb{Z}_3 \trianglelefteq D_3]^{\mathrm{Id}}_{2\mid 1}$ | 8 | 24. | ✗ | ✗ | ✗ | ✗ | ✗ | ✗ |
| $FR^{8,4}_{6}$ | $[\mathbb{Z}_3 \trianglelefteq \mathbb{Z}_6]^{\mathrm{Id}}_{1\mid 1}$ | 8 | 24. | ✓ | | | | | ✗ |
| $FR^{8,4}_{7}$ | $\mathrm{Fib} \times \mathbb{Z}_4$ | 8 | 14.4721 | ✓ | ✓ | ✓ | ✓ | ✓ | ✓ |
| $FR^{8,4}_{8}$ | | 8 | 20. | ✗ | | | | | ✗ |
| $FR^{8,4}_{9}$ | $\mathrm{Fib} \times \mathrm{TY}(\mathbb{Z}_3)$ | 8 | 21.7082 | ✓ | ✓ | ✓ | ✓ | ✗ | ✗ |
| $FR^{8,4}_{10}$ | $\mathrm{Fib} \times \mathrm{Fib}(\mathbb{Z}_3)$ | 8 | 30.0397 | ✓ | ✗ | ✗ | ✗ | ✗ | ✗ |
| $FR^{8,4}_{11}$ | | 8 | 27.3137 | ✓ | | | | | ✗ |
| $FR^{8,4}_{12}$ | | 8 | 27.3137 | ✓ | | | | | ✗ |
| $FR^{8,4}_{13}$ | $\mathbb{Z}_2 \times (\mathrm{Pseudo}\ \mathrm{PSU}(2)_6)$ | 8 | 27.3137 | ✓ | ✓ | ✓ | ✓ | ✗ | ✗ |
| $FR^{8,4}_{14}$ | | 8 | 47.6333 | ✓ | ✗ | ✗ | ✗ | ✗ | ✗ |
| $FR^{8,4}_{15}$ | $\mathrm{Fib} \times (\mathrm{Pseudo}\ \mathrm{PSU}(2)_6)$ | 8 | 49.411 | ✓ | ✓ | ✓ | ✓ | ✗ | ✗ |
| $FR^{8,4}_{16}$ | $[I \trianglelefteq \mathbb{Z}_4]^{\mathrm{Id}}_{1\mid 1}$ | 8 | 75.7771 | ✓ | | | | | ✗ |
| $FR^{8,4}_{17}$ | $[I \trianglelefteq \mathbb{Z}_2 \times \mathbb{Z}_2]^{\mathrm{Id}}_{2\mid 1}$ | 8 | 75.7771 | ✓ | | | | | ✗ |
| $FR^{8,4}_{18}$ | | 8 | 78.1637 | ✓ | ✗ | ✗ | ✗ | ✗ | ✗ |
| $FR^{8,6}_{1}$ | $Q$ | 8 | 8. | ✗ | ✓ | ✓ | ✓ | ✗ | ✗ |
| $FR^{8,6}_{2}$ | $\mathbb{Z}_8$ | 8 | 8. | ✓ | ✓ | ✓ | ✓ | ✓ | ✓ |
| $FR^{8,6}_{3}$ | $\mathrm{TY}(\mathbb{Z}_7)$ | 8 | 14. | ✓ | ✓ | ✓ | ✓ | ✗ | ✗ |
| $FR^{8,6}_{4}$ | $[\mathbb{Z}_7 \trianglelefteq \mathbb{Z}_7]^{\mathrm{Id}}_{1\mid 1}$ | 8 | 17.1926 | ✓ | ✗ | ✗ | ✗ | ✗ | ✗ |
| $FR^{8,6}_{5}$ | $[\mathbb{Z}_3 \trianglelefteq \mathbb{Z}_6]^{\mathrm{Id}}_{2\mid 0}$ | 8 | 12. | ✓ | | | | | ✗ |

Table 8.1: List of multiplicity-free fusion rings up to rank 9 (Continued)

| Name | Common Name | Rank | $\mathscr{D}_{FP}^2$ | Comm. | FC | PFC | UFC | BFC | MFC |
|---|---|---|---|---|---|---|---|---|---|
| $FR_6^{8,6}$ | $[\mathbb{Z}_3 \trianglelefteq \mathbb{Z}_6]_{2\|1}^{\text{Id}}$ | 8 | 24. | ✓ | | | | | ✗ |
| $FR_7^{8,6}$ | | 8 | 27.3137 | ✗ | | | | | ✗ |
| $FR_8^{8,6}$ | | 8 | 27.3137 | ✓ | ✗ | ✗ | ✗ | ✗ | ✗ |
| $FR_9^{8,6}$ | $[I \trianglelefteq \mathbb{Z}_4]_{2\|1}^{(3\ 4)}$ | 8 | 75.7771 | ✗ | | | | | ✗ |
| $FR_{10}^{8,6}$ | $[I \trianglelefteq \mathbb{Z}_4]_{3\|1}^{\text{Id}}$ | 8 | 75.7771 | ✓ | | | | | ✗ |
| $FR_1^{9,0}$ | $TY(\mathbb{Z}_2 \times \mathbb{Z}_2 \times \mathbb{Z}_2)$ | 9 | 16. | ✓ | ✓ | ✓ | ✓ | ✓ | ✗ |
| $FR_2^{9,0}$ | $[(\mathbb{Z}_2)^{\times 3} \trianglelefteq (\mathbb{Z}_2)^{\times 3}]_{1\|1}^{\text{Id}}$ | 9 | 19.3723 | ✓ | ✗ | ✗ | ✗ | ✗ | ✗ |
| $FR_3^{9,0}$ | Ising × Ising | 9 | 16. | ✓ | ✓ | ✓ | ✓ | ✓ | ✓ |
| $FR_4^{9,0}$ | Ising × Rep($D_3$) | 9 | 24. | ✓ | ✓ | ✓ | ✓ | ✓ | ✗ |
| $FR_5^{9,0}$ | | 9 | 24. | ✓ | | | | | ✗ |
| $FR_6^{9,0}$ | Rep($D_3$) × Rep($D_3$) | 9 | 36. | ✓ | ✓ | ✓ | ✓ | ✓ | ✗ |
| $FR_7^{9,0}$ | | 9 | 32. | ✓ | | | | | ✗ |
| $FR_8^{9,0}$ | | 9 | 38.7446 | ✓ | ✗ | ✗ | ✗ | ✗ | ✗ |
| $FR_9^{9,0}$ | | 9 | 37.8564 | ✓ | | | | | ✗ |
| $FR_{10}^{9,0}$ | | 9 | 48. | ✓ | | | | | ✗ |
| $FR_{11}^{9,0}$ | | 9 | 37.8564 | ✓ | ✗ | ✗ | ✗ | ✗ | ✗ |
| $FR_{12}^{9,0}$ | | 9 | 30. | ✓ | | | | | ✗ |
| $FR_{13}^{9,0}$ | | 9 | 49.551 | ✓ | ✗ | ✗ | ✗ | ✗ | ✗ |
| $FR_{14}^{9,0}$ | Ising × PSU(2)$_5$ | 9 | 37.1836 | ✓ | ✓ | ✓ | ✓ | ✓ | ✓ |
| $FR_{15}^{9,0}$ | | 9 | 60. | ✓ | | | | | ✗ |
| $FR_{16}^{9,0}$ | | 9 | 58.7386 | ✓ | ✗ | ✗ | ✗ | ✗ | ✗ |
| $FR_{17}^{9,0}$ | | 9 | 58.2213 | ✓ | ✗ | ✗ | ✗ | ✗ | ✗ |
| $FR_{18}^{9,0}$ | | 9 | 58.2213 | ✓ | ✗ | ✗ | ✗ | ✗ | ✗ |
| $FR_{19}^{9,0}$ | SO(11)$_2$ | 9 | 44. | ✓ | ✓ | ✓ | ✓ | ✓ | ✓ |
| $FR_{20}^{9,0}$ | Rep($D_3$ × PSU(2)$_5$) | 9 | 55.7754 | ✓ | ✓ | ✓ | ✓ | ✓ | ✗ |
| $FR_{21}^{9,0}$ | | 9 | 51.7082 | ✓ | ✗ | ✗ | ✗ | ✗ | ✗ |

Table 8.1: List of multiplicity-free fusion rings up to rank 9 (Continued)

| Name | Common Name | Rank | $\mathscr{D}_{FP}^2$ | Comm. | FC | PFC | UFC | BFC | MFC |
|---|---|---|---|---|---|---|---|---|---|
| $FR_{22}^{9,0}$ | | 9 | 72. | ✓ | | | | | ✗ |
| $FR_{23}^{9,0}$ | | 9 | 74.3672 | ✓ | | | | | ✗ |
| $FR_{24}^{9,0}$ | | 9 | 61.8564 | ✓ | ✗ | ✗ | ✗ | ✗ | ✗ |
| $FR_{25}^{9,0}$ | | 9 | 48. | ✓ | ✗ | ✗ | ✗ | ✗ | ✗ |
| $FR_{26}^{9,0}$ | | 9 | 87.1918 | ✓ | | | | | ✗ |
| $FR_{27}^{9,0}$ | $SU(2)_8$ | 9 | 52.3607 | ✓ | ✓ | ✓ | ✓ | ✓ | ✓ |
| $FR_{28}^{9,0}$ | | 9 | 60. | ✓ | ✗ | ✗ | ✗ | ✗ | ✗ |
| $FR_{29}^{9,0}$ | | 9 | 108.321 | ✓ | ✗ | ✗ | ✗ | ✗ | ✗ |
| $FR_{30}^{9,0}$ | | 9 | 70.2101 | ✓ | ✗ | ✗ | ✗ | ✗ | ✗ |
| $FR_{31}^{9,0}$ | | 9 | 88.108 | ✓ | ✗ | ✗ | ✗ | ✗ | ✗ |
| $FR_{32}^{9,0}$ | | 9 | 77.166 | ✓ | ✗ | ✗ | ✗ | ✗ | ✗ |
| $FR_{33}^{9,0}$ | | 9 | 77.166 | ✓ | ✗ | ✗ | ✗ | ✗ | ✗ |
| $FR_{34}^{9,0}$ | $PSU(2)_5 \times PSU(2)_5$ | 9 | 86.4137 | ✓ | ✓ | ✓ | ✓ | ✓ | ✓ |
| $FR_{35}^{9,0}$ | | 9 | 134.976 | ✓ | ✗ | ✗ | ✗ | ✗ | ✗ |
| $FR_{36}^{9,0}$ | | 9 | 97.9329 | ✓ | ✗ | ✗ | ✗ | ✗ | ✗ |
| $FR_{37}^{9,0}$ | | 9 | 100.467 | ✓ | ✗ | ✗ | ✗ | ✗ | ✗ |
| $FR_{38}^{9,0}$ | | 9 | 108.99 | ✓ | | ✗ | | | ✗ |
| $FR_{39}^{9,0}$ | | 9 | 110.912 | ✓ | ✗ | ✗ | ✗ | ✗ | ✗ |
| $FR_{40}^{9,0}$ | | 9 | 130.596 | ✓ | ✗ | ✗ | ✗ | ✗ | ✗ |
| $FR_{41}^{9,0}$ | $PSU(2)_{16}$ | 9 | 149.235 | ✓ | ✓ | ✓ | ✓ | ✓ | ✗ |
| $FR_{42}^{9,0}$ | | 9 | 137.082 | ✓ | ✗ | ✗ | ✗ | ✗ | ✗ |
| $FR_{43}^{9,0}$ | | 9 | 179.586 | ✓ | ✗ | ✗ | ✗ | ✗ | ✗ |
| $FR_{44}^{9,0}$ | $PSU(2)_{17}$ | 9 | 175.333 | ✓ | ✓ | ✓ | ✓ | ✓ | ✓ |
| $FR_{45}^{9,0}$ | | 9 | 227.519 | ✓ | ✗ | ✗ | ✗ | ✗ | ✗ |
| $FR_{46}^{9,0}$ | | 9 | 318.114 | ✓ | ✗ | ✗ | ✗ | ✗ | ✗ |
| $FR_1^{9,2}$ | $TY(D_4)$ | 9 | 16. | ✗ | ✗ | ✗ | ✗ | ✗ | ✗ |

Table 8.1: List of multiplicity-free fusion rings up to rank 9 (Continued)

| Name | Common Name | Rank | $\mathscr{D}_{FP}^2$ | Comm. | FC | PFC | UFC | BFC | MFC |
|---|---|---|---|---|---|---|---|---|---|
| $FR_2^{9,2}$ | $[D_4 \trianglelefteq D_4]_{\mathbf{1}\mid 1}^{\text{Id}}$ | 9 | 19.3723 | ✗ | ✗ | ✗ | ✗ | ✗ | ✗ |
| $FR_3^{9,2}$ | | 9 | 16. | ✓ | | | | | ✗ |
| $FR_4^{9,2}$ | | 9 | 29.67 | ✗ | | | | | ✗ |
| $FR_5^{9,2}$ | | 9 | 24. | ✓ | | | | | ✗ |
| $FR_6^{9,2}$ | | 9 | 24. | ✓ | | | | | ✗ |
| $FR_7^{9,2}$ | | 9 | 24. | ✗ | | | | | ✗ |
| $FR_8^{9,2}$ | | 9 | 24. | ✓ | | | | | ✗ |
| $FR_9^{9,2}$ | | 9 | 32. | ✓ | | | | | ✗ |
| $FR_{10}^{9,2}$ | | 9 | 32. | ✓ | | | | | ✗ |
| $FR_{11}^{9,2}$ | | 9 | 38.7446 | ✓ | ✗ | ✗ | ✗ | ✗ | ✗ |
| $FR_{12}^{9,2}$ | | 9 | 37.8564 | ✓ | | | | | ✗ |
| $FR_{13}^{9,2}$ | | 9 | 37.8564 | ✓ | | | | | ✗ |
| $FR_{14}^{9,2}$ | | 9 | 48. | ✓ | | | | | ✗ |
| $FR_{15}^{9,2}$ | | 9 | 48. | ✓ | | | | | ✗ |
| $FR_{16}^{9,2}$ | | 9 | 37.8564 | ✓ | ✗ | ✗ | ✗ | ✗ | ✗ |
| $FR_{17}^{9,2}$ | | 9 | 49.551 | ✓ | ✗ | ✗ | ✗ | ✗ | ✗ |
| $FR_{18}^{9,2}$ | | 9 | 49.551 | ✓ | ✗ | ✗ | ✗ | ✗ | ✗ |
| $FR_{19}^{9,2}$ | | 9 | 60. | ✓ | | | | | ✗ |
| $FR_{20}^{9,2}$ | | 9 | 60. | ✓ | | | | | ✗ |
| $FR_{21}^{9,2}$ | | 9 | 58.7386 | ✓ | ✗ | ✗ | ✗ | ✗ | ✗ |
| $FR_{22}^{9,2}$ | | 9 | 58.7386 | ✗ | | | | | ✗ |
| $FR_{23}^{9,2}$ | | 9 | 58.7386 | ✓ | ✗ | ✗ | ✗ | ✗ | ✗ |
| $FR_{24}^{9,2}$ | | 9 | 58.2213 | ✓ | ✗ | ✗ | ✗ | ✗ | ✗ |
| $FR_{25}^{9,2}$ | | 9 | 58.2213 | ✓ | ✗ | ✗ | ✗ | ✗ | ✗ |
| $FR_{26}^{9,2}$ | | 9 | 44. | ✓ | | | | | ✗ |
| $FR_{27}^{9,2}$ | | 9 | 74.3672 | ✓ | | | | | ✗ |

| Name | Common Name | Rank | $\mathcal{D}_{FP}^2$ | Comm. | FC | PFC | UFC | BFC | MFC |
|---|---|---|---|---|---|---|---|---|---|
| $FR_{28}^{9,2}$ | | 9 | 74.3672 | ✓ | | | | | ✗ |
| $FR_{29}^{9,2}$ | | 9 | 61.8564 | ✓ | ✗ | ✗ | ✗ | ✗ | ✗ |
| $FR_{30}^{9,2}$ | | 9 | 48. | ✓ | ✗ | ✗ | ✗ | ✗ | ✗ |
| $FR_{31}^{9,2}$ | | 9 | 87.1918 | ✓ | | | | | ✗ |
| $FR_{32}^{9,2}$ | | 9 | 60. | ✗ | | | | | ✗ |
| $FR_{33}^{9,2}$ | | 9 | 60. | ✓ | ✗ | ✗ | ✗ | ✗ | ✗ |
| $FR_{34}^{9,2}$ | | 9 | 77.166 | ✓ | ✗ | ✗ | ✗ | ✗ | ✗ |
| $FR_{35}^{9,2}$ | | 9 | 77.166 | ✓ | ✗ | ✗ | ✗ | ✗ | ✗ |
| $FR_{36}^{9,2}$ | | 9 | 134.976 | ✗ | | | | | ✗ |
| $FR_{37}^{9,2}$ | | 9 | 134.976 | ✗ | | | | | ✗ |
| $FR_{38}^{9,2}$ | | 9 | 100.467 | ✗ | | | | | ✗ |
| $FR_{39}^{9,2}$ | | 9 | 100.467 | ✓ | ✗ | ✗ | ✗ | ✗ | ✗ |
| $FR_{40}^{9,2}$ | | 9 | 130.596 | ✓ | ✗ | ✗ | ✗ | ✗ | ✗ |
| $FR_{41}^{9,2}$ | | 9 | 137.082 | ✗ | | | | | ✗ |
| $FR_{42}^{9,2}$ | | 9 | 137.082 | ✓ | ✗ | ✗ | ✗ | ✗ | ✗ |
| $FR_{43}^{9,2}$ | | 9 | 179.586 | ✓ | ✗ | ✗ | ✗ | ✗ | ✗ |
| $FR_{44}^{9,2}$ | | 9 | 227.519 | ✗ | | | | | ✗ |
| $FR_{45}^{9,2}$ | | 9 | 227.519 | ✓ | ✗ | ✗ | ✗ | ✗ | ✗ |
| $FR_1^{9,4}$ | $TY(\mathbb{Z}_2 \times \mathbb{Z}_4)$ | 9 | 16. | ✓ | ✓ | ✓ | ✓ | ✗ | ✗ |
| $FR_2^{9,4}$ | $[\mathbb{Z}_2 \times \mathbb{Z}_4 \trianglelefteq \mathbb{Z}_2 \times \mathbb{Z}_4]_{1\vert 1}^{\text{Id}}$ | 9 | 19.3723 | ✓ | ✗ | ✗ | ✗ | ✗ | ✗ |
| $FR_3^{9,4}$ | $[\mathbb{Z}_2 \trianglelefteq \mathbb{Z}_6]_{1\vert 0}^{(2\ 3)}$ | 9 | 12. | ✗ | ✓ | | | | ✗ |
| $FR_4^{9,4}$ | | 9 | 16. | ✗ | | | | | ✗ |
| $FR_5^{9,4}$ | | 9 | 16. | ✓ | | | | | ✗ |
| $FR_6^{9,4}$ | | 9 | 29.67 | ✗ | | | | | ✗ |
| $FR_7^{9,4}$ | | 9 | 29.67 | ✓ | ✗ | ✗ | ✗ | ✗ | ✗ |
| $FR_8^{9,4}$ | | 9 | 24. | ✗ | | | | | ✗ |

Table 8.1: List of multiplicity-free fusion rings up to rank 9 (Continued)

| Name | Common Name | Rank | $\mathscr{D}_{FP}^2$ | Comm. | FC | PFC | UFC | BFC | MFC |
|---|---|---|---|---|---|---|---|---|---|
| $FR_9^{9,4}$ | | 9 | 24. | ✗ | | | | | ✗ |
| $FR_{10}^{9,4}$ | | 9 | 24. | ✓ | | | | | ✗ |
| $FR_{11}^{9,4}$ | $[\mathbb{Z}_2 \trianglelefteq \mathbb{Z}_6]_{1\|1}^{(2\ 3)}$ | 9 | 44.054 | ✗ | | | | | ✗ |
| $FR_{12}^{9,4}$ | | 9 | 32. | ✓ | | | | | ✗ |
| $FR_{13}^{9,4}$ | | 9 | 37.8564 | ✓ | | | | | ✗ |
| $FR_{14}^{9,4}$ | | 9 | 48. | ✓ | | | | | ✗ |
| $FR_{15}^{9,4}$ | | 9 | 49.551 | ✓ | ✗ | ✗ | ✗ | ✗ | ✗ |
| $FR_{16}^{9,4}$ | | 9 | 60. | ✓ | | | | | ✗ |
| $FR_{17}^{9,4}$ | | 9 | 58.7386 | ✗ | | | | | ✗ |
| $FR_{18}^{9,4}$ | | 9 | 58.7386 | ✓ | ✗ | ✗ | ✗ | ✗ | ✗ |
| $FR_{19}^{9,4}$ | | 9 | 74.3672 | ✓ | | | | | ✗ |
| $FR_{20}^{9,4}$ | | 9 | 48. | ✗ | | | | | ✗ |
| $FR_{21}^{9,4}$ | | 9 | 48. | ✓ | ✗ | ✗ | ✗ | ✗ | ✗ |
| $FR_{22}^{9,4}$ | | 9 | 87.1918 | ✗ | | | | | ✗ |
| $FR_{23}^{9,4}$ | | 9 | 87.1918 | ✓ | | | | | ✗ |
| $FR_{24}^{9,4}$ | | 9 | 52.3607 | ✓ | | | | | ✗ |
| $FR_{25}^{9,4}$ | | 9 | 108.321 | ✓ | ✗ | ✗ | ✗ | ✗ | ✗ |
| $FR_{26}^{9,4}$ | | 9 | 77.166 | ✓ | ✗ | ✗ | ✗ | ✗ | ✗ |
| $FR_{27}^{9,4}$ | | 9 | 94.2873 | ✗ | | | | | ✗ |
| $FR_{28}^{9,4}$ | | 9 | 134.976 | ✓ | ✗ | ✗ | ✗ | ✗ | ✗ |
| $FR_{29}^{9,4}$ | | 9 | 134.976 | ✓ | ✗ | ✗ | ✗ | ✗ | ✗ |
| $FR_{30}^{9,4}$ | | 9 | 97.9329 | ✓ | ✗ | ✗ | ✗ | ✗ | ✗ |
| $FR_{31}^{9,4}$ | | 9 | 130.596 | ✗ | | | | | ✗ |
| $FR_{32}^{9,4}$ | | 9 | 130.596 | ✓ | ✗ | ✗ | ✗ | ✗ | ✗ |
| $FR_{33}^{9,4}$ | | 9 | 227.519 | ✗ | | | | | ✗ |
| $FR_1^{9,6}$ | TY(Q) | 9 | 16. | ✗ | ✗ | ✗ | ✗ | ✗ | ✗ |

| Name | Common Name | Rank | $\mathcal{D}_{FP}^2$ | Comm. | FC | PFC | UFC | BFC | MFC |
|---|---|---|---|---|---|---|---|---|---|
| $FR_2^{9,6}$ | $TY(\mathbb{Z}_8)$ | 9 | 16. | ✓ | ✓ | ✓ | ✓ | ✗ | ✗ |
| $FR_3^{9,6}$ | $[Q \trianglelefteq Q]_{1\|1}^{Id}$ | 9 | 19.3723 | ✗ | ✗ | ✗ | ✗ | ✗ | ✗ |
| $FR_4^{9,6}$ | $[\mathbb{Z}_8 \trianglelefteq \mathbb{Z}_8]_{1\|1}^{Id}$ | 9 | 19.3723 | ✓ | ✗ | ✗ | ✗ | ✗ | ✗ |
| $FR_5^{9,6}$ | Ising $\times \mathbb{Z}_3$ | 9 | 12. | ✓ | ✓ | ✓ | ✓ | ✓ | ✓ |
| $FR_6^{9,6}$ | $Rep(D_3) \times \mathbb{Z}_3$ | 9 | 18. | ✓ | ✓ | ✓ | ✓ | ✓ | ✗ |
| $FR_7^{9,6}$ | | 9 | 29.67 | ✓ | ✗ | ✗ | ✗ | ✗ | ✗ |
| $FR_8^{9,6}$ | | 9 | 28.3923 | ✓ | ✗ | ✗ | ✗ | ✗ | ✗ |
| $FR_9^{9,6}$ | $[\mathbb{Z}_2 \trianglelefteq \mathbb{Z}_6]_{1\|1}^{Id}$ | 9 | 44.054 | ✓ | ✗ | ✗ | ✗ | ✗ | ✗ |
| $FR_{10}^{9,6}$ | $PSU(2)_5 \times \mathbb{Z}_3$ | 9 | 27.8877 | ✓ | ✓ | ✓ | ✓ | ✓ | ✓ |
| $FR_{11}^{9,6}$ | | 9 | 55.1689 | ✓ | ✗ | ✗ | ✗ | ✗ | ✗ |
| $FR_{12}^{9,6}$ | | 9 | 94.2873 | ✓ | ✗ | ✗ | ✗ | ✗ | ✗ |
| $FR_{13}^{9,6}$ | | 9 | 134.976 | ✗ | | | | | ✗ |
| $FR_{14}^{9,6}$ | | 9 | 134.976 | ✓ | ✗ | ✗ | ✗ | ✗ | ✗ |
| $FR_{15}^{9,6}$ | | 9 | 127.95 | ✓ | ✗ | ✗ | ✗ | ✗ | ✗ |
| $FR_{16}^{9,6}$ | | 9 | 163.373 | ✓ | ✗ | ✗ | ✗ | ✗ | ✗ |
| $FR_1^{9,8}$ | $\mathbb{Z}_9$ | 9 | 9. | ✓ | ✓ | ✓ | ✓ | ✓ | ✓ |
| $FR_2^{9,8}$ | $\mathbb{Z}_3 \times \mathbb{Z}_3$ | 9 | 9. | ✓ | ✓ | ✓ | ✓ | ✓ | ✓ |

# Chapter 9

# List of Multiplicity-free Fusion Categories up to Rank 7

The following is a list of all the fusion categories for multiplicity-free fusion rings up to rank 7. The naming of the categories is described in section 4.5. The table has several visual guides that groups the categories by property, namely

- Categories with the same fusion ring are separated by a horizontal line.

- Categories of the same fusion ring with equivalent $F$-symbols are separated by a dashed horizontal line.

- Categories of the same fusion ring with equivalent $F$-symbols and equivalent $R$-symbols are grouped by white or gray background color. The background colors have no other meaning.

- The symbols , , , ,  denote that the category in question is respectively braided, unitary, spherical, ribbon, modular.

Table 9.1: List of multiplicity-free fusion categories up to rank 7

| Formal Name | Common Name | | | | | |
|---|---|---|---|---|---|---|
| $[\mathrm{FR}_1^{1,1,0}]_{1,1}^1$ | $[\mathrm{Trivial}]_{1,1}^1$ |  |  |  |  |  |
| $[\mathrm{FR}_1^{2,1,0}]_{1,1}^1$ | $[\mathbb{Z}_2]_{1,1}^1$ |  |  |  |  |  |
| $[\mathrm{FR}_1^{2,1,0}]_{1,1}^2$ | $[\mathbb{Z}_2]_{1,1}^2$ |  | |  |  |  |
| $[\mathrm{FR}_1^{2,1,0}]_{1,2}^1$ | $[\mathbb{Z}_2]_{1,2}^1$ |  |  |  |  |  |
| $[\mathrm{FR}_1^{2,1,0}]_{1,2}^2$ | $[\mathbb{Z}_2]_{1,2}^2$ |  | |  |  |  |
| $[\mathrm{FR}_1^{2,1,0}]_{2,1}^1$ | $[\mathbb{Z}_2]_{2,1}^1$ |  |  |  |  | |
| $[\mathrm{FR}_1^{2,1,0}]_{2,1}^2$ | $[\mathbb{Z}_2]_{2,1}^2$ |  | |  |  | |
| $[\mathrm{FR}_1^{2,1,0}]_{2,2}^1$ | $[\mathbb{Z}_2]_{2,2}^1$ |  |  |  |  | |

Table 9.1: List of multiplicity-free fusion categories up to rank 7 (Continued)

| Formal Name | Common Name | | | | | |
|---|---|---|---|---|---|---|
| $[FR_1^{2,1,0}]_{2,2}^2$ | $[\mathbb{Z}_2]_{2,2}^2$ | ✗ | | 🌐 | 🎀 | |
| $[FR_2^{2,1,0}]_{1,1}^1$ | $[Fib]_{1,1}^1$ | ✗ | † | 🌐 | 🎀 | ⭕ |
| $[FR_2^{2,1,0}]_{1,2}^1$ | $[Fib]_{1,2}^1$ | ✗ | † | 🌐 | 🎀 | ⭕ |
| $[FR_2^{2,1,0}]_{2,1}^1$ | $[Fib]_{2,1}^1$ | ✗ | | 🌐 | 🎀 | ⭕ |
| $[FR_2^{2,1,0}]_{2,2}^1$ | $[Fib]_{2,2}^1$ | ✗ | | 🌐 | 🎀 | ⭕ |
| $[FR_1^{3,1,0}]_{1,1}^1$ | $[Ising]_{1,1}^1$ | ✗ | † | 🌐 | 🎀 | ⭕ |
| $[FR_1^{3,1,0}]_{1,1}^2$ | $[Ising]_{1,1}^2$ | ✗ | | 🌐 | 🎀 | ⭕ |
| $[FR_1^{3,1,0}]_{1,2}^1$ | $[Ising]_{1,2}^1$ | ✗ | † | 🌐 | 🎀 | ⭕ |
| $[FR_1^{3,1,0}]_{1,2}^2$ | $[Ising]_{1,2}^2$ | ✗ | | 🌐 | 🎀 | ⭕ |
| $[FR_1^{3,1,0}]_{1,3}^1$ | $[Ising]_{1,3}^1$ | ✗ | † | 🌐 | 🎀 | ⭕ |
| $[FR_1^{3,1,0}]_{1,3}^2$ | $[Ising]_{1,3}^2$ | ✗ | | 🌐 | 🎀 | ⭕ |
| $[FR_1^{3,1,0}]_{1,4}^1$ | $[Ising]_{1,4}^1$ | ✗ | † | 🌐 | 🎀 | ⭕ |
| $[FR_1^{3,1,0}]_{1,4}^2$ | $[Ising]_{1,4}^2$ | ✗ | | 🌐 | 🎀 | ⭕ |
| $[FR_1^{3,1,0}]_{2,1}^1$ | $[Ising]_{2,1}^1$ | ✗ | † | 🌐 | 🎀 | ⭕ |
| $[FR_1^{3,1,0}]_{2,1}^2$ | $[Ising]_{2,1}^2$ | ✗ | | 🌐 | 🎀 | ⭕ |
| $[FR_1^{3,1,0}]_{2,2}^1$ | $[Ising]_{2,2}^1$ | ✗ | † | 🌐 | 🎀 | ⭕ |
| $[FR_1^{3,1,0}]_{2,2}^2$ | $[Ising]_{2,2}^2$ | ✗ | | 🌐 | 🎀 | ⭕ |
| $[FR_1^{3,1,0}]_{2,3}^1$ | $[Ising]_{2,3}^1$ | ✗ | † | 🌐 | 🎀 | ⭕ |
| $[FR_1^{3,1,0}]_{2,3}^2$ | $[Ising]_{2,3}^2$ | ✗ | | 🌐 | 🎀 | ⭕ |
| $[FR_1^{3,1,0}]_{2,4}^1$ | $[Ising]_{2,4}^1$ | ✗ | † | 🌐 | 🎀 | ⭕ |
| $[FR_1^{3,1,0}]_{2,4}^2$ | $[Ising]_{2,4}^2$ | ✗ | | 🌐 | 🎀 | ⭕ |
| $[FR_2^{3,1,0}]_{1,1}^1$ | $[Rep(D_3)]_{1,1}^1$ | ✗ | † | 🌐 | 🎀 | |
| $[FR_2^{3,1,0}]_{1,2}^1$ | $[Rep(D_3)]_{1,2}^1$ | ✗ | † | 🌐 | 🎀 | |
| $[FR_2^{3,1,0}]_{1,3}^1$ | $[Rep(D_3)]_{1,3}^1$ | ✗ | † | 🌐 | 🎀 | |
| $[FR_2^{3,1,0}]_{2,1}^1$ | $[Rep(D_3)]_{2,1}^1$ | | † | 🌐 | | |
| $[FR_2^{3,1,0}]_{3,1}^1$ | $[Rep(D_3)]_{3,1}^1$ | | † | 🌐 | | |
| $[FR_3^{3,1,0}]_{1,1}^1$ | $[PSU(2)_5]_{1,1}^1$ | ✗ | † | 🌐 | 🎀 | ⭕ |
| $[FR_3^{3,1,0}]_{1,2}^1$ | $[PSU(2)_5]_{1,2}^1$ | ✗ | † | 🌐 | 🎀 | ⭕ |
| $[FR_3^{3,1,0}]_{2,1}^1$ | $[PSU(2)_5]_{2,1}^1$ | ✗ | | 🌐 | 🎀 | ⭕ |

Table 9.1: List of multiplicity-free fusion categories up to rank 7 (Continued)

| Formal Name | Common Name | | | | | |
|---|---|---|---|---|---|---|
| $[\text{FR}_3^{3,1,0}]_{2,2}^1$ | $[\text{PSU}(2)_5]_{2,2}^1$ | ✕ | | 🌐 | 🎀 | ⚭ |
| $[\text{FR}_3^{3,1,0}]_{3,1}^1$ | $[\text{PSU}(2)_5]_{3,1}^1$ | ✕ | | 🌐 | 🎀 | ⚭ |
| $[\text{FR}_3^{3,1,0}]_{3,2}^1$ | $[\text{PSU}(2)_5]_{3,2}^1$ | ✕ | | 🌐 | 🎀 | ⚭ |
| $[\text{FR}_1^{3,1,2}]_{1,1}^1$ | $[\mathbb{Z}_3]_{1,1}^1$ | ✕ | † | 🌐 | 🎀 | ⚭ |
| $[\text{FR}_1^{3,1,2}]_{1,1}^2$ | $[\mathbb{Z}_3]_{1,1}^2$ | ✕ | | | | |
| $[\text{FR}_1^{3,1,2}]_{1,2}^1$ | $[\mathbb{Z}_3]_{1,2}^1$ | ✕ | † | 🌐 | 🎀 | ⚭ |
| $[\text{FR}_1^{3,1,2}]_{1,2}^2$ | $[\mathbb{Z}_3]_{1,2}^2$ | ✕ | | | | |
| $[\text{FR}_1^{3,1,2}]_{1,3}^1$ | $[\mathbb{Z}_3]_{1,3}^1$ | ✕ | † | 🌐 | 🎀 | |
| $[\text{FR}_1^{3,1,2}]_{1,3}^2$ | $[\mathbb{Z}_3]_{1,3}^2$ | ✕ | | | | |
| $[\text{FR}_1^{3,1,2}]_{2,1}^1$ | $[\mathbb{Z}_3]_{2,1}^1$ | | † | 🌐 | | |
| $[\text{FR}_1^{3,1,2}]_{2,1}^2$ | $[\mathbb{Z}_3]_{2,1}^2$ | | | | | |
| $[\text{FR}_1^{3,1,2}]_{3,1}^1$ | $[\mathbb{Z}_3]_{3,1}^1$ | | † | 🌐 | | |
| $[\text{FR}_1^{3,1,2}]_{3,1}^2$ | $[\mathbb{Z}_3]_{3,1}^2$ | | | | | |
| $[\text{FR}_1^{4,1,0}]_{1,1}^1$ | $[\mathbb{Z}_2 \times \mathbb{Z}_2]_{1,1}^1$ | ✕ | † | 🌐 | 🎀 | ⚭ |
| $[\text{FR}_1^{4,1,0}]_{1,1}^2$ | $[\mathbb{Z}_2 \times \mathbb{Z}_2]_{1,1}^2$ | ✕ | | 🌐 | 🎀 | ⚭ |
| $[\text{FR}_1^{4,1,0}]_{1,1}^3$ | $[\mathbb{Z}_2 \times \mathbb{Z}_2]_{1,1}^3$ | ✕ | | 🌐 | 🎀 | ⚭ |
| $[\text{FR}_1^{4,1,0}]_{1,2}^1$ | $[\mathbb{Z}_2 \times \mathbb{Z}_2]_{1,2}^1$ | ✕ | † | 🌐 | 🎀 | ⚭ |
| $[\text{FR}_1^{4,1,0}]_{1,2}^2$ | $[\mathbb{Z}_2 \times \mathbb{Z}_2]_{1,2}^2$ | ✕ | | 🌐 | 🎀 | ⚭ |
| $[\text{FR}_1^{4,1,0}]_{1,3}^1$ | $[\mathbb{Z}_2 \times \mathbb{Z}_2]_{1,3}^1$ | ✕ | † | 🌐 | 🎀 | |
| $[\text{FR}_1^{4,1,0}]_{1,3}^2$ | $[\mathbb{Z}_2 \times \mathbb{Z}_2]_{1,3}^2$ | ✕ | | 🌐 | 🎀 | |
| $[\text{FR}_1^{4,1,0}]_{1,4}^1$ | $[\mathbb{Z}_2 \times \mathbb{Z}_2]_{1,4}^1$ | ✕ | † | 🌐 | 🎀 | |
| $[\text{FR}_1^{4,1,0}]_{1,4}^2$ | $[\mathbb{Z}_2 \times \mathbb{Z}_2]_{1,4}^2$ | ✕ | | 🌐 | 🎀 | |
| $[\text{FR}_1^{4,1,0}]_{1,4}^3$ | $[\mathbb{Z}_2 \times \mathbb{Z}_2]_{1,4}^3$ | ✕ | | 🌐 | 🎀 | |
| $[\text{FR}_1^{4,1,0}]_{2,1}^1$ | $[\mathbb{Z}_2 \times \mathbb{Z}_2]_{2,1}^1$ | ✕ | † | 🌐 | 🎀 | ⚭ |
| $[\text{FR}_1^{4,1,0}]_{2,1}^2$ | $[\mathbb{Z}_2 \times \mathbb{Z}_2]_{2,1}^2$ | ✕ | | 🌐 | 🎀 | ⚭ |
| $[\text{FR}_1^{4,1,0}]_{2,1}^3$ | $[\mathbb{Z}_2 \times \mathbb{Z}_2]_{2,1}^3$ | ✕ | | 🌐 | 🎀 | ⚭ |
| $[\text{FR}_1^{4,1,0}]_{2,2}^1$ | $[\mathbb{Z}_2 \times \mathbb{Z}_2]_{2,2}^1$ | ✕ | † | 🌐 | 🎀 | ⚭ |
| $[\text{FR}_1^{4,1,0}]_{2,2}^2$ | $[\mathbb{Z}_2 \times \mathbb{Z}_2]_{2,2}^2$ | ✕ | | 🌐 | 🎀 | ⚭ |
| $[\text{FR}_1^{4,1,0}]_{2,2}^3$ | $[\mathbb{Z}_2 \times \mathbb{Z}_2]_{2,2}^3$ | ✕ | | 🌐 | 🎀 | ⚭ |

Table 9.1: List of multiplicity-free fusion categories up to rank 7 (Continued)

| Formal Name | Common Name | | | | | |
|---|---|---|---|---|---|---|
| $[FR_1^{4,1,0}]_{2,2}^4$ | $[\mathbb{Z}_2 \times \mathbb{Z}_2]_{2,2}^4$ | ✗ | | 🌐 | 🎀 | ⭕ |
| $[FR_1^{4,1,0}]_{2,3}^1$ | $[\mathbb{Z}_2 \times \mathbb{Z}_2]_{2,3}^1$ | ✗ | † | 🌐 | 🎀 | ⭕ |
| $[FR_1^{4,1,0}]_{2,3}^2$ | $[\mathbb{Z}_2 \times \mathbb{Z}_2]_{2,3}^2$ | ✗ | | 🌐 | 🎀 | ⭕ |
| $[FR_1^{4,1,0}]_{2,3}^3$ | $[\mathbb{Z}_2 \times \mathbb{Z}_2]_{2,3}^3$ | ✗ | | 🌐 | 🎀 | ⭕ |
| $[FR_1^{4,1,0}]_{2,4}^1$ | $[\mathbb{Z}_2 \times \mathbb{Z}_2]_{2,4}^1$ | ✗ | † | 🌐 | 🎀 | |
| $[FR_1^{4,1,0}]_{2,4}^2$ | $[\mathbb{Z}_2 \times \mathbb{Z}_2]_{2,4}^2$ | ✗ | | 🌐 | 🎀 | |
| $[FR_1^{4,1,0}]_{2,4}^3$ | $[\mathbb{Z}_2 \times \mathbb{Z}_2]_{2,4}^3$ | ✗ | | 🌐 | 🎀 | |
| $[FR_1^{4,1,0}]_{2,5}^1$ | $[\mathbb{Z}_2 \times \mathbb{Z}_2]_{2,5}^1$ | ✗ | † | 🌐 | 🎀 | |
| $[FR_1^{4,1,0}]_{2,5}^2$ | $[\mathbb{Z}_2 \times \mathbb{Z}_2]_{2,5}^2$ | ✗ | | 🌐 | 🎀 | |
| $[FR_1^{4,1,0}]_{2,5}^3$ | $[\mathbb{Z}_2 \times \mathbb{Z}_2]_{2,5}^3$ | ✗ | | 🌐 | 🎀 | |
| $[FR_1^{4,1,0}]_{2,5}^4$ | $[\mathbb{Z}_2 \times \mathbb{Z}_2]_{2,5}^4$ | ✗ | | 🌐 | 🎀 | |
| $[FR_1^{4,1,0}]_{2,6}^1$ | $[\mathbb{Z}_2 \times \mathbb{Z}_2]_{2,6}^1$ | ✗ | † | 🌐 | 🎀 | |
| $[FR_1^{4,1,0}]_{2,6}^2$ | $[\mathbb{Z}_2 \times \mathbb{Z}_2]_{2,6}^2$ | ✗ | | 🌐 | 🎀 | |
| $[FR_1^{4,1,0}]_{2,6}^3$ | $[\mathbb{Z}_2 \times \mathbb{Z}_2]_{2,6}^3$ | ✗ | | 🌐 | 🎀 | |
| $[FR_1^{4,1,0}]_{3,1}^1$ | $[\mathbb{Z}_2 \times \mathbb{Z}_2]_{3,1}^1$ | | † | 🌐 | | |
| $[FR_1^{4,1,0}]_{3,1}^2$ | $[\mathbb{Z}_2 \times \mathbb{Z}_2]_{3,1}^2$ | | | 🌐 | | |
| $[FR_1^{4,1,0}]_{3,1}^3$ | $[\mathbb{Z}_2 \times \mathbb{Z}_2]_{3,1}^3$ | | | 🌐 | | |
| $[FR_1^{4,1,0}]_{4,1}^1$ | $[\mathbb{Z}_2 \times \mathbb{Z}_2]_{4,1}^1$ | | † | 🌐 | | |
| $[FR_1^{4,1,0}]_{4,1}^2$ | $[\mathbb{Z}_2 \times \mathbb{Z}_2]_{4,1}^2$ | | | 🌐 | | |
| $[FR_2^{4,1,0}]_{1,1}^1$ | $[SU(2)_3]_{1,1}^1$ | ✗ | † | 🌐 | 🎀 | ⭕ |
| $[FR_2^{4,1,0}]_{1,1}^2$ | $[SU(2)_3]_{1,1}^2$ | ✗ | | 🌐 | 🎀 | ⭕ |
| $[FR_2^{4,1,0}]_{1,2}^1$ | $[SU(2)_3]_{1,2}^1$ | ✗ | † | 🌐 | 🎀 | ⭕ |
| $[FR_2^{4,1,0}]_{1,2}^2$ | $[SU(2)_3]_{1,2}^2$ | ✗ | | 🌐 | 🎀 | ⭕ |
| $[FR_2^{4,1,0}]_{1,3}^1$ | $[SU(2)_3]_{1,3}^1$ | ✗ | † | 🌐 | 🎀 | ⭕ |
| $[FR_2^{4,1,0}]_{1,3}^2$ | $[SU(2)_3]_{1,3}^2$ | ✗ | | 🌐 | 🎀 | ⭕ |
| $[FR_2^{4,1,0}]_{1,4}^1$ | $[SU(2)_3]_{1,4}^1$ | ✗ | † | 🌐 | 🎀 | ⭕ |
| $[FR_2^{4,1,0}]_{1,4}^2$ | $[SU(2)_3]_{1,4}^2$ | ✗ | | 🌐 | 🎀 | ⭕ |
| $[FR_2^{4,1,0}]_{2,1}^1$ | $[SU(2)_3]_{2,1}^1$ | ✗ | † | 🌐 | 🎀 | |
| $[FR_2^{4,1,0}]_{2,1}^2$ | $[SU(2)_3]_{2,1}^2$ | ✗ | | 🌐 | 🎀 | |

Table 9.1: List of multiplicity-free fusion categories up to rank 7 (Continued)

| Formal Name | Common Name | | | | | |
|---|---|---|---|---|---|---|
| $[FR_2^{4,1,0}]_{2,2}^1$ | $[SU(2)_3]_{2,2}^1$ | ✕ | † | 🌐 | 🎀 | |
| $[FR_2^{4,1,0}]_{2,2}^2$ | $[SU(2)_3]_{2,2}^2$ | ✕ | | 🌐 | 🎀 | |
| $[FR_2^{4,1,0}]_{2,3}^1$ | $[SU(2)_3]_{2,3}^1$ | ✕ | † | 🌐 | 🎀 | |
| $[FR_2^{4,1,0}]_{2,3}^2$ | $[SU(2)_3]_{2,3}^2$ | ✕ | | 🌐 | 🎀 | |
| $[FR_2^{4,1,0}]_{2,4}^1$ | $[SU(2)_3]_{2,4}^1$ | ✕ | † | 🌐 | 🎀 | |
| $[FR_2^{4,1,0}]_{2,4}^2$ | $[SU(2)_3]_{2,4}^2$ | ✕ | | 🌐 | 🎀 | |
| $[FR_2^{4,1,0}]_{3,1}^1$ | $[SU(2)_3]_{3,1}^1$ | ✕ | | 🌐 | 🎀 | ⚭ |
| $[FR_2^{4,1,0}]_{3,1}^2$ | $[SU(2)_3]_{3,1}^2$ | ✕ | | 🌐 | 🎀 | ⚭ |
| $[FR_2^{4,1,0}]_{3,2}^1$ | $[SU(2)_3]_{3,2}^1$ | ✕ | | 🌐 | 🎀 | ⚭ |
| $[FR_2^{4,1,0}]_{3,2}^2$ | $[SU(2)_3]_{3,2}^2$ | ✕ | | 🌐 | 🎀 | ⚭ |
| $[FR_2^{4,1,0}]_{3,3}^1$ | $[SU(2)_3]_{3,3}^1$ | ✕ | | 🌐 | 🎀 | ⚭ |
| $[FR_2^{4,1,0}]_{3,3}^2$ | $[SU(2)_3]_{3,3}^2$ | ✕ | | 🌐 | 🎀 | ⚭ |
| $[FR_2^{4,1,0}]_{3,4}^1$ | $[SU(2)_3]_{3,4}^1$ | ✕ | | 🌐 | 🎀 | ⚭ |
| $[FR_2^{4,1,0}]_{3,4}^2$ | $[SU(2)_3]_{3,4}^2$ | ✕ | | 🌐 | 🎀 | ⚭ |
| $[FR_2^{4,1,0}]_{4,1}^1$ | $[SU(2)_3]_{4,1}^1$ | ✕ | | 🌐 | 🎀 | |
| $[FR_2^{4,1,0}]_{4,1}^2$ | $[SU(2)_3]_{4,1}^2$ | ✕ | | 🌐 | 🎀 | |
| $[FR_2^{4,1,0}]_{4,2}^1$ | $[SU(2)_3]_{4,2}^1$ | ✕ | | 🌐 | 🎀 | |
| $[FR_2^{4,1,0}]_{4,2}^2$ | $[SU(2)_3]_{4,2}^2$ | ✕ | | 🌐 | 🎀 | |
| $[FR_2^{4,1,0}]_{4,3}^1$ | $[SU(2)_3]_{4,3}^1$ | ✕ | | 🌐 | 🎀 | |
| $[FR_2^{4,1,0}]_{4,3}^2$ | $[SU(2)_3]_{4,3}^2$ | ✕ | | 🌐 | 🎀 | |
| $[FR_2^{4,1,0}]_{4,4}^1$ | $[SU(2)_3]_{4,4}^1$ | ✕ | | 🌐 | 🎀 | |
| $[FR_2^{4,1,0}]_{4,4}^2$ | $[SU(2)_3]_{4,4}^2$ | ✕ | | 🌐 | 🎀 | |
| $[FR_3^{4,1,0}]_{1,1}^1$ | $[Rep(D_5)]_{1,1}^1$ | ✕ | † | 🌐 | 🎀 | |
| $[FR_3^{4,1,0}]_{1,2}^1$ | $[Rep(D_5)]_{1,2}^1$ | ✕ | † | 🌐 | 🎀 | |
| $[FR_3^{4,1,0}]_{1,3}^1$ | $[Rep(D_5)]_{1,3}^1$ | ✕ | † | 🌐 | 🎀 | |
| $[FR_3^{4,1,0}]_{2,1}^1$ | $[Rep(D_5)]_{2,1}^1$ | | † | 🌐 | | |
| $[FR_3^{4,1,0}]_{3,1}^1$ | $[Rep(D_5)]_{3,1}^1$ | | † | 🌐 | | |
| $[FR_4^{4,1,0}]_{1,1}^1$ | $[PSU(2)_6]_{1,1}^1$ | ✕ | † | 🌐 | 🎀 | |
| $[FR_4^{4,1,0}]_{2,1}^1$ | $[PSU(2)_6]_{2,1}^1$ | ✕ | | 🌐 | 🎀 | |

Table 9.1: List of multiplicity-free fusion categories up to rank 7 (Continued)

| Formal Name | Common Name | | | | | |
|---|---|---|---|---|---|---|
| $[FR_5^{4,1,0}]_{1,1}^1$ | $[Fib \times Fib]_{1,1}^1$ | ✕ | † | 🌐 | 🎀 | ⭕ |
| $[FR_5^{4,1,0}]_{1,2}^1$ | $[Fib \times Fib]_{1,2}^1$ | ✕ | † | 🌐 | 🎀 | ⭕ |
| $[FR_5^{4,1,0}]_{1,3}^1$ | $[Fib \times Fib]_{1,3}^1$ | ✕ | † | 🌐 | 🎀 | ⭕ |
| $[FR_5^{4,1,0}]_{2,1}^1$ | $[Fib \times Fib]_{2,1}^1$ | ✕ | | 🌐 | 🎀 | ⭕ |
| $[FR_5^{4,1,0}]_{2,2}^1$ | $[Fib \times Fib]_{2,2}^1$ | ✕ | | 🌐 | 🎀 | ⭕ |
| $[FR_5^{4,1,0}]_{2,3}^1$ | $[Fib \times Fib]_{2,3}^1$ | ✕ | | 🌐 | 🎀 | ⭕ |
| $[FR_5^{4,1,0}]_{2,4}^1$ | $[Fib \times Fib]_{2,4}^1$ | ✕ | | 🌐 | 🎀 | ⭕ |
| $[FR_5^{4,1,0}]_{3,1}^1$ | $[Fib \times Fib]_{3,1}^1$ | ✕ | | 🌐 | 🎀 | ⭕ |
| $[FR_5^{4,1,0}]_{3,2}^1$ | $[Fib \times Fib]_{3,2}^1$ | ✕ | | 🌐 | 🎀 | ⭕ |
| $[FR_5^{4,1,0}]_{3,3}^1$ | $[Fib \times Fib]_{3,3}^1$ | ✕ | | 🌐 | 🎀 | ⭕ |
| $[FR_6^{4,1,0}]_{1,1}^1$ | $[PSU(2)_7]_{1,1}^1$ | ✕ | † | 🌐 | 🎀 | ⭕ |
| $[FR_6^{4,1,0}]_{1,2}^1$ | $[PSU(2)_7]_{1,2}^1$ | ✕ | † | 🌐 | 🎀 | ⭕ |
| $[FR_6^{4,1,0}]_{2,1}^1$ | $[PSU(2)_7]_{2,1}^1$ | ✕ | | 🌐 | 🎀 | ⭕ |
| $[FR_6^{4,1,0}]_{2,2}^1$ | $[PSU(2)_7]_{2,2}^1$ | ✕ | | 🌐 | 🎀 | ⭕ |
| $[FR_6^{4,1,0}]_{3,1}^1$ | $[PSU(2)_7]_{3,1}^1$ | ✕ | | 🌐 | 🎀 | ⭕ |
| $[FR_6^{4,1,0}]_{3,2}^1$ | $[PSU(2)_7]_{3,2}^1$ | ✕ | | 🌐 | 🎀 | ⭕ |
| $[FR_1^{4,1,2}]_{1,1}^1$ | $[\mathbb{Z}_4]_{1,1}^1$ | ✕ | † | 🌐 | 🎀 | ⭕ |
| $[FR_1^{4,1,2}]_{1,1}^2$ | $[\mathbb{Z}_4]_{1,1}^2$ | ✕ | | 🌐 | 🎀 | ⭕ |
| $[FR_1^{4,1,2}]_{1,1}^3$ | $[\mathbb{Z}_4]_{1,1}^3$ | ✕ | | | | |
| $[FR_1^{4,1,2}]_{1,2}^1$ | $[\mathbb{Z}_4]_{1,2}^1$ | ✕ | † | 🌐 | 🎀 | ⭕ |
| $[FR_1^{4,1,2}]_{1,2}^2$ | $[\mathbb{Z}_4]_{1,2}^2$ | ✕ | | 🌐 | 🎀 | ⭕ |
| $[FR_1^{4,1,2}]_{1,2}^3$ | $[\mathbb{Z}_4]_{1,2}^3$ | ✕ | | | | |
| $[FR_1^{4,1,2}]_{1,3}^1$ | $[\mathbb{Z}_4]_{1,3}^1$ | ✕ | † | 🌐 | 🎀 | ⭕ |
| $[FR_1^{4,1,2}]_{1,3}^2$ | $[\mathbb{Z}_4]_{1,3}^2$ | ✕ | | 🌐 | 🎀 | ⭕ |
| $[FR_1^{4,1,2}]_{1,3}^3$ | $[\mathbb{Z}_4]_{1,3}^3$ | ✕ | | | | |
| $[FR_1^{4,1,2}]_{1,4}^1$ | $[\mathbb{Z}_4]_{1,4}^1$ | ✕ | † | 🌐 | 🎀 | ⭕ |
| $[FR_1^{4,1,2}]_{1,4}^2$ | $[\mathbb{Z}_4]_{1,4}^2$ | ✕ | | 🌐 | 🎀 | ⭕ |
| $[FR_1^{4,1,2}]_{1,4}^3$ | $[\mathbb{Z}_4]_{1,4}^3$ | ✕ | | | | |
| $[FR_1^{4,1,2}]_{2,1}^1$ | $[\mathbb{Z}_4]_{2,1}^1$ | ✕ | † | 🌐 | 🎀 | |

Table 9.1: List of multiplicity-free fusion categories up to rank 7 (Continued)

| Formal Name | Common Name | | | | |
|---|---|---|---|---|---|
| $[FR_1^{4,1,2}]_{2,1}^2$ | $[\mathbb{Z}_4]_{2,1}^2$ | ✕ | | 🌐 | 🎀 |
| $[FR_1^{4,1,2}]_{2,1}^3$ | $[\mathbb{Z}_4]_{2,1}^3$ | ✕ | | | |
| $[FR_1^{4,1,2}]_{2,2}^1$ | $[\mathbb{Z}_4]_{2,2}^1$ | ✕ | 🗡 | 🌐 | 🎀 |
| $[FR_1^{4,1,2}]_{2,2}^2$ | $[\mathbb{Z}_4]_{2,2}^2$ | ✕ | | 🌐 | 🎀 |
| $[FR_1^{4,1,2}]_{2,2}^3$ | $[\mathbb{Z}_4]_{2,2}^3$ | ✕ | | | |
| $[FR_1^{4,1,2}]_{2,3}^1$ | $[\mathbb{Z}_4]_{2,3}^1$ | ✕ | 🗡 | 🌐 | 🎀 |
| $[FR_1^{4,1,2}]_{2,3}^2$ | $[\mathbb{Z}_4]_{2,3}^2$ | ✕ | | 🌐 | 🎀 |
| $[FR_1^{4,1,2}]_{2,3}^3$ | $[\mathbb{Z}_4]_{2,3}^3$ | ✕ | | | |
| $[FR_1^{4,1,2}]_{2,4}^1$ | $[\mathbb{Z}_4]_{2,4}^1$ | ✕ | 🗡 | 🌐 | 🎀 |
| $[FR_1^{4,1,2}]_{2,4}^2$ | $[\mathbb{Z}_4]_{2,4}^2$ | ✕ | | 🌐 | 🎀 |
| $[FR_1^{4,1,2}]_{2,4}^3$ | $[\mathbb{Z}_4]_{2,4}^3$ | ✕ | | | |
| $[FR_1^{4,1,2}]_{3,1}^1$ | $[\mathbb{Z}_4]_{3,1}^1$ | | 🗡 | 🌐 | |
| $[FR_1^{4,1,2}]_{3,1}^2$ | $[\mathbb{Z}_4]_{3,1}^2$ | | | 🌐 | |
| $[FR_1^{4,1,2}]_{3,1}^3$ | $[\mathbb{Z}_4]_{3,1}^3$ | | | | |
| $[FR_1^{4,1,2}]_{4,1}^1$ | $[\mathbb{Z}_4]_{4,1}^1$ | | 🗡 | 🌐 | |
| $[FR_1^{4,1,2}]_{4,1}^2$ | $[\mathbb{Z}_4]_{4,1}^2$ | | | 🌐 | |
| $[FR_1^{4,1,2}]_{4,1}^3$ | $[\mathbb{Z}_4]_{4,1}^3$ | | | | |
| $[FR_2^{4,1,2}]_{1,1}^1$ | $[TY(\mathbb{Z}_3)]_{1,1}^1$ | | 🗡 | 🌐 | |
| $[FR_2^{4,1,2}]_{1,1}^2$ | $[TY(\mathbb{Z}_3)]_{1,1}^2$ | | | 🌐 | |
| $[FR_2^{4,1,2}]_{2,1}^1$ | $[TY(\mathbb{Z}_3)]_{2,1}^1$ | | 🗡 | 🌐 | |
| $[FR_2^{4,1,2}]_{2,1}^2$ | $[TY(\mathbb{Z}_3)]_{2,1}^2$ | | | 🌐 | |
| $[FR_2^{4,1,2}]_{3,1}^1$ | $[TY(\mathbb{Z}_3)]_{3,1}^1$ | | 🗡 | 🌐 | |
| $[FR_2^{4,1,2}]_{3,1}^2$ | $[TY(\mathbb{Z}_3)]_{3,1}^2$ | | | 🌐 | |
| $[FR_2^{4,1,2}]_{4,1}^1$ | $[TY(\mathbb{Z}_3)]_{4,1}^1$ | | 🗡 | 🌐 | |
| $[FR_2^{4,1,2}]_{4,1}^2$ | $[TY(\mathbb{Z}_3)]_{4,1}^2$ | | | 🌐 | |
| $[FR_4^{4,1,2}]_{1,1}^1$ | $[\text{Pseudo PSU}(2)_6]_{1,1}^1$ | | 🗡 | 🌐 | |
| $[FR_4^{4,1,2}]_{2,1}^1$ | $[\text{Pseudo PSU}(2)_6]_{2,1}^1$ | | 🗡 | 🌐 | |
| $[FR_4^{4,1,2}]_{3,1}^1$ | $[\text{Pseudo PSU}(2)_6]_{3,1}^1$ | | | 🌐 | |
| $[FR_4^{4,1,2}]_{4,1}^1$ | $[\text{Pseudo PSU}(2)_6]_{4,1}^1$ | | | 🌐 | |

Table 9.1: List of multiplicity-free fusion categories up to rank 7 (Continued)

| Formal Name | Common Name | | | | |
|---|---|---|---|---|---|
| $[FR_1^{5,1,0}]_{1,1}^1$ | $[Rep(D_4)]_{1,1}^1$ | ✕ | † | 🌐 | 🎗 |
| $[FR_1^{5,1,0}]_{1,1}^2$ | $[Rep(D_4)]_{1,1}^2$ | ✕ | | 🌐 | 🎗 |
| $[FR_1^{5,1,0}]_{1,2}^1$ | $[Rep(D_4)]_{1,2}^1$ | ✕ | † | 🌐 | 🎗 |
| $[FR_1^{5,1,0}]_{1,2}^2$ | $[Rep(D_4)]_{1,2}^2$ | ✕ | | 🌐 | 🎗 |
| $[FR_1^{5,1,0}]_{1,3}^1$ | $[Rep(D_4)]_{1,3}^1$ | ✕ | † | 🌐 | 🎗 |
| $[FR_1^{5,1,0}]_{1,3}^2$ | $[Rep(D_4)]_{1,3}^2$ | ✕ | | 🌐 | 🎗 |
| $[FR_1^{5,1,0}]_{1,4}^1$ | $[Rep(D_4)]_{1,4}^1$ | ✕ | † | 🌐 | 🎗 |
| $[FR_1^{5,1,0}]_{1,4}^2$ | $[Rep(D_4)]_{1,4}^2$ | ✕ | | 🌐 | 🎗 |
| $[FR_1^{5,1,0}]_{2,1}^1$ | $[Rep(D_4)]_{2,1}^1$ | ✕ | † | 🌐 | 🎗 |
| $[FR_1^{5,1,0}]_{2,1}^2$ | $[Rep(D_4)]_{2,1}^2$ | ✕ | | 🌐 | 🎗 |
| $[FR_1^{5,1,0}]_{2,2}^1$ | $[Rep(D_4)]_{2,2}^1$ | ✕ | † | 🌐 | 🎗 |
| $[FR_1^{5,1,0}]_{2,2}^2$ | $[Rep(D_4)]_{2,2}^2$ | ✕ | | 🌐 | 🎗 |
| $[FR_1^{5,1,0}]_{2,3}^1$ | $[Rep(D_4)]_{2,3}^1$ | ✕ | † | 🌐 | 🎗 |
| $[FR_1^{5,1,0}]_{2,3}^2$ | $[Rep(D_4)]_{2,3}^2$ | ✕ | | 🌐 | 🎗 |
| $[FR_1^{5,1,0}]_{2,4}^1$ | $[Rep(D_4)]_{2,4}^1$ | ✕ | † | 🌐 | 🎗 |
| $[FR_1^{5,1,0}]_{2,4}^2$ | $[Rep(D_4)]_{2,4}^2$ | ✕ | | 🌐 | 🎗 |
| $[FR_1^{5,1,0}]_{3,1}^1$ | $[Rep(D_4)]_{3,1}^1$ | ✕ | † | 🌐 | 🎗 |
| $[FR_1^{5,1,0}]_{3,1}^2$ | $[Rep(D_4)]_{3,1}^2$ | ✕ | | 🌐 | 🎗 |
| $[FR_1^{5,1,0}]_{3,2}^1$ | $[Rep(D_4)]_{3,2}^1$ | ✕ | † | 🌐 | 🎗 |
| $[FR_1^{5,1,0}]_{3,2}^2$ | $[Rep(D_4)]_{3,2}^2$ | ✕ | | 🌐 | 🎗 |
| $[FR_1^{5,1,0}]_{3,3}^1$ | $[Rep(D_4)]_{3,3}^1$ | ✕ | † | 🌐 | 🎗 |
| $[FR_1^{5,1,0}]_{3,3}^2$ | $[Rep(D_4)]_{3,3}^2$ | ✕ | | 🌐 | 🎗 |
| $[FR_1^{5,1,0}]_{3,4}^1$ | $[Rep(D_4)]_{3,4}^1$ | ✕ | † | 🌐 | 🎗 |
| $[FR_1^{5,1,0}]_{3,4}^2$ | $[Rep(D_4)]_{3,4}^2$ | ✕ | | 🌐 | 🎗 |
| $[FR_1^{5,1,0}]_{3,5}^1$ | $[Rep(D_4)]_{3,5}^1$ | ✕ | † | 🌐 | 🎗 |
| $[FR_1^{5,1,0}]_{3,5}^2$ | $[Rep(D_4)]_{3,5}^2$ | ✕ | | 🌐 | 🎗 |
| $[FR_1^{5,1,0}]_{3,6}^1$ | $[Rep(D_4)]_{3,6}^1$ | ✕ | † | 🌐 | 🎗 |
| $[FR_1^{5,1,0}]_{3,6}^2$ | $[Rep(D_4)]_{3,6}^2$ | ✕ | | 🌐 | 🎗 |
| $[FR_1^{5,1,0}]_{4,1}^1$ | $[Rep(D_4)]_{4,1}^1$ | ✕ | † | 🌐 | 🎗 |

Table 9.1: List of multiplicity-free fusion categories up to rank 7 (Continued)

| Formal Name | Common Name | | | | | |
|---|---|---|---|---|---|---|
| $[FR_1^{5,1,0}]_{4,1}^2$ | $[Rep(D_4)]_{4,1}^2$ | ✗ | | 🌐 | 🎗 | |
| $[FR_1^{5,1,0}]_{4,2}^1$ | $[Rep(D_4)]_{4,2}^1$ | ✗ | † | 🌐 | 🎗 | |
| $[FR_1^{5,1,0}]_{4,2}^2$ | $[Rep(D_4)]_{4,2}^2$ | ✗ | | 🌐 | 🎗 | |
| $[FR_1^{5,1,0}]_{4,3}^1$ | $[Rep(D_4)]_{4,3}^1$ | ✗ | † | 🌐 | 🎗 | |
| $[FR_1^{5,1,0}]_{4,3}^2$ | $[Rep(D_4)]_{4,3}^2$ | ✗ | | 🌐 | 🎗 | |
| $[FR_1^{5,1,0}]_{4,4}^1$ | $[Rep(D_4)]_{4,4}^1$ | ✗ | † | 🌐 | 🎗 | |
| $[FR_1^{5,1,0}]_{4,4}^2$ | $[Rep(D_4)]_{4,4}^2$ | ✗ | | 🌐 | 🎗 | |
| $[FR_1^{5,1,0}]_{4,5}^1$ | $[Rep(D_4)]_{4,5}^1$ | ✗ | † | 🌐 | 🎗 | |
| $[FR_1^{5,1,0}]_{4,5}^2$ | $[Rep(D_4)]_{4,5}^2$ | ✗ | | 🌐 | 🎗 | |
| $[FR_1^{5,1,0}]_{4,6}^1$ | $[Rep(D_4)]_{4,6}^1$ | ✗ | † | 🌐 | 🎗 | |
| $[FR_1^{5,1,0}]_{4,6}^2$ | $[Rep(D_4)]_{4,6}^2$ | ✗ | | 🌐 | 🎗 | |
| $[FR_3^{5,1,0}]_{1,1}^1$ | $[SU(2)_4]_{1,1}^1$ | ✗ | † | 🌐 | 🎗 | ⭕ |
| $[FR_3^{5,1,0}]_{1,1}^2$ | $[SU(2)_4]_{1,1}^2$ | ✗ | | 🌐 | 🎗 | ⭕ |
| $[FR_3^{5,1,0}]_{1,2}^1$ | $[SU(2)_4]_{1,2}^1$ | ✗ | † | 🌐 | 🎗 | ⭕ |
| $[FR_3^{5,1,0}]_{1,2}^2$ | $[SU(2)_4]_{1,2}^2$ | ✗ | | 🌐 | 🎗 | ⭕ |
| $[FR_3^{5,1,0}]_{2,1}^1$ | $[SU(2)_4]_{2,1}^1$ | ✗ | † | 🌐 | 🎗 | ⭕ |
| $[FR_3^{5,1,0}]_{2,1}^2$ | $[SU(2)_4]_{2,1}^2$ | ✗ | | 🌐 | 🎗 | ⭕ |
| $[FR_3^{5,1,0}]_{2,2}^1$ | $[SU(2)_4]_{2,2}^1$ | ✗ | † | 🌐 | 🎗 | ⭕ |
| $[FR_3^{5,1,0}]_{2,2}^2$ | $[SU(2)_4]_{2,2}^2$ | ✗ | | 🌐 | 🎗 | ⭕ |
| $[FR_4^{5,1,0}]_{1,1}^1$ | $[Rep(D_7)]_{1,1}^1$ | ✗ | † | 🌐 | 🎗 | |
| $[FR_4^{5,1,0}]_{1,2}^1$ | $[Rep(D_7)]_{1,2}^1$ | ✗ | † | 🌐 | 🎗 | |
| $[FR_4^{5,1,0}]_{1,3}^1$ | $[Rep(D_7)]_{1,3}^1$ | ✗ | † | 🌐 | 🎗 | |
| $[FR_4^{5,1,0}]_{2,1}^1$ | $[Rep(D_7)]_{2,1}^1$ | | † | 🌐 | | |
| $[FR_4^{5,1,0}]_{3,1}^1$ | $[Rep(D_7)]_{3,1}^1$ | | † | 🌐 | | |
| $[FR_4^{5,1,0}]_{4,1}^1$ | $[Rep(D_7)]_{4,1}^1$ | | † | 🌐 | | |
| $[FR_4^{5,1,0}]_{5,1}^1$ | $[Rep(D_7)]_{5,1}^1$ | | † | 🌐 | | |
| $[FR_4^{5,1,0}]_{6,1}^1$ | $[Rep(D_7)]_{6,1}^1$ | | † | 🌐 | | |
| $[FR_4^{5,1,0}]_{7,1}^1$ | $[Rep(D_7)]_{7,1}^1$ | | † | 🌐 | | |

Table 9.1: List of multiplicity-free fusion categories up to rank 7 (Continued)

| Formal Name | Common Name | | | | | |
|---|---|---|---|---|---|---|
| $[FR_6^{5,1,0}]_{1,1}^1$ | $[Rep(S_4)]_{1,1}^1$ | ✕ | † | 🌐 | 🎗 | |
| $[FR_6^{5,1,0}]_{2,1}^1$ | $[Rep(S_4)]_{2,1}^1$ | ✕ | † | 🌐 | 🎗 | |
| $[FR_7^{5,1,0}]_{1,1}^1$ | $[PSU(2)_8]_{1,1}^1$ | ✕ | † | 🌐 | 🎗 | |
| $[FR_7^{5,1,0}]_{1,2}^1$ | $[PSU(2)_8]_{1,2}^1$ | ✕ | † | 🌐 | 🎗 | |
| $[FR_7^{5,1,0}]_{2,1}^1$ | $[PSU(2)_8]_{2,1}^1$ | ✕ | | 🌐 | 🎗 | |
| $[FR_7^{5,1,0}]_{2,2}^1$ | $[PSU(2)_8]_{2,2}^1$ | ✕ | | 🌐 | 🎗 | |
| $[FR_{10}^{5,1,0}]_{1,1}^1$ | $[PSU(2)_9]_{1,1}^1$ | ✕ | † | 🌐 | 🎗 | ⭕ |
| $[FR_{10}^{5,1,0}]_{1,2}^1$ | $[PSU(2)_9]_{1,2}^1$ | ✕ | † | 🌐 | 🎗 | ⭕ |
| $[FR_{10}^{5,1,0}]_{2,1}^1$ | $[PSU(2)_9]_{2,1}^1$ | ✕ | | 🌐 | 🎗 | ⭕ |
| $[FR_{10}^{5,1,0}]_{2,2}^1$ | $[PSU(2)_9]_{2,2}^1$ | ✕ | | 🌐 | 🎗 | ⭕ |
| $[FR_{10}^{5,1,0}]_{3,1}^1$ | $[PSU(2)_9]_{3,1}^1$ | ✕ | | 🌐 | 🎗 | ⭕ |
| $[FR_{10}^{5,1,0}]_{3,2}^1$ | $[PSU(2)_9]_{3,2}^1$ | ✕ | | 🌐 | 🎗 | ⭕ |
| $[FR_{10}^{5,1,0}]_{4,1}^1$ | $[PSU(2)_9]_{4,1}^1$ | ✕ | | 🌐 | 🎗 | ⭕ |
| $[FR_{10}^{5,1,0}]_{4,2}^1$ | $[PSU(2)_9]_{4,2}^1$ | ✕ | | 🌐 | 🎗 | ⭕ |
| $[FR_{10}^{5,1,0}]_{5,1}^1$ | $[PSU(2)_9]_{5,1}^1$ | ✕ | | 🌐 | 🎗 | ⭕ |
| $[FR_{10}^{5,1,0}]_{5,2}^1$ | $[PSU(2)_9]_{5,2}^1$ | ✕ | | 🌐 | 🎗 | ⭕ |
| $[FR_1^{5,1,2}]_{1,1}^1$ | $[TY(\mathbb{Z}_4)]_{1,1}^1$ | | † | 🌐 | | |
| $[FR_1^{5,1,2}]_{1,1}^2$ | $[TY(\mathbb{Z}_4)]_{1,1}^2$ | | | 🌐 | | |
| $[FR_1^{5,1,2}]_{2,1}^1$ | $[TY(\mathbb{Z}_4)]_{2,1}^1$ | | † | 🌐 | | |
| $[FR_1^{5,1,2}]_{2,1}^2$ | $[TY(\mathbb{Z}_4)]_{2,1}^2$ | | | 🌐 | | |
| $[FR_1^{5,1,2}]_{3,1}^1$ | $[TY(\mathbb{Z}_4)]_{3,1}^1$ | | † | 🌐 | | |
| $[FR_1^{5,1,2}]_{3,1}^2$ | $[TY(\mathbb{Z}_4)]_{3,1}^2$ | | | 🌐 | | |
| $[FR_1^{5,1,2}]_{4,1}^1$ | $[TY(\mathbb{Z}_4)]_{4,1}^1$ | | † | 🌐 | | |
| $[FR_1^{5,1,2}]_{4,1}^2$ | $[TY(\mathbb{Z}_4)]_{4,1}^2$ | | | 🌐 | | |
| $[FR_3^{5,1,2}]_{1,1}^1$ | $[Pseudo\ SU(2)_4]_{1,1}^1$ | | † | 🌐 | | |
| $[FR_3^{5,1,2}]_{1,1}^2$ | $[Pseudo\ SU(2)_4]_{1,1}^2$ | | | 🌐 | | |
| $[FR_3^{5,1,2}]_{2,1}^1$ | $[Pseudo\ SU(2)_4]_{2,1}^1$ | | † | 🌐 | | |
| $[FR_3^{5,1,2}]_{2,1}^2$ | $[Pseudo\ SU(2)_4]_{2,1}^2$ | | | 🌐 | | |

Table 9.1: List of multiplicity-free fusion categories up to rank 7 (Continued)

| Formal Name | Common Name | | | | | |
|---|---|---|---|---|---|---|
| $[FR_4^{5,1,2}]_{1,1}^1$ | $[\text{Pseudo Rep}(S_4)]_{1,1}^1$ | | †| 🌐 | | |
| $[FR_4^{5,1,2}]_{2,1}^1$ | $[\text{Pseudo Rep}(S_4)]_{2,1}^1$ | | † | 🌐 | | |
| $[FR_1^{5,1,4}]_{1,1}^1$ | $[\mathbb{Z}_5]_{1,1}^1$ | ✗ | † | 🌐 | 🎗 | ⭕ |
| $[FR_1^{5,1,4}]_{1,1}^2$ | $[\mathbb{Z}_5]_{1,1}^2$ | ✗ | | | | |
| $[FR_1^{5,1,4}]_{1,1}^3$ | $[\mathbb{Z}_5]_{1,1}^3$ | ✗ | | | | |
| $[FR_1^{5,1,4}]_{1,2}^1$ | $[\mathbb{Z}_5]_{1,2}^1$ | ✗ | † | 🌐 | 🎗 | ⭕ |
| $[FR_1^{5,1,4}]_{1,2}^2$ | $[\mathbb{Z}_5]_{1,2}^2$ | ✗ | | | | |
| $[FR_1^{5,1,4}]_{1,2}^3$ | $[\mathbb{Z}_5]_{1,2}^3$ | ✗ | | | | |
| $[FR_1^{5,1,4}]_{1,3}^1$ | $[\mathbb{Z}_5]_{1,3}^1$ | ✗ | † | 🌐 | 🎗 | |
| $[FR_1^{5,1,4}]_{1,3}^2$ | $[\mathbb{Z}_5]_{1,3}^2$ | ✗ | | | | |
| $[FR_1^{5,1,4}]_{2,1}^1$ | $[\mathbb{Z}_5]_{2,1}^1$ | | † | 🌐 | | |
| $[FR_1^{5,1,4}]_{2,1}^2$ | $[\mathbb{Z}_5]_{2,1}^2$ | | | | | |
| $[FR_1^{5,1,4}]_{2,1}^3$ | $[\mathbb{Z}_5]_{2,1}^3$ | | | | | |
| $[FR_1^{5,1,4}]_{3,1}^1$ | $[\mathbb{Z}_5]_{3,1}^1$ | | † | 🌐 | | |
| $[FR_1^{5,1,4}]_{3,1}^2$ | $[\mathbb{Z}_5]_{3,1}^2$ | | | | | |
| $[FR_1^{5,1,4}]_{3,1}^3$ | $[\mathbb{Z}_5]_{3,1}^3$ | | | | | |
| $[FR_1^{6,1,0}]_{1,1}^1$ | $[\mathbb{Z}_2 \times \text{Ising}]_{1,1}^1$ | ✗ | † | 🌐 | 🎗 | ⭕ |
| $[FR_1^{6,1,0}]_{1,1}^2$ | $[\mathbb{Z}_2 \times \text{Ising}]_{1,1}^2$ | ✗ | | 🌐 | 🎗 | ⭕ |
| $[FR_1^{6,1,0}]_{1,1}^3$ | $[\mathbb{Z}_2 \times \text{Ising}]_{1,1}^3$ | ✗ | | 🌐 | 🎗 | ⭕ |
| $[FR_1^{6,1,0}]_{1,1}^4$ | $[\mathbb{Z}_2 \times \text{Ising}]_{1,1}^4$ | ✗ | | 🌐 | 🎗 | ⭕ |
| $[FR_1^{6,1,0}]_{1,2}^1$ | $[\mathbb{Z}_2 \times \text{Ising}]_{1,2}^1$ | ✗ | † | 🌐 | 🎗 | ⭕ |
| $[FR_1^{6,1,0}]_{1,2}^2$ | $[\mathbb{Z}_2 \times \text{Ising}]_{1,2}^2$ | ✗ | | 🌐 | 🎗 | ⭕ |
| $[FR_1^{6,1,0}]_{1,2}^3$ | $[\mathbb{Z}_2 \times \text{Ising}]_{1,2}^3$ | ✗ | | 🌐 | 🎗 | ⭕ |
| $[FR_1^{6,1,0}]_{1,2}^4$ | $[\mathbb{Z}_2 \times \text{Ising}]_{1,2}^4$ | ✗ | | 🌐 | 🎗 | ⭕ |
| $[FR_1^{6,1,0}]_{1,3}^1$ | $[\mathbb{Z}_2 \times \text{Ising}]_{1,3}^1$ | ✗ | † | 🌐 | 🎗 | ⭕ |
| $[FR_1^{6,1,0}]_{1,3}^2$ | $[\mathbb{Z}_2 \times \text{Ising}]_{1,3}^2$ | ✗ | | 🌐 | 🎗 | ⭕ |
| $[FR_1^{6,1,0}]_{1,3}^3$ | $[\mathbb{Z}_2 \times \text{Ising}]_{1,3}^3$ | ✗ | | 🌐 | 🎗 | ⭕ |
| $[FR_1^{6,1,0}]_{1,3}^4$ | $[\mathbb{Z}_2 \times \text{Ising}]_{1,3}^4$ | ✗ | | 🌐 | 🎗 | ⭕ |
| $[FR_1^{6,1,0}]_{1,4}^1$ | $[\mathbb{Z}_2 \times \text{Ising}]_{1,4}^1$ | ✗ | † | 🌐 | 🎗 | ⭕ |

Table 9.1: List of multiplicity-free fusion categories up to rank 7 (Continued)

| Formal Name | Common Name | | | | | |
|---|---|---|---|---|---|---|
| $[FR_1^{6,1,0}]_{1,4}^2$ | $[\mathbb{Z}_2 \times \text{Ising}]_{1,4}^2$ | ✕ | | 🌐 | 🎗 | ⭕ |
| $[FR_1^{6,1,0}]_{1,4}^3$ | $[\mathbb{Z}_2 \times \text{Ising}]_{1,4}^3$ | ✕ | | 🌐 | 🎗 | ⭕ |
| $[FR_1^{6,1,0}]_{1,4}^4$ | $[\mathbb{Z}_2 \times \text{Ising}]_{1,4}^4$ | ✕ | | 🌐 | 🎗 | ⭕ |
| $[FR_1^{6,1,0}]_{1,5}^1$ | $[\mathbb{Z}_2 \times \text{Ising}]_{1,5}^1$ | ✕ | † | 🌐 | 🎗 | ⭕ |
| $[FR_1^{6,1,0}]_{1,5}^2$ | $[\mathbb{Z}_2 \times \text{Ising}]_{1,5}^2$ | ✕ | | 🌐 | 🎗 | ⭕ |
| $[FR_1^{6,1,0}]_{1,5}^3$ | $[\mathbb{Z}_2 \times \text{Ising}]_{1,5}^3$ | ✕ | | 🌐 | 🎗 | ⭕ |
| $[FR_1^{6,1,0}]_{1,5}^4$ | $[\mathbb{Z}_2 \times \text{Ising}]_{1,5}^4$ | ✕ | | 🌐 | 🎗 | ⭕ |
| $[FR_1^{6,1,0}]_{1,6}^1$ | $[\mathbb{Z}_2 \times \text{Ising}]_{1,6}^1$ | ✕ | † | 🌐 | 🎗 | ⭕ |
| $[FR_1^{6,1,0}]_{1,6}^2$ | $[\mathbb{Z}_2 \times \text{Ising}]_{1,6}^2$ | ✕ | | 🌐 | 🎗 | ⭕ |
| $[FR_1^{6,1,0}]_{1,6}^3$ | $[\mathbb{Z}_2 \times \text{Ising}]_{1,6}^3$ | ✕ | | 🌐 | 🎗 | ⭕ |
| $[FR_1^{6,1,0}]_{1,6}^4$ | $[\mathbb{Z}_2 \times \text{Ising}]_{1,6}^4$ | ✕ | | 🌐 | 🎗 | ⭕ |
| $[FR_1^{6,1,0}]_{1,7}^1$ | $[\mathbb{Z}_2 \times \text{Ising}]_{1,7}^1$ | ✕ | † | 🌐 | 🎗 | ⭕ |
| $[FR_1^{6,1,0}]_{1,7}^2$ | $[\mathbb{Z}_2 \times \text{Ising}]_{1,7}^2$ | ✕ | | 🌐 | 🎗 | ⭕ |
| $[FR_1^{6,1,0}]_{1,7}^3$ | $[\mathbb{Z}_2 \times \text{Ising}]_{1,7}^3$ | ✕ | | 🌐 | 🎗 | ⭕ |
| $[FR_1^{6,1,0}]_{1,7}^4$ | $[\mathbb{Z}_2 \times \text{Ising}]_{1,7}^4$ | ✕ | | 🌐 | 🎗 | ⭕ |
| $[FR_1^{6,1,0}]_{1,8}^1$ | $[\mathbb{Z}_2 \times \text{Ising}]_{1,8}^1$ | ✕ | † | 🌐 | 🎗 | ⭕ |
| $[FR_1^{6,1,0}]_{1,8}^2$ | $[\mathbb{Z}_2 \times \text{Ising}]_{1,8}^2$ | ✕ | | 🌐 | 🎗 | ⭕ |
| $[FR_1^{6,1,0}]_{1,8}^3$ | $[\mathbb{Z}_2 \times \text{Ising}]_{1,8}^3$ | ✕ | | 🌐 | 🎗 | ⭕ |
| $[FR_1^{6,1,0}]_{1,8}^4$ | $[\mathbb{Z}_2 \times \text{Ising}]_{1,8}^4$ | ✕ | | 🌐 | 🎗 | ⭕ |
| $[FR_1^{6,1,0}]_{2,1}^1$ | $[\mathbb{Z}_2 \times \text{Ising}]_{2,1}^1$ | ✕ | † | 🌐 | 🎗 | |
| $[FR_1^{6,1,0}]_{2,1}^2$ | $[\mathbb{Z}_2 \times \text{Ising}]_{2,1}^2$ | ✕ | | 🌐 | 🎗 | |
| $[FR_1^{6,1,0}]_{2,1}^3$ | $[\mathbb{Z}_2 \times \text{Ising}]_{2,1}^3$ | ✕ | | 🌐 | 🎗 | |
| $[FR_1^{6,1,0}]_{2,1}^4$ | $[\mathbb{Z}_2 \times \text{Ising}]_{2,1}^4$ | ✕ | | 🌐 | 🎗 | |
| $[FR_1^{6,1,0}]_{2,2}^1$ | $[\mathbb{Z}_2 \times \text{Ising}]_{2,2}^1$ | ✕ | † | 🌐 | 🎗 | |
| $[FR_1^{6,1,0}]_{2,2}^2$ | $[\mathbb{Z}_2 \times \text{Ising}]_{2,2}^2$ | ✕ | | 🌐 | 🎗 | |
| $[FR_1^{6,1,0}]_{2,2}^3$ | $[\mathbb{Z}_2 \times \text{Ising}]_{2,2}^3$ | ✕ | | 🌐 | 🎗 | |
| $[FR_1^{6,1,0}]_{2,2}^4$ | $[\mathbb{Z}_2 \times \text{Ising}]_{2,2}^4$ | ✕ | | 🌐 | 🎗 | |
| $[FR_1^{6,1,0}]_{2,3}^1$ | $[\mathbb{Z}_2 \times \text{Ising}]_{2,3}^1$ | ✕ | † | 🌐 | 🎗 | |
| $[FR_1^{6,1,0}]_{2,3}^2$ | $[\mathbb{Z}_2 \times \text{Ising}]_{2,3}^2$ | ✕ | | 🌐 | 🎗 | |

Table 9.1: List of multiplicity-free fusion categories up to rank 7 (Continued)

| Formal Name | Common Name | | | | |
|---|---|---|---|---|---|
| $[FR_1^{6,1,0}]_{2,3}^3$ | $[\mathbb{Z}_2 \times \text{Ising}]_{2,3}^3$ | ✖ | | 🌐 | 🎗 |
| $[FR_1^{6,1,0}]_{2,4}^1$ | $[\mathbb{Z}_2 \times \text{Ising}]_{2,4}^1$ | ✖ | † | 🌐 | 🎗 |
| $[FR_1^{6,1,0}]_{2,4}^2$ | $[\mathbb{Z}_2 \times \text{Ising}]_{2,4}^2$ | ✖ | | 🌐 | 🎗 |
| $[FR_1^{6,1,0}]_{2,4}^3$ | $[\mathbb{Z}_2 \times \text{Ising}]_{2,4}^3$ | ✖ | | 🌐 | 🎗 |
| $[FR_1^{6,1,0}]_{2,5}^1$ | $[\mathbb{Z}_2 \times \text{Ising}]_{2,5}^1$ | ✖ | † | 🌐 | 🎗 |
| $[FR_1^{6,1,0}]_{2,5}^2$ | $[\mathbb{Z}_2 \times \text{Ising}]_{2,5}^2$ | ✖ | | 🌐 | 🎗 |
| $[FR_1^{6,1,0}]_{2,5}^3$ | $[\mathbb{Z}_2 \times \text{Ising}]_{2,5}^3$ | ✖ | | 🌐 | 🎗 |
| $[FR_1^{6,1,0}]_{2,6}^1$ | $[\mathbb{Z}_2 \times \text{Ising}]_{2,6}^1$ | ✖ | † | 🌐 | 🎗 |
| $[FR_1^{6,1,0}]_{2,6}^2$ | $[\mathbb{Z}_2 \times \text{Ising}]_{2,6}^2$ | ✖ | | 🌐 | 🎗 |
| $[FR_1^{6,1,0}]_{2,6}^3$ | $[\mathbb{Z}_2 \times \text{Ising}]_{2,6}^3$ | ✖ | | 🌐 | 🎗 |
| $[FR_1^{6,1,0}]_{3,1}^1$ | $[\mathbb{Z}_2 \times \text{Ising}]_{3,1}^1$ | ✖ | † | 🌐 | 🎗 |
| $[FR_1^{6,1,0}]_{3,1}^2$ | $[\mathbb{Z}_2 \times \text{Ising}]_{3,1}^2$ | ✖ | | 🌐 | 🎗 |
| $[FR_1^{6,1,0}]_{3,1}^3$ | $[\mathbb{Z}_2 \times \text{Ising}]_{3,1}^3$ | ✖ | | 🌐 | 🎗 |
| $[FR_1^{6,1,0}]_{3,1}^4$ | $[\mathbb{Z}_2 \times \text{Ising}]_{3,1}^4$ | ✖ | | 🌐 | 🎗 |
| $[FR_1^{6,1,0}]_{3,2}^1$ | $[\mathbb{Z}_2 \times \text{Ising}]_{3,2}^1$ | ✖ | † | 🌐 | 🎗 |
| $[FR_1^{6,1,0}]_{3,2}^2$ | $[\mathbb{Z}_2 \times \text{Ising}]_{3,2}^2$ | ✖ | | 🌐 | 🎗 |
| $[FR_1^{6,1,0}]_{3,2}^3$ | $[\mathbb{Z}_2 \times \text{Ising}]_{3,2}^3$ | ✖ | | 🌐 | 🎗 |
| $[FR_1^{6,1,0}]_{3,2}^4$ | $[\mathbb{Z}_2 \times \text{Ising}]_{3,2}^4$ | ✖ | | 🌐 | 🎗 |
| $[FR_1^{6,1,0}]_{3,3}^1$ | $[\mathbb{Z}_2 \times \text{Ising}]_{3,3}^1$ | ✖ | † | 🌐 | 🎗 |
| $[FR_1^{6,1,0}]_{3,3}^2$ | $[\mathbb{Z}_2 \times \text{Ising}]_{3,3}^2$ | ✖ | | 🌐 | 🎗 |
| $[FR_1^{6,1,0}]_{3,3}^3$ | $[\mathbb{Z}_2 \times \text{Ising}]_{3,3}^3$ | ✖ | | 🌐 | 🎗 |
| $[FR_1^{6,1,0}]_{3,4}^1$ | $[\mathbb{Z}_2 \times \text{Ising}]_{3,4}^1$ | ✖ | † | 🌐 | 🎗 |
| $[FR_1^{6,1,0}]_{3,4}^2$ | $[\mathbb{Z}_2 \times \text{Ising}]_{3,4}^2$ | ✖ | | 🌐 | 🎗 |
| $[FR_1^{6,1,0}]_{3,4}^3$ | $[\mathbb{Z}_2 \times \text{Ising}]_{3,4}^3$ | ✖ | | 🌐 | 🎗 |
| $[FR_1^{6,1,0}]_{3,5}^1$ | $[\mathbb{Z}_2 \times \text{Ising}]_{3,5}^1$ | ✖ | † | 🌐 | 🎗 |
| $[FR_1^{6,1,0}]_{3,5}^2$ | $[\mathbb{Z}_2 \times \text{Ising}]_{3,5}^2$ | ✖ | | 🌐 | 🎗 |
| $[FR_1^{6,1,0}]_{3,5}^3$ | $[\mathbb{Z}_2 \times \text{Ising}]_{3,5}^3$ | ✖ | | 🌐 | 🎗 |
| $[FR_1^{6,1,0}]_{3,6}^1$ | $[\mathbb{Z}_2 \times \text{Ising}]_{3,6}^1$ | ✖ | † | 🌐 | 🎗 |
| $[FR_1^{6,1,0}]_{3,6}^2$ | $[\mathbb{Z}_2 \times \text{Ising}]_{3,6}^2$ | ✖ | | 🌐 | 🎗 |

Table 9.1: List of multiplicity-free fusion categories up to rank 7 (Continued)

| Formal Name | Common Name | | | | |
|---|---|---|---|---|---|
| $[FR_1^{6,1,0}]_{3,6}^3$ | $[\mathbb{Z}_2 \times \text{Ising}]_{3,6}^3$ | ✖ | | 🌐 | 🎗 |
| $[FR_1^{6,1,0}]_{4,1}^1$ | $[\mathbb{Z}_2 \times \text{Ising}]_{4,1}^1$ | | † | 🌐 | |
| $[FR_1^{6,1,0}]_{4,1}^2$ | $[\mathbb{Z}_2 \times \text{Ising}]_{4,1}^2$ | | | 🌐 | |
| $[FR_1^{6,1,0}]_{4,1}^3$ | $[\mathbb{Z}_2 \times \text{Ising}]_{4,1}^3$ | | | 🌐 | |
| $[FR_1^{6,1,0}]_{4,1}^4$ | $[\mathbb{Z}_2 \times \text{Ising}]_{4,1}^4$ | | | 🌐 | |
| $[FR_1^{6,1,0}]_{5,1}^1$ | $[\mathbb{Z}_2 \times \text{Ising}]_{5,1}^1$ | | † | 🌐 | |
| $[FR_1^{6,1,0}]_{5,1}^2$ | $[\mathbb{Z}_2 \times \text{Ising}]_{5,1}^2$ | | | 🌐 | |
| $[FR_1^{6,1,0}]_{5,1}^3$ | $[\mathbb{Z}_2 \times \text{Ising}]_{5,1}^3$ | | | 🌐 | |
| $[FR_1^{6,1,0}]_{6,1}^1$ | $[\mathbb{Z}_2 \times \text{Ising}]_{6,1}^1$ | | † | 🌐 | |
| $[FR_1^{6,1,0}]_{6,1}^2$ | $[\mathbb{Z}_2 \times \text{Ising}]_{6,1}^2$ | | | 🌐 | |
| $[FR_1^{6,1,0}]_{6,1}^3$ | $[\mathbb{Z}_2 \times \text{Ising}]_{6,1}^3$ | | | 🌐 | |
| $[FR_2^{6,1,0}]_{1,1}^1$ | $[\mathbb{Z}_2 \times \text{Rep}(D_3)]_{1,1}^1$ | ✖ | † | 🌐 | 🎗 |
| $[FR_2^{6,1,0}]_{1,1}^2$ | $[\mathbb{Z}_2 \times \text{Rep}(D_3)]_{1,1}^2$ | ✖ | | 🌐 | 🎗 |
| $[FR_2^{6,1,0}]_{1,2}^1$ | $[\mathbb{Z}_2 \times \text{Rep}(D_3)]_{1,2}^1$ | ✖ | † | 🌐 | 🎗 |
| $[FR_2^{6,1,0}]_{1,2}^2$ | $[\mathbb{Z}_2 \times \text{Rep}(D_3)]_{1,2}^2$ | ✖ | | 🌐 | 🎗 |
| $[FR_2^{6,1,0}]_{1,3}^1$ | $[\mathbb{Z}_2 \times \text{Rep}(D_3)]_{1,3}^1$ | ✖ | † | 🌐 | 🎗 |
| $[FR_2^{6,1,0}]_{1,3}^2$ | $[\mathbb{Z}_2 \times \text{Rep}(D_3)]_{1,3}^2$ | ✖ | | 🌐 | 🎗 |
| $[FR_2^{6,1,0}]_{1,4}^1$ | $[\mathbb{Z}_2 \times \text{Rep}(D_3)]_{1,4}^1$ | ✖ | † | 🌐 | 🎗 |
| $[FR_2^{6,1,0}]_{1,4}^2$ | $[\mathbb{Z}_2 \times \text{Rep}(D_3)]_{1,4}^2$ | ✖ | | 🌐 | 🎗 |
| $[FR_2^{6,1,0}]_{1,5}^1$ | $[\mathbb{Z}_2 \times \text{Rep}(D_3)]_{1,5}^1$ | ✖ | † | 🌐 | 🎗 |
| $[FR_2^{6,1,0}]_{1,5}^2$ | $[\mathbb{Z}_2 \times \text{Rep}(D_3)]_{1,5}^2$ | ✖ | | 🌐 | 🎗 |
| $[FR_2^{6,1,0}]_{1,6}^1$ | $[\mathbb{Z}_2 \times \text{Rep}(D_3)]_{1,6}^1$ | ✖ | † | 🌐 | 🎗 |
| $[FR_2^{6,1,0}]_{1,6}^2$ | $[\mathbb{Z}_2 \times \text{Rep}(D_3)]_{1,6}^2$ | ✖ | | 🌐 | 🎗 |
| $[FR_2^{6,1,0}]_{2,1}^1$ | $[\mathbb{Z}_2 \times \text{Rep}(D_3)]_{2,1}^1$ | ✖ | † | 🌐 | 🎗 |
| $[FR_2^{6,1,0}]_{2,1}^2$ | $[\mathbb{Z}_2 \times \text{Rep}(D_3)]_{2,1}^2$ | ✖ | | 🌐 | 🎗 |
| $[FR_2^{6,1,0}]_{2,2}^1$ | $[\mathbb{Z}_2 \times \text{Rep}(D_3)]_{2,2}^1$ | ✖ | † | 🌐 | 🎗 |
| $[FR_2^{6,1,0}]_{2,2}^2$ | $[\mathbb{Z}_2 \times \text{Rep}(D_3)]_{2,2}^2$ | ✖ | | 🌐 | 🎗 |
| $[FR_2^{6,1,0}]_{2,3}^1$ | $[\mathbb{Z}_2 \times \text{Rep}(D_3)]_{2,3}^1$ | ✖ | † | 🌐 | 🎗 |
| $[FR_2^{6,1,0}]_{2,3}^2$ | $[\mathbb{Z}_2 \times \text{Rep}(D_3)]_{2,3}^2$ | ✖ | | 🌐 | 🎗 |

Table 9.1: List of multiplicity-free fusion categories up to rank 7 (Continued)

| Formal Name | Common Name | | | | | |
|---|---|---|---|---|---|---|
| $[FR_2^{6,1,0}]_{2,4}^1$ | $[\mathbb{Z}_2 \times \text{Rep}(D_3)]_{2,4}^1$ | ✗ | † | 🌐 | 🎗 | |
| $[FR_2^{6,1,0}]_{2,4}^2$ | $[\mathbb{Z}_2 \times \text{Rep}(D_3)]_{2,4}^2$ | ✗ | | 🌐 | 🎗 | |
| $[FR_2^{6,1,0}]_{2,5}^1$ | $[\mathbb{Z}_2 \times \text{Rep}(D_3)]_{2,5}^1$ | ✗ | † | 🌐 | 🎗 | |
| $[FR_2^{6,1,0}]_{2,5}^2$ | $[\mathbb{Z}_2 \times \text{Rep}(D_3)]_{2,5}^2$ | ✗ | | 🌐 | 🎗 | |
| $[FR_2^{6,1,0}]_{2,6}^1$ | $[\mathbb{Z}_2 \times \text{Rep}(D_3)]_{2,6}^1$ | ✗ | † | 🌐 | 🎗 | |
| $[FR_2^{6,1,0}]_{2,6}^2$ | $[\mathbb{Z}_2 \times \text{Rep}(D_3)]_{2,6}^2$ | ✗ | | 🌐 | 🎗 | |
| $[FR_2^{6,1,0}]_{3,1}^1$ | $[\mathbb{Z}_2 \times \text{Rep}(D_3)]_{3,1}^1$ | | † | 🌐 | | |
| $[FR_2^{6,1,0}]_{3,1}^2$ | $[\mathbb{Z}_2 \times \text{Rep}(D_3)]_{3,1}^2$ | | | 🌐 | | |
| $[FR_2^{6,1,0}]_{4,1}^1$ | $[\mathbb{Z}_2 \times \text{Rep}(D_3)]_{4,1}^1$ | | † | 🌐 | | |
| $[FR_2^{6,1,0}]_{4,1}^2$ | $[\mathbb{Z}_2 \times \text{Rep}(D_3)]_{4,1}^2$ | | | 🌐 | | |
| $[FR_2^{6,1,0}]_{5,1}^1$ | $[\mathbb{Z}_2 \times \text{Rep}(D_3)]_{5,1}^1$ | | † | 🌐 | | |
| $[FR_2^{6,1,0}]_{5,1}^2$ | $[\mathbb{Z}_2 \times \text{Rep}(D_3)]_{5,1}^2$ | | | 🌐 | | |
| $[FR_2^{6,1,0}]_{6,1}^1$ | $[\mathbb{Z}_2 \times \text{Rep}(D_3)]_{6,1}^1$ | | † | 🌐 | | |
| $[FR_2^{6,1,0}]_{6,1}^2$ | $[\mathbb{Z}_2 \times \text{Rep}(D_3)]_{6,1}^2$ | | | 🌐 | | |
| $[FR_4^{6,1,0}]_{1,1}^1$ | $[\text{TriCritIsing}]_{1,1}^1$ | ✗ | † | 🌐 | 🎗 | ⭕ |
| $[FR_4^{6,1,0}]_{1,1}^2$ | $[\text{TriCritIsing}]_{1,1}^2$ | ✗ | | 🌐 | 🎗 | ⭕ |
| $[FR_4^{6,1,0}]_{1,2}^1$ | $[\text{TriCritIsing}]_{1,2}^1$ | ✗ | † | 🌐 | 🎗 | ⭕ |
| $[FR_4^{6,1,0}]_{1,2}^2$ | $[\text{TriCritIsing}]_{1,2}^2$ | ✗ | | 🌐 | 🎗 | ⭕ |
| $[FR_4^{6,1,0}]_{1,3}^1$ | $[\text{TriCritIsing}]_{1,3}^1$ | ✗ | † | 🌐 | 🎗 | ⭕ |
| $[FR_4^{6,1,0}]_{1,3}^2$ | $[\text{TriCritIsing}]_{1,3}^2$ | ✗ | | 🌐 | 🎗 | ⭕ |
| $[FR_4^{6,1,0}]_{1,4}^1$ | $[\text{TriCritIsing}]_{1,4}^1$ | ✗ | † | 🌐 | 🎗 | ⭕ |
| $[FR_4^{6,1,0}]_{1,4}^2$ | $[\text{TriCritIsing}]_{1,4}^2$ | ✗ | | 🌐 | 🎗 | ⭕ |
| $[FR_4^{6,1,0}]_{1,5}^1$ | $[\text{TriCritIsing}]_{1,5}^1$ | ✗ | † | 🌐 | 🎗 | ⭕ |
| $[FR_4^{6,1,0}]_{1,5}^2$ | $[\text{TriCritIsing}]_{1,5}^2$ | ✗ | | 🌐 | 🎗 | ⭕ |
| $[FR_4^{6,1,0}]_{1,6}^1$ | $[\text{TriCritIsing}]_{1,6}^1$ | ✗ | † | 🌐 | 🎗 | ⭕ |
| $[FR_4^{6,1,0}]_{1,6}^2$ | $[\text{TriCritIsing}]_{1,6}^2$ | ✗ | | 🌐 | 🎗 | ⭕ |
| $[FR_4^{6,1,0}]_{1,7}^1$ | $[\text{TriCritIsing}]_{1,7}^1$ | ✗ | † | 🌐 | 🎗 | ⭕ |
| $[FR_4^{6,1,0}]_{1,7}^2$ | $[\text{TriCritIsing}]_{1,7}^2$ | ✗ | | 🌐 | 🎗 | ⭕ |
| $[FR_4^{6,1,0}]_{1,8}^1$ | $[\text{TriCritIsing}]_{1,8}^1$ | ✗ | † | 🌐 | 🎗 | ⭕ |

Table 9.1: List of multiplicity-free fusion categories up to rank 7 (Continued)

| Formal Name | Common Name | | | | | |
|---|---|---|---|---|---|---|
| $[FR_4^{6,1,0}]_{1,8}^2$ | $[TriCritIsing]_{1,8}^2$ | ✕ | | 🌐 | 🎗 | ⭕ |
| $[FR_4^{6,1,0}]_{2,1}^1$ | $[TriCritIsing]_{2,1}^1$ | ✕ | † | 🌐 | 🎗 | ⭕ |
| $[FR_4^{6,1,0}]_{2,1}^2$ | $[TriCritIsing]_{2,1}^2$ | ✕ | | 🌐 | 🎗 | ⭕ |
| $[FR_4^{6,1,0}]_{2,2}^1$ | $[TriCritIsing]_{2,2}^1$ | ✕ | † | 🌐 | 🎗 | ⭕ |
| $[FR_4^{6,1,0}]_{2,2}^2$ | $[TriCritIsing]_{2,2}^2$ | ✕ | | 🌐 | 🎗 | ⭕ |
| $[FR_4^{6,1,0}]_{2,3}^1$ | $[TriCritIsing]_{2,3}^1$ | ✕ | † | 🌐 | 🎗 | ⭕ |
| $[FR_4^{6,1,0}]_{2,3}^2$ | $[TriCritIsing]_{2,3}^2$ | ✕ | | 🌐 | 🎗 | ⭕ |
| $[FR_4^{6,1,0}]_{2,4}^1$ | $[TriCritIsing]_{2,4}^1$ | ✕ | † | 🌐 | 🎗 | ⭕ |
| $[FR_4^{6,1,0}]_{2,4}^2$ | $[TriCritIsing]_{2,4}^2$ | ✕ | | 🌐 | 🎗 | ⭕ |
| $[FR_4^{6,1,0}]_{2,5}^1$ | $[TriCritIsing]_{2,5}^1$ | ✕ | † | 🌐 | 🎗 | ⭕ |
| $[FR_4^{6,1,0}]_{2,5}^2$ | $[TriCritIsing]_{2,5}^2$ | ✕ | | 🌐 | 🎗 | ⭕ |
| $[FR_4^{6,1,0}]_{2,6}^1$ | $[TriCritIsing]_{2,6}^1$ | ✕ | † | 🌐 | 🎗 | ⭕ |
| $[FR_4^{6,1,0}]_{2,6}^2$ | $[TriCritIsing]_{2,6}^2$ | ✕ | | 🌐 | 🎗 | ⭕ |
| $[FR_4^{6,1,0}]_{2,7}^1$ | $[TriCritIsing]_{2,7}^1$ | ✕ | † | 🌐 | 🎗 | ⭕ |
| $[FR_4^{6,1,0}]_{2,7}^2$ | $[TriCritIsing]_{2,7}^2$ | ✕ | | 🌐 | 🎗 | ⭕ |
| $[FR_4^{6,1,0}]_{2,8}^1$ | $[TriCritIsing]_{2,8}^1$ | ✕ | † | 🌐 | 🎗 | ⭕ |
| $[FR_4^{6,1,0}]_{2,8}^2$ | $[TriCritIsing]_{2,8}^2$ | ✕ | | 🌐 | 🎗 | ⭕ |
| $[FR_4^{6,1,0}]_{3,1}^1$ | $[TriCritIsing]_{3,1}^1$ | ✕ | | 🌐 | 🎗 | ⭕ |
| $[FR_4^{6,1,0}]_{3,1}^2$ | $[TriCritIsing]_{3,1}^2$ | ✕ | | 🌐 | 🎗 | ⭕ |
| $[FR_4^{6,1,0}]_{3,2}^1$ | $[TriCritIsing]_{3,2}^1$ | ✕ | | 🌐 | 🎗 | ⭕ |
| $[FR_4^{6,1,0}]_{3,2}^2$ | $[TriCritIsing]_{3,2}^2$ | ✕ | | 🌐 | 🎗 | ⭕ |
| $[FR_4^{6,1,0}]_{3,3}^1$ | $[TriCritIsing]_{3,3}^1$ | ✕ | | 🌐 | 🎗 | ⭕ |
| $[FR_4^{6,1,0}]_{3,3}^2$ | $[TriCritIsing]_{3,3}^2$ | ✕ | | 🌐 | 🎗 | ⭕ |
| $[FR_4^{6,1,0}]_{3,4}^1$ | $[TriCritIsing]_{3,4}^1$ | ✕ | | 🌐 | 🎗 | ⭕ |
| $[FR_4^{6,1,0}]_{3,4}^2$ | $[TriCritIsing]_{3,4}^2$ | ✕ | | 🌐 | 🎗 | ⭕ |
| $[FR_4^{6,1,0}]_{3,5}^1$ | $[TriCritIsing]_{3,5}^1$ | ✕ | | 🌐 | 🎗 | ⭕ |
| $[FR_4^{6,1,0}]_{3,5}^2$ | $[TriCritIsing]_{3,5}^2$ | ✕ | | 🌐 | 🎗 | ⭕ |
| $[FR_4^{6,1,0}]_{3,6}^1$ | $[TriCritIsing]_{3,6}^1$ | ✕ | | 🌐 | 🎗 | ⭕ |
| $[FR_4^{6,1,0}]_{3,6}^2$ | $[TriCritIsing]_{3,6}^2$ | ✕ | | 🌐 | 🎗 | ⭕ |

Table 9.1: List of multiplicity-free fusion categories up to rank 7 (Continued)

| Formal Name | Common Name | | | | |
|---|---|---|---|---|---|
| $[FR_4^{6,1,0}]_{3,7}^1$ | $[TriCritIsing]_{3,7}^1$ | ✕ | 🌐 | 🎗 | ⭕ |
| $[FR_4^{6,1,0}]_{3,7}^2$ | $[TriCritIsing]_{3,7}^2$ | ✕ | 🌐 | 🎗 | ⭕ |
| $[FR_4^{6,1,0}]_{3,8}^1$ | $[TriCritIsing]_{3,8}^1$ | ✕ | 🌐 | 🎗 | ⭕ |
| $[FR_4^{6,1,0}]_{3,8}^2$ | $[TriCritIsing]_{3,8}^2$ | ✕ | 🌐 | 🎗 | ⭕ |
| $[FR_4^{6,1,0}]_{4,1}^1$ | $[TriCritIsing]_{4,1}^1$ | ✕ | 🌐 | 🎗 | ⭕ |
| $[FR_4^{6,1,0}]_{4,1}^2$ | $[TriCritIsing]_{4,1}^2$ | ✕ | 🌐 | 🎗 | ⭕ |
| $[FR_4^{6,1,0}]_{4,2}^1$ | $[TriCritIsing]_{4,2}^1$ | ✕ | 🌐 | 🎗 | ⭕ |
| $[FR_4^{6,1,0}]_{4,2}^2$ | $[TriCritIsing]_{4,2}^2$ | ✕ | 🌐 | 🎗 | ⭕ |
| $[FR_4^{6,1,0}]_{4,3}^1$ | $[TriCritIsing]_{4,3}^1$ | ✕ | 🌐 | 🎗 | ⭕ |
| $[FR_4^{6,1,0}]_{4,3}^2$ | $[TriCritIsing]_{4,3}^2$ | ✕ | 🌐 | 🎗 | ⭕ |
| $[FR_4^{6,1,0}]_{4,4}^1$ | $[TriCritIsing]_{4,4}^1$ | ✕ | 🌐 | 🎗 | ⭕ |
| $[FR_4^{6,1,0}]_{4,4}^2$ | $[TriCritIsing]_{4,4}^2$ | ✕ | 🌐 | 🎗 | ⭕ |
| $[FR_4^{6,1,0}]_{4,5}^1$ | $[TriCritIsing]_{4,5}^1$ | ✕ | 🌐 | 🎗 | ⭕ |
| $[FR_4^{6,1,0}]_{4,5}^2$ | $[TriCritIsing]_{4,5}^2$ | ✕ | 🌐 | 🎗 | ⭕ |
| $[FR_4^{6,1,0}]_{4,6}^1$ | $[TriCritIsing]_{4,6}^1$ | ✕ | 🌐 | 🎗 | ⭕ |
| $[FR_4^{6,1,0}]_{4,6}^2$ | $[TriCritIsing]_{4,6}^2$ | ✕ | 🌐 | 🎗 | ⭕ |
| $[FR_4^{6,1,0}]_{4,7}^1$ | $[TriCritIsing]_{4,7}^1$ | ✕ | 🌐 | 🎗 | ⭕ |
| $[FR_4^{6,1,0}]_{4,7}^2$ | $[TriCritIsing]_{4,7}^2$ | ✕ | 🌐 | 🎗 | ⭕ |
| $[FR_4^{6,1,0}]_{4,8}^1$ | $[TriCritIsing]_{4,8}^1$ | ✕ | 🌐 | 🎗 | ⭕ |
| $[FR_4^{6,1,0}]_{4,8}^2$ | $[TriCritIsing]_{4,8}^2$ | ✕ | 🌐 | 🎗 | ⭕ |
| $[FR_5^{6,1,0}]_{1,1}^1$ | $[Fib \times Rep(D_3)]_{1,1}^1$ | ✕ | 🗡 | 🌐 | 🎗 |
| $[FR_5^{6,1,0}]_{1,2}^1$ | $[Fib \times Rep(D_3)]_{1,2}^1$ | ✕ | 🗡 | 🌐 | 🎗 |
| $[FR_5^{6,1,0}]_{1,3}^1$ | $[Fib \times Rep(D_3)]_{1,3}^1$ | ✕ | 🗡 | 🌐 | 🎗 |
| $[FR_5^{6,1,0}]_{1,4}^1$ | $[Fib \times Rep(D_3)]_{1,4}^1$ | ✕ | 🗡 | 🌐 | 🎗 |
| $[FR_5^{6,1,0}]_{1,5}^1$ | $[Fib \times Rep(D_3)]_{1,5}^1$ | ✕ | 🗡 | 🌐 | 🎗 |
| $[FR_5^{6,1,0}]_{1,6}^1$ | $[Fib \times Rep(D_3)]_{1,6}^1$ | ✕ | 🗡 | 🌐 | 🎗 |
| $[FR_5^{6,1,0}]_{2,1}^1$ | $[Fib \times Rep(D_3)]_{2,1}^1$ | ✕ | | 🌐 | 🎗 |
| $[FR_5^{6,1,0}]_{2,2}^1$ | $[Fib \times Rep(D_3)]_{2,2}^1$ | ✕ | | 🌐 | 🎗 |
| $[FR_5^{6,1,0}]_{2,3}^1$ | $[Fib \times Rep(D_3)]_{2,3}^1$ | ✕ | | 🌐 | 🎗 |

Table 9.1: List of multiplicity-free fusion categories up to rank 7 (Continued)

| Formal Name | Common Name | | | | | | |
|---|---|---|---|---|---|---|---|
| $[FR_5^{6,1,0}]_{2,4}^1$ | $[\text{Fib} \times \text{Rep}(D_3)]_{2,4}^1$ | ✕ | | 🌐 | 🎗 | | |
| $[FR_5^{6,1,0}]_{2,5}^1$ | $[\text{Fib} \times \text{Rep}(D_3)]_{2,5}^1$ | ✕ | | 🌐 | 🎗 | | |
| $[FR_5^{6,1,0}]_{2,6}^1$ | $[\text{Fib} \times \text{Rep}(D_3)]_{2,6}^1$ | ✕ | | 🌐 | 🎗 | | |
| $[FR_6^{6,1,0}]_{1,1}^1$ | $[SU(2)_5]_{1,1}^1$ | ✕ | † | 🌐 | 🎗 | ⊚ | |
| $[FR_6^{6,1,0}]_{1,1}^2$ | $[SU(2)_5]_{1,1}^2$ | ✕ | | 🌐 | 🎗 | ⊚ | |
| $[FR_6^{6,1,0}]_{1,2}^1$ | $[SU(2)_5]_{1,2}^1$ | ✕ | † | 🌐 | 🎗 | ⊚ | |
| $[FR_6^{6,1,0}]_{1,2}^2$ | $[SU(2)_5]_{1,2}^2$ | ✕ | | 🌐 | 🎗 | ⊚ | |
| $[FR_6^{6,1,0}]_{1,3}^1$ | $[SU(2)_5]_{1,3}^1$ | ✕ | † | 🌐 | 🎗 | ⊚ | |
| $[FR_6^{6,1,0}]_{1,3}^2$ | $[SU(2)_5]_{1,3}^2$ | ✕ | | 🌐 | 🎗 | ⊚ | |
| $[FR_6^{6,1,0}]_{1,4}^1$ | $[SU(2)_5]_{1,4}^1$ | ✕ | † | 🌐 | 🎗 | ⊚ | |
| $[FR_6^{6,1,0}]_{1,4}^2$ | $[SU(2)_5]_{1,4}^2$ | ✕ | | 🌐 | 🎗 | ⊚ | |
| $[FR_6^{6,1,0}]_{2,1}^1$ | $[SU(2)_5]_{2,1}^1$ | ✕ | † | 🌐 | 🎗 | | |
| $[FR_6^{6,1,0}]_{2,1}^2$ | $[SU(2)_5]_{2,1}^2$ | ✕ | | 🌐 | 🎗 | | |
| $[FR_6^{6,1,0}]_{2,2}^1$ | $[SU(2)_5]_{2,2}^1$ | ✕ | † | 🌐 | 🎗 | | |
| $[FR_6^{6,1,0}]_{2,2}^2$ | $[SU(2)_5]_{2,2}^2$ | ✕ | | 🌐 | 🎗 | | |
| $[FR_6^{6,1,0}]_{2,3}^1$ | $[SU(2)_5]_{2,3}^1$ | ✕ | † | 🌐 | 🎗 | | |
| $[FR_6^{6,1,0}]_{2,3}^2$ | $[SU(2)_5]_{2,3}^2$ | ✕ | | 🌐 | 🎗 | | |
| $[FR_6^{6,1,0}]_{2,4}^1$ | $[SU(2)_5]_{2,4}^1$ | ✕ | † | 🌐 | 🎗 | | |
| $[FR_6^{6,1,0}]_{2,4}^2$ | $[SU(2)_5]_{2,4}^2$ | ✕ | | 🌐 | 🎗 | | |
| $[FR_6^{6,1,0}]_{3,1}^1$ | $[SU(2)_5]_{3,1}^1$ | ✕ | | 🌐 | 🎗 | ⊚ | |
| $[FR_6^{6,1,0}]_{3,1}^2$ | $[SU(2)_5]_{3,1}^2$ | ✕ | | 🌐 | 🎗 | ⊚ | |
| $[FR_6^{6,1,0}]_{3,2}^1$ | $[SU(2)_5]_{3,2}^1$ | ✕ | | 🌐 | 🎗 | ⊚ | |
| $[FR_6^{6,1,0}]_{3,2}^2$ | $[SU(2)_5]_{3,2}^2$ | ✕ | | 🌐 | 🎗 | ⊚ | |
| $[FR_6^{6,1,0}]_{3,3}^1$ | $[SU(2)_5]_{3,3}^1$ | ✕ | | 🌐 | 🎗 | ⊚ | |
| $[FR_6^{6,1,0}]_{3,3}^2$ | $[SU(2)_5]_{3,3}^2$ | ✕ | | 🌐 | 🎗 | ⊚ | |
| $[FR_6^{6,1,0}]_{3,4}^1$ | $[SU(2)_5]_{3,4}^1$ | ✕ | | 🌐 | 🎗 | ⊚ | |
| $[FR_6^{6,1,0}]_{3,4}^2$ | $[SU(2)_5]_{3,4}^2$ | ✕ | | 🌐 | 🎗 | ⊚ | |
| $[FR_6^{6,1,0}]_{4,1}^1$ | $[SU(2)_5]_{4,1}^1$ | ✕ | | 🌐 | 🎗 | ⊚ | |
| $[FR_6^{6,1,0}]_{4,1}^2$ | $[SU(2)_5]_{4,1}^2$ | ✕ | | 🌐 | 🎗 | ⊚ | |

Table 9.1: List of multiplicity-free fusion categories up to rank 7 (Continued)

| Formal Name | Common Name | | | | |
|---|---|---|---|---|---|
| $[FR_6^{6,1,0}]_{4,2}^1$ | $[SU(2)_5]_{4,2}^1$ | ✕ | 🌐 | 🎗 | ⭕⭕ |
| $[FR_6^{6,1,0}]_{4,2}^2$ | $[SU(2)_5]_{4,2}^2$ | ✕ | 🌐 | 🎗 | ⭕⭕ |
| $[FR_6^{6,1,0}]_{4,3}^1$ | $[SU(2)_5]_{4,3}^1$ | ✕ | 🌐 | 🎗 | ⭕⭕ |
| $[FR_6^{6,1,0}]_{4,3}^2$ | $[SU(2)_5]_{4,3}^2$ | ✕ | 🌐 | 🎗 | ⭕⭕ |
| $[FR_6^{6,1,0}]_{4,4}^1$ | $[SU(2)_5]_{4,4}^1$ | ✕ | 🌐 | 🎗 | ⭕⭕ |
| $[FR_6^{6,1,0}]_{4,4}^2$ | $[SU(2)_5]_{4,4}^2$ | ✕ | 🌐 | 🎗 | ⭕⭕ |
| $[FR_6^{6,1,0}]_{5,1}^1$ | $[SU(2)_5]_{5,1}^1$ | ✕ | 🌐 | 🎗 | |
| $[FR_6^{6,1,0}]_{5,1}^2$ | $[SU(2)_5]_{5,1}^2$ | ✕ | 🌐 | 🎗 | |
| $[FR_6^{6,1,0}]_{5,2}^1$ | $[SU(2)_5]_{5,2}^1$ | ✕ | 🌐 | 🎗 | |
| $[FR_6^{6,1,0}]_{5,2}^2$ | $[SU(2)_5]_{5,2}^2$ | ✕ | 🌐 | 🎗 | |
| $[FR_6^{6,1,0}]_{5,3}^1$ | $[SU(2)_5]_{5,3}^1$ | ✕ | 🌐 | 🎗 | |
| $[FR_6^{6,1,0}]_{5,3}^2$ | $[SU(2)_5]_{5,3}^2$ | ✕ | 🌐 | 🎗 | |
| $[FR_6^{6,1,0}]_{5,4}^1$ | $[SU(2)_5]_{5,4}^1$ | ✕ | 🌐 | 🎗 | |
| $[FR_6^{6,1,0}]_{5,4}^2$ | $[SU(2)_5]_{5,4}^2$ | ✕ | 🌐 | 🎗 | |
| $[FR_6^{6,1,0}]_{6,1}^1$ | $[SU(2)_5]_{6,1}^1$ | ✕ | 🌐 | 🎗 | |
| $[FR_6^{6,1,0}]_{6,1}^2$ | $[SU(2)_5]_{6,1}^2$ | ✕ | 🌐 | 🎗 | |
| $[FR_6^{6,1,0}]_{6,2}^1$ | $[SU(2)_5]_{6,2}^1$ | ✕ | 🌐 | 🎗 | |
| $[FR_6^{6,1,0}]_{6,2}^2$ | $[SU(2)_5]_{6,2}^2$ | ✕ | 🌐 | 🎗 | |
| $[FR_6^{6,1,0}]_{6,3}^1$ | $[SU(2)_5]_{6,3}^1$ | ✕ | 🌐 | 🎗 | |
| $[FR_6^{6,1,0}]_{6,3}^2$ | $[SU(2)_5]_{6,3}^2$ | ✕ | 🌐 | 🎗 | |
| $[FR_6^{6,1,0}]_{6,4}^1$ | $[SU(2)_5]_{6,4}^1$ | ✕ | 🌐 | 🎗 | |
| $[FR_6^{6,1,0}]_{6,4}^2$ | $[SU(2)_5]_{6,4}^2$ | ✕ | 🌐 | 🎗 | |
| $[FR_7^{6,1,0}]_{1,1}^1$ | $[\mathrm{Rep}(\mathbb{Z}_3 \rtimes D_3)]_{1,1}^1$ | ✕ | † 🌐 | 🎗 | |
| $[FR_7^{6,1,0}]_{1,2}^1$ | $[\mathrm{Rep}(\mathbb{Z}_3 \rtimes D_3)]_{1,2}^1$ | ✕ | † 🌐 | 🎗 | |
| $[FR_7^{6,1,0}]_{1,3}^1$ | $[\mathrm{Rep}(\mathbb{Z}_3 \rtimes D_3)]_{1,3}^1$ | ✕ | † 🌐 | 🎗 | |
| $[FR_7^{6,1,0}]_{1,4}^1$ | $[\mathrm{Rep}(\mathbb{Z}_3 \rtimes D_3)]_{1,4}^1$ | ✕ | † 🌐 | 🎗 | |
| $[FR_7^{6,1,0}]_{1,5}^1$ | $[\mathrm{Rep}(\mathbb{Z}_3 \rtimes D_3)]_{1,5}^1$ | ✕ | † 🌐 | 🎗 | |
| $[FR_7^{6,1,0}]_{2,1}^1$ | $[\mathrm{Rep}(\mathbb{Z}_3 \rtimes D_3)]_{2,1}^1$ | | † 🌐 | | |
| $[FR_7^{6,1,0}]_{3,1}^1$ | $[\mathrm{Rep}(\mathbb{Z}_3 \rtimes D_3)]_{3,1}^1$ | | † 🌐 | | |

Table 9.1: List of multiplicity-free fusion categories up to rank 7 (Continued)

| Formal Name | Common Name | | | | | |
|---|---|---|---|---|---|---|
| $[FR_7^{6,1,0}]_{4,1}^1$ | $[\text{Rep}(\mathbb{Z}_3 \rtimes D_3)]_{4,1}^1$ | | † | 🌐 | | |
| $[FR_7^{6,1,0}]_{5,1}^1$ | $[\text{Rep}(\mathbb{Z}_3 \rtimes D_3)]_{5,1}^1$ | | † | 🌐 | | |
| $[FR_8^{6,1,0}]_{1,1}^1$ | $[\text{Rep}(D_9)]_{1,1}^1$ | ✕ | † | 🌐 | 🎗 | |
| $[FR_8^{6,1,0}]_{1,2}^1$ | $[\text{Rep}(D_9)]_{1,2}^1$ | ✕ | † | 🌐 | 🎗 | |
| $[FR_8^{6,1,0}]_{1,3}^1$ | $[\text{Rep}(D_9)]_{1,3}^1$ | ✕ | † | 🌐 | 🎗 | |
| $[FR_8^{6,1,0}]_{1,4}^1$ | $[\text{Rep}(D_9)]_{1,4}^1$ | ✕ | † | 🌐 | 🎗 | |
| $[FR_8^{6,1,0}]_{1,5}^1$ | $[\text{Rep}(D_9)]_{1,5}^1$ | ✕ | † | 🌐 | 🎗 | |
| $[FR_8^{6,1,0}]_{2,1}^1$ | $[\text{Rep}(D_9)]_{2,1}^1$ | | † | 🌐 | | |
| $[FR_8^{6,1,0}]_{3,1}^1$ | $[\text{Rep}(D_9)]_{3,1}^1$ | | † | 🌐 | | |
| $[FR_8^{6,1,0}]_{4,1}^1$ | $[\text{Rep}(D_9)]_{4,1}^1$ | | † | 🌐 | | |
| $[FR_8^{6,1,0}]_{5,1}^1$ | $[\text{Rep}(D_9)]_{5,1}^1$ | | † | 🌐 | | |
| $[FR_9^{6,1,0}]_{1,1}^1$ | $[SO(5)_2]_{1,1}^1$ | ✕ | † | 🌐 | 🎗 | ⚭ |
| $[FR_9^{6,1,0}]_{1,1}^2$ | $[SO(5)_2]_{1,1}^2$ | ✕ | | 🌐 | 🎗 | ⚭ |
| $[FR_9^{6,1,0}]_{2,1}^1$ | $[SO(5)_2]_{2,1}^1$ | ✕ | † | 🌐 | 🎗 | ⚭ |
| $[FR_9^{6,1,0}]_{2,1}^2$ | $[SO(5)_2]_{2,1}^2$ | ✕ | | 🌐 | 🎗 | ⚭ |
| $[FR_9^{6,1,0}]_{3,1}^1$ | $[SO(5)_2]_{3,1}^1$ | ✕ | † | 🌐 | 🎗 | ⚭ |
| $[FR_9^{6,1,0}]_{3,1}^2$ | $[SO(5)_2]_{3,1}^2$ | ✕ | | 🌐 | 🎗 | ⚭ |
| $[FR_9^{6,1,0}]_{4,1}^1$ | $[SO(5)_2]_{4,1}^1$ | ✕ | † | 🌐 | 🎗 | ⚭ |
| $[FR_9^{6,1,0}]_{4,1}^2$ | $[SO(5)_2]_{4,1}^2$ | ✕ | | 🌐 | 🎗 | ⚭ |
| $[FR_{14}^{6,1,0}]_{1,1}^1$ | $[\text{Fib} \times PSU(2)_5]_{1,1}^1$ | ✕ | † | 🌐 | 🎗 | ⚭ |
| $[FR_{14}^{6,1,0}]_{1,2}^1$ | $[\text{Fib} \times PSU(2)_5]_{1,2}^1$ | ✕ | † | 🌐 | 🎗 | ⚭ |
| $[FR_{14}^{6,1,0}]_{1,3}^1$ | $[\text{Fib} \times PSU(2)_5]_{1,3}^1$ | ✕ | † | 🌐 | 🎗 | ⚭ |
| $[FR_{14}^{6,1,0}]_{1,4}^1$ | $[\text{Fib} \times PSU(2)_5]_{1,4}^1$ | ✕ | † | 🌐 | 🎗 | ⚭ |
| $[FR_{14}^{6,1,0}]_{2,1}^1$ | $[\text{Fib} \times PSU(2)_5]_{2,1}^1$ | ✕ | | 🌐 | 🎗 | ⚭ |
| $[FR_{14}^{6,1,0}]_{2,2}^1$ | $[\text{Fib} \times PSU(2)_5]_{2,2}^1$ | ✕ | | 🌐 | 🎗 | ⚭ |
| $[FR_{14}^{6,1,0}]_{2,3}^1$ | $[\text{Fib} \times PSU(2)_5]_{2,3}^1$ | ✕ | | 🌐 | 🎗 | ⚭ |
| $[FR_{14}^{6,1,0}]_{2,4}^1$ | $[\text{Fib} \times PSU(2)_5]_{2,4}^1$ | ✕ | | 🌐 | 🎗 | ⚭ |
| $[FR_{14}^{6,1,0}]_{3,1}^1$ | $[\text{Fib} \times PSU(2)_5]_{3,1}^1$ | ✕ | | 🌐 | 🎗 | ⚭ |

Table 9.1: List of multiplicity-free fusion categories up to rank 7 (Continued)

| Formal Name | Common Name | | | | | |
|---|---|---|---|---|---|---|
| $[\mathrm{FR}_{14}^{6,1,0}]_{3,2}^{1}$ | $[\mathrm{Fib} \times \mathrm{PSU}(2)_5]_{3,2}^{1}$ | 🧬 | | 🌐 | 🎗 | ⭕ |
| $[\mathrm{FR}_{14}^{6,1,0}]_{3,3}^{1}$ | $[\mathrm{Fib} \times \mathrm{PSU}(2)_5]_{3,3}^{1}$ | 🧬 | | 🌐 | 🎗 | ⭕ |
| $[\mathrm{FR}_{14}^{6,1,0}]_{3,4}^{1}$ | $[\mathrm{Fib} \times \mathrm{PSU}(2)_5]_{3,4}^{1}$ | 🧬 | | 🌐 | 🎗 | ⭕ |
| $[\mathrm{FR}_{14}^{6,1,0}]_{4,1}^{1}$ | $[\mathrm{Fib} \times \mathrm{PSU}(2)_5]_{4,1}^{1}$ | 🧬 | | 🌐 | 🎗 | ⭕ |
| $[\mathrm{FR}_{14}^{6,1,0}]_{4,2}^{1}$ | $[\mathrm{Fib} \times \mathrm{PSU}(2)_5]_{4,2}^{1}$ | 🧬 | | 🌐 | 🎗 | ⭕ |
| $[\mathrm{FR}_{14}^{6,1,0}]_{4,3}^{1}$ | $[\mathrm{Fib} \times \mathrm{PSU}(2)_5]_{4,3}^{1}$ | 🧬 | | 🌐 | 🎗 | ⭕ |
| $[\mathrm{FR}_{14}^{6,1,0}]_{4,4}^{1}$ | $[\mathrm{Fib} \times \mathrm{PSU}(2)_5]_{4,4}^{1}$ | 🧬 | | 🌐 | 🎗 | ⭕ |
| $[\mathrm{FR}_{14}^{6,1,0}]_{5,1}^{1}$ | $[\mathrm{Fib} \times \mathrm{PSU}(2)_5]_{5,1}^{1}$ | 🧬 | | 🌐 | 🎗 | ⭕ |
| $[\mathrm{FR}_{14}^{6,1,0}]_{5,2}^{1}$ | $[\mathrm{Fib} \times \mathrm{PSU}(2)_5]_{5,2}^{1}$ | 🧬 | | 🌐 | 🎗 | ⭕ |
| $[\mathrm{FR}_{14}^{6,1,0}]_{5,3}^{1}$ | $[\mathrm{Fib} \times \mathrm{PSU}(2)_5]_{5,3}^{1}$ | 🧬 | | 🌐 | 🎗 | ⭕ |
| $[\mathrm{FR}_{14}^{6,1,0}]_{5,4}^{1}$ | $[\mathrm{Fib} \times \mathrm{PSU}(2)_5]_{5,4}^{1}$ | 🧬 | | 🌐 | 🎗 | ⭕ |
| $[\mathrm{FR}_{14}^{6,1,0}]_{6,1}^{1}$ | $[\mathrm{Fib} \times \mathrm{PSU}(2)_5]_{6,1}^{1}$ | 🧬 | | 🌐 | 🎗 | ⭕ |
| $[\mathrm{FR}_{14}^{6,1,0}]_{6,2}^{1}$ | $[\mathrm{Fib} \times \mathrm{PSU}(2)_5]_{6,2}^{1}$ | 🧬 | | 🌐 | 🎗 | ⭕ |
| $[\mathrm{FR}_{14}^{6,1,0}]_{6,3}^{1}$ | $[\mathrm{Fib} \times \mathrm{PSU}(2)_5]_{6,3}^{1}$ | 🧬 | | 🌐 | 🎗 | ⭕ |
| $[\mathrm{FR}_{14}^{6,1,0}]_{6,4}^{1}$ | $[\mathrm{Fib} \times \mathrm{PSU}(2)_5]_{6,4}^{1}$ | 🧬 | | 🌐 | 🎗 | ⭕ |
| $[\mathrm{FR}_{16}^{6,1,0}]_{1,1}^{1}$ | $[\mathrm{PSU}(2)_{10}]_{1,1}^{1}$ | 🧬 | 🗡 | 🌐 | 🎗 | |
| $[\mathrm{FR}_{16}^{6,1,0}]_{1,2}^{1}$ | $[\mathrm{PSU}(2)_{10}]_{1,2}^{1}$ | 🧬 | 🗡 | 🌐 | 🎗 | |
| $[\mathrm{FR}_{16}^{6,1,0}]_{2,1}^{1}$ | $[\mathrm{PSU}(2)_{10}]_{2,1}^{1}$ | 🧬 | | 🌐 | 🎗 | |
| $[\mathrm{FR}_{16}^{6,1,0}]_{2,2}^{1}$ | $[\mathrm{PSU}(2)_{10}]_{2,2}^{1}$ | 🧬 | | 🌐 | 🎗 | |
| $[\mathrm{FR}_{18}^{6,1,0}]_{1,1}^{1}$ | $[\mathrm{PSU}(2)_{11}]_{1,1}^{1}$ | 🧬 | 🗡 | 🌐 | 🎗 | ⭕ |
| $[\mathrm{FR}_{18}^{6,1,0}]_{1,2}^{1}$ | $[\mathrm{PSU}(2)_{11}]_{1,2}^{1}$ | 🧬 | 🗡 | 🌐 | 🎗 | ⭕ |
| $[\mathrm{FR}_{18}^{6,1,0}]_{2,1}^{1}$ | $[\mathrm{PSU}(2)_{11}]_{2,1}^{1}$ | 🧬 | | 🌐 | 🎗 | ⭕ |
| $[\mathrm{FR}_{18}^{6,1,0}]_{2,2}^{1}$ | $[\mathrm{PSU}(2)_{11}]_{2,2}^{1}$ | 🧬 | | 🌐 | 🎗 | ⭕ |
| $[\mathrm{FR}_{18}^{6,1,0}]_{3,1}^{1}$ | $[\mathrm{PSU}(2)_{11}]_{3,1}^{1}$ | 🧬 | | 🌐 | 🎗 | ⭕ |
| $[\mathrm{FR}_{18}^{6,1,0}]_{3,2}^{1}$ | $[\mathrm{PSU}(2)_{11}]_{3,2}^{1}$ | 🧬 | | 🌐 | 🎗 | ⭕ |
| $[\mathrm{FR}_{18}^{6,1,0}]_{4,1}^{1}$ | $[\mathrm{PSU}(2)_{11}]_{4,1}^{1}$ | 🧬 | | 🌐 | 🎗 | ⭕ |
| $[\mathrm{FR}_{18}^{6,1,0}]_{4,2}^{1}$ | $[\mathrm{PSU}(2)_{11}]_{4,2}^{1}$ | 🧬 | | 🌐 | 🎗 | ⭕ |
| $[\mathrm{FR}_{18}^{6,1,0}]_{5,1}^{1}$ | $[\mathrm{PSU}(2)_{11}]_{5,1}^{1}$ | 🧬 | | 🌐 | 🎗 | ⭕ |

Table 9.1: List of multiplicity-free fusion categories up to rank 7 (Continued)

| Formal Name | Common Name | | | | |
|---|---|---|---|---|---|
| $[FR_{18}^{6,1,0}]_{5,2}^1$ | $[PSU(2)_{11}]_{5,2}^1$ | ✕ | 🌐 | 🎀 | ⭕ |
| $[FR_{18}^{6,1,0}]_{6,1}^1$ | $[PSU(2)_{11}]_{6,1}^1$ | ✕ | 🌐 | 🎀 | ⭕ |
| $[FR_{18}^{6,1,0}]_{6,2}^1$ | $[PSU(2)_{11}]_{6,2}^1$ | ✕ | 🌐 | 🎀 | ⭕ |
| $[FR_1^{6,1,2}]_{1,1}^1$ | $[D_3]_{1,1}^1$ | † | 🌐 | | |
| $[FR_1^{6,1,2}]_{1,1}^2$ | $[D_3]_{1,1}^2$ | | 🌐 | | |
| $[FR_1^{6,1,2}]_{2,1}^1$ | $[D_3]_{2,1}^1$ | † | 🌐 | | |
| $[FR_1^{6,1,2}]_{2,1}^2$ | $[D_3]_{2,1}^2$ | | 🌐 | | |
| $[FR_1^{6,1,2}]_{3,1}^1$ | $[D_3]_{3,1}^1$ | † | 🌐 | | |
| $[FR_1^{6,1,2}]_{3,1}^2$ | $[D_3]_{3,1}^2$ | | 🌐 | | |
| $[FR_1^{6,1,2}]_{4,1}^1$ | $[D_3]_{4,1}^1$ | † | 🌐 | | |
| $[FR_1^{6,1,2}]_{4,1}^2$ | $[D_3]_{4,1}^2$ | | 🌐 | | |
| $[FR_1^{6,1,2}]_{5,1}^1$ | $[D_3]_{5,1}^1$ | † | 🌐 | | |
| $[FR_1^{6,1,2}]_{5,1}^2$ | $[D_3]_{5,1}^2$ | | 🌐 | | |
| $[FR_1^{6,1,2}]_{6,1}^1$ | $[D_3]_{6,1}^1$ | † | 🌐 | | |
| $[FR_1^{6,1,2}]_{6,1}^2$ | $[D_3]_{6,1}^2$ | | 🌐 | | |
| $[FR_2^{6,1,2}]_{1,1}^1$ | $[[\mathbb{Z}_2 \trianglelefteq \mathbb{Z}_4]_{\mathbf{1}|0}^{\text{Id}}]_{1,1}^1$ | † | 🌐 | | |
| $[FR_2^{6,1,2}]_{1,1}^2$ | $[[\mathbb{Z}_2 \trianglelefteq \mathbb{Z}_4]_{\mathbf{1}|0}^{\text{Id}}]_{1,1}^2$ | | 🌐 | | |
| $[FR_2^{6,1,2}]_{1,1}^3$ | $[[\mathbb{Z}_2 \trianglelefteq \mathbb{Z}_4]_{\mathbf{1}|0}^{\text{Id}}]_{1,1}^3$ | | 🌐 | | |
| $[FR_2^{6,1,2}]_{2,1}^1$ | $[[\mathbb{Z}_2 \trianglelefteq \mathbb{Z}_4]_{\mathbf{1}|0}^{\text{Id}}]_{2,1}^1$ | † | 🌐 | | |
| $[FR_2^{6,1,2}]_{2,1}^2$ | $[[\mathbb{Z}_2 \trianglelefteq \mathbb{Z}_4]_{\mathbf{1}|0}^{\text{Id}}]_{2,1}^2$ | | 🌐 | | |
| $[FR_2^{6,1,2}]_{2,1}^3$ | $[[\mathbb{Z}_2 \trianglelefteq \mathbb{Z}_4]_{\mathbf{1}|0}^{\text{Id}}]_{2,1}^3$ | | 🌐 | | |
| $[FR_2^{6,1,2}]_{3,1}^1$ | $[[\mathbb{Z}_2 \trianglelefteq \mathbb{Z}_4]_{\mathbf{1}|0}^{\text{Id}}]_{3,1}^1$ | † | 🌐 | | |
| $[FR_2^{6,1,2}]_{3,1}^2$ | $[[\mathbb{Z}_2 \trianglelefteq \mathbb{Z}_4]_{\mathbf{1}|0}^{\text{Id}}]_{3,1}^2$ | | 🌐 | | |
| $[FR_2^{6,1,2}]_{3,1}^3$ | $[[\mathbb{Z}_2 \trianglelefteq \mathbb{Z}_4]_{\mathbf{1}|0}^{\text{Id}}]_{3,1}^3$ | | 🌐 | | |
| $[FR_2^{6,1,2}]_{3,1}^4$ | $[[\mathbb{Z}_2 \trianglelefteq \mathbb{Z}_4]_{\mathbf{1}|0}^{\text{Id}}]_{3,1}^4$ | | 🌐 | | |
| $[FR_2^{6,1,2}]_{4,1}^1$ | $[[\mathbb{Z}_2 \trianglelefteq \mathbb{Z}_4]_{\mathbf{1}|0}^{\text{Id}}]_{4,1}^1$ | † | 🌐 | | |
| $[FR_2^{6,1,2}]_{4,1}^2$ | $[[\mathbb{Z}_2 \trianglelefteq \mathbb{Z}_4]_{\mathbf{1}|0}^{\text{Id}}]_{4,1}^2$ | | 🌐 | | |
| $[FR_2^{6,1,2}]_{4,1}^3$ | $[[\mathbb{Z}_2 \trianglelefteq \mathbb{Z}_4]_{\mathbf{1}|0}^{\text{Id}}]_{4,1}^3$ | | 🌐 | | |

Table 9.1: List of multiplicity-free fusion categories up to rank 7 (Continued)

| Formal Name | Common Name | | | | |
|---|---|---|---|---|---|
| $[FR_2^{6,1,2}]_{5,1}^1$ | $[[\mathbb{Z}_2 \trianglelefteq \mathbb{Z}_4]_{1|0}^{Id}]_{5,1}^1$ | | † | 🌐 | |
| $[FR_2^{6,1,2}]_{5,1}^2$ | $[[\mathbb{Z}_2 \trianglelefteq \mathbb{Z}_4]_{1|0}^{Id}]_{5,1}^2$ | | | 🌐 | |
| $[FR_2^{6,1,2}]_{5,1}^3$ | $[[\mathbb{Z}_2 \trianglelefteq \mathbb{Z}_4]_{1|0}^{Id}]_{5,1}^3$ | | | 🌐 | |
| $[FR_2^{6,1,2}]_{6,1}^1$ | $[[\mathbb{Z}_2 \trianglelefteq \mathbb{Z}_4]_{1|0}^{Id}]_{6,1}^1$ | | † | 🌐 | |
| $[FR_2^{6,1,2}]_{6,1}^2$ | $[[\mathbb{Z}_2 \trianglelefteq \mathbb{Z}_4]_{1|0}^{Id}]_{6,1}^2$ | | | 🌐 | |
| $[FR_2^{6,1,2}]_{6,1}^3$ | $[[\mathbb{Z}_2 \trianglelefteq \mathbb{Z}_4]_{1|0}^{Id}]_{6,1}^3$ | | | 🌐 | |
| $[FR_2^{6,1,2}]_{6,1}^4$ | $[[\mathbb{Z}_2 \trianglelefteq \mathbb{Z}_4]_{1|0}^{Id}]_{6,1}^4$ | | | 🌐 | |
| $[FR_3^{6,1,2}]_{1,1}^1$ | $[[\mathbb{Z}_2 \trianglelefteq \mathbb{Z}_2 \times \mathbb{Z}_2]_{3|0}^{Id}]_{1,1}^1$ | ✕ | † | 🌐 | 🎀 |
| $[FR_3^{6,1,2}]_{1,1}^2$ | $[[\mathbb{Z}_2 \trianglelefteq \mathbb{Z}_2 \times \mathbb{Z}_2]_{3|0}^{Id}]_{1,1}^2$ | ✕ | | 🌐 | 🎀 |
| $[FR_3^{6,1,2}]_{1,1}^3$ | $[[\mathbb{Z}_2 \trianglelefteq \mathbb{Z}_2 \times \mathbb{Z}_2]_{3|0}^{Id}]_{1,1}^3$ | ✕ | | | |
| $[FR_3^{6,1,2}]_{1,2}^1$ | $[[\mathbb{Z}_2 \trianglelefteq \mathbb{Z}_2 \times \mathbb{Z}_2]_{3|0}^{Id}]_{1,2}^1$ | ✕ | † | 🌐 | 🎀 |
| $[FR_3^{6,1,2}]_{1,2}^2$ | $[[\mathbb{Z}_2 \trianglelefteq \mathbb{Z}_2 \times \mathbb{Z}_2]_{3|0}^{Id}]_{1,2}^2$ | ✕ | | 🌐 | 🎀 |
| $[FR_3^{6,1,2}]_{1,2}^3$ | $[[\mathbb{Z}_2 \trianglelefteq \mathbb{Z}_2 \times \mathbb{Z}_2]_{3|0}^{Id}]_{1,2}^3$ | ✕ | | | |
| $[FR_3^{6,1,2}]_{1,3}^1$ | $[[\mathbb{Z}_2 \trianglelefteq \mathbb{Z}_2 \times \mathbb{Z}_2]_{3|0}^{Id}]_{1,3}^1$ | ✕ | † | 🌐 | 🎀 |
| $[FR_3^{6,1,2}]_{1,3}^2$ | $[[\mathbb{Z}_2 \trianglelefteq \mathbb{Z}_2 \times \mathbb{Z}_2]_{3|0}^{Id}]_{1,3}^2$ | ✕ | | 🌐 | 🎀 |
| $[FR_3^{6,1,2}]_{1,3}^3$ | $[[\mathbb{Z}_2 \trianglelefteq \mathbb{Z}_2 \times \mathbb{Z}_2]_{3|0}^{Id}]_{1,3}^3$ | ✕ | | | |
| $[FR_3^{6,1,2}]_{1,4}^1$ | $[[\mathbb{Z}_2 \trianglelefteq \mathbb{Z}_2 \times \mathbb{Z}_2]_{3|0}^{Id}]_{1,4}^1$ | ✕ | † | 🌐 | 🎀 |
| $[FR_3^{6,1,2}]_{1,4}^2$ | $[[\mathbb{Z}_2 \trianglelefteq \mathbb{Z}_2 \times \mathbb{Z}_2]_{3|0}^{Id}]_{1,4}^2$ | ✕ | | 🌐 | 🎀 |
| $[FR_3^{6,1,2}]_{1,4}^3$ | $[[\mathbb{Z}_2 \trianglelefteq \mathbb{Z}_2 \times \mathbb{Z}_2]_{3|0}^{Id}]_{1,4}^3$ | ✕ | | | |
| $[FR_3^{6,1,2}]_{1,5}^1$ | $[[\mathbb{Z}_2 \trianglelefteq \mathbb{Z}_2 \times \mathbb{Z}_2]_{3|0}^{Id}]_{1,5}^1$ | ✕ | † | 🌐 | 🎀 |
| $[FR_3^{6,1,2}]_{1,5}^2$ | $[[\mathbb{Z}_2 \trianglelefteq \mathbb{Z}_2 \times \mathbb{Z}_2]_{3|0}^{Id}]_{1,5}^2$ | ✕ | | 🌐 | 🎀 |
| $[FR_3^{6,1,2}]_{1,5}^3$ | $[[\mathbb{Z}_2 \trianglelefteq \mathbb{Z}_2 \times \mathbb{Z}_2]_{3|0}^{Id}]_{1,5}^3$ | ✕ | | | |
| $[FR_3^{6,1,2}]_{1,6}^1$ | $[[\mathbb{Z}_2 \trianglelefteq \mathbb{Z}_2 \times \mathbb{Z}_2]_{3|0}^{Id}]_{1,6}^1$ | ✕ | † | 🌐 | 🎀 |
| $[FR_3^{6,1,2}]_{1,6}^2$ | $[[\mathbb{Z}_2 \trianglelefteq \mathbb{Z}_2 \times \mathbb{Z}_2]_{3|0}^{Id}]_{1,6}^2$ | ✕ | | 🌐 | 🎀 |
| $[FR_3^{6,1,2}]_{1,6}^3$ | $[[\mathbb{Z}_2 \trianglelefteq \mathbb{Z}_2 \times \mathbb{Z}_2]_{3|0}^{Id}]_{1,6}^3$ | ✕ | | | |
| $[FR_3^{6,1,2}]_{1,7}^1$ | $[[\mathbb{Z}_2 \trianglelefteq \mathbb{Z}_2 \times \mathbb{Z}_2]_{3|0}^{Id}]_{1,7}^1$ | ✕ | † | 🌐 | 🎀 |
| $[FR_3^{6,1,2}]_{1,7}^2$ | $[[\mathbb{Z}_2 \trianglelefteq \mathbb{Z}_2 \times \mathbb{Z}_2]_{3|0}^{Id}]_{1,7}^2$ | ✕ | | 🌐 | 🎀 |
| $[FR_3^{6,1,2}]_{1,7}^3$ | $[[\mathbb{Z}_2 \trianglelefteq \mathbb{Z}_2 \times \mathbb{Z}_2]_{3|0}^{Id}]_{1,7}^3$ | ✕ | | | |

Table 9.1: List of multiplicity-free fusion categories up to rank 7 (Continued)

| Formal Name | Common Name | | | | |
|---|---|---|---|---|---|
| $[FR_3^{6,1,2}]_{1,8}^1$ | $[[\mathbb{Z}_2 \trianglelefteq \mathbb{Z}_2 \times \mathbb{Z}_2]_{3\|0}^{\text{Id}}]_{1,8}^1$ | ✕ | † | ● | ⚘ |
| $[FR_3^{6,1,2}]_{1,8}^2$ | $[[\mathbb{Z}_2 \trianglelefteq \mathbb{Z}_2 \times \mathbb{Z}_2]_{3\|0}^{\text{Id}}]_{1,8}^2$ | ✕ | | ● | ⚘ |
| $[FR_3^{6,1,2}]_{1,8}^3$ | $[[\mathbb{Z}_2 \trianglelefteq \mathbb{Z}_2 \times \mathbb{Z}_2]_{3\|0}^{\text{Id}}]_{1,8}^3$ | ✕ | | | |
| $[FR_3^{6,1,2}]_{2,1}^1$ | $[[\mathbb{Z}_2 \trianglelefteq \mathbb{Z}_2 \times \mathbb{Z}_2]_{3\|0}^{\text{Id}}]_{2,1}^1$ | | † | ● | |
| $[FR_3^{6,1,2}]_{2,1}^2$ | $[[\mathbb{Z}_2 \trianglelefteq \mathbb{Z}_2 \times \mathbb{Z}_2]_{3\|0}^{\text{Id}}]_{2,1}^2$ | | | ● | |
| $[FR_3^{6,1,2}]_{2,1}^3$ | $[[\mathbb{Z}_2 \trianglelefteq \mathbb{Z}_2 \times \mathbb{Z}_2]_{3\|0}^{\text{Id}}]_{2,1}^3$ | | | | |
| $[FR_4^{6,1,2}]_{1,1}^1$ | $[\text{Rep}(\text{Dic}_{12})]_{1,1}^1$ | ✕ | † | ● | ⚘ |
| $[FR_4^{6,1,2}]_{1,1}^2$ | $[\text{Rep}(\text{Dic}_{12})]_{1,1}^2$ | ✕ | | ● | ⚘ |
| $[FR_4^{6,1,2}]_{1,2}^1$ | $[\text{Rep}(\text{Dic}_{12})]_{1,2}^1$ | ✕ | † | ● | ⚘ |
| $[FR_4^{6,1,2}]_{1,2}^2$ | $[\text{Rep}(\text{Dic}_{12})]_{1,2}^2$ | ✕ | | ● | ⚘ |
| $[FR_4^{6,1,2}]_{1,3}^1$ | $[\text{Rep}(\text{Dic}_{12})]_{1,3}^1$ | ✕ | † | ● | ⚘ |
| $[FR_4^{6,1,2}]_{1,3}^2$ | $[\text{Rep}(\text{Dic}_{12})]_{1,3}^2$ | ✕ | | ● | ⚘ |
| $[FR_4^{6,1,2}]_{1,4}^1$ | $[\text{Rep}(\text{Dic}_{12})]_{1,4}^1$ | ✕ | † | ● | ⚘ |
| $[FR_4^{6,1,2}]_{1,4}^2$ | $[\text{Rep}(\text{Dic}_{12})]_{1,4}^2$ | ✕ | | ● | ⚘ |
| $[FR_4^{6,1,2}]_{1,5}^1$ | $[\text{Rep}(\text{Dic}_{12})]_{1,5}^1$ | ✕ | † | ● | ⚘ |
| $[FR_4^{6,1,2}]_{1,5}^2$ | $[\text{Rep}(\text{Dic}_{12})]_{1,5}^2$ | ✕ | | ● | ⚘ |
| $[FR_4^{6,1,2}]_{1,6}^1$ | $[\text{Rep}(\text{Dic}_{12})]_{1,6}^1$ | ✕ | † | ● | ⚘ |
| $[FR_4^{6,1,2}]_{1,6}^2$ | $[\text{Rep}(\text{Dic}_{12})]_{1,6}^2$ | ✕ | | ● | ⚘ |
| $[FR_4^{6,1,2}]_{2,1}^1$ | $[\text{Rep}(\text{Dic}_{12})]_{2,1}^1$ | ✕ | † | ● | ⚘ |
| $[FR_4^{6,1,2}]_{2,1}^2$ | $[\text{Rep}(\text{Dic}_{12})]_{2,1}^2$ | ✕ | | ● | ⚘ |
| $[FR_4^{6,1,2}]_{2,2}^1$ | $[\text{Rep}(\text{Dic}_{12})]_{2,2}^1$ | ✕ | † | ● | ⚘ |
| $[FR_4^{6,1,2}]_{2,2}^2$ | $[\text{Rep}(\text{Dic}_{12})]_{2,2}^2$ | ✕ | | ● | ⚘ |
| $[FR_4^{6,1,2}]_{2,3}^1$ | $[\text{Rep}(\text{Dic}_{12})]_{2,3}^1$ | ✕ | † | ● | ⚘ |
| $[FR_4^{6,1,2}]_{2,3}^2$ | $[\text{Rep}(\text{Dic}_{12})]_{2,3}^2$ | ✕ | | ● | ⚘ |
| $[FR_4^{6,1,2}]_{2,4}^1$ | $[\text{Rep}(\text{Dic}_{12})]_{2,4}^1$ | ✕ | † | ● | ⚘ |
| $[FR_4^{6,1,2}]_{2,4}^2$ | $[\text{Rep}(\text{Dic}_{12})]_{2,4}^2$ | ✕ | | ● | ⚘ |
| $[FR_4^{6,1,2}]_{2,5}^1$ | $[\text{Rep}(\text{Dic}_{12})]_{2,5}^1$ | ✕ | † | ● | ⚘ |
| $[FR_4^{6,1,2}]_{2,5}^2$ | $[\text{Rep}(\text{Dic}_{12})]_{2,5}^2$ | ✕ | | ● | ⚘ |
| $[FR_4^{6,1,2}]_{2,6}^1$ | $[\text{Rep}(\text{Dic}_{12})]_{2,6}^1$ | ✕ | † | ● | ⚘ |

Table 9.1: List of multiplicity-free fusion categories up to rank 7 (Continued)

| Formal Name | Common Name | | | | | |
|---|---|---|---|---|---|---|
| $[FR_4^{6,1,2}]_{2,6}^2$ | $[\text{Rep}(\text{Dic}_{12})]_{2,6}^2$ | ✕ | | ● | ♀ | |
| $[FR_4^{6,1,2}]_{3,1}^1$ | $[\text{Rep}(\text{Dic}_{12})]_{3,1}^1$ | | † | ● | | |
| $[FR_4^{6,1,2}]_{3,1}^2$ | $[\text{Rep}(\text{Dic}_{12})]_{3,1}^2$ | | | ● | | |
| $[FR_4^{6,1,2}]_{4,1}^1$ | $[\text{Rep}(\text{Dic}_{12})]_{4,1}^1$ | | † | ● | | |
| $[FR_4^{6,1,2}]_{4,1}^2$ | $[\text{Rep}(\text{Dic}_{12})]_{4,1}^2$ | | | ● | | |
| $[FR_4^{6,1,2}]_{5,1}^1$ | $[\text{Rep}(\text{Dic}_{12})]_{5,1}^1$ | | † | ● | | |
| $[FR_4^{6,1,2}]_{5,1}^2$ | $[\text{Rep}(\text{Dic}_{12})]_{5,1}^2$ | | | ● | | |
| $[FR_4^{6,1,2}]_{6,1}^1$ | $[\text{Rep}(\text{Dic}_{12})]_{6,1}^1$ | | † | ● | | |
| $[FR_4^{6,1,2}]_{6,1}^2$ | $[\text{Rep}(\text{Dic}_{12})]_{6,1}^2$ | | | ● | | |
| $[FR_7^{6,1,2}]_{1,1}^1$ | $[\text{Pseudo SO}(5)_2]_{1,1}^1$ | | † | ● | | |
| $[FR_7^{6,1,2}]_{1,1}^2$ | $[\text{Pseudo SO}(5)_2]_{1,1}^2$ | | | ● | | |
| $[FR_7^{6,1,2}]_{2,1}^1$ | $[\text{Pseudo SO}(5)_2]_{2,1}^1$ | | † | ● | | |
| $[FR_7^{6,1,2}]_{2,1}^2$ | $[\text{Pseudo SO}(5)_2]_{2,1}^2$ | | | ● | | |
| $[FR_8^{6,1,2}]_{1,1}^1$ | $[\text{HI}(\mathbb{Z}_3)]_{1,1}^1$ | | † | ● | | |
| $[FR_8^{6,1,2}]_{2,1}^1$ | $[\text{HI}(\mathbb{Z}_3)]_{2,1}^1$ | | † | ● | | |
| $[FR_8^{6,1,2}]_{3,1}^1$ | $[\text{HI}(\mathbb{Z}_3)]_{3,1}^1$ | | | ● | | |
| $[FR_8^{6,1,2}]_{4,1}^1$ | $[\text{HI}(\mathbb{Z}_3)]_{4,1}^1$ | | | ● | | |
| $[FR_1^{6,1,4}]_{1,1}^1$ | $[\mathbb{Z}_6]_{1,1}^1$ | ✕ | † | ● | ♀ | ⊚ |
| $[FR_1^{6,1,4}]_{1,1}^2$ | $[\mathbb{Z}_6]_{1,1}^2$ | ✕ | | ● | ♀ | ⊚ |
| $[FR_1^{6,1,4}]_{1,1}^3$ | $[\mathbb{Z}_6]_{1,1}^3$ | ✕ | | | | |
| $[FR_1^{6,1,4}]_{1,1}^4$ | $[\mathbb{Z}_6]_{1,1}^4$ | ✕ | | | | |
| $[FR_1^{6,1,4}]_{1,2}^1$ | $[\mathbb{Z}_6]_{1,2}^1$ | ✕ | † | ● | ♀ | ⊚ |
| $[FR_1^{6,1,4}]_{1,2}^2$ | $[\mathbb{Z}_6]_{1,2}^2$ | ✕ | | ● | ♀ | ⊚ |
| $[FR_1^{6,1,4}]_{1,2}^3$ | $[\mathbb{Z}_6]_{1,2}^3$ | ✕ | | | | |
| $[FR_1^{6,1,4}]_{1,2}^4$ | $[\mathbb{Z}_6]_{1,2}^4$ | ✕ | | | | |
| $[FR_1^{6,1,4}]_{1,3}^1$ | $[\mathbb{Z}_6]_{1,3}^1$ | ✕ | † | ● | ♀ | ⊚ |
| $[FR_1^{6,1,4}]_{1,3}^2$ | $[\mathbb{Z}_6]_{1,3}^2$ | ✕ | | ● | ♀ | ⊚ |
| $[FR_1^{6,1,4}]_{1,3}^3$ | $[\mathbb{Z}_6]_{1,3}^3$ | ✕ | | | | |

Table 9.1: List of multiplicity-free fusion categories up to rank 7 (Continued)

| Formal Name | Common Name | | | | | |
|---|---|---|---|---|---|---|
| $[FR_1^{6,1,4}]_{1,3}^4$ | $[\mathbb{Z}_6]_{1,3}^4$ | ✗ | | | | |
| $[FR_1^{6,1,4}]_{1,4}^1$ | $[\mathbb{Z}_6]_{1,4}^1$ | ✗ | † | 🌐 | 🎗 | ⚭ |
| $[FR_1^{6,1,4}]_{1,4}^2$ | $[\mathbb{Z}_6]_{1,4}^2$ | ✗ | | 🌐 | 🎗 | ⚭ |
| $[FR_1^{6,1,4}]_{1,4}^3$ | $[\mathbb{Z}_6]_{1,4}^3$ | ✗ | | | | |
| $[FR_1^{6,1,4}]_{1,4}^4$ | $[\mathbb{Z}_6]_{1,4}^4$ | ✗ | | | | |
| $[FR_1^{6,1,4}]_{1,5}^1$ | $[\mathbb{Z}_6]_{1,5}^1$ | ✗ | † | 🌐 | 🎗 | |
| $[FR_1^{6,1,4}]_{1,5}^2$ | $[\mathbb{Z}_6]_{1,5}^2$ | ✗ | | 🌐 | 🎗 | |
| $[FR_1^{6,1,4}]_{1,5}^3$ | $[\mathbb{Z}_6]_{1,5}^3$ | ✗ | | | | |
| $[FR_1^{6,1,4}]_{1,5}^4$ | $[\mathbb{Z}_6]_{1,5}^4$ | ✗ | | | | |
| $[FR_1^{6,1,4}]_{1,6}^1$ | $[\mathbb{Z}_6]_{1,6}^1$ | ✗ | † | 🌐 | 🎗 | |
| $[FR_1^{6,1,4}]_{1,6}^2$ | $[\mathbb{Z}_6]_{1,6}^2$ | ✗ | | 🌐 | 🎗 | |
| $[FR_1^{6,1,4}]_{1,6}^3$ | $[\mathbb{Z}_6]_{1,6}^3$ | ✗ | | | | |
| $[FR_1^{6,1,4}]_{1,6}^4$ | $[\mathbb{Z}_6]_{1,6}^4$ | ✗ | | | | |
| $[FR_1^{6,1,4}]_{2,1}^1$ | $[\mathbb{Z}_6]_{2,1}^1$ | ✗ | † | 🌐 | 🎗 | |
| $[FR_1^{6,1,4}]_{2,1}^2$ | $[\mathbb{Z}_6]_{2,1}^2$ | ✗ | | 🌐 | 🎗 | |
| $[FR_1^{6,1,4}]_{2,1}^3$ | $[\mathbb{Z}_6]_{2,1}^3$ | ✗ | | | | |
| $[FR_1^{6,1,4}]_{2,1}^4$ | $[\mathbb{Z}_6]_{2,1}^4$ | ✗ | | | | |
| $[FR_1^{6,1,4}]_{2,2}^1$ | $[\mathbb{Z}_6]_{2,2}^1$ | ✗ | † | 🌐 | 🎗 | |
| $[FR_1^{6,1,4}]_{2,2}^2$ | $[\mathbb{Z}_6]_{2,2}^2$ | ✗ | | 🌐 | 🎗 | |
| $[FR_1^{6,1,4}]_{2,2}^3$ | $[\mathbb{Z}_6]_{2,2}^3$ | ✗ | | | | |
| $[FR_1^{6,1,4}]_{2,2}^4$ | $[\mathbb{Z}_6]_{2,2}^4$ | ✗ | | | | |
| $[FR_1^{6,1,4}]_{2,3}^1$ | $[\mathbb{Z}_6]_{2,3}^1$ | ✗ | † | 🌐 | 🎗 | |
| $[FR_1^{6,1,4}]_{2,3}^2$ | $[\mathbb{Z}_6]_{2,3}^2$ | ✗ | | 🌐 | 🎗 | |
| $[FR_1^{6,1,4}]_{2,3}^3$ | $[\mathbb{Z}_6]_{2,3}^3$ | ✗ | | | | |
| $[FR_1^{6,1,4}]_{2,3}^4$ | $[\mathbb{Z}_6]_{2,3}^4$ | ✗ | | | | |
| $[FR_1^{6,1,4}]_{2,4}^1$ | $[\mathbb{Z}_6]_{2,4}^1$ | ✗ | † | 🌐 | 🎗 | |
| $[FR_1^{6,1,4}]_{2,4}^2$ | $[\mathbb{Z}_6]_{2,4}^2$ | ✗ | | 🌐 | 🎗 | |
| $[FR_1^{6,1,4}]_{2,4}^3$ | $[\mathbb{Z}_6]_{2,4}^3$ | ✗ | | | | |
| $[FR_1^{6,1,4}]_{2,4}^4$ | $[\mathbb{Z}_6]_{2,4}^4$ | ✗ | | | | |

Table 9.1: List of multiplicity-free fusion categories up to rank 7 (Continued)

| Formal Name | Common Name | | | | |
|---|---|---|---|---|---|
| $[FR_1^{6,1,4}]_{2,5}^1$ | $[\mathbb{Z}_6]_{2,5}^1$ | ⚕ | † | 🌐 | 🎗 |
| $[FR_1^{6,1,4}]_{2,5}^2$ | $[\mathbb{Z}_6]_{2,5}^2$ | ⚕ | | 🌐 | 🎗 |
| $[FR_1^{6,1,4}]_{2,5}^3$ | $[\mathbb{Z}_6]_{2,5}^3$ | ⚕ | | | |
| $[FR_1^{6,1,4}]_{2,5}^4$ | $[\mathbb{Z}_6]_{2,5}^4$ | ⚕ | | | |
| $[FR_1^{6,1,4}]_{2,6}^1$ | $[\mathbb{Z}_6]_{2,6}^1$ | ⚕ | † | 🌐 | 🎗 |
| $[FR_1^{6,1,4}]_{2,6}^2$ | $[\mathbb{Z}_6]_{2,6}^2$ | ⚕ | | 🌐 | 🎗 |
| $[FR_1^{6,1,4}]_{2,6}^3$ | $[\mathbb{Z}_6]_{2,6}^3$ | ⚕ | | | |
| $[FR_1^{6,1,4}]_{2,6}^4$ | $[\mathbb{Z}_6]_{2,6}^4$ | ⚕ | | | |
| $[FR_1^{6,1,4}]_{3,1}^1$ | $[\mathbb{Z}_6]_{3,1}^1$ | | † | 🌐 | |
| $[FR_1^{6,1,4}]_{3,1}^2$ | $[\mathbb{Z}_6]_{3,1}^2$ | | | 🌐 | |
| $[FR_1^{6,1,4}]_{3,1}^3$ | $[\mathbb{Z}_6]_{3,1}^3$ | | | | |
| $[FR_1^{6,1,4}]_{3,1}^4$ | $[\mathbb{Z}_6]_{3,1}^4$ | | | | |
| $[FR_1^{6,1,4}]_{4,1}^1$ | $[\mathbb{Z}_6]_{4,1}^1$ | | † | 🌐 | |
| $[FR_1^{6,1,4}]_{4,1}^2$ | $[\mathbb{Z}_6]_{4,1}^2$ | | | 🌐 | |
| $[FR_1^{6,1,4}]_{4,1}^3$ | $[\mathbb{Z}_6]_{4,1}^3$ | | | | |
| $[FR_1^{6,1,4}]_{4,1}^4$ | $[\mathbb{Z}_6]_{4,1}^4$ | | | | |
| $[FR_1^{6,1,4}]_{5,1}^1$ | $[\mathbb{Z}_6]_{5,1}^1$ | | † | 🌐 | |
| $[FR_1^{6,1,4}]_{5,1}^2$ | $[\mathbb{Z}_6]_{5,1}^2$ | | | 🌐 | |
| $[FR_1^{6,1,4}]_{5,1}^3$ | $[\mathbb{Z}_6]_{5,1}^3$ | | | | |
| $[FR_1^{6,1,4}]_{5,1}^4$ | $[\mathbb{Z}_6]_{5,1}^4$ | | | | |
| $[FR_1^{6,1,4}]_{6,1}^1$ | $[\mathbb{Z}_6]_{6,1}^1$ | | † | 🌐 | |
| $[FR_1^{6,1,4}]_{6,1}^2$ | $[\mathbb{Z}_6]_{6,1}^2$ | | | 🌐 | |
| $[FR_1^{6,1,4}]_{6,1}^3$ | $[\mathbb{Z}_6]_{6,1}^3$ | | | | |
| $[FR_1^{6,1,4}]_{6,1}^4$ | $[\mathbb{Z}_6]_{6,1}^4$ | | | | |
| $[FR_2^{6,1,4}]_{1,1}^1$ | $[MR_6]_{1,1}^1$ | | † | 🌐 | |
| $[FR_2^{6,1,4}]_{1,1}^2$ | $[MR_6]_{1,1}^2$ | | | 🌐 | |
| $[FR_2^{6,1,4}]_{1,1}^3$ | $[MR_6]_{1,1}^3$ | | | | |
| $[FR_2^{6,1,4}]_{2,1}^1$ | $[MR_6]_{2,1}^1$ | | † | 🌐 | |
| $[FR_2^{6,1,4}]_{2,1}^2$ | $[MR_6]_{2,1}^2$ | | | 🌐 | |

Table 9.1: List of multiplicity-free fusion categories up to rank 7 (Continued)

| Formal Name | Common Name | | | | | |
|---|---|---|---|---|---|---|
| $[FR_2^{6,1,4}]_{2,1}^3$ | $[MR_6]_{2,1}^3$ | | | | | |
| $[FR_2^{6,1,4}]_{3,1}^1$ | $[MR_6]_{3,1}^1$ | | † | 🌐 | | |
| $[FR_2^{6,1,4}]_{3,1}^2$ | $[MR_6]_{3,1}^2$ | | | 🌐 | | |
| $[FR_2^{6,1,4}]_{3,1}^3$ | $[MR_6]_{3,1}^3$ | | | | | |
| $[FR_2^{6,1,4}]_{4,1}^1$ | $[MR_6]_{4,1}^1$ | | † | 🌐 | | |
| $[FR_2^{6,1,4}]_{4,1}^2$ | $[MR_6]_{4,1}^2$ | | | 🌐 | | |
| $[FR_2^{6,1,4}]_{4,1}^3$ | $[MR_6]_{4,1}^3$ | | | | | |
| $[FR_3^{6,1,4}]_{1,1}^1$ | $[TY(\mathbb{Z}_5)]_{1,1}^1$ | | † | 🌐 | | |
| $[FR_3^{6,1,4}]_{1,1}^2$ | $[TY(\mathbb{Z}_5)]_{1,1}^2$ | | | 🌐 | | |
| $[FR_3^{6,1,4}]_{2,1}^1$ | $[TY(\mathbb{Z}_5)]_{2,1}^1$ | | † | 🌐 | | |
| $[FR_3^{6,1,4}]_{2,1}^2$ | $[TY(\mathbb{Z}_5)]_{2,1}^2$ | | | 🌐 | | |
| $[FR_3^{6,1,4}]_{3,1}^1$ | $[TY(\mathbb{Z}_5)]_{3,1}^1$ | | † | 🌐 | | |
| $[FR_3^{6,1,4}]_{3,1}^2$ | $[TY(\mathbb{Z}_5)]_{3,1}^2$ | | | 🌐 | | |
| $[FR_3^{6,1,4}]_{4,1}^1$ | $[TY(\mathbb{Z}_5)]_{4,1}^1$ | | † | 🌐 | | |
| $[FR_3^{6,1,4}]_{4,1}^2$ | $[TY(\mathbb{Z}_5)]_{4,1}^2$ | | | 🌐 | | |
| $[FR_5^{6,1,4}]_{1,1}^1$ | $[Fib \times \mathbb{Z}_3]_{1,1}^1$ | ✕ | † | 🌐 | 🎗 | ⊚ |
| $[FR_5^{6,1,4}]_{1,1}^2$ | $[Fib \times \mathbb{Z}_3]_{1,1}^2$ | ✕ | | | | |
| $[FR_5^{6,1,4}]_{1,2}^1$ | $[Fib \times \mathbb{Z}_3]_{1,2}^1$ | ✕ | † | 🌐 | 🎗 | ⊚ |
| $[FR_5^{6,1,4}]_{1,2}^2$ | $[Fib \times \mathbb{Z}_3]_{1,2}^2$ | ✕ | | | | |
| $[FR_5^{6,1,4}]_{1,3}^1$ | $[Fib \times \mathbb{Z}_3]_{1,3}^1$ | ✕ | † | 🌐 | 🎗 | ⊚ |
| $[FR_5^{6,1,4}]_{1,3}^2$ | $[Fib \times \mathbb{Z}_3]_{1,3}^2$ | ✕ | | | | |
| $[FR_5^{6,1,4}]_{1,4}^1$ | $[Fib \times \mathbb{Z}_3]_{1,4}^1$ | ✕ | † | 🌐 | 🎗 | ⊚ |
| $[FR_5^{6,1,4}]_{1,4}^2$ | $[Fib \times \mathbb{Z}_3]_{1,4}^2$ | ✕ | | | | |
| $[FR_5^{6,1,4}]_{1,5}^1$ | $[Fib \times \mathbb{Z}_3]_{1,5}^1$ | ✕ | † | 🌐 | 🎗 | |
| $[FR_5^{6,1,4}]_{1,5}^2$ | $[Fib \times \mathbb{Z}_3]_{1,5}^2$ | ✕ | | | | |
| $[FR_5^{6,1,4}]_{1,6}^1$ | $[Fib \times \mathbb{Z}_3]_{1,6}^1$ | ✕ | † | 🌐 | 🎗 | |
| $[FR_5^{6,1,4}]_{1,6}^2$ | $[Fib \times \mathbb{Z}_3]_{1,6}^2$ | ✕ | | | | |
| $[FR_5^{6,1,4}]_{2,1}^1$ | $[Fib \times \mathbb{Z}_3]_{2,1}^1$ | ✕ | | 🌐 | 🎗 | ⊚ |
| $[FR_5^{6,1,4}]_{2,1}^2$ | $[Fib \times \mathbb{Z}_3]_{2,1}^2$ | ✕ | | | | |

Table 9.1: List of multiplicity-free fusion categories up to rank 7 (Continued)

| Formal Name | Common Name | | | | | |
|---|---|---|---|---|---|---|
| $[FR_5^{6,1,4}]_{2,2}^1$ | $[Fib \times \mathbb{Z}_3]_{2,2}^1$ | ✗ | | 🌐 | 🎗 | ⭕ |
| $[FR_5^{6,1,4}]_{2,2}^2$ | $[Fib \times \mathbb{Z}_3]_{2,2}^2$ | ✗ | | | | |
| $[FR_5^{6,1,4}]_{2,3}^1$ | $[Fib \times \mathbb{Z}_3]_{2,3}^1$ | ✗ | | 🌐 | 🎗 | ⭕ |
| $[FR_5^{6,1,4}]_{2,3}^2$ | $[Fib \times \mathbb{Z}_3]_{2,3}^2$ | ✗ | | | | |
| $[FR_5^{6,1,4}]_{2,4}^1$ | $[Fib \times \mathbb{Z}_3]_{2,4}^1$ | ✗ | | 🌐 | 🎗 | ⭕ |
| $[FR_5^{6,1,4}]_{2,4}^2$ | $[Fib \times \mathbb{Z}_3]_{2,4}^2$ | ✗ | | | | |
| $[FR_5^{6,1,4}]_{2,5}^1$ | $[Fib \times \mathbb{Z}_3]_{2,5}^1$ | ✗ | | 🌐 | 🎗 | |
| $[FR_5^{6,1,4}]_{2,5}^2$ | $[Fib \times \mathbb{Z}_3]_{2,5}^2$ | ✗ | | | | |
| $[FR_5^{6,1,4}]_{2,6}^1$ | $[Fib \times \mathbb{Z}_3]_{2,6}^1$ | ✗ | | 🌐 | 🎗 | |
| $[FR_5^{6,1,4}]_{2,6}^2$ | $[Fib \times \mathbb{Z}_3]_{2,6}^2$ | ✗ | | | | |
| $[FR_5^{6,1,4}]_{3,1}^1$ | $[Fib \times \mathbb{Z}_3]_{3,1}^1$ | | 🗡 | 🌐 | | |
| $[FR_5^{6,1,4}]_{3,1}^2$ | $[Fib \times \mathbb{Z}_3]_{3,1}^2$ | | | | | |
| $[FR_5^{6,1,4}]_{4,1}^1$ | $[Fib \times \mathbb{Z}_3]_{4,1}^1$ | | 🗡 | 🌐 | | |
| $[FR_5^{6,1,4}]_{4,1}^2$ | $[Fib \times \mathbb{Z}_3]_{4,1}^2$ | | | | | |
| $[FR_5^{6,1,4}]_{5,1}^1$ | $[Fib \times \mathbb{Z}_3]_{5,1}^1$ | | | 🌐 | | |
| $[FR_5^{6,1,4}]_{5,1}^2$ | $[Fib \times \mathbb{Z}_3]_{5,1}^2$ | | | | | |
| $[FR_5^{6,1,4}]_{6,1}^1$ | $[Fib \times \mathbb{Z}_3]_{6,1}^1$ | | | 🌐 | | |
| $[FR_5^{6,1,4}]_{6,1}^2$ | $[Fib \times \mathbb{Z}_3]_{6,1}^2$ | | | | | |
| $[FR_1^{7,1,0}]_{1,1}^1$ | $[Adj(SO(16)_2)]_{1,1}^1$ | ✗ | 🗡 | 🌐 | 🎗 | |
| $[FR_1^{7,1,0}]_{1,1}^2$ | $[Adj(SO(16)_2)]_{1,1}^2$ | ✗ | | 🌐 | 🎗 | |
| $[FR_1^{7,1,0}]_{1,2}^1$ | $[Adj(SO(16)_2)]_{1,2}^1$ | ✗ | 🗡 | 🌐 | 🎗 | |
| $[FR_1^{7,1,0}]_{1,2}^2$ | $[Adj(SO(16)_2)]_{1,2}^2$ | ✗ | | 🌐 | 🎗 | |
| $[FR_1^{7,1,0}]_{1,3}^1$ | $[Adj(SO(16)_2)]_{1,3}^1$ | ✗ | 🗡 | 🌐 | 🎗 | |
| $[FR_1^{7,1,0}]_{1,3}^2$ | $[Adj(SO(16)_2)]_{1,3}^2$ | ✗ | | 🌐 | 🎗 | |
| $[FR_1^{7,1,0}]_{1,4}^1$ | $[Adj(SO(16)_2)]_{1,4}^1$ | ✗ | 🗡 | 🌐 | 🎗 | |
| $[FR_1^{7,1,0}]_{1,4}^2$ | $[Adj(SO(16)_2)]_{1,4}^2$ | ✗ | | 🌐 | 🎗 | |
| $[FR_1^{7,1,0}]_{1,5}^1$ | $[Adj(SO(16)_2)]_{1,5}^1$ | ✗ | 🗡 | 🌐 | 🎗 | |
| $[FR_1^{7,1,0}]_{1,5}^2$ | $[Adj(SO(16)_2)]_{1,5}^2$ | ✗ | | 🌐 | 🎗 | |
| $[FR_1^{7,1,0}]_{1,6}^1$ | $[Adj(SO(16)_2)]_{1,6}^1$ | ✗ | 🗡 | 🌐 | 🎗 | |

Table 9.1: List of multiplicity-free fusion categories up to rank 7 (Continued)

| Formal Name | Common Name | | | | |
|---|---|---|---|---|---|
| $[FR_1^{7,1,0}]_{1,6}^2$ | $[Adj(SO(16)_2)]_{1,6}^2$ | ✕ | | 🌐 | 🎗 |
| $[FR_1^{7,1,0}]_{1,7}^1$ | $[Adj(SO(16)_2)]_{1,7}^1$ | ✕ | † | 🌐 | 🎗 |
| $[FR_1^{7,1,0}]_{1,7}^2$ | $[Adj(SO(16)_2)]_{1,7}^2$ | ✕ | | 🌐 | 🎗 |
| $[FR_1^{7,1,0}]_{1,8}^1$ | $[Adj(SO(16)_2)]_{1,8}^1$ | ✕ | † | 🌐 | 🎗 |
| $[FR_1^{7,1,0}]_{1,8}^2$ | $[Adj(SO(16)_2)]_{1,8}^2$ | ✕ | | 🌐 | 🎗 |
| $[FR_1^{7,1,0}]_{2,1}^1$ | $[Adj(SO(16)_2)]_{2,1}^1$ | ✕ | † | 🌐 | 🎗 |
| $[FR_1^{7,1,0}]_{2,1}^2$ | $[Adj(SO(16)_2)]_{2,1}^2$ | ✕ | | 🌐 | 🎗 |
| $[FR_1^{7,1,0}]_{2,2}^1$ | $[Adj(SO(16)_2)]_{2,2}^1$ | ✕ | † | 🌐 | 🎗 |
| $[FR_1^{7,1,0}]_{2,2}^2$ | $[Adj(SO(16)_2)]_{2,2}^2$ | ✕ | | 🌐 | 🎗 |
| $[FR_1^{7,1,0}]_{2,3}^1$ | $[Adj(SO(16)_2)]_{2,3}^1$ | ✕ | † | 🌐 | 🎗 |
| $[FR_1^{7,1,0}]_{2,3}^2$ | $[Adj(SO(16)_2)]_{2,3}^2$ | ✕ | | 🌐 | 🎗 |
| $[FR_1^{7,1,0}]_{2,4}^1$ | $[Adj(SO(16)_2)]_{2,4}^1$ | ✕ | † | 🌐 | 🎗 |
| $[FR_1^{7,1,0}]_{2,4}^2$ | $[Adj(SO(16)_2)]_{2,4}^2$ | ✕ | | 🌐 | 🎗 |
| $[FR_1^{7,1,0}]_{2,5}^1$ | $[Adj(SO(16)_2)]_{2,5}^1$ | ✕ | † | 🌐 | 🎗 |
| $[FR_1^{7,1,0}]_{2,5}^2$ | $[Adj(SO(16)_2)]_{2,5}^2$ | ✕ | | 🌐 | 🎗 |
| $[FR_1^{7,1,0}]_{2,6}^1$ | $[Adj(SO(16)_2)]_{2,6}^1$ | ✕ | † | 🌐 | 🎗 |
| $[FR_1^{7,1,0}]_{2,6}^2$ | $[Adj(SO(16)_2)]_{2,6}^2$ | ✕ | | 🌐 | 🎗 |
| $[FR_1^{7,1,0}]_{2,7}^1$ | $[Adj(SO(16)_2)]_{2,7}^1$ | ✕ | † | 🌐 | 🎗 |
| $[FR_1^{7,1,0}]_{2,7}^2$ | $[Adj(SO(16)_2)]_{2,7}^2$ | ✕ | | 🌐 | 🎗 |
| $[FR_1^{7,1,0}]_{2,8}^1$ | $[Adj(SO(16)_2)]_{2,8}^1$ | ✕ | † | 🌐 | 🎗 |
| $[FR_1^{7,1,0}]_{2,8}^2$ | $[Adj(SO(16)_2)]_{2,8}^2$ | ✕ | | 🌐 | 🎗 |
| $[FR_1^{7,1,0}]_{3,1}^1$ | $[Adj(SO(16)_2)]_{3,1}^1$ | ✕ | † | 🌐 | 🎗 |
| $[FR_1^{7,1,0}]_{3,1}^2$ | $[Adj(SO(16)_2)]_{3,1}^2$ | ✕ | | 🌐 | 🎗 |
| $[FR_1^{7,1,0}]_{3,2}^1$ | $[Adj(SO(16)_2)]_{3,2}^1$ | ✕ | † | 🌐 | 🎗 |
| $[FR_1^{7,1,0}]_{3,2}^2$ | $[Adj(SO(16)_2)]_{3,2}^2$ | ✕ | | 🌐 | 🎗 |
| $[FR_1^{7,1,0}]_{3,3}^1$ | $[Adj(SO(16)_2)]_{3,3}^1$ | ✕ | † | 🌐 | 🎗 |
| $[FR_1^{7,1,0}]_{3,3}^2$ | $[Adj(SO(16)_2)]_{3,3}^2$ | ✕ | | 🌐 | 🎗 |
| $[FR_1^{7,1,0}]_{3,4}^1$ | $[Adj(SO(16)_2)]_{3,4}^1$ | ✕ | † | 🌐 | 🎗 |
| $[FR_1^{7,1,0}]_{3,4}^2$ | $[Adj(SO(16)_2)]_{3,4}^2$ | ✕ | | 🌐 | 🎗 |

Table 9.1: List of multiplicity-free fusion categories up to rank 7 (Continued)

| Formal Name | Common Name | | | | | |
|---|---|---|---|---|---|---|
| $[FR_1^{7,1,0}]_{4,1}^1$ | $[Adj(SO(16)_2)]_{4,1}^1$ | ✕ | † | 🌐 | 🎀 | |
| $[FR_1^{7,1,0}]_{4,1}^2$ | $[Adj(SO(16)_2)]_{4,1}^2$ | ✕ | | 🌐 | 🎀 | |
| $[FR_1^{7,1,0}]_{4,2}^1$ | $[Adj(SO(16)_2)]_{4,2}^1$ | ✕ | † | 🌐 | 🎀 | |
| $[FR_1^{7,1,0}]_{4,2}^2$ | $[Adj(SO(16)_2)]_{4,2}^2$ | ✕ | | 🌐 | 🎀 | |
| $[FR_1^{7,1,0}]_{4,3}^1$ | $[Adj(SO(16)_2)]_{4,3}^1$ | ✕ | † | 🌐 | 🎀 | |
| $[FR_1^{7,1,0}]_{4,3}^2$ | $[Adj(SO(16)_2)]_{4,3}^2$ | ✕ | | 🌐 | 🎀 | |
| $[FR_1^{7,1,0}]_{4,4}^1$ | $[Adj(SO(16)_2)]_{4,4}^1$ | ✕ | † | 🌐 | 🎀 | |
| $[FR_1^{7,1,0}]_{4,4}^2$ | $[Adj(SO(16)_2)]_{4,4}^2$ | ✕ | | 🌐 | 🎀 | |
| $[FR_1^{7,1,0}]_{5,1}^1$ | $[Adj(SO(16)_2)]_{5,1}^1$ | | † | 🌐 | | |
| $[FR_1^{7,1,0}]_{5,1}^2$ | $[Adj(SO(16)_2)]_{5,1}^2$ | | | 🌐 | | |
| $[FR_1^{7,1,0}]_{6,1}^1$ | $[Adj(SO(16)_2)]_{6,1}^1$ | | † | 🌐 | | |
| $[FR_1^{7,1,0}]_{6,1}^2$ | $[Adj(SO(16)_2)]_{6,1}^2$ | | | 🌐 | | |
| $[FR_1^{7,1,0}]_{7,1}^1$ | $[Adj(SO(16)_2)]_{7,1}^1$ | | † | 🌐 | | |
| $[FR_1^{7,1,0}]_{7,1}^2$ | $[Adj(SO(16)_2)]_{7,1}^2$ | | | 🌐 | | |
| $[FR_1^{7,1,0}]_{8,1}^1$ | $[Adj(SO(16)_2)]_{8,1}^1$ | | † | 🌐 | | |
| $[FR_1^{7,1,0}]_{8,1}^2$ | $[Adj(SO(16)_2)]_{8,1}^2$ | | | 🌐 | | |
| $[FR_6^{7,1,0}]_{1,1}^1$ | $[Adj(SO(11)_2)]_{1,1}^1$ | ✕ | † | 🌐 | 🎀 | |
| $[FR_6^{7,1,0}]_{1,2}^1$ | $[Adj(SO(11)_2)]_{1,2}^1$ | ✕ | † | 🌐 | 🎀 | |
| $[FR_6^{7,1,0}]_{1,3}^1$ | $[Adj(SO(11)_2)]_{1,3}^1$ | ✕ | † | 🌐 | 🎀 | |
| $[FR_6^{7,1,0}]_{2,1}^1$ | $[Adj(SO(11)_2)]_{2,1}^1$ | | † | 🌐 | | |
| $[FR_6^{7,1,0}]_{3,1}^1$ | $[Adj(SO(11)_2)]_{3,1}^1$ | | † | 🌐 | | |
| $[FR_7^{7,1,0}]_{1,1}^1$ | $[SU(2)_6]_{1,1}^1$ | ✕ | † | 🌐 | 🎀 | ⭕ |
| $[FR_7^{7,1,0}]_{1,1}^2$ | $[SU(2)_6]_{1,1}^2$ | ✕ | | 🌐 | 🎀 | ⭕ |
| $[FR_7^{7,1,0}]_{1,2}^1$ | $[SU(2)_6]_{1,2}^1$ | ✕ | † | 🌐 | 🎀 | ⭕ |
| $[FR_7^{7,1,0}]_{1,2}^2$ | $[SU(2)_6]_{1,2}^2$ | ✕ | | 🌐 | 🎀 | ⭕ |
| $[FR_7^{7,1,0}]_{1,3}^1$ | $[SU(2)_6]_{1,3}^1$ | ✕ | † | 🌐 | 🎀 | ⭕ |
| $[FR_7^{7,1,0}]_{1,3}^2$ | $[SU(2)_6]_{1,3}^2$ | ✕ | | 🌐 | 🎀 | ⭕ |
| $[FR_7^{7,1,0}]_{1,4}^1$ | $[SU(2)_6]_{1,4}^1$ | ✕ | † | 🌐 | 🎀 | ⭕ |
| $[FR_7^{7,1,0}]_{1,4}^2$ | $[SU(2)_6]_{1,4}^2$ | ✕ | | 🌐 | 🎀 | ⭕ |

Table 9.1: List of multiplicity-free fusion categories up to rank 7 (Continued)

| Formal Name | Common Name | | | | | |
|---|---|---|---|---|---|---|
| $[FR_7^{7,1,0}]_{2,1}^1$ | $[SU(2)_6]_{2,1}^1$ | ✗ | | 🌐 | 🎀 | ⭕ |
| $[FR_7^{7,1,0}]_{2,1}^2$ | $[SU(2)_6]_{2,1}^2$ | ✗ | | 🌐 | 🎀 | ⭕ |
| $[FR_7^{7,1,0}]_{2,2}^1$ | $[SU(2)_6]_{2,2}^1$ | ✗ | | 🌐 | 🎀 | ⭕ |
| $[FR_7^{7,1,0}]_{2,2}^2$ | $[SU(2)_6]_{2,2}^2$ | ✗ | | 🌐 | 🎀 | ⭕ |
| $[FR_7^{7,1,0}]_{2,3}^1$ | $[SU(2)_6]_{2,3}^1$ | ✗ | | 🌐 | 🎀 | ⭕ |
| $[FR_7^{7,1,0}]_{2,3}^2$ | $[SU(2)_6]_{2,3}^2$ | ✗ | | 🌐 | 🎀 | ⭕ |
| $[FR_7^{7,1,0}]_{2,4}^1$ | $[SU(2)_6]_{2,4}^1$ | ✗ | | 🌐 | 🎀 | ⭕ |
| $[FR_7^{7,1,0}]_{2,4}^2$ | $[SU(2)_6]_{2,4}^2$ | ✗ | | 🌐 | 🎀 | ⭕ |
| $[FR_8^{7,1,0}]_{1,1}^1$ | $[SO(7)_2]_{1,1}^1$ | ✗ | † | 🌐 | 🎀 | ⭕ |
| $[FR_8^{7,1,0}]_{1,1}^2$ | $[SO(7)_2]_{1,1}^2$ | ✗ | | 🌐 | 🎀 | ⭕ |
| $[FR_8^{7,1,0}]_{1,2}^1$ | $[SO(7)_2]_{1,2}^1$ | ✗ | † | 🌐 | 🎀 | ⭕ |
| $[FR_8^{7,1,0}]_{1,2}^2$ | $[SO(7)_2]_{1,2}^2$ | ✗ | | 🌐 | 🎀 | ⭕ |
| $[FR_8^{7,1,0}]_{2,1}^1$ | $[SO(7)_2]_{2,1}^1$ | ✗ | † | 🌐 | 🎀 | ⭕ |
| $[FR_8^{7,1,0}]_{2,1}^2$ | $[SO(7)_2]_{2,1}^2$ | ✗ | | 🌐 | 🎀 | ⭕ |
| $[FR_8^{7,1,0}]_{2,2}^1$ | $[SO(7)_2]_{2,2}^1$ | ✗ | † | 🌐 | 🎀 | ⭕ |
| $[FR_8^{7,1,0}]_{2,2}^2$ | $[SO(7)_2]_{2,2}^2$ | ✗ | | 🌐 | 🎀 | ⭕ |
| $[FR_{14}^{7,1,0}]_{1,1}^1$ | $[PSU(2)_{12}]_{1,1}^1$ | ✗ | † | 🌐 | 🎀 | |
| $[FR_{14}^{7,1,0}]_{1,2}^1$ | $[PSU(2)_{12}]_{1,2}^1$ | ✗ | † | 🌐 | 🎀 | |
| $[FR_{14}^{7,1,0}]_{2,1}^1$ | $[PSU(2)_{12}]_{2,1}^1$ | ✗ | | 🌐 | 🎀 | |
| $[FR_{14}^{7,1,0}]_{2,2}^1$ | $[PSU(2)_{12}]_{2,2}^1$ | ✗ | | 🌐 | 🎀 | |
| $[FR_{14}^{7,1,0}]_{3,1}^1$ | $[PSU(2)_{12}]_{3,1}^1$ | ✗ | | 🌐 | 🎀 | |
| $[FR_{14}^{7,1,0}]_{3,2}^1$ | $[PSU(2)_{12}]_{3,2}^1$ | ✗ | | 🌐 | 🎀 | |
| $[FR_{17}^{7,1,0}]_{1,1}^1$ | $[PSU(2)_{13}]_{1,1}^1$ | ✗ | † | 🌐 | 🎀 | ⭕ |
| $[FR_{17}^{7,1,0}]_{1,2}^1$ | $[PSU(2)_{13}]_{1,2}^1$ | ✗ | † | 🌐 | 🎀 | ⭕ |
| $[FR_{17}^{7,1,0}]_{2,1}^1$ | $[PSU(2)_{13}]_{2,1}^1$ | ✗ | | 🌐 | 🎀 | ⭕ |
| $[FR_{17}^{7,1,0}]_{2,2}^1$ | $[PSU(2)_{13}]_{2,2}^1$ | ✗ | | 🌐 | 🎀 | ⭕ |
| $[FR_{17}^{7,1,0}]_{3,1}^1$ | $[PSU(2)_{13}]_{3,1}^1$ | ✗ | | 🌐 | 🎀 | ⭕ |
| $[FR_{17}^{7,1,0}]_{3,2}^1$ | $[PSU(2)_{13}]_{3,2}^1$ | ✗ | | 🌐 | 🎀 | ⭕ |
| $[FR_{17}^{7,1,0}]_{4,1}^1$ | $[PSU(2)_{13}]_{4,1}^1$ | ✗ | | 🌐 | 🎀 | ⭕ |

Table 9.1: List of multiplicity-free fusion categories up to rank 7 (Continued)

| Formal Name | Common Name | | | | |
|---|---|---|---|---|---|
| $[FR_{17}^{7,1,0}]_{4,2}^{1}$ | $[PSU(2)_{13}]_{4,2}^{1}$ | ✕ | | 🌐 | 🎗 ⚭ |
| $[FR_{3}^{7,1,2}]_{1,1}^{1}$ | | | † | 🌐 | |
| $[FR_{3}^{7,1,2}]_{1,1}^{2}$ | | | | 🌐 | |
| $[FR_{3}^{7,1,2}]_{2,1}^{1}$ | | | † | 🌐 | |
| $[FR_{3}^{7,1,2}]_{2,1}^{2}$ | | | | 🌐 | |
| $[FR_{3}^{7,1,2}]_{3,1}^{1}$ | | | † | 🌐 | |
| $[FR_{3}^{7,1,2}]_{3,1}^{2}$ | | | | 🌐 | |
| $[FR_{3}^{7,1,2}]_{4,1}^{1}$ | | | † | 🌐 | |
| $[FR_{3}^{7,1,2}]_{4,1}^{2}$ | | | | 🌐 | |
| $[FR_{3}^{7,1,2}]_{5,1}^{1}$ | | | † | 🌐 | |
| $[FR_{3}^{7,1,2}]_{5,1}^{2}$ | | | | 🌐 | |
| $[FR_{3}^{7,1,2}]_{6,1}^{1}$ | | | † | 🌐 | |
| $[FR_{3}^{7,1,2}]_{6,1}^{2}$ | | | | 🌐 | |
| $[FR_{3}^{7,1,2}]_{7,1}^{1}$ | | | † | 🌐 | |
| $[FR_{3}^{7,1,2}]_{7,1}^{2}$ | | | | 🌐 | |
| $[FR_{3}^{7,1,2}]_{8,1}^{1}$ | | | † | 🌐 | |
| $[FR_{3}^{7,1,2}]_{8,1}^{2}$ | | | | 🌐 | |
| $[FR_{4}^{7,1,2}]_{1,1}^{1}$ | | ✕ | † | 🌐 | 🎗 |
| $[FR_{4}^{7,1,2}]_{1,1}^{2}$ | | ✕ | | 🌐 | 🎗 |
| $[FR_{4}^{7,1,2}]_{1,2}^{1}$ | | ✕ | † | 🌐 | 🎗 |
| $[FR_{4}^{7,1,2}]_{1,2}^{2}$ | | ✕ | | 🌐 | 🎗 |
| $[FR_{4}^{7,1,2}]_{1,3}^{1}$ | | ✕ | † | 🌐 | 🎗 |
| $[FR_{4}^{7,1,2}]_{1,3}^{2}$ | | ✕ | | 🌐 | 🎗 |
| $[FR_{4}^{7,1,2}]_{1,4}^{1}$ | | ✕ | † | 🌐 | 🎗 |
| $[FR_{4}^{7,1,2}]_{1,4}^{2}$ | | ✕ | | 🌐 | 🎗 |
| $[FR_{4}^{7,1,2}]_{1,5}^{1}$ | | ✕ | † | 🌐 | 🎗 |
| $[FR_{4}^{7,1,2}]_{1,5}^{2}$ | | ✕ | | 🌐 | 🎗 |
| $[FR_{4}^{7,1,2}]_{1,6}^{1}$ | | ✕ | † | 🌐 | 🎗 |
| $[FR_{4}^{7,1,2}]_{1,6}^{2}$ | | ✕ | | 🌐 | 🎗 |

Table 9.1: List of multiplicity-free fusion categories up to rank 7 (Continued)

| Formal Name | Common Name | | | | |
|---|---|---|---|---|---|
| $[FR_4^{7,1,2}]_{1,7}^1$ | | ✕ | † | 🌐 | 🎗 |
| $[FR_4^{7,1,2}]_{1,7}^2$ | | ✕ | | 🌐 | 🎗 |
| $[FR_4^{7,1,2}]_{1,8}^1$ | | ✕ | † | 🌐 | 🎗 |
| $[FR_4^{7,1,2}]_{1,8}^2$ | | ✕ | | 🌐 | 🎗 |
| $[FR_4^{7,1,2}]_{2,1}^1$ | | | † | 🌐 | |
| $[FR_4^{7,1,2}]_{2,1}^2$ | | | | 🌐 | |
| $[FR_4^{7,1,2}]_{3,1}^1$ | | | † | 🌐 | |
| $[FR_4^{7,1,2}]_{3,1}^2$ | | | | 🌐 | |
| $[FR_4^{7,1,2}]_{4,1}^1$ | | | † | 🌐 | |
| $[FR_4^{7,1,2}]_{4,1}^2$ | | | | 🌐 | |
| $[FR_{12}^{7,1,2}]_{1,1}^1$ | | | † | 🌐 | |
| $[FR_{12}^{7,1,2}]_{1,1}^2$ | | | | 🌐 | |
| $[FR_{12}^{7,1,2}]_{2,1}^1$ | | | † | 🌐 | |
| $[FR_{12}^{7,1,2}]_{2,1}^2$ | | | | 🌐 | |
| $[FR_1^{7,1,4}]_{1,1}^1$ | $[TY(\mathbb{Z}_2 \times \mathbb{Z}_3)]_{1,1}^1$ | | † | 🌐 | |
| $[FR_1^{7,1,4}]_{1,1}^2$ | $[TY(\mathbb{Z}_2 \times \mathbb{Z}_3)]_{1,1}^2$ | | | 🌐 | |
| $[FR_1^{7,1,4}]_{2,1}^1$ | $[TY(\mathbb{Z}_2 \times \mathbb{Z}_3)]_{2,1}^1$ | | † | 🌐 | |
| $[FR_1^{7,1,4}]_{2,1}^2$ | $[TY(\mathbb{Z}_2 \times \mathbb{Z}_3)]_{2,1}^2$ | | | 🌐 | |
| $[FR_1^{7,1,4}]_{3,1}^1$ | $[TY(\mathbb{Z}_2 \times \mathbb{Z}_3)]_{3,1}^1$ | | † | 🌐 | |
| $[FR_1^{7,1,4}]_{3,1}^2$ | $[TY(\mathbb{Z}_2 \times \mathbb{Z}_3)]_{3,1}^2$ | | | 🌐 | |
| $[FR_1^{7,1,4}]_{4,1}^1$ | $[TY(\mathbb{Z}_2 \times \mathbb{Z}_3)]_{4,1}^1$ | | † | 🌐 | |
| $[FR_1^{7,1,4}]_{4,1}^2$ | $[TY(\mathbb{Z}_2 \times \mathbb{Z}_3)]_{4,1}^2$ | | | 🌐 | |
| $[FR_3^{7,1,4}]_{1,1}^1$ | | | † | 🌐 | |
| $[FR_3^{7,1,4}]_{1,1}^2$ | | | | 🌐 | |
| $[FR_3^{7,1,4}]_{2,1}^1$ | | | † | 🌐 | |
| $[FR_3^{7,1,4}]_{2,1}^2$ | | | | 🌐 | |
| $[FR_3^{7,1,4}]_{3,1}^1$ | | | † | 🌐 | |
| $[FR_3^{7,1,4}]_{3,1}^2$ | | | | 🌐 | |

Table 9.1: List of multiplicity-free fusion categories up to rank 7 (Continued)

| Formal Name | Common Name | | | | | |
|---|---|---|---|---|---|---|
| $[FR_3^{7,1,4}]_{4,1}^1$ | | |  |  | | |
| $[FR_3^{7,1,4}]_{4,1}^2$ | | | |  | | |
| $[FR_1^{7,1,6}]_{1,1}^1$ | $[\mathbb{Z}_7]_{1,1}^1$ |  |  |  |  |  |
| $[FR_1^{7,1,6}]_{1,1}^2$ | $[\mathbb{Z}_7]_{1,1}^2$ |  | | | | |
| $[FR_1^{7,1,6}]_{1,1}^3$ | $[\mathbb{Z}_7]_{1,1}^3$ |  | | | | |
| $[FR_1^{7,1,6}]_{1,1}^4$ | $[\mathbb{Z}_7]_{1,1}^4$ |  | | | | |
| $[FR_1^{7,1,6}]_{1,2}^1$ | $[\mathbb{Z}_7]_{1,2}^1$ |  |  |  |  |  |
| $[FR_1^{7,1,6}]_{1,2}^2$ | $[\mathbb{Z}_7]_{1,2}^2$ |  | | | | |
| $[FR_1^{7,1,6}]_{1,2}^3$ | $[\mathbb{Z}_7]_{1,2}^3$ |  | | | | |
| $[FR_1^{7,1,6}]_{1,2}^4$ | $[\mathbb{Z}_7]_{1,2}^4$ |  | | | | |
| $[FR_1^{7,1,6}]_{1,3}^1$ | $[\mathbb{Z}_7]_{1,3}^1$ |  |  |  |  | |
| $[FR_1^{7,1,6}]_{1,3}^2$ | $[\mathbb{Z}_7]_{1,3}^2$ |  | | | | |
| $[FR_1^{7,1,6}]_{2,1}^1$ | $[\mathbb{Z}_7]_{2,1}^1$ | |  |  | | |
| $[FR_1^{7,1,6}]_{2,1}^2$ | $[\mathbb{Z}_7]_{2,1}^2$ | | | | | |
| $[FR_1^{7,1,6}]_{2,1}^3$ | $[\mathbb{Z}_7]_{2,1}^3$ | | | | | |
| $[FR_1^{7,1,6}]_{2,1}^4$ | $[\mathbb{Z}_7]_{2,1}^4$ | | | | | |
| $[FR_1^{7,1,6}]_{3,1}^1$ | $[\mathbb{Z}_7]_{3,1}^1$ | |  |  | | |
| $[FR_1^{7,1,6}]_{3,1}^2$ | $[\mathbb{Z}_7]_{3,1}^2$ | | | | | |
| $[FR_1^{7,1,6}]_{3,1}^3$ | $[\mathbb{Z}_7]_{3,1}^3$ | | | | | |
| $[FR_1^{7,1,6}]_{3,1}^4$ | $[\mathbb{Z}_7]_{3,1}^4$ | | | | | |

# Chapter 10

# Solutions For Specific Graph-Braid Models

## 10.1 Solutions for the Ising Model

The Ising fusion ring has 3 particles, $1, \psi, \sigma$ subject to the multiplication rules

$$1 \times a = a \times 1 = a, \quad \forall\, a \in \{1, \psi, \sigma\}, \tag{10.1}$$

$$\psi \times \psi = 1, \quad \sigma \times \psi = \psi \times \sigma = \sigma, \quad \sigma \times \sigma = 1 + \psi. \tag{10.2}$$

In the following section we will list the solutions to various equations for the Ising model. To save space we omit any well-defined symbol equal to 1. By well-defined we mean that the fusion tree corresponding to the symbol exists.

### 10.1.1 Solutions to the Pentagon Equations

There are two solutions to the pentagon equations for the Ising fusion ring. Both solutions share the same values for the following $F$-symbols

$$[F_\sigma^{\psi\sigma\psi}]_\sigma^\sigma = [F_1^{\sigma\psi\sigma}]_\sigma^\sigma = [F_1^{\sigma\sigma\psi}]_\sigma^\psi = [F_\psi^{\sigma\sigma\psi}]_\sigma^1 = -1, \tag{10.3}$$

but have a different sign for the $F$-matrix

$$[F_\sigma^{\sigma\sigma\sigma}] = \pm \frac{1}{\sqrt{2}} \begin{bmatrix} 1 & -1 \\ 1 & 1 \end{bmatrix}. \tag{10.4}$$

We will denote these solutions by $\mathscr{F}_\kappa$ where $\kappa = \pm 1$. Note that some of the $F$-symbols in these solutions are gauge dependent and so they may differ from those in other works. In, e.g. [81] a gauge is used such that

$$[F_\sigma^{\sigma\sigma\sigma}] = \pm \frac{1}{\sqrt{2}} \begin{bmatrix} 1 & 1 \\ 1 & -1 \end{bmatrix}. \tag{10.5}$$

As a matter of fact, the *F*-matrices of the Ising model that are currently stored in Anyonica are also in a diagonal gauge, but they weren't at the time we did these calculations. Obviously this has no qualitative consequences for the results.

### 10.1.2 Solutions to the Planar Hexagon Equations

Each set of *F*-symbols allows four solutions to the planar hexagon equations. They can be parameterized as follows:

$$R_1^{\psi\psi} = -1, R_\sigma^{\psi\sigma} = R_\sigma^{\sigma\psi} = \varepsilon_1 i, R_1^{\sigma\sigma} = \left(\frac{\kappa\varepsilon_1}{i}\right)^{\frac{1+\kappa}{2}} i^{\varepsilon_2+1} e^{-\varepsilon_1 \frac{i\pi}{8}}, R_1^{\sigma\sigma} = \left(\frac{\kappa\varepsilon_1}{i}\right)^{\frac{1-\kappa}{2}} i^{\varepsilon_2+1} e^{-\varepsilon_1 \frac{i\pi}{8}},$$

where $\varepsilon_i \in \{-1, 1\}$.

### 10.1.3 Solutions to the Trijunction Equations

#### 10.1.3.1 Three Particles

For three particles each solution to the pentagon equation gives rise to two classes of solutions to the trijunction equations. Each class of solutions are parameterized by two complex phases $z_1, z_2$. To save space we will not denote the symbols $P_{ed}^{ab1} \equiv Q_{ed}^{ab1}$ since these are equal to the *R*-symbols $R_e^{ab}$. The four combinations of *F*- and *R*-symbols have the following form

$$
\begin{aligned}
&P_{1\psi}^{\psi\psi\psi} = \tfrac{1}{z_1}, & &P_{1\sigma}^{\psi\psi\sigma} = -1, & &P_{\sigma\sigma}^{\psi\sigma\psi} = \tfrac{-\varepsilon i}{z_1}, & &P_{\sigma 1}^{\psi\sigma\sigma} = -\varepsilon i z_1, & &P_{\sigma\psi}^{\psi\sigma\sigma} = \varepsilon i, \\
&P_{\sigma\sigma}^{\sigma\psi\psi} = \varepsilon i, & &P_{\sigma 1}^{\sigma\psi\sigma} = \varepsilon i, & &P_{\sigma\psi}^{\sigma\psi\sigma} = \varepsilon i, & &P_{1\psi}^{\sigma\sigma\psi} = z_2, & &P_{\psi 1}^{\sigma\sigma\psi} = \varepsilon i z_2, \\
&P_{1\sigma}^{\sigma\sigma\sigma} = \tfrac{\kappa}{z_2} e^{-\frac{\varepsilon i\pi}{4}}, & &P_{\psi\sigma}^{\sigma\sigma\sigma} = \tfrac{\kappa}{z_2} e^{\frac{\varepsilon i\pi}{4}}, \\
&Q_{1\psi}^{\psi\psi\psi} = \tfrac{1}{z_1}, & &Q_{1\sigma}^{\psi\psi\sigma} = -1, & &Q_{\sigma\sigma}^{\psi\sigma\psi} = \varepsilon i, & &Q_{\sigma 1}^{\psi\sigma\sigma} = \varepsilon i, \\
&Q_{\sigma\psi}^{\psi\sigma\sigma} = \varepsilon i, & &Q_{\sigma\sigma}^{\sigma\psi\psi} = \tfrac{-\varepsilon i}{z_1}, & &Q_{\sigma 1}^{\sigma\psi\sigma} = -\varepsilon i z_1, & &Q_{\sigma\psi}^{\sigma\psi\sigma} = \varepsilon i, & &Q_{1\psi}^{\sigma\sigma\psi} = z_2, \\
&Q_{\psi 1}^{\sigma\sigma\psi} = \varepsilon i z_2, & &Q_{1\sigma}^{\sigma\sigma\sigma} = \tfrac{\kappa}{z_2} e^{-\frac{\varepsilon i\pi}{4}}, & &Q_{\psi\sigma}^{\sigma\sigma\sigma} = \tfrac{\kappa}{z_2} e^{\frac{\varepsilon i\pi}{4}}, \\
&R_1^{\psi\psi} = z_1, & &R_\sigma^{\psi\sigma} = \varepsilon i, & &R_\sigma^{\sigma\psi} = \varepsilon i, & &R_1^{\sigma\sigma} = z_2, & &R_\psi^{\sigma\sigma} = \varepsilon i z_2,
\end{aligned}
$$

where $\varepsilon \in \{-1, 1\}$. We will label each solution by $\mathcal{T}_{\kappa,\varepsilon}^{(3)}$.

#### 10.1.3.2 Four Particles

For four particles, the trijunction equations can only be satisfied if $z_1 = -1$ (which implies $R_1^{\psi\psi} = -1$) and therefore the solutions have the property that $P \equiv Q$. Note that this does not necessarily imply that $P \equiv R$, i.e. that the solutions are planar. The solutions are then described by adding to the $\mathcal{T}_{\kappa,\varepsilon}^{(3)}$ the respective values of the *A*, *B*, *X*, and *Y*-symbols which we will describe here. All symbols with a 1 as the third or fourth top label are *P*, *Q*, or *R*-symbols and will therefore not be listed.

For the Ising model, it turns out that all symbols with the same labels are equal to each other. We can thus write the solutions in terms of the symbol $M$, where $M$ could be any of $A, B, X, Y$. The solutions then read

$$M^{\psi\psi cd}_{fge} \equiv -1 \tag{10.6}$$

$$M^{\sigma\psi cd}_{fge} \equiv M^{\psi\sigma cd}_{fge} \equiv \varepsilon i \tag{10.7}$$

$$M^{\sigma\sigma cd}_{fge} = \begin{cases} z_2 & \text{if } c = d \text{ and } f = 1 \\ \varepsilon i z_2 & \text{if } c = d \text{ and } f = \psi \\ \frac{\kappa}{z_2} \exp\left(\frac{-\varepsilon i \pi}{4}\right) & \text{if } c \neq d \text{ and } f = 1 \\ \frac{\kappa}{z_2} \exp\left(\frac{\varepsilon i \pi}{4}\right) & \text{if } c \neq d \text{ and } f = \psi \end{cases} \tag{10.8}$$

where $c, d \in \{\psi, \sigma\}$, $f, g, e \in \{1, \psi, \sigma\}$, and the value of $\kappa$ and $\varepsilon$ are fixed by the choice of $\mathcal{T}^{(3)}_{\kappa,\varepsilon}$.

### 10.1.4 Solutions to the Lollipop Equations

#### 10.1.4.1 Lollipop Trijunction

For the Ising model on the trijunction, on the lollipop, the demand that $P^{abc}_{ed} \equiv R^{ab}_{e}$ implies that the solutions must be planar. In particular, the solutions are the four solutions for the planar hexagon equations with the addition of the $P$-and $Q$-symbols, which obey $P^{abc}_{ed} \equiv Q^{abc}_{ed} \equiv R^{ab}_{e}$.

#### 10.1.4.2 Circle Solutions

There are sixteen solutions to the circle equations for each set of $F$-symbols. They can be written as

$$D^{1\psi}_{\psi} = -1, \; D^{1\sigma}_{\sigma} = e^{i\pi \frac{-2-\nu_1+4\nu_2-2\kappa}{8}}, D^{\psi\sigma}_{\sigma} = -\nu_1 e^{i\pi \frac{2-\nu_1+4\nu_2-2\kappa}{8}},$$
$$D^{\sigma\psi}_{\sigma} = \nu_1 i, \; D^{\sigma\sigma}_{1} = \nu_3, \qquad\qquad D^{\sigma\sigma}_{\psi} = \nu_4 i,$$

where the $\nu_i \in \{-1, 1\}$ and $\kappa$ is fixed by the choice of $F$-symbols. In particular we find that, per set of $F$-symbols, there are four possible values for the generalized topological spins. These coincide with the values of the topological spins for planar Ising anyons.

#### 10.1.4.3 Full Lollipop Solutions

There are 32 solutions to the full lollipop equations per set of $F$-symbols. Because there is no gauge freedom left after fixing a set of $F$-symbols, for a given set of $F$-symbols any solution can be found by combining a solution to the lollipop trijunction equations with matching label $\kappa$ with a solution to the circle equations with matching label $\kappa$.

## 10.2 Solutions for the Quantum Double of $\mathbb{Z}_2$

The quantum double of $\mathbb{Z}_2$ is a model with four anyons $1, e, m, \varepsilon$ that follow the fusion rules of $\mathbb{Z}_2 \times \mathbb{Z}_2$ (via, e.g., the identification $1 = (0,0), e = (1,0), m = (0,1), \varepsilon = (1,1)$) and for which $[F_d^{abc}]_f^e \equiv 1$ for each well-defined $F$-symbol. This model arises as the excitations in the Toric code model with gauge group $\mathbb{Z}_2$ [55].

### 10.2.1 Solutions to the Planar Hexagon Equations

There are eight gauge-independent planar hexagon solutions:

$$R_1^{\varepsilon\varepsilon} = \nu_1, \quad R_m^{\varepsilon e} = \nu_2, \quad R_e^{\varepsilon m} = \nu_1 \nu_2, \quad R_1^{ee} = \nu_3,$$
$$R_\varepsilon^{em} = \nu_3, \quad R_e^{m\varepsilon} = \nu_1, \quad R_\varepsilon^{me} = \nu_2 \nu_3, \quad R_1^{mm} = \nu_1 \nu_2 \nu_3,$$

where $\nu_i \in \{-1, 1\}$.

### 10.2.2 Solutions to the Trijunction Equations

The trijunction equations that impose constraints on any of the $R$-symbols are trivially satisfied. Therefore, we find that all non-trivial $R$-symbols are free parameters

$$R_1^{\varepsilon\varepsilon} = z_1, \quad R_m^{\varepsilon e} = z_2, \quad R_e^{\varepsilon m} = z_3, \quad R_m^{ee} = z_4, \quad R_1^{ee} = z_5,$$
$$R_\varepsilon^{em} = z_6, \quad R_e^{m\varepsilon} = z_7, \quad R_\varepsilon^{me} = z_8, \quad R_1^{mm} = z_9,$$

and all other symbols can be expressed in terms of these free parameters. To save space we will omit the symbols $P_{ed}^{ab1} \equiv Q_{ed}^{ab1}$ since these are equal to the $R$-symbols $R_e^{ab}$. The $P$-and $Q$-symbols are the following

$$
\begin{array}{llllllll}
P_{1\varepsilon}^{\varepsilon\varepsilon\varepsilon} = \frac{1}{z_1}, & P_{me}^{\varepsilon\varepsilon\varepsilon} = \frac{1}{z_4}, & P_{em}^{m\varepsilon\varepsilon} = \frac{1}{z_7}, & Q_{1\varepsilon}^{\varepsilon\varepsilon\varepsilon} = \frac{1}{z_1}, & Q_{me}^{\varepsilon\varepsilon\varepsilon} = \frac{z_7}{z_1}, & Q_{em}^{m\varepsilon\varepsilon} = \frac{z_4}{z_1}, \\
P_{1e}^{\varepsilon\varepsilon e} = \frac{z_3}{z_2}, & P_{me}^{\varepsilon e\varepsilon} = \frac{z_6}{z_5}, & P_{e1}^{mee} = \frac{z_9}{z_8}, & Q_{1e}^{\varepsilon\varepsilon\varepsilon} = \frac{z_7}{z_4}, & Q_{m\varepsilon}^{\varepsilon\varepsilon} = \frac{1}{z_4}, & Q_{e1}^{m\varepsilon e} = \frac{z_1}{z_4}, \\
P_{1m}^{\varepsilon\varepsilon m} = \frac{z_2}{z_3}, & P_{m1}^{\varepsilon em} = \frac{z_5}{z_6}, & P_{ee}^{m\varepsilon m} = \frac{z_8}{z_9}, & Q_{1m}^{\varepsilon\varepsilon m} = \frac{z_4}{z_7}, & Q_{m1}^{\varepsilon em} = \frac{z_1}{z_7}, & Q_{e\varepsilon}^{mem} = \frac{1}{z_7}, \\
P_{me}^{\varepsilon e\varepsilon} = \frac{z_3}{z_1}, & P_{1\varepsilon}^{ee\varepsilon} = \frac{z_6}{z_4}, & P_{\varepsilon 1}^{me\varepsilon} = \frac{z_9}{z_7}, & Q_{me}^{\varepsilon\varepsilon\varepsilon} = \frac{1}{z_2}, & Q_{1\varepsilon}^{ee\varepsilon} = \frac{z_8}{z_2}, & Q_{\varepsilon 1}^{me\varepsilon} = \frac{z_5}{z_2}, \\
P_{m\varepsilon}^{\varepsilon ee} = \frac{1}{z_2}, & P_{1e}^{eee} = \frac{1}{z_5}, & P_{\varepsilon m}^{mee} = \frac{1}{z_8}, & Q_{m\varepsilon}^{\varepsilon ee} = \frac{z_8}{z_5}, & Q_{1e}^{eee} = \frac{1}{z_5}, & Q_{\varepsilon m}^{mee} = \frac{z_2}{z_5}, \\
P_{m1}^{\varepsilon em} = \frac{z_1}{z_3}, & P_{1m}^{eem} = \frac{z_4}{z_6}, & P_{\varepsilon e}^{mem} = \frac{z_7}{z_9}, & Q_{m1}^{\varepsilon em} = \frac{z_5}{z_8}, & Q_{1m}^{eem} = \frac{z_2}{z_8}, & Q_{\varepsilon e}^{mem} = \frac{1}{z_8}, \\
P_{em}^{\varepsilon m\varepsilon} = \frac{z_2}{z_1}, & P_{\varepsilon 1}^{em\varepsilon} = \frac{z_5}{z_4}, & P_{1\varepsilon}^{mm\varepsilon} = \frac{z_8}{z_7}, & Q_{em}^{\varepsilon m\varepsilon} = \frac{1}{z_3}, & Q_{\varepsilon 1}^{em\varepsilon} = \frac{z_9}{z_3}, & Q_{1\varepsilon}^{mm\varepsilon} = \frac{z_6}{z_3}, \\
P_{e1}^{\varepsilon me} = \frac{z_1}{z_2}, & P_{\varepsilon m}^{eme} = \frac{z_4}{z_5}, & P_{1e}^{mme} = \frac{z_7}{z_8}, & Q_{e1}^{\varepsilon me} = \frac{z_9}{z_6}, & Q_{\varepsilon m}^{eme} = \frac{1}{z_6}, & Q_{1e}^{mme} = \frac{z_3}{z_6}, \\
P_{e\varepsilon}^{\varepsilon mm} = \frac{1}{z_3}, & P_{\varepsilon e}^{emm} = \frac{1}{z_6}, & P_{1m}^{mmm} = \frac{1}{z_9}, & Q_{e\varepsilon}^{\varepsilon mm} = \frac{z_6}{z_9}, & Q_{\varepsilon e}^{emm} = \frac{z_3}{z_9}, & Q_{1m}^{mmm} = \frac{1}{z_9}.
\end{array}
$$

We can observe some interesting features in this table, namely when all of the particles are of the same type we find $P^{aaa} = Q^{aaa}$.

### 10.2.2.1 Four Particles

For four particles we have the following solutions.

| | Value of $M$ | | | | | Value of $M$ | | | | | Value of $M$ | | | |
|---|---|---|---|---|---|---|---|---|---|---|---|---|---|---|
| | $A$ | $B$ | $X$ | $Y$ | | $A$ | $B$ | $X$ | $Y$ | | $A$ | $B$ | $X$ | $Y$ |
| $M^{\varepsilon\varepsilon\varepsilon\varepsilon}_{1\varepsilon 1}$ | $z_1$ | $z_1$ | $z_1$ | $z_1$ | $M^{\varepsilon\varepsilon\varepsilon\varepsilon}_{mem}$ | $\frac{z_1}{z_7}$ | $\frac{z_1}{z_7}$ | $z_4$ | $z_4$ | $M^{m\varepsilon\varepsilon\varepsilon}_{eme}$ | $\frac{z_1}{z_4}$ | $\frac{z_1}{z_4}$ | $z_7$ | $z_7$ |
| $M^{\varepsilon\varepsilon\varepsilon\varepsilon}_{1\varepsilon m}$ | $\frac{z_2}{z_3}$ | $\frac{z_4}{z_7}$ | $\frac{z_2}{z_3}$ | $\frac{z_4}{z_7}$ | $M^{\varepsilon\varepsilon\varepsilon\varepsilon}_{me1}$ | $\frac{z_2 z_9}{z_3 z_8}$ | $z_4$ | $\frac{z_5}{z_6}$ | $\frac{z_1}{z_7}$ | $M^{m\varepsilon\varepsilon\varepsilon}_{em\varepsilon}$ | $\frac{z_2 z_6}{z_3 z_5}$ | $\frac{z_4}{z_1}$ | $\frac{z_8}{z_9}$ | $\frac{1}{z_7}$ |
| $M^{\varepsilon\varepsilon\varepsilon m}_{1\varepsilon e}$ | $\frac{z_3}{z_2}$ | $\frac{z_7}{z_4}$ | $\frac{z_3}{z_2}$ | $\frac{z_7}{z_4}$ | $M^{\varepsilon\varepsilon\varepsilon m}_{me\varepsilon}$ | $\frac{z_3 z_8}{z_2 z_9}$ | $\frac{z_7}{z_1}$ | $\frac{z_6}{z_5}$ | $\frac{1}{z_4}$ | $M^{m\varepsilon\varepsilon m}_{em1}$ | $\frac{z_3 z_5}{z_2 z_6}$ | $z_7$ | $\frac{z_9}{z_8}$ | $\frac{z_1}{z_4}$ |
| $M^{\varepsilon\varepsilon\varepsilon\varepsilon}_{1em}$ | $\frac{z_4}{z_7}$ | $\frac{z_2}{z_3}$ | $\frac{z_2}{z_3}$ | $\frac{z_4}{z_7}$ | $M^{\varepsilon\varepsilon\varepsilon\varepsilon}_{me1}$ | $z_4$ | $\frac{z_2 z_9}{z_3 z_8}$ | $\frac{z_5}{z_6}$ | $\frac{z_1}{z_7}$ | $M^{m\varepsilon\varepsilon\varepsilon}_{e1\varepsilon}$ | $\frac{z_4}{z_1}$ | $\frac{z_2 z_6}{z_3 z_5}$ | $\frac{z_8}{z_9}$ | $\frac{1}{z_7}$ |
| $M^{\varepsilon\varepsilon\varepsilon\varepsilon}_{1e1}$ | $\frac{z_5 z_9}{z_6 z_8}$ | $\frac{z_5 z_9}{z_6 z_8}$ | $z_1$ | $z_1$ | $M^{\varepsilon\varepsilon\varepsilon\varepsilon}_{mem}$ | $\frac{z_5}{z_6}$ | $\frac{z_5}{z_6}$ | $z_4$ | $z_4$ | $M^{m\varepsilon\varepsilon\varepsilon}_{e1e}$ | $\frac{z_3 z_5}{z_2 z_6}$ | $\frac{z_3 z_5}{z_2 z_6}$ | $z_7$ | $z_7$ |
| $M^{\varepsilon\varepsilon\varepsilon m}_{1e\varepsilon}$ | $\frac{z_6 z_8}{z_5 z_9}$ | $\frac{z_6 z_8}{z_5 z_9}$ | $\frac{1}{z_1}$ | $\frac{1}{z_1}$ | $M^{\varepsilon\varepsilon\varepsilon m}_{me\varepsilon}$ | $\frac{z_6}{z_5}$ | $\frac{z_3 z_8}{z_2 z_9}$ | $\frac{1}{z_4}$ | $\frac{z_7}{z_1}$ | $M^{m\varepsilon\varepsilon m}_{e1m}$ | $\frac{z_2 z_6}{z_3 z_5}$ | $\frac{z_8}{z_9}$ | $\frac{1}{z_7}$ | $\frac{z_4}{z_1}$ |
| $M^{\varepsilon\varepsilon m\varepsilon}_{1me}$ | $\frac{z_7}{z_4}$ | $\frac{z_3}{z_2}$ | $\frac{z_3}{z_2}$ | $\frac{z_7}{z_4}$ | $M^{\varepsilon\varepsilon m\varepsilon}_{m1\varepsilon}$ | $\frac{z_7}{z_1}$ | $\frac{z_3 z_8}{z_2 z_9}$ | $\frac{z_6}{z_5}$ | $\frac{1}{z_4}$ | $M^{m\varepsilon m\varepsilon}_{ee1}$ | $z_7$ | $\frac{z_3 z_5}{z_2 z_6}$ | $\frac{z_9}{z_8}$ | $\frac{z_1}{z_4}$ |
| $M^{\varepsilon\varepsilon m e}_{1m\varepsilon}$ | $\frac{z_6 z_8}{z_5 z_9}$ | $\frac{z_6 z_8}{z_5 z_9}$ | $\frac{1}{z_1}$ | $\frac{1}{z_1}$ | $M^{\varepsilon\varepsilon m e}_{m1e}$ | $\frac{z_3 z_8}{z_2 z_9}$ | $\frac{z_6}{z_5}$ | $\frac{1}{z_4}$ | $\frac{z_7}{z_1}$ | $M^{m e m e}_{eem}$ | $\frac{z_8}{z_9}$ | $\frac{z_2 z_6}{z_3 z_5}$ | $\frac{1}{z_7}$ | $\frac{z_4}{z_1}$ |
| $M^{\varepsilon\varepsilon m m}_{1m1}$ | $\frac{z_5 z_9}{z_6 z_8}$ | $\frac{z_5 z_9}{z_6 z_8}$ | $z_1$ | $z_1$ | $M^{\varepsilon\varepsilon m m}_{m1m}$ | $\frac{z_2 z_9}{z_3 z_8}$ | $\frac{z_2 z_9}{z_3 z_8}$ | $z_4$ | $z_4$ | $M^{m\varepsilon m m}_{ee\varepsilon}$ | $\frac{z_9}{z_8}$ | $\frac{z_9}{z_8}$ | $z_7$ | $z_7$ |
| $M^{\varepsilon\varepsilon\varepsilon\varepsilon}_{mem}$ | $\frac{z_1}{z_3}$ | $\frac{z_1}{z_3}$ | $z_2$ | $z_2$ | $M^{\varepsilon\varepsilon\varepsilon\varepsilon}_{1\varepsilon 1}$ | $\frac{z_1 z_9}{z_3 z_7}$ | $\frac{z_1 z_9}{z_3 z_7}$ | $z_5$ | $z_5$ | $M^{m\varepsilon\varepsilon\varepsilon}_{\varepsilon 1\varepsilon}$ | $\frac{z_1 z_6}{z_3 z_4}$ | $\frac{z_1 z_6}{z_3 z_4}$ | $z_8$ | $z_8$ |
| $M^{\varepsilon\varepsilon\varepsilon\varepsilon}_{me1}$ | $z_2$ | $\frac{z_4 z_9}{z_6 z_7}$ | $\frac{z_1}{z_3}$ | $\frac{z_5}{z_8}$ | $M^{\varepsilon\varepsilon\varepsilon\varepsilon}_{1\varepsilon m}$ | $\frac{z_2}{z_8}$ | $\frac{z_4}{z_6}$ | $\frac{z_4}{z_6}$ | $\frac{z_2}{z_8}$ | $M^{m\varepsilon\varepsilon\varepsilon}_{\varepsilon 1e}$ | $\frac{z_2}{z_5}$ | $\frac{z_3 z_4}{z_1 z_6}$ | $\frac{z_7}{z_9}$ | $\frac{1}{z_8}$ |
| $M^{\varepsilon\varepsilon\varepsilon m}_{me\varepsilon}$ | $\frac{z_3}{z_1}$ | $\frac{z_6 z_7}{z_4 z_9}$ | $\frac{1}{z_2}$ | $\frac{z_8}{z_5}$ | $M^{\varepsilon\varepsilon\varepsilon m}_{1\varepsilon e}$ | $\frac{z_3 z_7}{z_1 z_9}$ | $\frac{z_3 z_7}{z_1 z_9}$ | $\frac{1}{z_5}$ | $\frac{1}{z_5}$ | $M^{m\varepsilon\varepsilon m}_{\varepsilon 1m}$ | $\frac{z_3 z_4}{z_1 z_6}$ | $\frac{z_7}{z_9}$ | $\frac{1}{z_8}$ | $\frac{z_2}{z_5}$ |
| $M^{\varepsilon\varepsilon\varepsilon\varepsilon}_{me1}$ | $\frac{z_4 z_9}{z_6 z_7}$ | $z_2$ | $\frac{z_1}{z_3}$ | $\frac{z_5}{z_8}$ | $M^{\varepsilon\varepsilon\varepsilon\varepsilon}_{1em}$ | $\frac{z_4}{z_6}$ | $\frac{z_2}{z_8}$ | $\frac{z_4}{z_6}$ | $\frac{z_2}{z_8}$ | $M^{m\varepsilon\varepsilon\varepsilon}_{\varepsilon me}$ | $\frac{z_3 z_4}{z_1 z_6}$ | $\frac{z_2}{z_5}$ | $\frac{z_7}{z_9}$ | $\frac{1}{z_8}$ |
| $M^{\varepsilon\varepsilon\varepsilon\varepsilon}_{mem}$ | $\frac{z_5}{z_8}$ | $\frac{z_5}{z_8}$ | $z_2$ | $z_2$ | $M^{\varepsilon\varepsilon\varepsilon\varepsilon}_{1e1}$ | $z_5$ | $z_5$ | $z_5$ | $z_5$ | $M^{m\varepsilon\varepsilon\varepsilon}_{\varepsilon m\varepsilon}$ | $\frac{z_5}{z_2}$ | $\frac{z_5}{z_2}$ | $z_8$ | $z_8$ |
| $M^{\varepsilon\varepsilon\varepsilon m}_{me\varepsilon}$ | $\frac{z_6 z_7}{z_4 z_9}$ | $\frac{z_8}{z_5}$ | $\frac{z_3}{z_1}$ | $\frac{1}{z_2}$ | $M^{\varepsilon\varepsilon\varepsilon m}_{1e\varepsilon}$ | $\frac{z_6}{z_4}$ | $\frac{z_8}{z_2}$ | $\frac{z_6}{z_4}$ | $\frac{z_8}{z_2}$ | $M^{m\varepsilon\varepsilon m}_{\varepsilon m1}$ | $\frac{z_1 z_6}{z_3 z_4}$ | $z_8$ | $\frac{z_9}{z_7}$ | $\frac{z_5}{z_2}$ |
| $M^{\varepsilon\varepsilon m\varepsilon}_{m1\varepsilon}$ | $\frac{z_6 z_7}{z_4 z_9}$ | $\frac{z_3}{z_1}$ | $\frac{1}{z_2}$ | $\frac{z_8}{z_5}$ | $M^{\varepsilon\varepsilon m\varepsilon}_{1me}$ | $\frac{z_3 z_7}{z_1 z_9}$ | $\frac{z_3 z_7}{z_1 z_9}$ | $\frac{1}{z_5}$ | $\frac{1}{z_5}$ | $M^{m\varepsilon m\varepsilon}_{\varepsilon em}$ | $\frac{z_7}{z_9}$ | $\frac{z_3 z_4}{z_1 z_6}$ | $\frac{1}{z_8}$ | $\frac{z_2}{z_5}$ |
| $M^{\varepsilon\varepsilon me}_{m1e}$ | $\frac{z_8}{z_5}$ | $\frac{z_6 z_7}{z_4 z_9}$ | $\frac{z_3}{z_1}$ | $\frac{1}{z_2}$ | $M^{\varepsilon\varepsilon me}_{1m\varepsilon}$ | $\frac{z_8}{z_2}$ | $\frac{z_6}{z_4}$ | $\frac{z_6}{z_4}$ | $\frac{z_8}{z_2}$ | $M^{meme}_{\varepsilon e 1}$ | $z_8$ | $\frac{z_1 z_6}{z_3 z_4}$ | $\frac{z_9}{z_7}$ | $\frac{z_5}{z_2}$ |
| $M^{\varepsilon\varepsilon m m}_{m1m}$ | $\frac{z_4 z_9}{z_6 z_7}$ | $\frac{z_4 z_9}{z_6 z_7}$ | $z_2$ | $z_2$ | $M^{\varepsilon\varepsilon m m}_{1m1}$ | $\frac{z_1 z_9}{z_3 z_7}$ | $\frac{z_1 z_9}{z_3 z_7}$ | $z_5$ | $z_5$ | $M^{m emm}_{\varepsilon e\varepsilon}$ | $\frac{z_9}{z_7}$ | $\frac{z_9}{z_7}$ | $z_8$ | $z_8$ |
| $M^{\varepsilon m\varepsilon\varepsilon}_{eme}$ | $\frac{z_1}{z_2}$ | $\frac{z_1}{z_2}$ | $z_3$ | $z_3$ | $M^{\varepsilon m\varepsilon\varepsilon}_{\varepsilon 1\varepsilon}$ | $\frac{z_1 z_8}{z_2 z_7}$ | $\frac{z_1 z_8}{z_2 z_7}$ | $z_6$ | $z_6$ | $M^{mm\varepsilon\varepsilon}_{1\varepsilon 1}$ | $\frac{z_1 z_5}{z_2 z_4}$ | $\frac{z_1 z_5}{z_2 z_4}$ | $z_9$ | $z_9$ |
| $M^{\varepsilon m\varepsilon\varepsilon}_{em\varepsilon}$ | $\frac{z_2}{z_1}$ | $\frac{z_4 z_8}{z_5 z_7}$ | $\frac{1}{z_3}$ | $\frac{z_6}{z_9}$ | $M^{\varepsilon m\varepsilon\varepsilon}_{\varepsilon 1e}$ | $\frac{z_2 z_7}{z_1 z_8}$ | $\frac{z_4}{z_5}$ | $\frac{1}{z_6}$ | $\frac{z_3}{z_9}$ | $M^{mm\varepsilon\varepsilon}_{1\varepsilon m}$ | $\frac{z_2 z_4}{z_1 z_5}$ | $\frac{z_2 z_4}{z_1 z_5}$ | $\frac{1}{z_9}$ | $\frac{1}{z_9}$ |
| $M^{\varepsilon m\varepsilon m}_{em1}$ | $z_3$ | $\frac{z_5 z_7}{z_4 z_8}$ | $\frac{z_1}{z_2}$ | $\frac{z_9}{z_6}$ | $M^{\varepsilon m\varepsilon m}_{\varepsilon 1m}$ | $\frac{z_3}{z_9}$ | $\frac{z_2 z_7}{z_1 z_8}$ | $\frac{z_4}{z_5}$ | $\frac{1}{z_6}$ | $M^{mm\varepsilon m}_{1\varepsilon e}$ | $\frac{z_3}{z_6}$ | $\frac{z_7}{z_8}$ | $\frac{z_7}{z_8}$ | $\frac{z_3}{z_6}$ |
| $M^{\varepsilon m e\varepsilon}_{e1\varepsilon}$ | $\frac{z_4 z_8}{z_5 z_7}$ | $\frac{z_2}{z_1}$ | $\frac{1}{z_3}$ | $\frac{z_6}{z_9}$ | $M^{\varepsilon m e\varepsilon}_{\varepsilon me}$ | $\frac{z_4}{z_5}$ | $\frac{z_2 z_7}{z_1 z_8}$ | $\frac{1}{z_6}$ | $\frac{z_3}{z_9}$ | $M^{mm\varepsilon\varepsilon}_{1em}$ | $\frac{z_2 z_4}{z_1 z_5}$ | $\frac{z_2 z_4}{z_1 z_5}$ | $\frac{1}{z_9}$ | $\frac{1}{z_9}$ |
| $M^{\varepsilon m ee}_{e1e}$ | $\frac{z_5 z_7}{z_4 z_8}$ | $\frac{z_5 z_7}{z_4 z_8}$ | $z_3$ | $z_3$ | $M^{\varepsilon m ee}_{\varepsilon m\varepsilon}$ | $\frac{z_5}{z_4}$ | $\frac{z_5}{z_4}$ | $z_6$ | $z_6$ | $M^{mm ee}_{1e1}$ | $\frac{z_1 z_5}{z_2 z_4}$ | $\frac{z_1 z_5}{z_2 z_4}$ | $z_9$ | $z_9$ |
| $M^{\varepsilon m em}_{e1m}$ | $\frac{z_6}{z_9}$ | $\frac{z_4 z_8}{z_5 z_7}$ | $\frac{z_2}{z_1}$ | $\frac{1}{z_3}$ | $M^{\varepsilon m em}_{\varepsilon m1}$ | $z_6$ | $\frac{z_1 z_8}{z_2 z_7}$ | $\frac{z_5}{z_4}$ | $\frac{z_9}{z_3}$ | $M^{mm em}_{1ee}$ | $\frac{z_6}{z_3}$ | $\frac{z_8}{z_7}$ | $\frac{z_8}{z_7}$ | $\frac{z_6}{z_3}$ |
| $M^{\varepsilon m m\varepsilon}_{ee1}$ | $\frac{z_5 z_7}{z_4 z_8}$ | $z_3$ | $\frac{z_1}{z_2}$ | $\frac{z_9}{z_6}$ | $M^{\varepsilon m m\varepsilon}_{\varepsilon em}$ | $\frac{z_2 z_7}{z_1 z_8}$ | $\frac{z_3}{z_9}$ | $\frac{z_4}{z_5}$ | $\frac{1}{z_6}$ | $M^{mm m\varepsilon}_{1me}$ | $\frac{z_7}{z_8}$ | $\frac{z_3}{z_6}$ | $\frac{z_7}{z_8}$ | $\frac{z_3}{z_6}$ |
| $M^{\varepsilon m m e}_{eem}$ | $\frac{z_4 z_8}{z_5 z_7}$ | $\frac{z_6}{z_9}$ | $\frac{z_2}{z_1}$ | $\frac{1}{z_3}$ | $M^{\varepsilon m m e}_{\varepsilon e 1}$ | $\frac{z_1 z_8}{z_2 z_7}$ | $z_6$ | $\frac{z_5}{z_4}$ | $\frac{z_9}{z_3}$ | $M^{mm m e}_{1m\varepsilon}$ | $\frac{z_8}{z_7}$ | $\frac{z_6}{z_3}$ | $\frac{z_8}{z_7}$ | $\frac{z_6}{z_3}$ |
| $M^{\varepsilon m m m}_{ee\varepsilon}$ | $\frac{z_9}{z_6}$ | $\frac{z_9}{z_6}$ | $z_3$ | $z_3$ | $M^{\varepsilon m m m}_{\varepsilon e\varepsilon}$ | $\frac{z_9}{z_3}$ | $\frac{z_9}{z_3}$ | $z_6$ | $z_6$ | $M^{mmmm}_{1m1}$ | $z_9$ | $z_9$ | $z_9$ | $z_9$ |

We can notice here again, when all of the particles are the same type the graph braid symbols are equal, i.e. $X^{aaaa} = Y^{aaaa} = A^{aaaa} = B^{aaaa}$.

### 10.2.3 Solutions to the Lollipop Equations

#### 10.2.3.1 Lollipop trijunction solutions

In contrast to the Ising model, demanding that $P_{ed}^{abc} \equiv R_e^{ab}$ does not necessarily imply that the solutions must be planar. There are 32 solutions in total which can be presented as follows:

$$R_1^{22} = \nu_1, \quad R_4^{23} = \nu_2, \quad R_3^{24} = \nu_1\nu_2, \quad R_4^{32} = \nu_3, \quad R_1^{33} = \nu_4,$$
$$R_2^{34} = \nu_3\nu_4, \quad R_3^{42} = -1, \quad R_2^{43} = \nu_5, \quad R_1^{44} = -\nu_5$$

and

|  | Value of $M$ | |  | Value of $M$ | |  | Value of $M$ | |
|---|---|---|---|---|---|---|---|---|
|  | $P$ | $Q$ |  | $P$ | $Q$ |  | $P$ | $Q$ |
| $M_{1\varepsilon}^{\varepsilon\varepsilon\varepsilon}$ | $\nu_1$ | $\nu_1$ | $M_{me}^{\varepsilon\varepsilon\varepsilon}$ | $\nu_3$ | $-\nu_1$ | $M_{em}^{m\varepsilon\varepsilon}$ | $-1$ | $\nu_1\nu_3$ |
| $M_{1e}^{\varepsilon\varepsilon e}$ | $\nu_1$ | $-\nu_3$ | $M_{m\varepsilon}^{\varepsilon\varepsilon e}$ | $\nu_3$ | $\nu_3$ | $M_{e1}^{m\varepsilon e}$ | $-1$ | $\nu_1\nu_3$ |
| $M_{1m}^{\varepsilon\varepsilon m}$ | $\nu_1$ | $-\nu_3$ | $M_{m1}^{\varepsilon\varepsilon m}$ | $\nu_3$ | $-\nu_1$ | $M_{e\varepsilon}^{m\varepsilon m}$ | $-1$ | $-1$ |
| $M_{me}^{\varepsilon e\varepsilon}$ | $\nu_2$ | $\nu_2$ | $M_{1\varepsilon}^{e\varepsilon\varepsilon}$ | $\nu_4$ | $\nu_2\nu_5$ | $M_{\varepsilon 1}^{me\varepsilon}$ | $\nu_5$ | $\nu_2\nu_4$ |
| $M_{m\varepsilon}^{\varepsilon e e}$ | $\nu_2$ | $\nu_4\nu_5$ | $M_{1e}^{e\varepsilon e}$ | $\nu_4$ | $\nu_4$ | $M_{\varepsilon m}^{mee}$ | $\nu_5$ | $\nu_2\nu_4$ |
| $M_{m1}^{\varepsilon em}$ | $\nu_2$ | $\nu_4\nu_5$ | $M_{1m}^{e\varepsilon m}$ | $\nu_4$ | $\nu_2\nu_5$ | $M_{\varepsilon e}^{mem}$ | $\nu_5$ | $\nu_5$ |
| $M_{em}^{\varepsilon m\varepsilon}$ | $\nu_1\nu_2$ | $\nu_1\nu_2$ | $M_{\varepsilon 1}^{em\varepsilon}$ | $\nu_3\nu_4$ | $-\nu_1\nu_2\nu_5$ | $M_{1\varepsilon}^{mm\varepsilon}$ | $-\nu_5$ | $\nu_1\nu_2\nu_3\nu_4$ |
| $M_{e1}^{\varepsilon me}$ | $\nu_1\nu_2$ | $-\nu_3\nu_4\nu_5$ | $M_{\varepsilon m}^{eme}$ | $\nu_3\nu_4$ | $\nu_3\nu_4$ | $M_{1e}^{mme}$ | $-\nu_5$ | $\nu_1\nu_2\nu_3\nu_4$ |
| $M_{e\varepsilon}^{\varepsilon mm}$ | $\nu_1\nu_2$ | $-\nu_3\nu_4\nu_5$ | $M_{\varepsilon e}^{emm}$ | $\nu_3\nu_4$ | $-\nu_1\nu_2\nu_5$ | $M_{1m}^{mmm}$ | $-\nu_5$ | $-\nu_5$ |

where $\nu_i \in \{-1, 1\}$. Demanding planarity then comes down to demanding that $\nu_1 = -\nu_3$ and $\nu_2 = \nu_4\nu_5$. The loss of two binary degrees of freedom thus implies that only one out of four solutions are planar. We find here again $P^{aaa} = Q^{aaa}$.

#### 10.2.3.2 Circle Solutions

There are 128 solutions to the circle equations. They can be presented as follows:

$$D_\varepsilon^{1\varepsilon} = \mu_1, \quad D_e^{1e} = \mu_2, \quad D_m^{1m} = \mu_3, \quad D_1^{\varepsilon\varepsilon} = \mu_4, \quad D_m^{ee} = \mu_5, \quad D_e^{\varepsilon m} = \mu_6,$$
$$D_m^{e\varepsilon} = \mu_3\mu_5, \quad D_1^{ee} = \mu_7, \quad D_\varepsilon^{em} = -\mu_1, \quad D_e^{m\varepsilon} = \mu_2\mu_6, \quad D_\varepsilon^{me} = -1, \quad D_1^{mm} = \mu_4\mu_7,$$
(10.9)

where $\mu_i \in \{-1, 1\}$. The twist factors are the same as in the planar case.

### 10.2.3.3 Full Lollipop Solutions

In contrast to the Ising model, after fixing the $F$-symbols, there is a discrete $\mathbb{Z}_2$ gauge symmetry left that has the following form:

$$M_m^{\varepsilon e} \mapsto -M_m^{\varepsilon e}, \quad M_e^{\varepsilon m} \mapsto -M_e^{\varepsilon m}, \quad M_m^{e\varepsilon} \mapsto -M_m^{e\varepsilon},$$
$$M_\varepsilon^{em} \mapsto -M_\varepsilon^{em}, \quad M_e^{m\varepsilon} \mapsto -M_e^{m\varepsilon}, \quad M_\varepsilon^{me} \mapsto -M_\varepsilon^{me},$$

for $M = R$ and $M = D$, and

$$M_{md}^{\varepsilon ec} \mapsto -M_{md}^{\varepsilon ec}, \quad M_{ed}^{\varepsilon mc} \mapsto -M_{ed}^{\varepsilon mc}, \quad M_{md}^{e\varepsilon c} \mapsto -M_{md}^{e\varepsilon c},$$
$$M_{\varepsilon d}^{emc} \mapsto -M_{\varepsilon d}^{emc}, \quad M_{ed}^{m\varepsilon c} \mapsto -M_{ed}^{m\varepsilon c}, \quad M_{\varepsilon d}^{mec} \mapsto -M_{\varepsilon d}^{mec},$$

for $M = P$ and $M = Q$. For the solutions to the lollipop trijunction equations and the circle equations, described in Section 10.2.3.1 and 10.2.3.2, this gauge symmetry has been removed. To construct the full solution set to the lollipop equations one should therefore re-introduce these gauge equivalent solutions, construct all products between solutions to the trijunction lollipop equations and circle equations, and finally remove this gauge symmetry again. This set has twice the size of the product set of gauge-inequivalent solutions. It can be constructed by taking the product of the lollipop trijunction solutions and the following set of non-reduced solutions to the circle equations

$$D_\varepsilon^{1\varepsilon} = \mu_1, \quad D_e^{1e} = \mu_2, \quad D_m^{1m} = \mu_3, \quad D_1^{\varepsilon\varepsilon} = \mu_4, \quad D_m^{\varepsilon e} = \sigma\mu_5, \quad D_e^{\varepsilon m} = \sigma\mu_6,$$
$$D_m^{e\varepsilon} = \sigma\mu_3\mu_5, \quad D_1^{ee} = \mu_7, \quad D_\varepsilon^{em} = -\sigma\mu_1, \quad D_e^{m\varepsilon} = \sigma\mu_2\mu_6, \quad D_\varepsilon^{me} = -\sigma, \quad D_1^{mm} = \mu_4\mu_7,$$
(10.10)

where $\sigma \in \{-1, 1\}$ reintroduces the $\mathbb{Z}_2$ gauge freedom.

## 10.3 Solutions for the $TY(\mathbb{Z}_3)$ Model

The $TY(\mathbb{Z}_3)$ fusion ring has 4 particles, $1, \psi_1, \psi_2, \sigma$, where $\{1, \psi_1, \psi_2\}$ form a $\mathbb{Z}_3$ subgroup, and

$$1 \times a = a \times 1 = a, \quad \forall a \in \{1, \psi_1, \psi_2, \sigma\}, \tag{10.11}$$
$$\sigma \times a = a \times \sigma = \sigma, \quad \forall a \in \{1, \psi_1, \psi_2\}, \tag{10.12}$$
$$\sigma \times \sigma = 1 + \psi_1 + \psi_2. \tag{10.13}$$

In the following sections we will list the solutions to the pentagon equations as well as the circle equations. All other equations admit no solutions. We omit any well-defined symbol equal to 1.

### 10.3.1 Solutions to the Pentagon Equations

There are four solutions to the pentagon equations which can be presented as follows:

$$[F^{\sigma\sigma\psi_1}_{\psi_1}]^1_\sigma = \kappa_1, \qquad [F^{\sigma\sigma\psi_2}_{\psi_2}]^1_\sigma = \kappa_1, \qquad [F^{\sigma\psi_1\sigma}_{\psi_1}]^\sigma_\sigma = e^{-\frac{2}{3}i\pi\kappa_1\kappa_2},$$

$$[F^{\sigma\psi_1\sigma}_{\psi_2}]^\sigma_\sigma = e^{\frac{2}{3}i\pi\kappa_1\kappa_2}, \qquad [F^{\sigma\psi_1\psi_2}_{\sigma}]^\sigma_1 = \kappa_1, \qquad [F^{\sigma\psi_2\sigma}_{\psi_1}]^\sigma_\sigma = e^{\frac{2}{3}i\pi\kappa_1\kappa_2},$$

$$[F^{\sigma\psi_2\sigma}_{\psi_2}]^\sigma_\sigma = e^{-\frac{2}{3}i\pi\kappa_1\kappa_2}, \qquad [F^{\sigma\psi_2\psi_1}_{\sigma}]^\sigma_1 = \kappa_1, \qquad [F^{\psi_1\sigma\sigma}_{\psi_1}]^\sigma_1 = \kappa_1,$$

$$[F^{\psi_1\sigma\psi_1}_{\sigma}]^\sigma_\sigma = e^{-\frac{2}{3}i\pi\kappa_1\kappa_2}, \qquad [F^{\psi_1\sigma\psi_2}_{\sigma}]^\sigma_\sigma = e^{\frac{2}{3}i\pi\kappa_1\kappa_2}, \qquad [F^{\psi_1\psi_1\psi_2}_{\psi_1}]^{\psi_2} = \kappa_1,$$

$$[F^{\psi_1\psi_2\sigma}_{\sigma}]^1_\sigma = \kappa_1, \qquad [F^{\psi_1\psi_2\psi_2}_{\psi_2}]^1_{\psi_1} = \kappa_1, \qquad [F^{\psi_2\sigma\sigma}_{\psi_2}]^\sigma_1 = \kappa_1,$$

$$[F^{\psi_2\sigma\psi_1}_{\sigma}]^\sigma_\sigma = e^{\frac{2}{3}i\pi\kappa_1\kappa_2}, \qquad [F^{\psi_2\sigma\psi_2}_{\sigma}]^\sigma_\sigma = e^{-\frac{2}{3}i\pi\kappa_1\kappa_2}, \qquad [F^{\psi_2\psi_1\sigma}_{\sigma}]^1_\sigma = \kappa_1,$$

$$[F^{\psi_2\psi_1\psi_1}_{\psi_1}]^1_{\psi_2} = \kappa_1, \qquad [F^{\psi_2\psi_2\psi_1}_{\psi_2}]^{\psi_1}_1 = \kappa_1,$$

$$[F^{\sigma\sigma\sigma}_\sigma] = \frac{1}{\sqrt{3}}\begin{bmatrix} \kappa_1 & 1 & 1 \\ 1 & e^{i\pi\left(\frac{\kappa_1}{6}+\frac{1}{2}\right)\kappa_2} & e^{-i\pi\left(\frac{\kappa_1}{6}+\frac{1}{2}\right)\kappa_2} \\ 1 & e^{-i\pi\left(\frac{\kappa_1}{6}+\frac{1}{2}\right)\kappa_2} & e^{i\pi\left(\frac{\kappa_1}{6}+\frac{1}{2}\right)\kappa_2} \end{bmatrix},$$

where $\kappa_1, \kappa_2 \in \{-1, 1\}$ and the matrix indices of $[F^{\sigma\sigma\sigma}_\sigma]$ range over $(1, \psi_1, \psi_2)$.

### 10.3.2 Solutions to the Circle Equations

In contrast to the planar hexagon equations, we now find there are 48 solutions, per set of $F$-symbols, to the circle equations. Let $\varepsilon_i \in \{-1, 1\}$ and $\nu \in \{0, 1, 2\}$, then they can be presented as follows.

If $(\kappa_1, \kappa_2) = (-1, -1)$ then

$$D^{1\sigma}_\sigma = \varepsilon_1 e^{\frac{i\pi}{12}(7-2\nu(\nu+1))}, \qquad D^{1\psi_1}_{\psi_1} = e^{-\frac{2i\pi}{3}},$$

$$D^{1\psi_2}_{\psi_2} = e^{-\frac{2i\pi}{3}}, \qquad D^{\sigma\sigma}_1 = \varepsilon_2,$$

$$D^{\sigma\sigma}_{\psi_1} = e^{i\pi\left(\frac{\varepsilon_3}{2}+\frac{1}{6}\right)}, \qquad D^{\sigma\sigma}_{\psi_2} = e^{i\pi\left(\frac{\varepsilon_4}{2}+\frac{1}{6}\right)},$$

$$D^{\sigma\psi_1}_\sigma = e^{\frac{2i\pi}{3}\left((\nu-1)^2\varepsilon_1-1\right)}, \qquad D^{\sigma\psi_2}_\sigma = e^{-\frac{2i\pi}{3}\left((\nu-1)^2\varepsilon_1+1\right)},$$

$$D^{\psi_1\sigma}_\sigma = e^{-\frac{i\pi}{12}\left(2\nu^2+2\nu-9+2\varepsilon_1\left(4\nu^2-8\nu+1\right)\right)}, \qquad D^{\psi_1\psi_1}_{\psi_2} = e^{\frac{2i\pi}{3}},$$

$$D^{\psi_2\sigma}_\sigma = e^{-\frac{i\pi}{12}\left(2\nu^2-10\nu+3+\varepsilon_1\left(4\nu^2-8\nu-2\right)\right)}, \qquad D^{\psi_2\psi_2}_{\psi_1} = e^{\frac{2i\pi}{3}}.$$

If $(\kappa_1, \kappa_2) = (-1, 1)$ then

$$D^{1\sigma}_\sigma = \varepsilon_1 e^{\frac{i\pi}{12}(5+2\nu(\nu+1))}, \qquad D^{1\psi_1}_{\psi_1} = e^{\frac{2i\pi}{3}},$$

$$D^{1\psi_2}_{\psi_2} = e^{\frac{2i\pi}{3}}, \qquad D^{\sigma\sigma}_1 = \varepsilon_2,$$

$$D^{\sigma\sigma}_{\psi_1} = e^{-i\pi\left(\frac{1}{6}-\frac{\varepsilon_3}{2}\right)}, \qquad D^{\sigma\sigma}_{\psi_2} = e^{-i\pi\left(\frac{1}{6}-\frac{\varepsilon_4}{2}\right)},$$

$$D^{\sigma\psi_1}_\sigma = e^{\frac{2i\pi}{3}(3-2\nu)}, \qquad D^{\sigma\psi_2}_\sigma = e^{\frac{2i\pi}{3}(2\nu-1)},$$

$$D^{\psi_1\sigma}_\sigma = e^{\frac{i\pi}{12}\left(2\nu^2-6\nu-1+6\varepsilon_1\right)}, \qquad D^{\psi_1\psi_1}_{\psi_2} = e^{-\frac{2i\pi}{3}},$$

$$D^{\psi_2\sigma}_\sigma = e^{\frac{i\pi}{12}\left(2\nu^2-2\nu-5-6\varepsilon_1\left(2\nu^2-4\nu+1\right)\right)}, \qquad D^{\psi_2\psi_2}_{\psi_1} = e^{-\frac{2i\pi}{3}}.$$

If $(\kappa_1, \kappa_2) = (1, -1)$ then

$$D_\sigma^{1\sigma} = \varepsilon_1 e^{-\frac{i\pi}{12}(-1+2\nu(\nu+1))}, \qquad D_{\psi_1}^{1\psi_1} = e^{-\frac{2i\pi}{3}},$$
$$D_{\psi_2}^{1\psi_2} = e^{-\frac{2i\pi}{3}}, \qquad D_1^{\sigma\sigma} = \varepsilon_2,$$
$$D_{\psi_1}^{\sigma\sigma} = e^{i\pi\left(\frac{\varepsilon_3}{2}+\frac{1}{6}\right)}, \qquad D_{\psi_2}^{\sigma\sigma} = e^{i\pi\left(\frac{\varepsilon_4}{2}+\frac{1}{6}\right)},$$
$$D_\sigma^{\sigma\psi_1} = e^{\frac{2i\pi}{3}\left((\nu-1)^2\varepsilon_1-1\right)}, \qquad D_\sigma^{\sigma\psi_2} = e^{\frac{2i\pi}{3}\left((\nu-1)^2(-\varepsilon_1)-1\right)},$$
$$D_\sigma^{\psi_1\sigma} = e^{-\frac{i\pi}{12}\left(2\nu^2-10\nu-3-\varepsilon_1\left(4\nu^2-8\nu-2\right)\right)}, \qquad D_{\psi_2}^{\psi_1\psi_1} = e^{\frac{2i\pi}{3}},$$
$$D_\sigma^{\psi_2\sigma} = e^{-\frac{i\pi}{12}\left(2\nu^2+2\nu-15-2\varepsilon_1\left(4\nu^2-8\nu+1\right)\right)}, \qquad D_{\psi_1}^{\psi_2\psi_2} = e^{\frac{2i\pi}{3}}.$$

If $(\kappa_1, \kappa_2) = (1, 1)$ then

$$D_\sigma^{1\sigma} = \varepsilon_1 e^{\frac{i\pi}{12}(-1+2\nu(\nu+1))}, \qquad D_{\psi_1}^{1\psi_1} = e^{\frac{2i\pi}{3}},$$
$$D_{\psi_2}^{1\psi_2} = e^{\frac{2i\pi}{3}}, \qquad D_1^{\sigma\sigma} = \varepsilon_2,$$
$$D_{\psi_1}^{\sigma\sigma} = e^{-i\pi\left(\frac{1}{6}-\frac{\varepsilon_3}{2}\right)}, \qquad D_{\psi_2}^{\sigma\sigma} = e^{-i\pi\left(\frac{1}{6}-\frac{\varepsilon_4}{2}\right)},$$
$$D_\sigma^{\sigma\psi_1} = e^{\frac{2i\pi}{3}(3-2\nu)}, \qquad D_\sigma^{\sigma\psi_2} = e^{\frac{2i\pi}{3}(2\nu-1)},$$
$$D_\sigma^{\psi_1\sigma} = e^{\frac{i\pi}{12}\left(2\nu^2-6\nu+5-6\varepsilon_1\right)}, \qquad D_{\psi_2}^{\psi_1\psi_1} = e^{-\frac{2i\pi}{3}},$$
$$D_\sigma^{\psi_2\sigma} = e^{\frac{i\pi}{12}\left(2\nu^2-2\nu+1+6\varepsilon_1\left(2\nu^2-4\nu+1\right)\right)}, \qquad D_{\psi_1}^{\psi_2\psi_2} = e^{-\frac{2i\pi}{3}}.$$

# Bibliography

[1] Willie Aboumrad. *Quantum computing with anyons: an F-matrix and braid calculator*. en. arXiv:2212.00831 [quant-ph]. Dec. 2022.

[2] S. M. Albrecht et al. "Exponential protection of zero modes in Majorana islands". en. In: *Nature* 531.7593 (Mar. 2016), pp. 206–209. DOI: `10.1038/nature17162`.

[3] Max A. Alekseyev et al. *Classification of modular data of integral modular fusion categories up to rank 12*. en. arXiv:2302.01613 [math-ph]. May 2023.

[4] Byung Hee An and Tomasz Maciazek. "Geometric Presentations of Braid Groups for Particles on a Graph". en. In: *Communications in Mathematical Physics* 384.2 (June 2021), pp. 1109–1140. DOI: `10.1007/s00220-021-04095-x`.

[5] Eddy Ardonne. *alatc*. May 2023.

[6] Eddy Ardonne and Joost Slingerland. "Clebsch–Gordan and 6$j$-coefficients for rank 2 quantum groups". en. In: *Journal of Physics A: Mathematical and Theoretical* 43.39 (Oct. 2010), p. 395205. DOI: `10.1088/1751-8113/43/39/395205`.

[7] E. Artin. "Theory of Braids". In: *Annals of Mathematics* 48.1 (1947). Publisher: Annals of Mathematics, pp. 101–126.

[8] Bojko Bakalov and A.A. Kirillov. "Lectures on tensor categories and modular functors". In: *Amer. Math. Soc. Univ. Lect. Ser.* 21 (Jan. 2001).

[9] John W. Barrett and Bruce W. Westbury. *Invariants of Piecewise-Linear 3-Manifolds*. en. arXiv:hep-th/9311155. Aug. 1995.

[10] Parsa Hassan Bonderson. "Non-Abelian Anyons and Interferometry". en. In: (2007).

[11] B. Buchberger. "A theoretical basis for the reduction of polynomials to canonical forms". en. In: *ACM SIGSAM Bulletin* 10.3 (Aug. 1976), pp. 19–29. DOI: `10.1145/1088216.1088219`.

[12] Jacob C. Bridgeman. *smallRankUnitaryFusionData*. June 2023.

[13] Frank Calegari, Scott Morrison, and Noah Snyder. "Cyclotomic Integers, Fusion Categories, and Subfactors". en. In: *Communications in Mathematical Physics* 303.3 (May 2011), pp. 845–896. DOI: `10.1007/s00220-010-1136-2`.

[14] H. O. H. Churchill et al. "Superconductor-nanowire devices from tunneling to the multichannel regime: Zero-bias oscillations and magnetoconductance crossover". In: *Phys. Rev. B* 87.24 (June 2013). Publisher: American Physical Society, p. 241401. DOI: 10.1103/PhysRevB.87.241401.

[15] A. Conlon and J. K. Slingerland. "Compatibility of braiding and fusion on wire networks". In: *Phys. Rev. B* 108.3 (July 2023). Publisher: American Physical Society, p. 035150. DOI: 10.1103/PhysRevB.108.035150.

[16] Louis Crane, Louis H. Kauffman, and David N. Yetter. "State-Sum Invariants of 4-Manifolds". In: *Journal of Knot Theory and Its Ramifications* 06.02 (1997). _eprint: https://doi.org/10.1142/S0218216597000145, pp. 177–234. DOI: 10.1142/S0218216597000145.

[17] Thomas Creutzig. "Fusion categories for affine vertex algebras at admissible levels". en. In: *Selecta Mathematica* 25.2 (June 2019), p. 27. DOI: 10.1007/s00029-019-0479-6.

[18] Shawn X. Cui and Zhenghan Wang. "State sum invariants of three manifolds from spherical multi-fusion categories". en. In: *Journal of Knot Theory and Its Ramifications* 26.14 (Dec. 2017), p. 1750104. DOI: 10.1142/S0218216517501048.

[19] Orit Davidovich, Tobias Hagge, and Zhenghan Wang. *On Arithmetic Modular Categories*. en. arXiv:1305.2229 [math]. May 2013.

[20] Colleen Delaney, Eric C. Rowell, and Zhenghan Wang. "Local unitary representations of the braid group and their applications to quantum computing". en. In: *Revista Colombiana de Matemáticas* 50.2 (Jan. 2017), p. 211. DOI: 10.15446/recolma.v50n2.62211.

[21] Colleen Delaney et al. "Braided zesting and its applications". en. In: *Communications in Mathematical Physics* 386.1 (Aug. 2021). arXiv:2005.05544 [math], pp. 1–55. DOI: 10.1007/s00220-021-04002-4.

[22] M. T. Deng et al. "Anomalous Zero-Bias Conductance Peak in a Nb–InSb Nanowire–Nb Hybrid Device". en. In: *Nano Letters* 12.12 (Dec. 2012), pp. 6414–6419. DOI: 10.1021/nl303758w.

[23] Lukas Devos. *CategoryData*. June 2024.

[24] Thomas W. Dubé. "The Structure of Polynomial Ideals and Gröbner Bases". In: *SIAM Journal on Computing* 19.4 (1990). _eprint: https://doi.org/10.1137/0219053, pp. 750–773. DOI: 10.1137/0219053.

[25] P. I. Etingof et al., eds. *Tensor categories*. en. Mathematical surveys and monographs volume 205. Providence, Rhode Island: American Mathematical Society, 2015.

[26] Pavel Etingof, Dmitri Nikshych, and Victor Ostrik. "Weakly group-theoretical and solvable fusion categories". en. In: *Advances in Mathematics* 226.1 (Jan. 2011), pp. 176–205. DOI: 10.1016/j.aim.2010.06.009.

[27] Pavel Etingof, Dmitri Nikshych, and Viktor Ostrik. *On fusion categories*. en. arXiv:math/0203060. Apr. 2017.

[28] Pavel Etingof and Victor Ostrik. *On semisimplification of tensor categories*. 2019. arXiv: `1801.04409 [math.RT]`.

[29] Pavel Etingof et al. *Tensor categories*. en. Mathematical surveys and monographs volume 205. Providence, Rhode Island: American Mathematical Society, 2015.

[30] David E. Evans and Terry Gannon. *Near-group fusion categories and their doubles*. en. arXiv:1208.1500 [hep-th]. Aug. 2012.

[31] Daniel Farley and Lucas Sabalka. "Discrete Morse theory and graph braid groups". en. In: *Algebraic & Geometric Topology* 5.3 (Aug. 2005), pp. 1075–1109. DOI: `10.2140/agt.2005.5.1075`.

[32] Daniel Farley and Lucas Sabalka. "Presentations of graph braid groups". In: *Forum Mathematicum* 24.4 (2012), pp. 827–859. DOI: `doi:10.1515/form.2011.086`.

[33] Jean Charles Faugère. "A New Efficient Algorithm for Computing Gröbner Bases without Reduction to Zero (F5)". In: *Proceedings of the 2002 International Symposium on Symbolic and Algebraic Computation*. ISSAC '02. event-place: Lille, France. New York, NY, USA: Association for Computing Machinery, 2002, pp. 75–83. DOI: `10.1145/780506.780516`.

[34] Jean-Charles Faugère. "A new e cient algorithm for computing Grobner bases (F4)". en. In: *Journal of Pure and Applied Algebra* (1999).

[35] Richard P. Feynman. "Simulating physics with computers". In: *International Journal of Theoretical Physics* 21.6 (June 1982), pp. 467–488. DOI: `10.1007/BF02650179`.

[36] A. D. K. Finck et al. "Anomalous Modulation of a Zero-Bias Peak in a Hybrid Nanowire-Superconductor Device". en. In: *Physical Review Letters* 110.12 (Mar. 2013), p. 126406. DOI: `10.1103/PhysRevLett.110.126406`.

[37] Michael H. Freedman et al. *Topological Quantum Computation*. en. arXiv:quant-ph/0101025. Sept. 2002.

[38] Jürg Fröhlich and Thomas Kerler. *Quantum groups, quantum categories, and quantum field theory*. en. Lecture notes in mathematics 1542. Berlin : New York: Springer-Verlag, 1993.

[39] Jürgen Fuchs. "Fusion Rules in Conformal Field Theory". en. In: *Fortschritte der Physik/Progress of Physics* 42.1 (1994), pp. 1–48. DOI: `10.1002/prop.2190420102`.

[40] César Galindo. "On Braided and Ribbon Unitary Fusion Categories". en. In: *Canadian Mathematical Bulletin* 57.3 (Sept. 2014), pp. 506–510. DOI: `10.4153/CMB-2013-017-5`.

[41] Doron Gepner and Anton Kapustin. "On the classification of fusion rings". en. In: *Physics Letters B* 349.1-2 (Apr. 1995), pp. 71–75. DOI: 10.1016/0370-2693(95)00172-H.

[42] Google Quantum AI and Collaborators et al. "Non-Abelian braiding of graph vertices in a superconducting processor". en. In: *Nature* 618.7964 (June 2023), pp. 264–269. DOI: 10.1038/s41586-023-05954-4.

[43] Pinhas Grossman and Noah Snyder. "Quantum Subgroups of the Haagerup Fusion Categories". en. In: *Communications in Mathematical Physics* 311.3 (May 2012), pp. 617–643. DOI: 10.1007/s00220-012-1427-x.

[44] Pinhas Grossman et al. *The Extended Haagerup fusion categories*. en. arXiv:1810.06076 [math]. Oct. 2018.

[45] Lov K. Grover. "Quantum Mechanics Helps in Searching for a Needle in a Haystack". en. In: *Physical Review Letters* 79.2 (July 1997), pp. 325–328. DOI: 10.1103/PhysRevLett.79.325.

[46] Jutho Haegeman. *TensorKit*. "DOI: 10.5281/zenodo.10574897". July 2024.

[47] Tobias Hagge and Matthew Titsworth. *Geometric Invariants for Fusion Categories*. en. arXiv:1509.03275 [math]. Sept. 2015.

[48] André Henriques and David Penneys. *Bicommutant categories from fusion categories*. en. arXiv:1511.05226 [math]. Dec. 2016.

[49] Mohsin Iqbal et al. "Non-Abelian topological order and anyons on a trapped-ion processor". en. In: *Nature* 626.7999 (Feb. 2024), pp. 505–511. DOI: 10.1038/s41586-023-06934-4.

[50] Corey Jones and David Penneys. "Operator algebras in rigid $C^*$-tensor categories". en. In: *Communications in Mathematical Physics* 355.3 (Nov. 2017). arXiv:1611.04620 [math], pp. 1121–1188. DOI: 10.1007/s00220-017-2964-0.

[51] Vaughan F. R. Jones, Scott Morrison, and Noah Snyder. "The classification of subfactors of index at most 5". en. In: *Bulletin of the American Mathematical Society* 51.2 (Dec. 2013). arXiv:1304.6141 [math], pp. 277–327. DOI: 10.1090/S0273-0979-2013-01442-3.

[52] A. Joyal and R. Street. "Braided Tensor Categories". In: *Advances in Mathematics* 102.1 (1993), pp. 20–78. DOI: https://doi.org/10.1006/aima.1993.1055.

[53] André Joyal and Ross Street. "The geometry of tensor calculus, I". en. In: *Advances in Mathematics* 88.1 (July 1991), pp. 55–112. DOI: 10.1016/0001-8708(91)90003-P.

[54] Christian Kassel. *Quantum Groups*. en. Vol. 155. Graduate Texts in Mathematics. New York, NY: Springer New York, 1995. DOI: 10.1007/978-1-4612-0783-2.

[55] A.Yu. Kitaev. "Fault-tolerant quantum computation by anyons". en. In: *Annals of Physics* 303.1 (Jan. 2003), pp. 2–30. DOI: 10.1016/S0003-4916(02)00018-0.

[56] Alexei Kitaev. "Anyons in an exactly solved model and beyond". en. In: *Annals of Physics* 321.1 (Jan. 2006). arXiv:cond-mat/0506438, pp. 2–111. DOI: 10.1016/j.aop.2005.10.005.

[57] Liang Kong and Xiao-Gang Wen. *Braided fusion categories, gravitational anomalies, and the mathematical framework for topological orders in any dimensions*. en. arXiv:1405.5858 [cond-mat, physics:hep-th]. May 2014.

[58] Michaël Krajecki, Christophe Jaillet, and Alain Bui. "Parallel tree search for combinatorial problems: a comparative study between openMP and MPI". en. In: (2005).

[59] Vitaliy Kurlin. "Computing braid groups of graphs with applications to robot motion planning". en. In: *Homology, Homotopy and Applications* 14.1 (2012), pp. 159–180. DOI: 10.4310/HHA.2012.v14.n1.a8.

[60] J. M. Leinaas and J. Myrheim. "On the theory of identical particles". In: *Il Nuovo Cimento B (1971-1996)* 37.1 (Jan. 1977), pp. 1–23. DOI: 10.1007/BF02727953.

[61] Michael A. Levin and Xiao-Gang Wen. "String-net condensation: A physical mechanism for topological phases". en. In: *Physical Review B* 71.4 (Jan. 2005), p. 045110. DOI: 10.1103/PhysRevB.71.045110.

[62] Zhengwei Liu, Scott Morrison, and David Penneys. "1-supertransitive subfactors with index at most 6+1/5". en. In: *Communications in Mathematical Physics* 334.2 (Mar. 2015). arXiv:1310.8566 [math], pp. 889–922. DOI: 10.1007/s00220-014-2160-4.

[63] Zhengwei Liu, Sebastien Palcoux, and Yunxiang Ren. "Classification of Grothendieck rings of complex fusion categories of multiplicity one up to rank six". en. In: *Letters in Mathematical Physics* 112.3 (June 2022). arXiv:2010.10264 [math], p. 54. DOI: 10.1007/s11005-022-01542-1.

[64] Zhengwei Liu, Sebastien Palcoux, and Yunxiang Ren. *Triangular Prism Equations and Categorification*. en. arXiv:2203.06522 [math-ph]. May 2023.

[65] Zhengwei Liu, Sebastien Palcoux, and Jinsong Wu. *Fusion Bialgebras and Fourier Analysis*. en. arXiv:1910.12059 [math]. June 2021.

[66] Seth Lloyd. "Universal Quantum Simulators". en. In: *Science* 273.5278 (Aug. 1996), pp. 1073–1078. DOI: 10.1126/science.273.5278.1073.

[67] Saunders Mac Lane. *Categories for the Working Mathematician*. en. Vol. 5. Graduate Texts in Mathematics. New York, NY: Springer New York, 1978. DOI: 10.1007/978-1-4757-4721-8.

[68] Tomasz Maciazek et al. *Extending the planar theory of anyons to quantum wire networks*. en. arXiv:2301.06590 [cond-mat, physics:math-ph, physics:quant-ph]. Jan. 2023.

[69] Fabian Mäurer. "Developing a category theory framework in Julia". MA thesis. University of Kaiserslautern, Jan. 2021.

[70] Fabian Mäurer and Ulrich Thiel. *An algorithm for computing the center of a fusion category*. In preparation.

[71] Fabian Mäurer and Ulrich Thiel. *TensorCategories*.

[72] Ernst W Mayr and Albert R Meyer. "The complexity of the word problems for commutative semigroups and polynomial ideals". en. In: *Advances in Mathematics* 46.3 (Dec. 1982), pp. 305–329. DOI: 10.1016/0001-8708(82)90048-2.

[73] Catherine Meusburger. "State sum models with defects based on spherical fusion categories". en. In: *Advances in Mathematics* 429 (Sept. 2023), p. 109177. DOI: 10.1016/j.aim.2023.109177.

[74] Roger Mong et al. "Fibonacci anyons and charge density order in the 12/5 and 13/5 quantum Hall plateaus". In: *Physical Review B* 95 (Mar. 2017). DOI: 10.1103/PhysRevB.95.115136.

[75] Gregory Moore and Nathan Seiberg. "Classical and quantum conformal field theory". en. In: *Communications in Mathematical Physics* 123.2 (June 1989), pp. 177–254. DOI: 10.1007/BF01238857.

[76] V. Mourik et al. "Signatures of Majorana Fermions in Hybrid Superconductor-Semiconductor Nanowire Devices". en. In: *Science* 336.6084 (May 2012), pp. 1003–1007. DOI: 10.1126/science.1222360.

[77] Michael Müger. "From subfactors to categories and topology II: The quantum double of tensor categories and subfactors". en. In: *Journal of Pure and Applied Algebra* 180.1-2 (May 2003), pp. 159–219. DOI: 10.1016/S0022-4049(02)00248-7.

[78] Deepak Naidu and Eric C. Rowell. "A Finiteness Property for Braided Fusion Categories". en. In: *Algebras and Representation Theory* 14.5 (Oct. 2011), pp. 837–855. DOI: 10.1007/s10468-010-9219-5.

[79] Chetan Nayak et al. "Non-Abelian Anyons and Topological Quantum Computation". en. In: *Reviews of Modern Physics* 80.3 (Sept. 2008). arXiv:0707.1889 [cond-mat], pp. 1083–1159. DOI: 10.1103/RevModPhys.80.1083.

[80] Siu-Hung Ng, Eric C. Rowell, and Xiao-Gang Wen. *Classification of modular data up to rank 11*. en. arXiv:2308.09670 [cond-mat]. Aug. 2023.

[81] Jiannis K. Pachos. *Introduction to Topological Quantum Computation*. Cambridge University Press, 2012.

[82] John Preskill. "Lecture Notes for Physics 219: Quantum Computation". In: (Jan. 1999).

[83] N. Reshetikhin and V. G. Turaev. "Invariants of 3-manifolds via link polynomials and quantum groups". en. In: *Inventiones Mathematicae* 103.1 (Dec. 1991), pp. 547–597. DOI: 10.1007/BF01239527.

[84] N. Y. Reshetikhin and V. G. Turaev. "Ribbon graphs and their invaraints derived from quantum groups". en. In: *Communications in Mathematical Physics* 127.1 (Jan. 1990), pp. 1–26. DOI: 10.1007/BF02096491.

[85] David Reutter. "Uniqueness of Unitary Structure for Unitarizable Fusion Categories". en. In: *Communications in Mathematical Physics* 397.1 (Jan. 2023), pp. 37–52. DOI: 10.1007/s00220-022-04425-7.

[86] Leonid P. Rokhinson, Xinyu Liu, and Jacek K. Furdyna. "The fractional a.c. Josephson effect in a semiconductor–superconductor nanowire as a signature of Majorana particles". en. In: *Nature Physics* 8.11 (Nov. 2012), pp. 795–799. DOI: 10.1038/nphys2429.

[87] E C Rowell. "An Invitation to the Mathematics of Topological Quantum Computation". en. In: *Journal of Physics: Conference Series* 698 (Mar. 2016), p. 012012. DOI: 10.1088/1742-6596/698/1/012012.

[88] Eric Rowell and Zhenghan Wang. "Mathematics of topological quantum computing". en. In: *Bulletin of the American Mathematical Society* 55.2 (Jan. 2018), pp. 183–238. DOI: 10.1090/bull/1605.

[89] Eric C. Rowell. *From Quantum Groups to Unitary Modular Tensor Categories*. en. arXiv:math/0503226. Mar. 2006.

[90] Eric C. Rowell and Zhenghan Wang. "Degeneracy and non-Abelian statistics". In: *Physical Review A* 93.3 (Mar. 2016). DOI: 10.1103/physreva.93.030102.

[91] Eric C. Rowell and Zhenghan Wang. *Mathematics of Topological Quantum Computing*. en. arXiv:1705.06206 [cond-mat, physics:math-ph, physics:quant-ph]. Dec. 2017.

[92] Andrew Schopieray. *Lie Theory for Fusion Categories: a Research Primer*. en. arXiv:1810.09055 [math]. Oct. 2018.

[93] Peter W. Shor. "Algorithms for quantum computation: discrete logarithms and factoring". In: *Proceedings 35th Annual Symposium on Foundations of Computer Science* (1994), pp. 124–134.

[94] Daisuke Tambara and Shigeru Yamagami. "Tensor Categories with Fusion Rules of Self-Duality for Finite Abelian Groups". en. In: *Journal of Algebra* 209.2 (Nov. 1998), pp. 692–707. DOI: 10.1006/jabr.1998.7558.

[95] Josiah E Thornton. "Generalized near-group categories". en. In: (2012).

[96] V. G. Turaev. *Quantum invariants of knots and 3-manifolds*. en. 2nd rev. ed. De Gruyter studies in mathematics 18. Berlin ; New York: De Gruyter, 2010.

[97] V.G. Turaev and O.Y. Viro. "State sum invariants of 3-manifolds and quantum 6j-symbols". en. In: *Topology* 31.4 (Oct. 1992), pp. 865–902. DOI: `10.1016/0040-9383(92)90015-A`.

[98] Vladimir Turaev and Alexis Virelizier. *Monoidal Categories and Topological Field Theory*. en. Vol. 322. Progress in Mathematics. Cham: Springer International Publishing, 2017. DOI: `10.1007/978-3-319-49834-8`.

[99] Cumrun Vafa. "Toward classification of conformal theories". en. In: *Physics Letters B* 206.3 (May 1988), pp. 421–426. DOI: `10.1016/0370-2693(88)91603-6`.

[100] Gert Vercleyen. *Anyonica*. "DOI: 10.5281/zenodo.10686860". Feb. 2024.

[101] Gert Vercleyen. "The Mathematical Structure of Tensor Networks". en. In: (2018).

[102] Gert Vercleyen and Joost Slingerland. *On Low Rank Fusion Rings*. en. arXiv:2205.15637 [math-ph]. Mar. 2023.

[103] Zhenghan Wang. "Topological Quantum Computation". en. In: (2010).

[104] Zhenghan Wang. "Topologization of electron liquids with Chern-Simons theory and quantum computation". en. In: *Differential Geometry and Physics*. Tianjin, China: WORLD SCIENTIFIC, Dec. 2006, pp. 106–120. DOI: `10.1142/9789812772527_0005`.

[105] Shigeru Yamagami. "Frobenius duality in $C^*$-tensor categories". en. In: (2004).

[106] Shigeru Yamagami. "Polygonal presentations of semisimple tensor categories". en. In: *Journal of the Mathematical Society of Japan* 54.1 (Jan. 2002). DOI: `10.2969/jmsj/1191593955`.